# Domain walls in spin-valve nanotracks: characterisation and applications

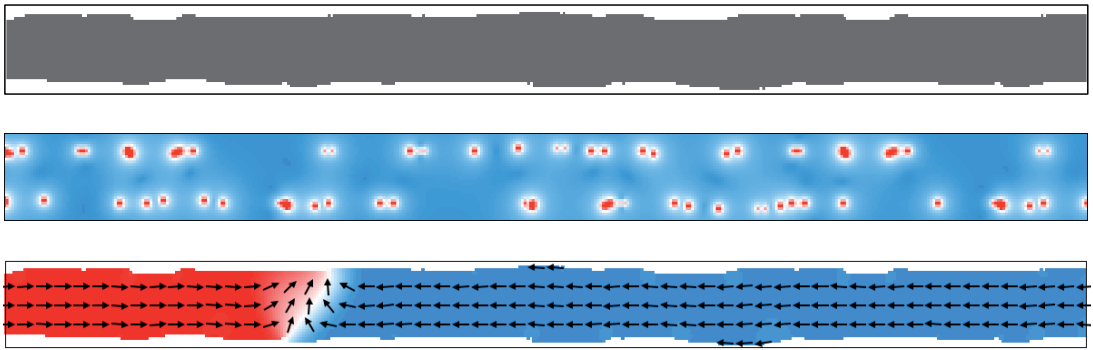

## João Miguel Ramos Melo Sampaio

October 2011

*Except where specific reference is made to the work of others, this work is original and has not been already submitted either wholly or in part to satisfy any degree requirement at this or any other university.*

João Sampaio

# Abstract

Magnetic systems based on the manipulation of domain walls (DWs) in nanometre-scaled tracks have been shown to store data at high density, perform complex logic operations, and even mechanically manipulate magnetic beads. The magnetic nano-track has also been an indispensable model system to study fundamental magnetic and magneto-electronic phenomena, such as field induced DW propagation, spin-transfer torque, and other micromagnetic properties. Its value to fundamental research and the breath of potentially useful applications have made this class of systems the focus of wide research in the area of nanomagnetism and spintronics.

This thesis focuses on DW manipulation and DW-based devices in spin-valve nanotracks. The spin-valve is a metallic multi-layered spintronic structure, wherein the electrical resistance varies greatly with the magnetisation of its layers. In comparison to monolayer tracks, the spin-valve track enables more sensitive and versatile measurements, as well as demonstrating electronic output of DW-based devices, an achievement of crucial interest to technological applications. However, these multi-layered tracks introduce new, potentially disruptive magnetic interactions, as well as fabrication challenges.

In this thesis, the DW propagation in spin-valve nanotracks of different compositions was studied, and a system with DW propagation properties comparable to the state-of-the-art in monolayer tracks was demonstrated, down to an unprecedented lateral size of 33 nm.

Several DW logic devices of variable complexity were demonstrated and studied, namely a turn-counting DW spiral, a DW gate, multiple DW logic NOT gates, and a DW-DW interactor. It was found that, where the comparison was possible, the overall magnetic behaviour of these devices was analogous to that of monolayer structures, and the device performance, as defined by the range of field wherein they function desirably, was found to be comparable, albeit inferior, to that of their monolayer counterparts. The interaction between DWs in adjacent tracks was studied and new phenomena were observed and characterised, such as DW depinning induced by a static or travelling adjacent DW.

The contribution of different physical mechanisms to electrical current induced depinning were quantified, and it was found that the Oersted field, typically negligible in monolayer tracks, was responsible for large variations in depinning field in SV tracks, and that the strength of spin-transfer effect was similar in magnitude to that reported in monolayer tracks. Finally, current induced ferromagnetic resonance was measured, and the domain uniform resonant mode was observed, in very good agreement to Kittel theory and simulations.

# Acknowledgments

I would like to firstly thank my supervisor, Prof Russell Cowburn, for his inspiring ideas and helpful discussions, without whom I would not have started —or finished— this thesis. I am also immensely thankful for the friendship, discussions, and precious help of my colleagues Laura Thevenard, Dorothée Petit, Dan Read, Vanessa Jausovec, Liam O'Brien, Emma Lewis, Huang Zeng, Amalio Pacheco, Anthony Beguivin, and Julie Kite.

This thesis was made with the help of the European Union's Marie Curie project *SpinSwitch*, to which I am very thankful for the possibility of collaborating with other research groups. I would like to thank the people at the CNRS/Thales UMR with whom I had the pleasure to work, in particular to Drs Vincent Cros and Julie Grollier. I also enjoyed very much working with Drs Kerry O'Shea and Steve McVitie at the University of Glasgow. And I am deeply grateful to Drs Susana Freitas and Ricardo Ferreira, at INESC-MN, who produced the spin valve stacks I used in this work.

Finally, I would like to thank especially those who helped me in the writing of this thesis: Drs Dorothée Petit, Liam O'Brien, Laura Thevenard, and Russell Cowburn who read and corrected my thesis, and Dr Will Branford who was my supervisor during the writing period.

# Contents

**[4] Domain wall logic**

**[5] Effects of electric current**

**[6] Conclusion and outlook**

**[A] Sample details**

**[B] Additional data (Chapter 5)**

# Figures

## [1] Introduction

**Fig. 1-1** Flux closure domains.

**Fig. 1-2** Magnetisation in nanotracks.

**Fig. 1-3** DW structure in thin nanotracks.

**Fig. 1-4** DW velocity versus applied field (according to 1-D analytical model).

**Fig. 1-5** Retrograde motion above Walker breakdown.

**Fig. 1-6** DW velocity vs. current effective velocity (simulation).

**Fig. 1-7** Pinning potential and magnetisation configuration of a TDW pinned at a notch.

**Fig. 1-8** Stoner Wohlfarth astroid.

**Fig. 1-9** DW racetrack (schematic).

**Fig. 1-10** DW logic elements.

**Fig. 1-11** DW logic 5-bit shift register.

**Fig. 1-12** Resistor model for a single FM layer.

**Fig. 1-13** Resistor model for a triple layer FM/NM/FM.

**Fig. 1-14** MR ratio of Fe/Cr multilayers.

**Fig. 1-15** RKKY coupling.

**Fig. 1-16** Néel and magnetostatic coupling, in a triple layer FM/NM/FM.

**Fig. 1-17** Meiklejohn & Bean model for exchange bias.

**Fig. 1-18** GMR in a SV.

**Fig. 1-19** Effect of Co interfacial layers.

**Fig. 1-20** DW propagation and pinning in a SV nanotrack.

**Fig. 1-21** Measurement of DW velocity.

**Fig. 1-22** Direct observation of Walker breakdown on a SV nanotrack.

**Fig. 1-23** Current induced DW propagation in SV nanotracks.

**Fig. 1-24** DW depinning stochasticity and temperature dependence.

**Fig. 1-25** Variation of depinning field with DW polarity.

**Fig. 1-26** Turn counter.

## [2] Fabrication and measurement of spin valve nanotracks

## [3] Nucleation and domain wall propagation in spin valve tracks

applied field (simulation).

## [4] Domain wall logic

## [5] Effects of electric current

## [B] Additional data (Chapter 5)

# Acronyms & symbols

*a–b* ...............................range from *a* to *b* (wherever unambiguous)

[*a*, *b*] ...........................closed interval from *a* to *b*

(*x*, *y*) ...........................vector with coordinates *x* and *y*

CIP.................................current in plane (spin valve geometry)

CVD ...............................chemical vapour deposition

CW, CCW.......................clockwise, counter-clockwise

DW ...............................domain wall

EBL ...............................electron beam lithography

FM, NM, AFM ...............ferromagnetic, non-magnetic, antiferromagnetic.

FMR ..............................ferromagnetic resonance

SV .................................spin-valve

GMR .............................giant magnetoresistance

HF.................................high frequency

HH, TT...........................head-to-head and tail-to-tail (domain wall)

high/low-R..................high and low resistance states (of a spin valve)

$H_{NUC}$ ............................nucleation field

$H_{PR}$...............................propagation field

$H_{SH}$...............................coupling shift field

$H_{Pin}$ ..............................depinning field

IBS .................................ion-beam assisted sputtering

MR.................................magnetoresistance (ratio)

MRAM ...........................magnetic random access memory

OOMMF ........................object oriented micromagnetic framework (simulation code)

PVD ..............................physical vapour deposition

P, AP..............................parallel, anti-parallel

$R_0$ .................................resistance of the low resistance state (of a spin valve)

SAF ...............................synthetic anti-ferromagnet

SEM ..............................scanning electron microscopy

SNR ...............................signal to noise ratio

SV .................................spin valve

TDW ..............................transverse DW

TEM ..............................transmission electron microscopy

# Publications & Communications

# [1] Introduction

This thesis is focused on the study of the propagation and manipulation of magnetic domain walls (DWs) in spin-valve (SV) nanotracks, with an interest in the field of DW logic. It is then based on two related fields of research. The first, *micromagnetism*, deals with magnetisation of nm to µm-scaled objects. The second, *spintronics*, studies the interaction of magnetically polarised conduction electrons with the material magnetisation, including the giant magnetoresistive effect (GMR) present in SV structures.

The magnetic nanotrack is a magnet with a nanometre-scaled cross-section, a thickness smaller than the exchange length, and a much longer length, made of a soft ferromagnetic material (e.g. Permalloy, widely used in nanotrack studies). As will be seen below, it is an information-bearing system with remarkable technological possibilities. It holds magnetic domains that are binary-stable and separated by structurally simple DWs. These can be injected, moved, and pinned inside the material by external fields and currents, enabling the manipulation of the information they encode. From a fabrication technology point of view, when compared with current nanoelectronics technologies, they are simple to make and scale to lower dimensions. They are also good model systems to study physical phenomena, ranging from DW dynamics to spintronic effects.

The SV track, simply enough, is a similarly shaped nanotrack consisting of multiple layers of magnetic and non-magnetic metals. As it will be explained in more detail below, in a SV track, the electrical resistance can be used to determine the magnetic domain configuration of the track, and specifically the position of a DW in a track with only two domains. The SV nanotrack then serves both as a tool for the study of the micromagnetism of nanotracks and as a path to practical integration of future DW-based digital devices with electronic components. It is, however, a distinct system from the magnetic nanotrack, as the presence of multiple magnetic layers adds new magnetostatic and spintronic interactions. The magnetic behaviour of the SV nanotrack and the control of these interactions are the object of this thesis.

In chapter 1, the fundamentals of micromagnetism and spintronics will be briefly presented, as well as previous work on DW logic and spin-valve tracks. Chapter 2 describes the main fabrication and measurement methods, developed for or used in this work. Chapter 3 presents our results on DW propagation in a SV track, and how it is affected by SV composition and by geometrical parameters. In chapter 4, we demonstrate several complex DW logic structures, and use them to study the behaviour of DWs in these systems. Chapter 5 presents some results on the interaction of electrical current and the track magnetisation, namely current induced DW depinning and current induced ferromagnetic resonance (FMR).

**Table of contents**

# 1-1. Micromagnetism

## 1-1.1. Micromagnetic theory

Micromagnetic theory attempts to explain the magnetisation of ferromagnetic materials at the sub-$\mu$m scale. This scale is far larger than the individual composing magnetic moments, which occur at the atomic scale, allowing for the semi-classical approximation of the local magnetic moment average density by a continuous vector field $\mathbf{M}$ ([1]), called the (local) magnetisation. In order for this approximation to hold, we must assume $\mathbf{M}$ varies slowly across the distance of several individual spins, which is indeed the case enforced by the exchange interaction in ferromagnets, as we shall see. This assumption underpins the micromagnetic model, introduced in the first half of the 20th century, mainly in the work by Landau & Lifshitz [Landau & Lifshitz 1935] and later developed by Brown [Brown 1963].

The **magnetisation M** can be written as

$$\mathbf{M} = M_S \cdot \mathbf{m}(\mathbf{r}, t) \qquad \text{(eq. 1-1)}$$

where $M_S$ is the saturation magnetisation, a scalar constant of the material [2], and $\mathbf{m}$ a unit-sized vector field representing the direction of the magnetisation, called normalised (local) magnetisation. All the energy terms relevant to the magnetic interactions are defined as functions of the field $\mathbf{M}$, and the stable configurations of $\mathbf{M}$ are found by finding the local minima of the total energy. In this way, the intractable problem of calculating the mutual interactions of ~$10^{15}$ spins is reduced to a simpler problem of multivariate calculus.

There are several energy terms that must be taken into account to describe adequately the magnetisation: the exchange interaction, magnetocrystalline anisotropy, Zeeman (or magnetostatic) energy, and demagnetising energy (or shape anisotropy).

**Exchange interaction**

The direct exchange interaction is a quantum-mechanical electrostatic phenomenon between electron pairs with overlapping wave functions, which gives rise to a difference in energy between the parallel and anti-parallel spin states. This interaction in a many-body system is described by the Heisenberg Hamiltonian [Blundell 2001]

$$\mathcal{H} = -\sum_{i \neq j} J_{i,j} \, \mathbf{S_i} \cdot \mathbf{S_j} \qquad \text{(eq. 1-2)}$$

where $\mathbf{S_i}$ are the individual spin vectors, and $J_{i,j}$ is the exchange constant (or integral) which quantifies the overlap of the two electron wavefunctions and parameterises the strength of the interaction. The range of this interaction is limited to neighbouring atoms where the wavefunction overlap is significant [3]. Thus, $J_{i,j}$ is negligible for all but the immediate neighbours. In ferromagnetic materials, $J > 0$, favouring the parallel alignment of spins. In antiferromagnetic materials, where adjacent spins are anti-parallel aligned, $J < 0$.

---

[1] In this chapter, we use bold letters for vector or tensorial variables, and italics for scalars. The amplitude of vector quantities are also written with italics (e.g. $|\mathbf{M}| = M$ ).

[2] $M_S$ is typically considered as a function of temperature. At the temperatures studied in this thesis (room temperature, or, at maximum, some tens of K above it), there is little change in $M_S$ in the used materials.

Using the micromagnetic approximation for **m**, the exchange energy density term can be expressed as [4]

$$\varepsilon_{ex} = A\left(|\nabla m_X|^2 + |\nabla m_Y|^2 + |\nabla m_Z|^2\right) \qquad \text{(eq. 1-3)}$$

where $A$, the *exchange stiffness* (in units of J/m), is a parameter of the material proportional to $J$ ([5]). In Permalloy ($Ni_{81}Fe_{19}$), $A \approx 1.3 \times 10^{11}$ J/m [Aharoni 2000]. It can be seen here that the exchange interaction favours regions of uniform **m**.

**Magnetocrystalline anisotropy**

Magnetocrystalline anisotropy is the property of some materials to favour energetically the magnetisation along certain crystalline axes. It is caused by the spin-orbit interaction [Aharoni 2000]. The electron orbits are linked to the crystalline structure and, via the spin-orbit interaction, so are the atomic spins, favouring some spin orientations along certain crystalline axes. This defines **easy axes** (and, by opposition, *hard axes*). The number and orientation of these axes depend on the crystalline symmetry.

The simplest case, of a single easy axis, is called **uniaxial anisotropy**, and the energy density may be expressed phenomenologically as a series of powers of $\sin^2 \theta$

$$\varepsilon_a = K_{u1} \sin^2 \theta + K_{u2} \sin^4 \theta + \cdots \qquad \text{(eq. 1-4)}$$

where $K_{un}$ are the (material specific) *anisotropy constants* (unit: J/m³), and $\theta$ is the angle between the magnetisation and the easy axis. Note that magnetocrystalline anisotropy is symmetric for opposite magnetisations (e.g. **m** = +**e**z and **m** = -**e**z). For small $\theta$, such as in e.g. FMR, the effect of this anisotropy may be approximated by an effective *anisotropy field*,

$$H_K = 2\,K_{u1}/\mu_0 M_S \qquad \text{(eq. 1-5).}$$

Polycrystalline Permalloy (Py; of which the free layer of the SV studied in this thesis is made) is a soft ferromagnet, i.e. one where the magnetocrystalline anisotropy is negligible (e.g. [Tannous & Gieraltowski 2008]). This formalism will still be used to understand shape anisotropy, which is described further below.

---

[3] This form of the exchange interaction is called direct exchange, as the spins are directly overlapping. Other forms of exchange exist, mediated by other electrons, such as the RKKY interaction described below.

[4] The derivation can be seen in, e.g., [Blundell 2001].

[5] The relation is $A = 2\,J\,S^2\,z/a$ , where $a$ is the nearest neighbour distance and $z$ the number of sites in the unit cell.

**Zeeman energy**

Zeeman energy accounts for the classical energy of a magnetised body in an external field $\mathbf{H}_{EXT}$. The Zeeman energy density is given by

$$\varepsilon_{Zeeman} = -\mu_0 \, \mathbf{M} \cdot \mathbf{H}_{EXT} \qquad \text{(eq. 1-6)}.$$

It favours the alignment of $\mathbf{M}$ along the externally applied field.

**Demagnetising energy**

The demagnetising energy $E_D$ accounts for the dipolar interaction between the individual spins. The demagnetising energy density in the ferromagnet can be written as a function of the *demagnetising field* $\mathbf{H}_D$, in an equation similar to that of $\varepsilon_{Zeeman}$ [Aharoni 2000]:

$$\varepsilon_D = -\frac{\mu_0}{2} \, \mathbf{M} \cdot \mathbf{H}_D \qquad \text{(eq. 1-7)}.$$

The ½ factor is included to avoid double-counting the dipole pairs. $\mathbf{H}_D$ can be deduced from Maxwell's equations [Aharoni 2000]

$$\nabla \cdot \mathbf{M} = -\nabla \cdot \mathbf{H}_D \, . \qquad \text{(eq. 1-8)}$$

In a uniformly magnetised infinite magnet, $\nabla \cdot \mathbf{M} = 0$ everywhere and $\mathbf{H}_D$ vanishes. In a uniformly magnetised *finite* magnet, it is common to approximate $\mathbf{H}_D$ by a constant field given by $\mathbf{H}_D = -\boldsymbol{\mathcal{N}} \cdot \mathbf{M}$, where $\boldsymbol{\mathcal{N}}$ is the 3×3 *demagnetisation tensor*. This yields an expression for the demagnetising energy density:

$$\varepsilon_D = +\frac{\mu_0}{2} \, M_S^2 \, (\mathbf{m} \cdot \boldsymbol{\mathcal{N}} \cdot \mathbf{m}) \, , \qquad \text{(eq. 1-9)}$$

which is a function of magnetisation angle only. Thus, in uniformly magnetised particles, the demagnetisation energy is analogous to the magnetocrystalline energy seen above, and is responsible for the so-called **shape anisotropy**, and the existence of shape-defined easy and hard axes.

An important class of geometries is the ellipsoid and its limit shapes, the infinite plane and the infinite cylinder. In these geometries, a uniform $\mathbf{M}$ produces in fact a constant $\mathbf{H}_D$. In addition, in these cases, $\boldsymbol{\mathcal{N}}$ is diagonal in the coordinate system of the ellipsoidal axes, and its components ($\mathcal{N}_{XX}$, $\mathcal{N}_{YY}$, and $\mathcal{N}_{ZZ}$) are called the *demagnetisation factors*.

Generally, $\boldsymbol{\mathcal{N}}$ must be calculated numerically. Analytical solutions do exist for some important special cases, such as ellipsoids [Osborn 1945] and rectangular prisms (such as nanotracks) [Aharoni 1998].

Using Maxwell's equations, $\mathbf{H}_D$ may be re-written as a gradient of a scalar potential, $\mathbf{H}_D = -\nabla U$, where $\nabla^2 U = \nabla \cdot \mathbf{M}$. This leads to a useful and intuitive mathematical picture, which is to consider $\mathbf{H}_D$ as the field created by **volume and surface magnetic charges** analogous to electrostatics. The magnetic charge density is defined as

$\rho = -\mu_0\,\boldsymbol{\nabla} \cdot \mathbf{M}$       (volume charge density)

$\sigma = -\mu_0\,\mathbf{n} \cdot \mathbf{M}$       (surface charge density)         (eqs. 1-10)

where $\mathbf{n}$ is the normal to the surface. These charges are only a mathematical object, similar to the concept of bound electrical charges in polarised media. The expressions for the scalar potential $U$ and the $E_D$ obtained from these charges are [Aharoni 2000] :

$U = \frac{1}{4\pi\mu_0}\left(\int \frac{\rho}{|\mathbf{r}-\mathbf{r}'|}dV' + \int \frac{\sigma}{|\mathbf{r}-\mathbf{r}'|}dS'\right)$

$E_D = \int \rho\,U\,dV + \int \sigma\,U\,dS$         (eqs. 1-11)

### Domains and DWS

It can be seen directly from eqs. 1-11 above that the existence of magnetic charges increases the demagnetisation energy, and that the system will favour the reduction and separation of magnetic charges, unless opposed by another energy term. The amount of charges (and $E_D$) can be minimised by the formation of **magnetic domains**. The domain is a region of uniform magnetisation, therefore with no volume charges and minimal exchange energy, preferentially aligned so to minimise anisotropy energy. It has, however, surface charges that still generate a demagnetising field, FIG. 1-1A. The demagnetising energy can be minimised by the formation of **flux closure domains**, with several adjacent domains of opposite magnetisation, FIG. 1-1B.

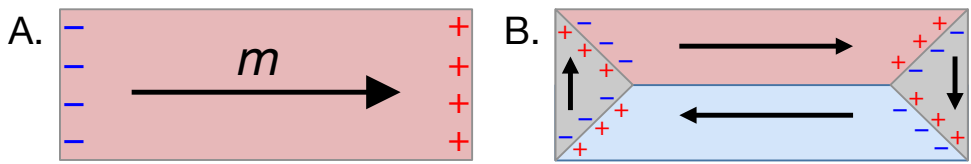

**FIG. 1-1 Flux closure domains. A.** Single domain state, and its surface charges in the left and right sides. **B.** Flux closure domains, with four domains. The positive charges of each domain are adjacent to the negative charges of another domain, minimising the total $H_D$.

Across the regions between the uniformly magnetised domains, the magnetisation is forced to change directions with an associated energy cost. These regions are called **domain walls** (DWs). The DWs are typically characterised by the angular difference between the domains they separate (e.g. 90°, 180°, or even 360° DWs), and their internal structure (e.g. the axis and sense of magnetisation rotation). Their structure and extension are the product of the balance between the exchange energy, which favours thick DWs with smoothly varying magnetisation, and the shape or magnetocrystalline anisotropy, which favours thin DWs that maximise the alignment to the easy axis. The different types of DWs in nanotracks will be presented further below.

The scale in which the magnetisation changes is called the **exchange length**, typically expressed by [Aharoni 2000],

$$l_{ex} = \sqrt{\frac{A}{K_{eff}}} \qquad \text{(eq. 1-12)}$$

where $A$ is the exchange stiffness, and $K_{eff}$ is the effective anisotropy of the ferromagnet, describing both the magnetocrystalline and the shape anisotropy. The $K$ constant for shape anisotropy can be found by equating the expression for $\mathbf{H_K}$ (eq. 1-5) and the maximum value of $\mathbf{H_D}$. As a general case, the maximum $\mathbf{H_D}$ is $M_S$, and $K_{Shape} = \mu_0 M_S^2/2$. In a material with negligible crystalline anisotropy, such as Py, $l_{ex}$ is then

$$l_{ex} = \sqrt{\frac{2\,A}{\mu_0\,M_S^2}} \qquad \text{(eq. 1-13)},$$

which is ~5 nm for Py.

In this thesis, we will study DWs in nanotracks, which will be discussed further below.

## 1-1.2. Magnetisation dynamics

An isolated magnetic moment $\boldsymbol{\mu}$ in an external field $\mathbf{H_0}$ is subject to a torque $\boldsymbol{\mu} \times \mathbf{H_0}$, leading to the gyroscopic equation [Blundell 2001]

$$\frac{\partial \boldsymbol{\mu}}{\partial t} = \gamma_0\, \boldsymbol{\mu} \times \mathbf{H_0} \qquad \text{(eq. 1-14)}$$

Where $\gamma_0$ is the gyromagnetic ratio, which for the electron is $-\gamma_0 = -\mu_0\,\mu_B\,g/\hbar$ (where $g$ is the Landé factor and $\mu_B$ is the Bohr magnetron). The magnetisation inside the ferromagnet is subject to other interactions besides the external field, as we saw above. The same expression can still be applied, but using an effective field $\mathbf{H_{eff}}$ [Aharoni 2000]

$$\frac{\partial \mathbf{M}}{\partial t} = -\gamma_0\, \mathbf{M} \times \mathbf{H_{eff}} \qquad \mathbf{H_{eff}} = -\frac{1}{\mu_0}\frac{\partial \varepsilon_T}{\partial \mathbf{M}} \qquad \text{(eqs. 1-15)}$$

where $\varepsilon_T$ is the total energy density, and $\partial \varepsilon_T / \partial \mathbf{M}$ is shorthand for $\sum_{i=x,y,z}\frac{\partial \varepsilon_T}{\partial m_i}\mathbf{e}_i$.

Note that in stationary solutions $\mathbf{M}$ is parallel to $\mathbf{H}_{eff}$, which leads to Brown's static equations [Aharoni 2000] :

$$\mathbf{M} \times \mathbf{H}_{eff} = 0 \quad \text{(eq. 1-16)}.$$

For $\mathbf{M}$ not parallel to $\mathbf{H}_{eff}$, this equation describes an undamped precessing magnetisation, while in reality it is observed that magnetisation oscillations do decay. The damping is included as phenomenological damping term (*Gilbert damping*), leading to the **Landau Lifshitz Gilbert equation** (LLG) [Landau & Lifshitz 1935; Aharoni 2000; Gilbert 2004]:

$$\frac{\partial \mathbf{m}}{\partial t} = -\gamma_0 \, \mathbf{m} \times \mathbf{H}_{eff} + \alpha \, \mathbf{m} \times \frac{\partial \mathbf{m}}{\partial t} \qquad \text{(eq. 1-17)}$$

where $\alpha$ is the (material dependent) Gilbert damping constant, and $\mathbf{M}$ was substituted by $M_s \, \mathbf{m}$. The damping torque rotates the magnetisation towards the direction of $\mathbf{H}_{eff}$, and towards a stationary solution. In Py, $\alpha$ was experimentally determined to be 0.008 [Rantschler et al. 2007], a low value compared to other common materials.

Generally, the determination of either the stationary magnetisation or the dynamical response cannot be solved analytically. In this thesis, we used the OOMMF package [Donahue & Porter 1999] to numerically integrate the LLG equation. A brief description of the simulation parameters and of OOMMF's integration methods will be given in Chapter 2.

## Ferromagnetic resonance (FMR)

The eq. 1-17 predicts a magnetisation precessional state with frequency $\omega_0 = \gamma_0 \, H_{eff}$, which can be resonantly excited by an external field. This excitation is called **ferromagnetic resonance**. If the ferromagnet is uniformly magnetised, and all the spins precess in phase with the same amplitude, this is called the **uniform mode**, as opposed to non-uniform modes such as spin-wave excitations.

For a small ellipsoidal soft ferromagnet, $\mathbf{H}_{eff}$ contains contributions only from $\mathbf{H}_0$ and the demagnetising field $\mathbf{H}_D = -\mathcal{N} \cdot \mathbf{M}$, and the resonance frequency can be written as [Kittel 2005]:

$$\omega_0^2 = \gamma_0^2 \left[ H_0 + (\mathcal{N}_{YY} - \mathcal{N}_{ZZ}) \, M \right] \left[ H_0 + (\mathcal{N}_{XX} - \mathcal{N}_{ZZ}) \, M \right] \qquad \text{(eq. 1-18)}$$

where the z-axis was chosen to lie along $\mathbf{H}_0$.

The study of FMR as a function of $H_0$ magnitude and angle, and of excitation frequency, is called **FMR spectroscopy**, and is used to experimentally measure many material parameters, such as $\alpha$ [Rantschler et al. 2007] or magnetocrystalline anisotropy constants.

## 1-2. Magnetisation of nanotracks

In this thesis, we will mostly study the magnetisation of the free layer of a SV nanotrack. This layer is made of soft ferromagnetic material with negligible magnetocrystalline anisotropy (Py [6]). The thickness ($t$) of the free layer in this thesis is in the order of 10 nm, the width ($w$) from 30 to 300 nm, and the length ($l$) was of several μm. The thickness of these nanotracks are comparable to the exchange length (which is ~5 nm in Py), while its length is several orders-of-magnitude greater. The reduced dimensions along with negligible magnetocrystalline anisotropy result in the magnetisation being governed mostly by the element shape. The extreme shape anisotropy forces the magnetisation to lie length-wise, while the reduced cross-section dimensions favour a quasi-single domain state: the magnetic domain occupies the whole cross-section of the nanotrack, while more than one domain may exist length-wise, FIG. 1-2. The domains are separated by DWs, which have a width similar to the track width (these are called 180° DWs as they separate domains of opposite magnetisation). These DWs have a net magnetic charge, +2$Q$ (head-to-head DW; HH DW) or −2$Q$ (tail-to-tail DW; TT DW), where $Q = \mu_0 \, S \, M_S$ is the characteristic charge of the track, and $S$ is the cross-section area (cf. eqs. 1-10).

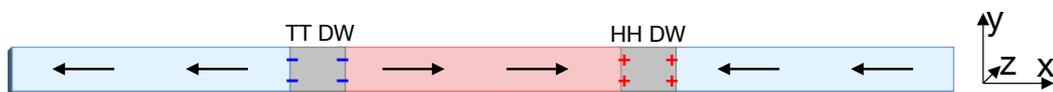

**FIG. 1-2 Magnetisation in nanotracks.**

### DW structure in nanotracks

Depending on the track cross-section, there can be several possible DW structures. We will focus here on the DWs in thin nanotracks, i.e. tracks where $t < w \ll l$, which is the case of all studied structures in this thesis. In these tracks, micromagnetic simulations [McMichael & Donahue 1997; Nakatani et al. 2005] predict three structure types: the **transverse DW** (TDW), the **asymmetric TDW**, and the **vortex DW** (see FIG. 1-3A). TDW occur in thinner and narrower tracks. Here, the magnetisation rotates while remaining in the plane of the strip, and forms a typical V-shape. The vortex DW occurs in thicker and wider tracks. Here the magnetisation curls in-plane about a perpendicular-to-plane core. The vortex DW allows some flux closure, lowering the demagnetisation energy, at the expense of an increase in the exchange energy. The asymmetric TDW occurs in intermediate thickness and width. It is similar to the TDW but with a length-wise (X direction) asymmetry. All these structures types have been observed experimentally with magnetic imaging techniques (see [Kläui 2008] for a review).

---

[6] In some structures, the free layer is a Py/CoFe double layer, also with negligible crystalline anisotropy.

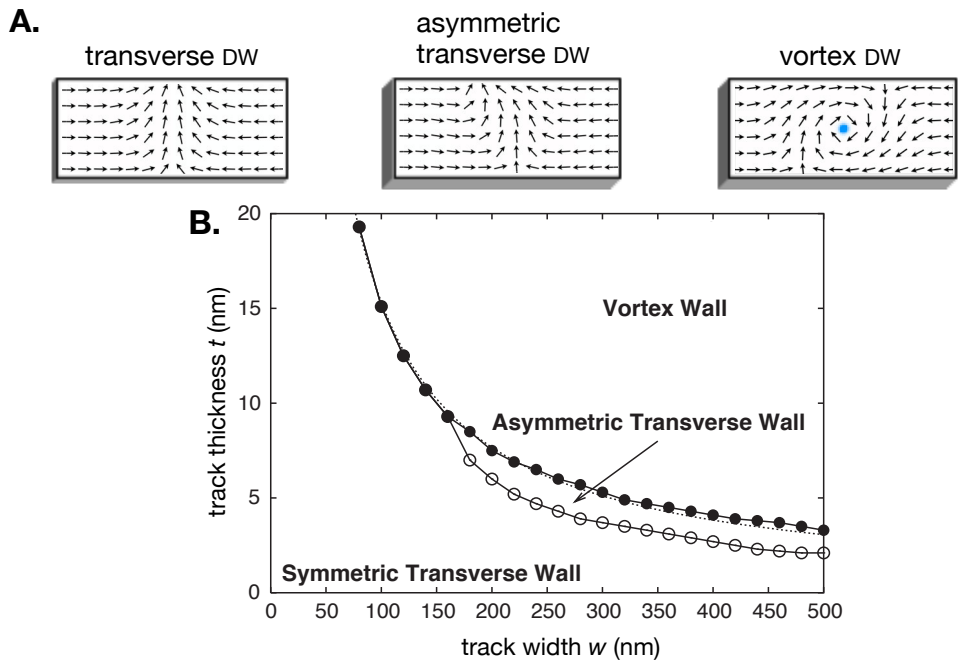

**FIG. 1-3 DW structure in thin nanotracks. A.** Micromagnetic simulation of the three HH DW types in a Py strip ($w$ = 250 nm, $t$ = 3 nm for the TDW, t = 10 nm for the rest). The colour scale represents $M_Z$, white for 0, blue for >200 kA/m. **B.** Phase diagram of stable DW types, from [Nakatani et al. 2005].

The **phase diagram** has been determined by simulation [McMichael & Donahue 1997; Nakatani et al. 2005] and experimentally [Kläui et al. 2004] ([7]), and is shown in FIG. 1-3B. The lines show the phase boundaries between each DW type. The vortex to TDW (or asymmetric TDW) is a first order transition, and the (asymmetric) TDWs are meta-stable above the transition line (such is the case in FIG. 1-3A). The coexistence of vortex and transverse DWs has been observed experimentally, as well as the thermally activated transformation from the meta-stable transverse to vortex structure [Laufenberg et al. 2006]. On the other hand, the asymmetric to symmetric TDW transition is of second order, and symmetric and asymmetric TDWs cannot coexist in the same track.

---

[7] see [Kläui 2008] for a comparison of both.

Moreover, for any DW polarity (HH or TT), each DW structure possesses several **symmetric sub-types**. If no symmetry breaking field or track deformation is present, these are degenerate states, meaning their energies are the same. For the TDW, there are two such states: one with the central magnetisation pointing up, or down (FIG. 1-3A shows the up case). The asymmetric TDW has one extra degree of freedom, with the direction of the length-wise asymmetry (FIG. 1-3A shows the up, left case). The vortex has also two degrees of freedom, the vortex curl direction (clockwise or counter-clockwise), and the z-alignment of the vortex core (up or down).

## 1-2.1. DW propagation

If an external field is applied parallel to the length of the track, the Zeeman energy favours the DW displacement in the direction that reduces the domain anti-parallel to the field. In a real track, the field needs to be sufficiently large to overcome the pinning by local defects. This is not the case for a perfectly smooth track. A one dimensional (1-D) analytical model of the TDW introduced by Walker and Schryer [Schryer & Walker 1974; Thiaville & Nakatani 2006] shows that, for sufficiently small fields, the DW moves with a velocity proportional to the applied field, $v = \frac{\gamma_0}{\alpha} H_0 \Delta^*$, where $H_0$ is the applied field and $\Delta^*$ is the dynamic DW width. However, as the field increases, so does the deformation it induces to the DW. For a TDW in a thin track, the main distortion is an out-of-plane tilt of the central magnetisation under the action of the torque of $H_0$. For low enough $H_0$, this torque is balanced by the shape anisotropy field, and the central magnetisation has a stable tilt. Above a certain threshold, called the **Walker breakdown field**, $H_W$, this balance breaks, the DW structure is no longer static, and its central magnetisation precesses. When this occurs, the DW velocity oscillates, with retrograde motion during part of its cycle. The mean velocity lowers drastically and is only increased again at much higher fields. In Walker's 1-D model, $H_W$ is given by

$$H_W = \frac{\alpha}{2} H_K,$$ 
<div align="right">(eq. 1-19)</div>

where $H_K$ is perpendicular-to-plane shape anisotropy field (i.e. $\mathcal{N}_{ZZ} M_S$). The associated velocity is called Walker velocity, $v_W$. FIG. 1-4 shows the (1D model) mean DW velocity versus $H_0$ for tracks with different shape anisotropy.

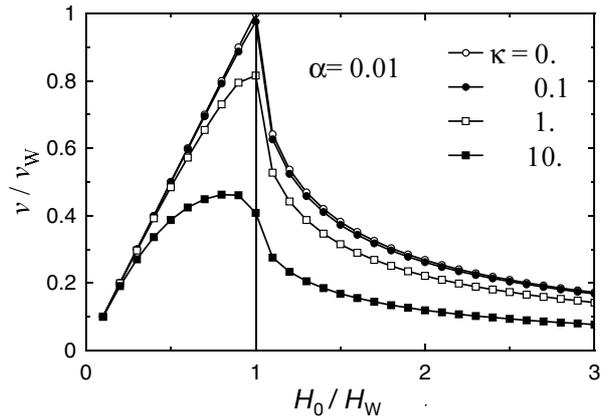

**FIG. 1-4 DW velocity versus applied field** (according to 1-D analytical model). κ is the ratio of the transversal and perpendicular-to-plane shape anisotropy constants. From [Thiaville & Nakatani 2006].

For TDWs, micromagnetic simulations predict that Walker breakdown cyclically reverses the DWs central magnetisation via an intermediate anti-vortex or vortex state [Lee et al. 2007]. Whether the transition occurs via a vortex or an anti-vortex depends on the track cross-section, applied field, and on the shape of any existing border defect. FIG. 1-5 shows the anti-vortex case. The (anti-) vortex is injected at the track border, propagates transversely across the DW, and is annihilated at the opposite border. During the presence of the (anti-) vortex core, the DW moves backwards (FIG. 1-5 bottom).

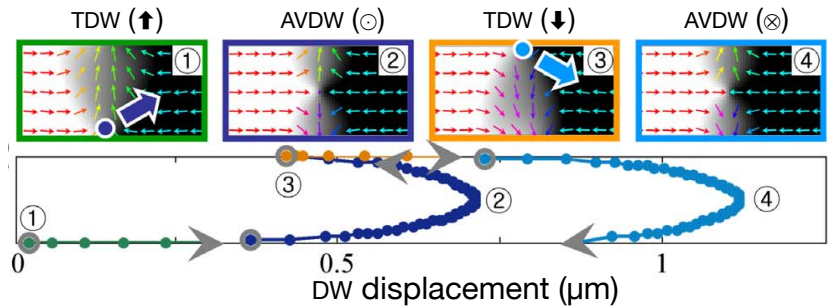

**FIG. 1-5 Retrograde motion above Walker breakdown. Top.** DW structure at various moments during the Walker cycle (AVDW is an anti-vortex DW; the circles and arrows mark the nucleation of an anti-vortex core). **Bottom.** DW position during same Walker cycle. The Y axis is the transverse position of the vortex core. From [Lee et al. 2007].

Experimental measurements of DW velocity are consistent with the Walker breakdown model, showing the predicted velocity oscillations [Hayashi et al. 2006b; Glathe et al. 2008]. In tracks with many borders defects, however, the Walker process was found to be non-periodic and stochastic [Glathe et al. 2008] ([8]). Though the experimental $v_W$ depends on the track material, dimensions, and border defects, it typically ranges from few 100s to 1000 m/s [Hayashi et al. 2006b; Glathe et al. 2008; Lewis et al. 2010].

---

[8] These results are partially shown in FIG. 1-22.

## Current induced DW propagation

When an electric current is applied across a DW, the spin-polarised conducting electrons interact with the DW magnetisation, transferring some of its angular momentum. The resulting torque on the magnetisation is called **spin-transfer torque**. This torque can deform and propagate the DW, an effect first predicted and experimentally demonstrated by Berger and colleagues in continuous thin films [Berger 1984; Freitas & Berger 1985], and later observed by many groups in nanotracks (e.g. [Kläui et al. 2003; Vernier et al. 2004; Hayashi et al. 2006a]; see also references compiled in [Kläui 2008]).

The effect of the length-wise current on the DW magnetisation in a nanotrack can be accounted for by adding two phenomenological terms to the LLG equation (eq. 1-17) [Beach 2008]:

$$\frac{\partial \mathbf{m}}{\partial t} = -\gamma_0 \, \mathbf{m} \times \mathbf{H}_{\text{eff}} + \alpha \, \mathbf{m} \times \frac{\partial \mathbf{m}}{\partial t} - (\mathbf{u} \cdot \nabla)\mathbf{m} + \beta \, \mathbf{m} \times [(\mathbf{u} \cdot \nabla) \, \mathbf{m}] \qquad \text{(eq. 1-20)}$$

$$\mathbf{u} = \frac{g \, \mu_{\text{B}} \, P}{2 \, e \, M_S} \mathbf{j}$$

where $\mathbf{u}$ is the effective velocity (in the direction of the electron velocity), $P$ the spin polarisation, $\mathbf{j}$ the current density, and $\beta$ the non-adiabaticity parameter. The first added term, $-(\mathbf{u} \cdot \nabla)\mathbf{m}$, corresponds to the **adiabatic STT**. This corresponds to the torque generated by the conduction electron on the magnetisation as its spin follows the local DW magnetisation. The second added term, $\beta \, \mathbf{m} \times [(\mathbf{u} \cdot \nabla) \, \mathbf{m}]$, corresponds to the **non-adiabatic STT**, and acts as an additional effective field onto the DW. The physical origin of this term is still subject to much debate, and several origins have been suggested, such as linear momentum transfer, spin mis-tracking, spin-flip scattering, or even having the same origin as adiabatic STT (see [Burrowes et al. 2009] and references therein).

The effect of current on the propagation of DWs [9] depends on the value of $\beta$. If $\beta = 0$, the DW is distorted but static at low current densities, even for a perfect track. Above a current density threshold $j_C$, the DW propagates with a Walker-like precession of its structure, and $v \propto \left(j^2 - j_C^2\right)^{1/2}$ [Thiaville et al. 2005]. For a finite $\beta$, $j_C = 0$, and the DW propagates with a $v \propto u$, until it reaches the Walker breakdown, with the accompanying velocity decrease (see FIG. 1-6).

---

[9] Here, as in this whole section, we focus on DWs in nanotracks, with length-wise currents.

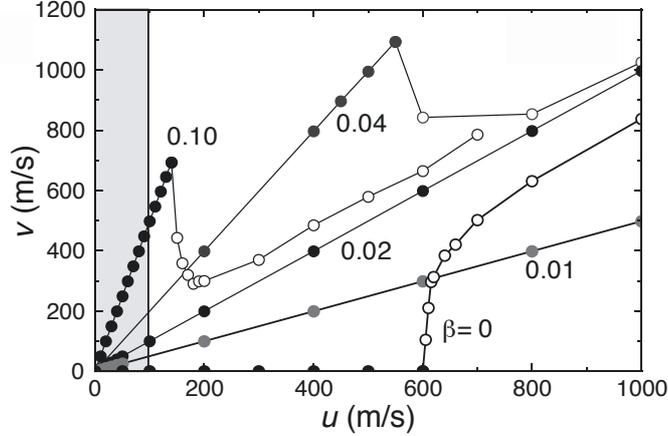

**FIG. 1-6 DW velocity vs. current effective velocity** (simulation), for various values of β. The shaded region denotes experimentally accessible values of u. From [Thiaville et al. 2005].

The two cases, $\beta = 0$ and $\beta \neq 0$, are however harder to distinguish in experiment than what the above description suggests. The presence of natural pinning defects introduces a so-called *extrinsic* current threshold proportional to the defect pinning field, hard to distinguish from the intrinsic threshold of the $\beta = 0$ case, and generating qualitatively similar DW velocity as a function of current for both hypotheses. Experimental studies have reported different values for $\beta$, though always in the same order of magnitude of $\alpha$ [Beach 2008].

For relatively small defects, the field-induced depinning also changes linearly with applied current [Vernier et al. 2004; Thiaville et al. 2005; Berger 2006; Parkin et al. 2008]. For a perfect wire, the change to the depinning field (H_P) induced by the current is

$$\Delta H_P = \frac{\beta}{\alpha \, \mu} |u| \qquad \qquad \text{(eq. 1-21)}$$

where $\mu = \partial v \, / \, \partial H$ is the DW mobility.

Experimentally, the variation of $H_P$ with current density in Py nanotracks is about $5 \times 10^{12}$ Oe·A$^{-1}$·m$^2$ [Vernier et al. 2004; Parkin et al. 2008]. In SV tracks, some studies report $H_P$ variations with current similar to the observed in Py tracks [10] [Jiang et al. 2011; Mihai et al. 2011], while others report variations orders-of-magnitude greater [Grollier et al. 2004; Ravelosona et al. 2007; Pizzini et al. 2009]. The reasons for this discrepancy are as yet unknown.

---

[10] Including our own study, in Chapter 5.

## 1-2.2. DW pinning

DW pinning occurs when there is an energy variation with DW position, which stabilises the DW position in a local energetic minimum. The variation may be caused by geometric defects on the track borders, film roughness, or by changes of material parameters. In any situation, it is useful to treat the DW as a quasi-particle with a well-defined position inside a **pinning potential landscape** $U_P(x)$ [Kläui 2008]. The energy of a pinned DW under an applied field $H_0$ is then $E = E_{Zeeman} + U_P(x) = -2\,x\,Q\,H_0 + U_P(x)$ [11], where $Q$ is the characteristic DW magnetic charge defined before. According to this model, depinning occurs when the energy gradient $\partial E/\partial x$ is negative everywhere, thus defining the static **depinning field** $H_{Pin}$: $2\,Q\,H_{Pin} = \max(\partial U_P/\partial x)$. In this 0 K model, pinning depends on the potential gradient and not on the depth of the potential minimum. For finite temperatures, though, depinning is a stochastic phenomenon that can occur even at $H_0 < H_{Pin}|_{T=0\,K}$, when a pinning potential barrier remains, with the probability of depinning increasing with decreasing barrier height. Depinning is thus very dependent on the profile of the pinning potential. The difference between $H_{Pin}$ at 0 K and room temperature depends on the pinning potential, with differences in the order of 1/2 for the typically studied artificial notch [Himeno et al. 2005b].

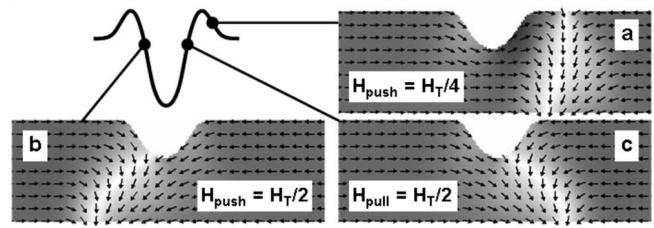

**FIG. 1-7 Pinning potential and magnetisation configuration of a TDW pinned at a notch.** The different points (**a, b, c**) correspond to different applied external fields ($H_{push}$ and $H_{pull}$ refer to the direction of applied field, left and rightwards respectively. $H_T$ is the depinning field from the bottom of the potential well). From [Petit et al. 2008a].

Artificial pinning traps are commonly created by patterning shapes on the nanotrack lateral border (e.g. [Parkin et al. 2008; Petit et al. 2008a]), as shown in FIG. 1-7. Shape defects generate pinning potentials by altering the DW internal exchange or demagnetisation energies. Pinning potentials are thus generally highly dependent on the DW structure [Petit et al. 2008a]. Shape defects can also alter the pinning potential by generating magnetostatic fields that interact with the DW charge [Petit et al. 2008b]. This is especially true in SV nanotracks, where a border defect generates intense magnetostatic fields emanating from the several magnetic layers that compose the SV [Briones et al. 2008].

---

[11] The internal energy of the DW is ignored, except its variations with $x$, which must be included in $U_P$.

Experimentally, the pinning potential can be characterised qualitatively by measuring $H_{Pin}$ as a function of the direction and amplitude of the field used to propagate the DW towards the pinning centre [Petit et al. 2008a]. FIG. 1-7 shows the potential shape of a TDW pinned at a notch determined with this technique. Other techniques used to characterise the pinning potential include direct imaging of the pinned DW under an applied field (e.g. [Kläui et al. 2005a; Petit et al. 2010]), analysis of the resonance of a pinned DW [Bedau et al. 2008], and analysis of $H_{Pin}$ versus temperature [Himeno et al. 2005b].

### 1-2.3. Reversal by nucleation

When a magnetic field is (slowly [12]) applied to a nanotrack, the two possible magnetic domain directions have different Zeeman energies ($\propto \mathbf{M} \cdot \mathbf{H}_0$). The shape anisotropy, however, prevents an unfavoured domain from spontaneously rotating towards the energy minimum. If a DW is present, and the applied field is strong enough to depin it, the propagation of the DW will reverse the unfavoured domain. If no DW is present, a strong enough field may still reverse it.

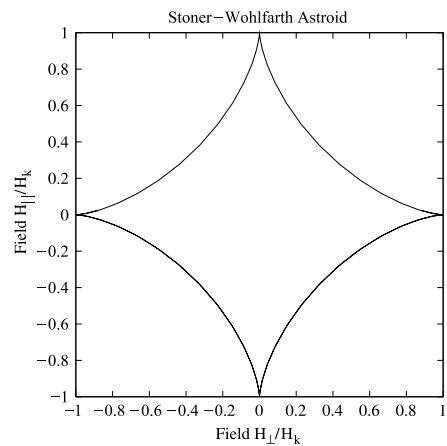

FIG. 1-8 **Stoner Wohlfarth astroid.** The x and y axis are the applied field components, normalised to $H_K$. From [Tannous & Gieraltowski 2008].

---

[12] We consider here the quasi-static regime, i.e. the applied field ramping rate is slow enough that any precessional effects can be safely ignored.

In small magnets, where the exchange energy enforces a uniform magnetisation, this rotation is uniform. This case was studied by Stoner and Wohlfarth [Stoner & Wohlfarth 1948], by considering the anisotropy and Zeeman energy terms as a function of magnetisation and field angles ($\theta_M$ and $\theta_H$):

$$E_{SW} = V\,K\sin^2(\theta_M) - V\,\mu_0\,M_S\,H_0\cos(\theta_H - \theta_M), \qquad \text{(eq. 1-22)}$$

where $V$ is the particle volume, $K = \frac{1}{2}\mu_0\,M_S^2\,(\mathcal{N}_\parallel - \mathcal{N}_\perp)$ the effective shape anisotropy constant, and $\mathcal{N}_\parallel$ & $\mathcal{N}_\perp$ the in-plane demagnetisation factors. Below a certain $H_{Critical}(\theta_H)$, two local minima exist, corresponding to two stable domain orientations, while above it only one exists. FIG. 1-8 shows the plot of $H_{Critical}$ as a function of applied field components (the so-called *Stoner Wohlfarth astroid*). Note that $H_{Critical}$ is proportional to $H_K$.

In a long nanotrack, the magnetisation is not forced to be uniform length-wise, nor is the demagnetisation field uniform. As such, also the rotation of the magnetisation will not be uniform, and will be easier where anisotropy is weaker: typically near the track ends, or in some edge defect. Therefore, reversal in tracks occurs via rotation of a small region with injection of a DW, and then by DW propagation. The critical field needed to reverse a nanotrack initially in a single domain state is called **nucleation field**, $H_{NUC}$.

The difference between the field amplitude needed for DW propagation and that needed for nucleation of a new domain is the basis of the concept of **DW conduit**. A DW conduit is a magnetic nanotrack in which that difference exists and is, preferentially, wide. In such tracks, it is possible to apply an external field that propagates existing DWs while not creating new ones. As we shall see below, this property is often a basic requirement for proposed information devices.

## 1-3. Digital devices

The ability of magnetic nanotracks to contain a large number of bi-stable domains in a relatively small and simple structure has led to the idea that these could be used in information-storing devices of higher density than current semiconductor technologies [Chappert et al. 2007; ITRS 2009; Kryder & Kim 2009]. Furthermore, the discovery of GMR (discussed further below), which allows the probing of the magnetisation with an electric current, introduced a way to retrieve electronically the information stored in these magnetic tracks, while the STT effect and the application of magnetic fields can be used to manipulate the track magnetisation.

We will present below two proposed schemes for using magnetic nanotracks in logic or data storage devices. Though there are many differences between the two, they can be distinguished firstly by the mechanism used for DW manipulation: the DW racetrack uses electric current pulses while the other, which we shall call here DW logic circuits, uses external uniform magnetic fields. Also, the racetrack is a data storing device, while the DW logic circuits are able to implement many logic functions, including data storage devices.

### The DW racetrack

The working principle of the DW racetrack, proposed by Parkin et al. [Parkin 2004; Parkin et al. 2008], is schematised in FIG. 1-9. A Permalloy nanotrack contains information encoded in the orientation of its magnetic domains. This nanotrack may be fabricated vertically in order to increase the information density of the device, or horizontally for easier fabrication. Current pulses, injected in the nanotrack, are used to shift the domains along the track. Several artificial traps may be patterned in order to stabilise the DW positions. In order to write information to the device, the localised Oersted field of a current line is applied somewhere along the track, injecting a new domain. The domains are then shifted (using injected current pulses), so that the process may be repeated. To read the information, the domains are shifted likewise, and a magnetoresistive sensor (e.g. a magnetic tunnel junction), coupled somewhere to the track, serially reads the orientation of the domains. This device is analogous to an electronic shift register, where information is fed and read at the ends of a series of information bearing simpler devices (the flip-flops in the electronic shift register, and the nanotrack segments here).

As presented above and further below in this chapter, many of the required components of the DW racetrack have been demonstrated: DW propagation with current pulses, local domain reversal with Oersted fields, and magnetoresistive reading of the nanotrack magnetisation. However, there are still many issues to be solved before such device can be truly demonstrated. The correct operation of the racetrack depends crucially on the controlled and reliable DW propagation by current pulses, as otherwise the domains could change their size, or even be annihilated, corrupting the encoded information. However, experiments so far showed that the distance travelled by the DW, its velocity, and the critical depinning current, vary stochastically and are strongly dependent on hard to control parameters, such as DW structure and border defects (see e.g. [Nakatani et al. 2003; Kläui et al. 2005b]). There are issues with the high current densities needed for DW depinning and propagation, as well as fabrication difficulties in realising a vertical magnetic nanotrack with good conduit properties.

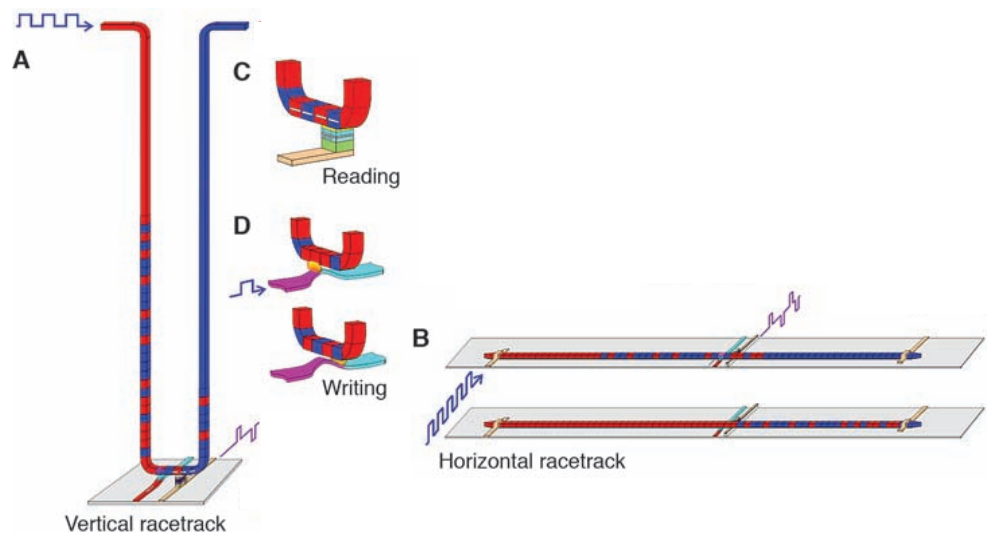

**FIG. 1-9 DW racetrack** (schematic). Two geometries are shown, a vertical racetrack (**A**) and an in-plane racetrack (**B**). In both, current pulses injected in the nanowire push the DWs through the wire. A magnetoresistive sensor, such as a magnetic tunnel junction, is coupled to the track in order to read the magnetisation of the travelling domains (**C**). Writing is accomplished with the Oersted field of a nearby current line, or the stray field of a nearby DW (**D**). From [Parkin et al. 2008].

## DW logic circuits

By manipulating the configuration of the nanotrack, along with the patterning of artificial defects, Cowburn et al. demonstrated an all-magnetic system of inter-connectable **DW logic gates**, which could be used to implement complex logic circuits [Allwood et al. 2005]. In this system, the DWs are propagated by a rotating external field, and the wire geometry is used to control their movements. Some of the demonstrated logic gates are shown in FIG. 1-10, along with a ring oscillator circuit. One remarkable characteristic of these logic circuits is that the DW propagation, and the operation of the logic gates, are synchronised by the external rotating field, analogously to the clock signal of traditional electronic logic.

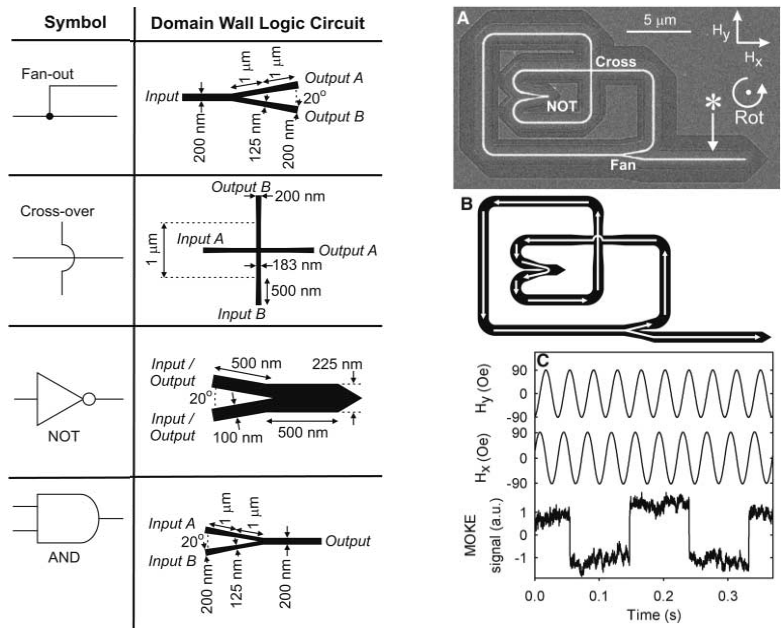

**Fig. 1-10 DW logic elements. Left.** Some DW logic gates. **Right.** Example of a DW logic circuit with a NOT gate, a cross-over and a fan-out. **A.** SEM image, **B.** magnetisation schematic, and **C.** applied field and measured magnetisation (on the region marked with an * in A.). From [Allwood et al. 2005].

The **DW NOT gate** [Allwood et al. 2002] (shown in Fig. 1-10) consists in a cusp, where a incoming DW, driven by a rotating field, enters the cusp and exits on the opposite side half a field rotation later, with its polarity (HH/TT) reversed. If several NOT gates are connected in series they will form a shift register, capable of storing data. A DW will only traverse the gate series one gate at a time, every half a field turn. Moreover, a group of DWs in such a structure is reliably propagated in synchrony with the rotating field, without ever meeting each other. Allwood et al. demonstrated a data storing shift register where information could be input [Allwood et al. 2005], Fig. 1-11. This was achieved by altering one of the NOT gates so that new domains could be injected in it when the field exceeded a certain threshold. Since then, Huang et al. [Zeng et al. 2010] demonstrated that similar shift registers could have high information densities, by redesigning the shape of the NOT gates so that they could be densely packed.

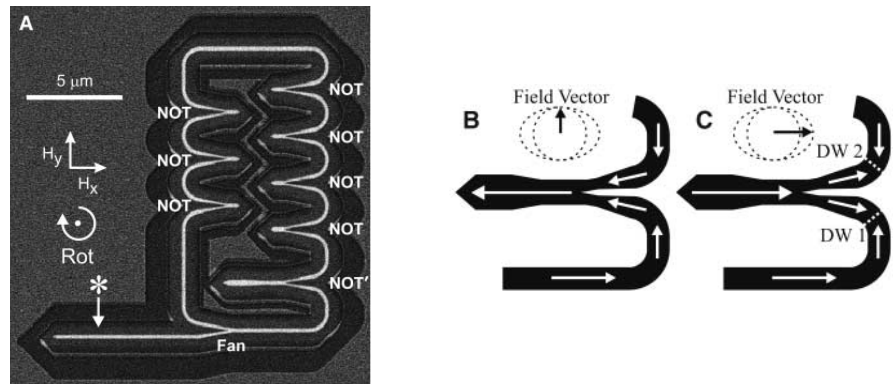

**FIG. 1-11 DW logic 5-bit shift register. A.** SEM image. **B,C.** Mechanism for inputting information. If the rotating field is kept within a certain threshold, the enlarged gate operates as a NOT gate. It the field exceeds the threshold, the cusp reverses and injects two DWs. From [Allwood et al. 2005].

# 1-4. Giant magnetoresistance

GMR, discovered independently by Fert et al. and Grünberg et al. [Baibich et al. 1988; Binasch et al. 1989], is a magnetoresistive effect, by which the change in relative orientation of the magnetisation of two ferromagnets induces a change in electrical resistance. It was first measured in metallic multilayers, and is the main magnetoresistive effect in the (multi-layered) SV nanotracks studied in this thesis.

## 1-4.1. Principles of GMR

To understand the basic mechanism behind GMR in metallic multilayers, we must first consider the transport of electrons inside a ferromagnet. When the frequency of spin-flipping scattering events is low, such as is the case of transition metals at low temperature, the spin up and down conduction electrons can be thought as carrying the electric current in two parallel channels, $I = I_\uparrow + I_\downarrow$ (this is called Mott's **two current model** [Mott 1936]). In a ferromagnet, the two current channels have very different scattering rates, regardless of the particular nature of the scattering mechanism [Fert & Campbell 1976; Dieny 2004]. In Py, for example, the mean free path of the two spin channels differs by more than five times. This difference is a consequence of large difference between the two spin orientations of the density of states at the Fermi level into which the electrons can be scattered, as predicted by the Stoner band model of ferromagnetism [Stoner 1938]. The two channels will have then two resistivities, typically parameterised as

$$\rho_\uparrow = \frac{2\rho}{1+\beta'} \qquad \rho_\downarrow = \frac{2\rho}{1-\beta'} \qquad \text{(eq. 1-23)}$$

where $\rho$ is the macroscopic resistivity, and $\beta'$ is dimensionless parameter ([13]) between -1 and 1 ($|\beta'| \approx 0.7$ in NiFe at 4 K [Dieny 2004] and 0 in non-magnetic materials).

Spin-flipping scattering events have the effect of mixing the two spin channels. The length scale over which the electron maintains its orientation is quantified by the spin-diffusion length, $l_{SF}$, which is ~5 nm in Py and ~140 nm in copper [Dieny 2004]. Consequently, a current injected in a ferromagnet, even if initially balanced between ↑ and ↓ electrons, will become **spin polarised**. This is quantified by the polarisation factor $P = (j_+ - j_-)/(j_+ + j_-) = |\beta'|$, where we replaced the ↑/↓ notation by +/– to mean the majority/minority spin channels.

The bulk spin-dependent scattering described above is the principal mechanism of GMR in the SV studied in this thesis. To understand how spin-dependent scattering leads to GMR, it is useful to consider the following **resistor network model**, shown in FIG. 1-12 for current injected in a single ferromagnetic layer. The two spin current channels ($j_\uparrow$ and $j_\downarrow$) are separated, each subject to different resistivities. The two channels are joined at the beginning and end, representing the spin relaxation in the non-magnetic electrical contacts. When the magnetisation is reversed, the resistances of the two spin channels switch: if at first the spin ↑ had resistance $R_-$, after reversal it has $R_+$, and vice-versa [14]. The total (macroscopic) resistance is, though, still the same.

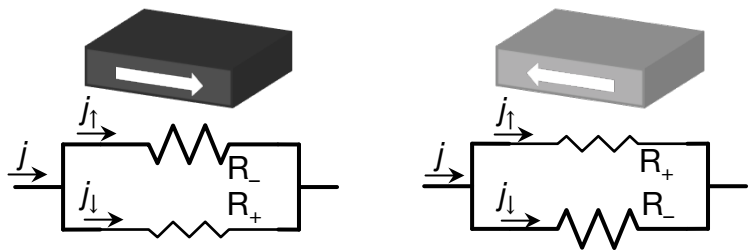

FIG. 1-12 **Resistor model for a single FM layer.**

Now, considering a triple layer FM/NM/FM, in the limit where the thickness of all layers is much smaller than the electron mean free path, and the thickness of the NM layer is negligible. In these conditions, this model is valid both for currents injected in the layer plane (known as current-in-plane/CIP geometry) and for currents injected perpendicularly to the plane (CPP geometry). The resistor model is shown in FIG. 1-13. In the case where the magnetisation of both layers are parallel (P), one spin channel (↑) is strongly scattered in both FM layers, and thus represented by two large resistors ($2 R_-$). The other channel (↓) is weakly scattered, and thus has low resistance ($2 R_+$). The total resistance is $R_P = (2R_- R_+)/(R_- + R_+)$. The anti-parallel case (AP), both spins are strongly scattered in one layer and weakly scattered in the other. As a result, both their resistances are $R_- + R_+$, and the total resistance is $R_{AP} = (R_- + R_+)/2$. There is thus a difference in total resistance ($R_{AP} > R_P$), and this resistance variation is called GMR. Resistance variation is typically quantified by the **magnetoresistance ratio** (MR) defined as $MR = \Delta R/R_{Min,}$, which in this case is

$$\text{MR} = \frac{R_{AP} - R_P}{R_P} = \frac{(R_+ - R_-)^2}{4\,R_+\,R_-} = \frac{(\rho_+ - \rho_-)^2}{4\,\rho_+\,\rho_-} = \frac{\beta'^2}{1 - \beta'^2} \qquad \text{(eq. 1-24)}.$$

---

[13] This β' parameter is unrelated to the earlier mentioned spin-transfer torque $\beta$ parameter.

[14] Where R+ is proportional to $\rho_+$, with a constant that can be derived from the layer geometry.

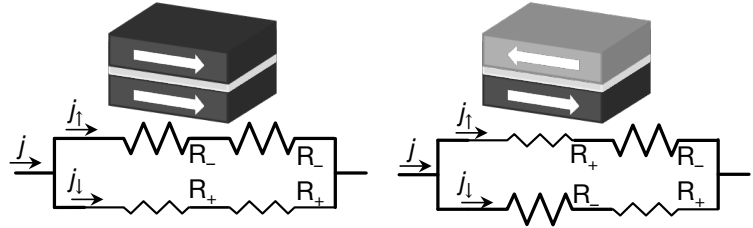

FIG. 1-13 Resistor model for a triple layer FM/NM/FM.

Using the values of β' at 4 K of NiFe (0.7) and Co (0.4) [Dieny 2004], this result predicts an MR ratio of 30% to 96% respectively. Some of the first measurements of such large MR ratios are shown in FIG. 1-14. The measured MR ratios, though, are rarely as high as predicted by this model, as they are affected by a variety of issues, such as the non-negligible layer thickness, current shunting by other metallic layers, or material limitations (inter-diffusion, alloying, etc.). Also important, at room temperature, the electron mean free path is reduced (coming farther away from the thin-layer limit), and the effects of magnon spin-flipping scattering are larger, mixing both current channels. Both these factors lower the MR ratio at room temperature (typically a factor of 2–3x [Tsymbal & Pettifor 2001]). Finally, the resistor model fails to predict several effects, such as the increase of MR with increasing number of interfaces, as shown in the results of FIG. 1-14. This happens because it does not consider the spin-dependent transmission and scattering at the interfaces, which also contribute to MR (though with a minor role in CIP SVs).

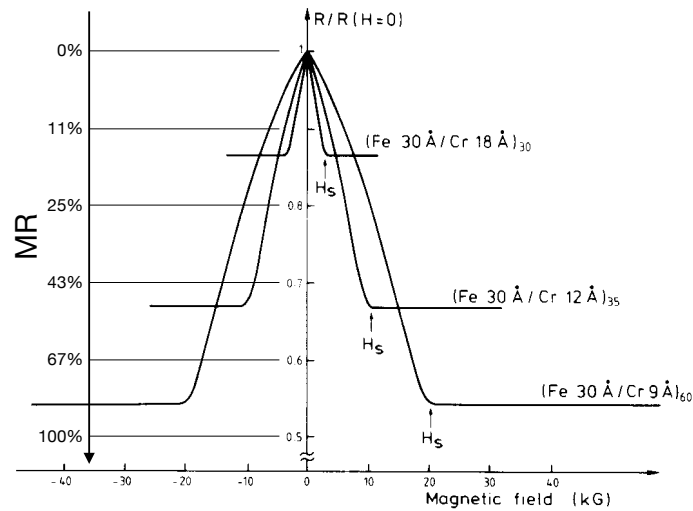

**FIG. 1-14 MR ratio of Fe/Cr multilayers.** Showing R/R(H=0) (central axis) versus applied field. The left axis corresponds to the MR ratio (same curves). The magnetic layers are anti-parallel at H=0, and parallel for |H|>|H_S|. From [Baibich et al. 1988].

The shortcomings of the resistor model for finite-thickness multilayers are addressed by a more advanced theory, the **semi-classical model**, first proposed by Camley and Barnaś [Camley & Barnaś 1989]. It expands the classical Boltzmann transport model, using some quantum physics results such as the Fermi distribution for electron momentum. It takes into account the non-uniform, spin dependent electron distributions across the multilayers, the effects of interface scattering and specular reflection, and of layer band mismatching [Tsymbal & Pettifor 2001]. It reproduces quantitatively many of the trends of MR as a function of the thickness of the layers observed experimentally, such as the existence of an optimum thickness of the FM layer, and the sub-exponential decay of MR with the thickness of the NM layer, both with a characteristic scale proportional to the electron mean free path.

## 1-4.2. Spin valves

SVs are multilayer systems which present GMR at very low applied fields, first introduced in a series of papers by Dieny and colleagues [Dieny et al. 1991a; 1991b; 1991c]. They are characterised by the very weak coupling between the participating magnetic layers, in opposition to the strongly coupled multi-layered systems where GMR was discovered. This weak coupling allowed the design of SV magnetic sensors, where small fields were capable of inducing large changes in the angle of the magnetisation of one of the layers, thus producing large changes in the resistance of the SV due to the GMR effect. They represent an important technological application of GMR, one that revolutionised the field of magnetic data storage when IBM introduced the first SV hard-disk drive read heads in 1997 [Chappert et al. 2007]. The very high field sensitivity (i.e. $\partial R/\partial H_0$) of SVs introduced them in many other technological applications, such as MRAM elements, sensors of electric currents in microchips, magnetic compasses, or sensors of magnetic markers in biomedical electronic devices [Johnson 2004].

In order to better understand multi-layered systems such as the SV, we will start by briefly reviewing the most relevant coupling mechanisms between magnetic layers.

### Interlayer coupling

Arguably, the simplest coupling mechanism is **direct exchange coupling** between two adjacent FM layers. This coupling mechanism is identical to the exchange coupling inside a ferromagnet that was studied before. It can also arise when two magnetic layers are separated by an imperfect spacer layer that has physical holes, through which the two magnetic layers are in direct contact.

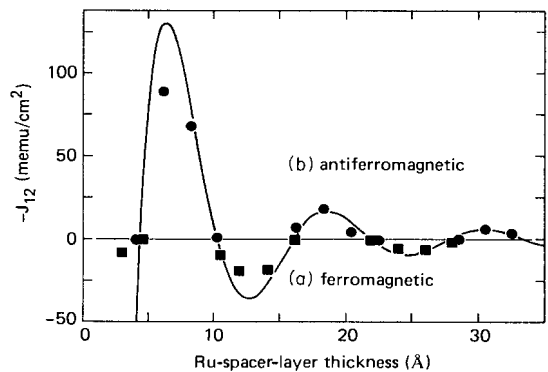

**FIG. 1-15 RKKY coupling.** Inter-layer RKKY coupling energy as a function of NM layer thickness, in $Ni_{80}Co_{20}/Ru/Ni_{80}Co_{20}$ multilayers. From [Parkin & Mauri 1991].

The **RKKY interaction** (named after its discovers [Ruderman & Kittel 1954; Kasuya 1956; Yosida 1957]) is a type of exchange interaction between spins in conducting materials, similar to the direct exchange coupling referred before. A localised moment in a metal couples to and polarises the conduction electrons in its vicinity. A second moment at some distance from the first, will be similarly coupled to the conduction electrons. The two moments then become *indirectly* coupled via spin-polarised conduction electrons [Blundell 2001]. The RKKY coupling constant is oscillatory and decaying with distance $r$: $J_{RKKY}(r) \propto \frac{\cos(2\,k_F\,r)}{r^3}$, where $k_F$ is the Fermi wave vector [Blundell 2001]. In a metallic triple layer FM/NM/FM, RKKY induces an alternating ferromagnetic/anti-ferromagnetic coupling as function of the NM thickness, shown in FIG. 1-15.

**Magnetostatic inter-layer coupling** typically occurs in two forms. **Néel coupling** [Néel 1962] (also known as *orange peel coupling*) arises from the magnetostatic charges created by surface roughness, see schematic in FIG. 1-16A. As the surface roughness is usually correlated throughout the multilayer, the charges from the two layers occur in the same position, generating a ferromagnetic interlayer coupling. The other important magnetostatic coupling effect occurs in patterned multilayers. The magnetic charges, which occur wherever the magnetisation is not parallel to the border, induce an anti-ferromagnetic inter-layer coupling (FIG. 1-16B). In a SV nanotrack, this effect occurs at the track ends, at any border defects (both artificial structures or natural roughness), and whenever a DW is present.

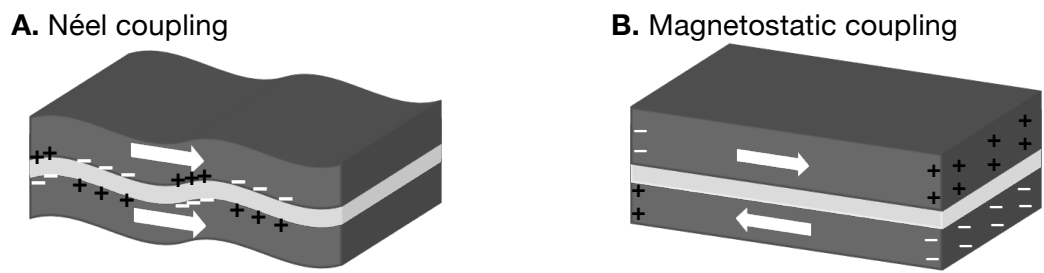

**A.** Néel coupling   **B.** Magnetostatic coupling

FIG. 1-16 **Néel and magnetostatic coupling,** in a triple layer FM/NM/FM. The arrows are the magnetisation directions and the **+/−** the magnetic charges.

**Exchange anisotropy or exchange bias** refers to the shift in switching fields and the increase in coercivity observed in a ferromagnetic layer adjacent to an anti-ferromagnetic layer (AFM). It is used in SVs to fix the magnetisation of a FM layer in a certain direction, which is then said to be *exchange pinned*. Though discovered in the 60s, it is not yet quantitatively understood [Dieny 2004]. The first and simplest model, introduced by its discoverers Meiklejohn & Bean [Meiklejohn 1962], is schematised in FIG. 1-17. It is based on the direct exchange coupling between the ferromagnetic spins and the last layer of the anti-ferromagnet. This coupling lowers the energy when the FM is parallel to the last layer of AFM spins, and increases the energy when otherwise, thus shifting the switching fields. Though this model gives an intuitive picture of the basic pinning mechanism, its quantitative predictions are several orders of magnitude off from experimental measurements. Furthermore, it fails to account for surface roughness. More advanced models, which take into account surface roughness, grain structure, and anisotropy distribution in the AFM, among other aspects, are reviewed in [Berkowitz & Takano 1999]. Phenomenologically, the exchange anisotropy is modelled by $E_{EB} = -J_{EB}\,A\cos\theta = -J_{EB}\,V/t_{FM}\cos\theta$, where $J_{EB}$ is a coupling constant (~0.1–0.4 erg/cm² for typical SV structures), $A$ and $V$ are the area and volume of the FM layer, and $t_{FM}$ its thickness [Dieny 2004]. The shift in switching fields is then given $H_{EB} = J_{EB}/\mu_0\,M_S\,t_{FM}$.

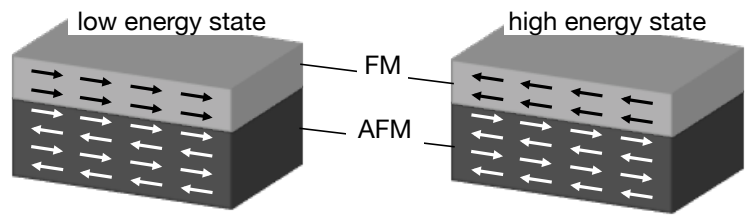

**FIG. 1-17 Meiklejohn & Bean model for exchange bias.** The arrows represent the magnetic spins.

## Spin valves

A SV contains two FM layers separated by a NM layer, such that the two FM layers are relatively uncoupled, and that the reversal field of one of the FM layers is much lower than that of the other FM layer. The layer that is easily reversed is called the *free layer*, the other the *reference* (or *pinned*) *layer*. The magnetisation of the reference layer is fixed (or 'pinned') by coupling to an adjacent anti-ferromagnetic (AFM) layer by exchange bias. In the so-called **pseudo-SVs**, the reference layer is fixed instead by using a FM layer of higher coercivity. In either case, as an external field is applied, the magnetisation of the free layer reverses while the reference layer stays fixed, and the resistance changes by the full GMR ratio.

FIG. 1-18 shows the hysteresis and MR loops for a Py/Cu SV. The pinning AFM is the FeMn layer, and the Ta layers serve to protect the stack from oxidation (*cap layer*) and to promote smooth layer growth (*seed layer*). The effects of exchange bias on the reference layer are readily observable: the hysteresis loop is shifted (420 Oe) and the coercivity increased (100 Oe). The free layer, instead, has very low coercivity (~1 Oe) and a small shift (5 Oe). This ferromagnetic shift is due to Néel coupling between the free and reference layers, as the RKKY coupling is negligible at a $t_{Cu}$ = 2.2 nm.

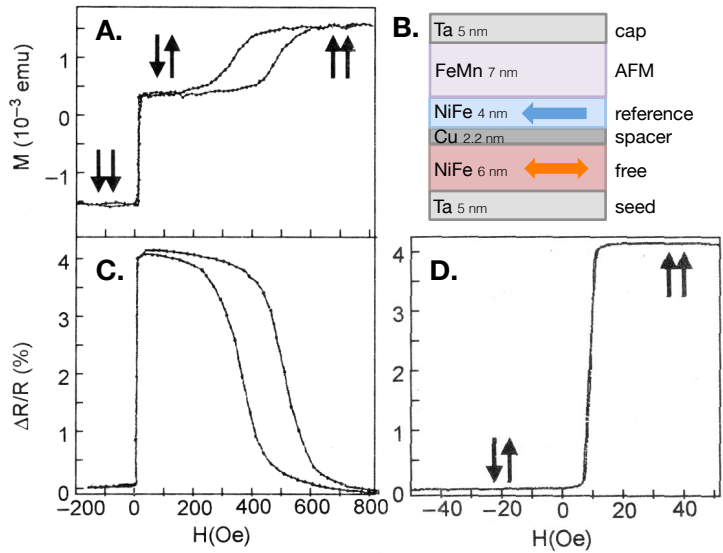

**FIG. 1-18 GMR in a SV. A.** Average magnetisation (the arrows represent the reference and free layer orientations). **B.** SV composition. **C, D.** MR ratio, as a function of external applied field (full and partial loop). From [Dieny 2004].

In order to increase the pinning strength of the reference layer, and to reduce the free to pinned magnetostatic coupling in patterned SVs, sometimes a **synthetic anti-ferromagnet** (SAF) reference layer is used instead of a single-material layer. This consists in a multilayer AFM/FM/NM/FM where the NM thickness is chosen so the two FM layers are strongly and anti-ferromagnetically coupled, and the two FM thicknesses are chosen so the total magnetic momentum is negligible. A typical composition is a Ru NM layer of $t_{Ru} = 0.5–1$ nm, and either NiFe or CoFe of $t = 1.5–3$ nm as the FM layers [Dieny 2004]. In such SVs, reference layers pinned at fields in >1000 Oe are typically observed.

The free layer may also be composed of several magnetic layers. One important instance is the use of Co-rich ultra-thin layers between a Py free layer and the Cu spacer layer to increase the MR ratio. FIG. 1-19 shows the MR versus field loop for a SV with and without Co interfacial layers, showing an increase of ~3x in MR. This effect is mostly due to Co (or CoFe) being a good diffusion barrier between the Cu and NiFe layers, preventing the creation of a NiFeCu interfacial layer with much reduced $l_{SF}$ [Dieny 2004].

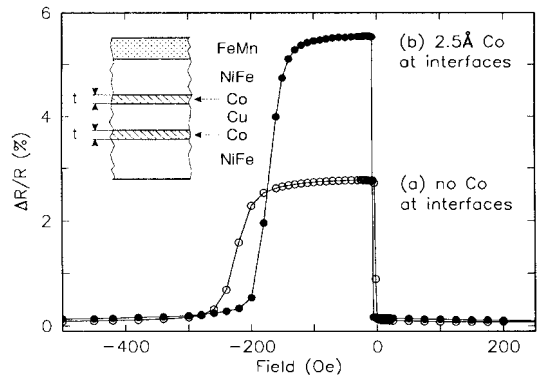

**FɪG. 1-19 Effect of Co interfacial layers.** From [Parkin 1993].

## 1-4.3. Anisotropic magnetoresistance

In ferromagnets, the resistivity depends on the angle between current and magnetisation, the so-called *anisotropic magnetoresistance*. In NiFe alloys, the resistivity is lowest by up to 5% when the magnetisation and current are perpendicular [McGuire & Potter 1975]. This effect has its origin in the spin-orbit interaction, which links the magnetisation direction to the anisotropic electron scattering probabilities.

In a SV track, the AMR effect does not change the resistance of the parallel and anti-parallel states, but lowers the resistance of the state with the free layer at 90° by about 1% [Li et al. 2001].

# 1-5. SV nanotracks: state of the art

We will now consider SV nanotracks, where the magnetisation of the reference layer is pinned length-wise. The resistance of such SV nanotrack increases linearly with the inter-contact length occupied by domains in the free layer magnetisation[15] that are anti-parallel to the reference layer. In a nanotrack with only two domains, the resistance may then be linearly mapped to DW position.

**Measuring DW position using GMR**

This idea was first applied by Ono et al. [Ono et al. 1998], where it was used to study the propagation and pinning at an artificial notch of DWs in both layers of a Py/Cu pseudo-SV, using a nucleation pad for DW injection, FIG. 1-20. They observed several resistance plateaus as they swept an external field, which corresponded linearly to the position of the traps.

---

[15] For sake of brevity, in this thesis, frequently we will refer to the magnetisation of the SV track when we mean the magnetisation of the free layer of the SV track.

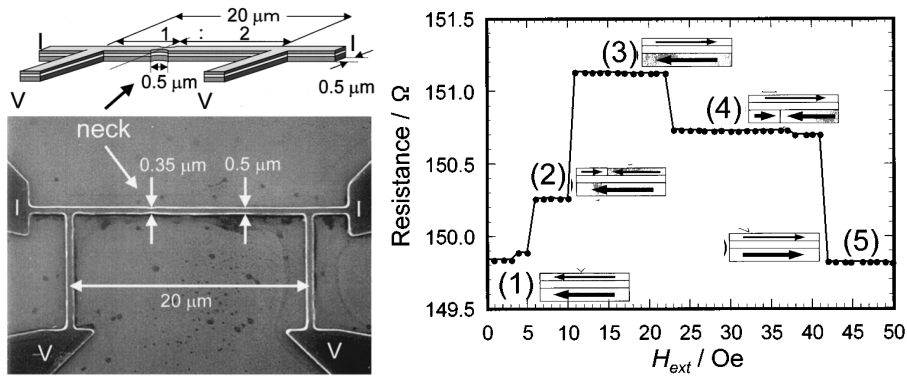

**FIG. 1-20 DW propagation and pinning in a SV nanotrack.** Schematic and SEM image of the SV track (left), and resistance versus external horizontal field. From [Ono et al. 1998].

## Using SV tracks to measure DW velocity

The direct electrical measurement of DW position also made these systems ideal for studying DW propagation phenomena, including the measurement of DW velocity and of the Walker breakdown process. The first of such studies was performed by Ono et al. [Ono et al. 1999] on a straight pseudo-SV nanotrack, FIG. 1-21. There, the position of the propagating DW was monitored, and the velocity could be measured. As the reversal field showed a large stochastic variation, the velocity as a function of applied field could also be measured.

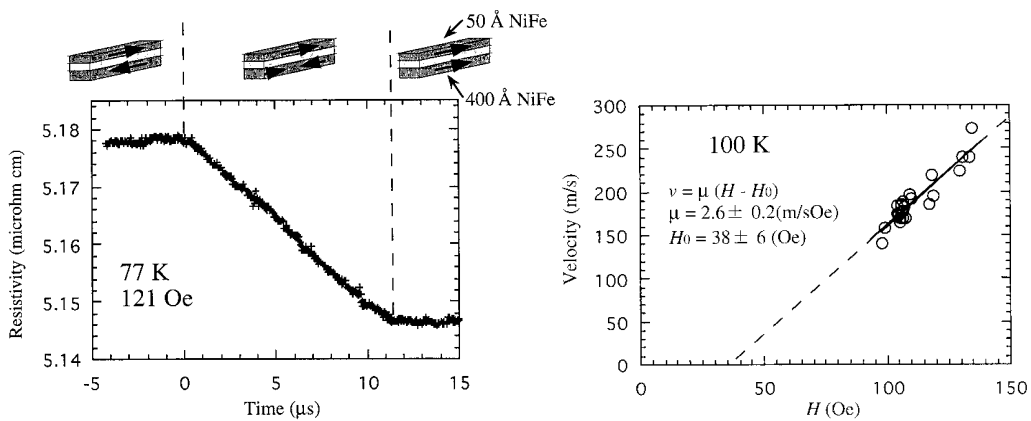

**FIG. 1-21 Measurement of DW velocity.** From [Ono et al. 1999].

Glathe et al. [Glathe et al. 2008] reported a similar experiment, on a straight SV nanotrack, where they were able to directly observe the oscillations of DW velocity due to the Walker breakdown process, FIG. 1-22. This allowed them to study how transverse applied fields affected the Walker breakdown, and that the predicted periodicity of the breakdown process was not found in SV nanotracks.

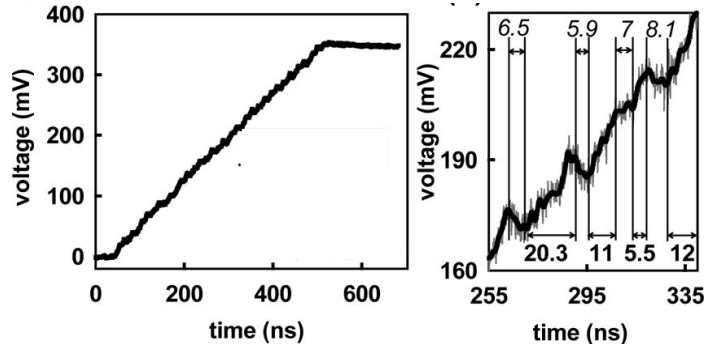

**FIG. 1-22 Direct observation of Walker breakdown on a SV nanotrack.** Voltage versus time, with the whole track reversal on the left, and a zoomed portion on the right, revealing velocity oscillations characteristic of Walker breakdown. The track was 1 µm wide and 45 µm long. From [Glathe et al. 2008].

## Current induced DW propagation in SV tracks

Current induced DW propagation was also studied with SV nanotracks. Grollier et al. [Grollier et al. 2003] first reported the observation of DW depinning and propagation with current, between two natural pinning defects, in a straight SV track, FIG. 1-23. Moreover, they observed that the current density needed to induce DW propagation was orders-of-magnitude lower than the value needed in monolayer tracks. As referred earlier, since this first report, other studies have confirmed this enhanced current induced propagation [Ravelosona et al. 2007; Pizzini et al. 2009], while others found efficiencies similar to monolayer tracks [Jiang et al. 2011; Mihai et al. 2011].

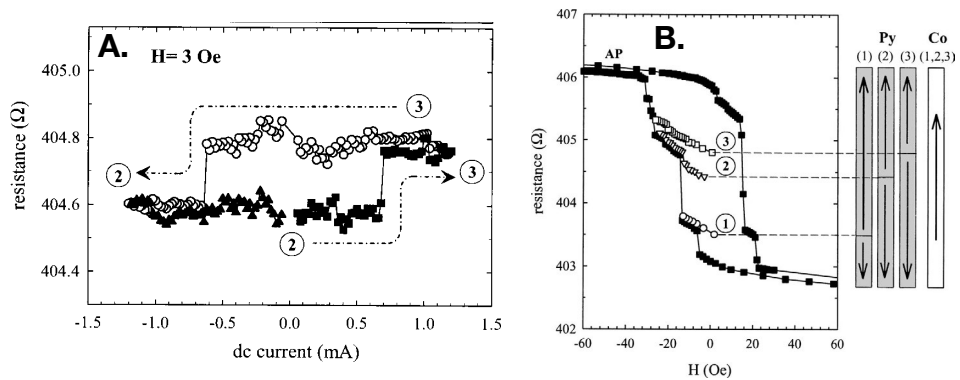

**FIG. 1-23 Current induced DW propagation in SV nanotracks. A.** Forward and backward DW motion under current. The DW moved between natural defects *2* and *3*. **B.** Resistance versus field, for complete and partial hysteretic loops, revealing several natural defects. From [Grollier et al. 2003].

## Artificial traps in SV tracks

Of importance to the field of DW logic is the study of artificial DW traps. Ono and colleagues studied several aspects of DW pinning by notches in SV nanotracks since their first study mentioned earlier (FIG. 1-20). Using a straight pseudo-SV, and now injecting DW with the Oersted field generated by an overlaid current line, FIG. 1-24A, they observed the stochastic variation of depinning field and its variation with trap neck width [Himeno et al. 2003] (FIG. 1-24B), and also the temperature variation of the depinning field [Himeno et al. 2005b]. Using a similar setup, traps with a more complex design were also demonstrated. Himeno et al. showed how a SV nanotrack patterned with a series of asymmetric notches presented different depinning fields for DWs travelling forward and backwards [Himeno et al. 2005a].

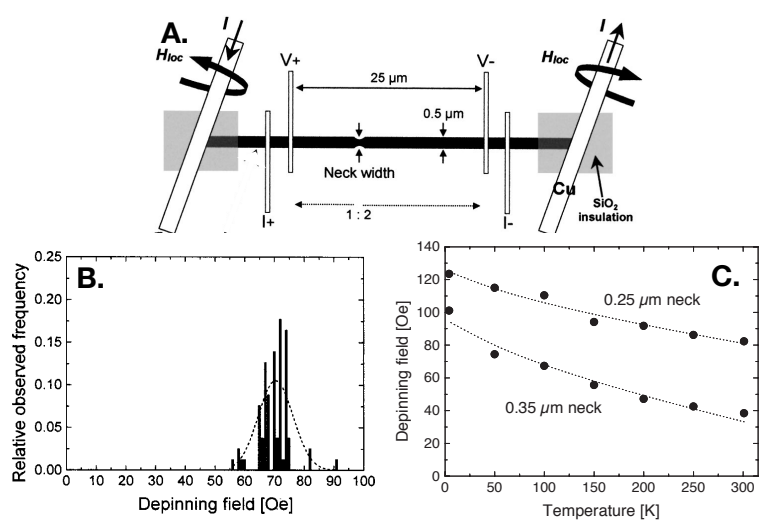

**FIG. 1-24 DW depinning stochasticity and temperature dependence. A.** Schematic of the SV track and electrical contacts. **B.** Stochastic distribution of depinning field. **C.** Dependence of depinning field with neck width and temperature. A,B from [Himeno et al. 2003] and C from [Himeno et al. 2005b].

In many of these studies, it was observed that the SV tracks presented a large number of significant natural pinning sites when compared to monolayer tracks. Also, the observed depinning fields presented large stochastic variations, far larger than those reported in Py nanotracks. Though this is partly attributable on one hand to the different patterning techniques used to fabricate SV nanotracks and, on the other hand, to a poorer control of the injected DW structure in these experiments, a study by Briones et al. [Briones et al. 2008] revealed an important difference between border defects in monolayer and SV tracks. Using a notched SV track, Briones et al. showed that the depinning field on SV tracks was dependent on DW polarity, FIG. 1-25. Using micromagnetic simulations, they showed how the stray field from the reference layer was very significant at any border defects, and how that localised field affected differently the pinning configuration of the two DW polarities (FIG. 1-25c).

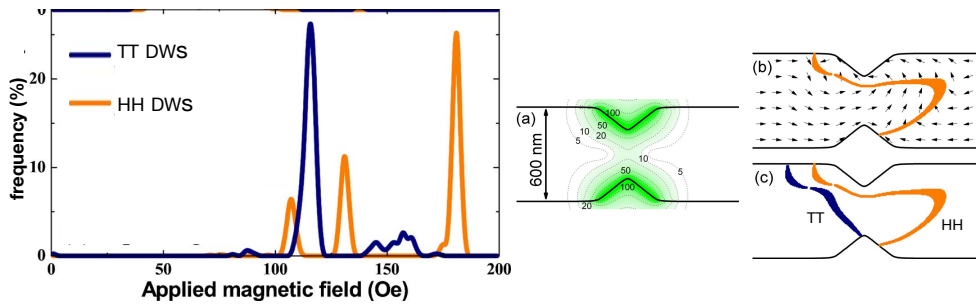

**FIG. 1-25 Variation of depinning field with DW polarity. Left.** Depinning field frequency distribution. **(a)** amplitude of the stray field from the reference layer. **(b)&(c)** configuration of a pinned HH and TT DWs (simulation). Colour represents regions of transverse (Y) magnetisation. From [Briones et al. 2008].

## Digital DW devices in SV tracks

A digital DW device was also demonstrated using SV nanotracks. Using a spiral SV nanotrack with a nucleation pad, FIG. 1-26, Mattheis et al. demonstrated a device capable of registering the number of rotations performed by an external field, which could then be read electronically [Diegel & Mattheis 2005; Mattheis et al. 2006]. As the field rotates, DWs are injected from the pad into the spiral, and those DWs progressively propagate through the spiral. In doing this, they reverse the magnetisation of the spiral segments, which is measured by the GMR effect. In order to obtain an increased signal and to remove thermal resistance variations, the authors measured the resistance in four spirals with a Wheatstone bridge configuration, with two clockwise and two counter-clockwise spirals, FIG. 1-26B.

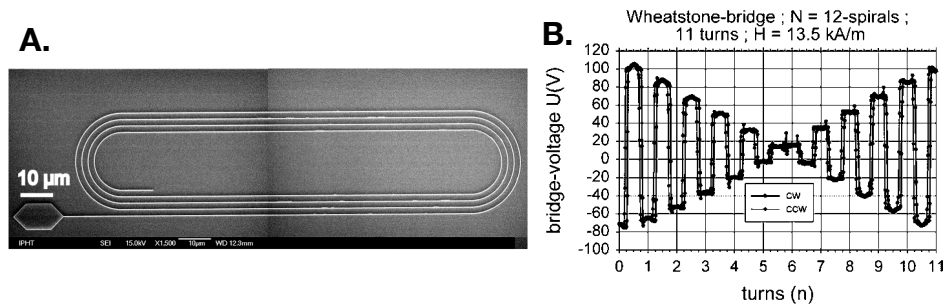

**FIG. 1-26 Turn counter. A.** SEM image of the spiral track. **B.** Voltage versus number of external field turns. From [Mattheis et al. 2006].

# 1-6. Conclusion

We started by introducing the micromagnetic theory, and the different energy terms that control the magnetisation of small magnets, and showed how domains and DWs are formed. We then studied the LLG equation, which governs the magnetisation dynamics. Afterwards, we examined the magnetisation of nanotracks, particularly the DW structure, propagation, and pinning. We then briefly presented some of the studies on the field of DW digital devices. Next, we reviewed some of the principles of spintronics and GMR, and described the resistor model to explain the GMR of thin multilayers. We then presented the SV, and the inter-layer coupling mechanisms most important to its understanding.

Finally, we reviewed some of the reported studies of SV nanotracks. These studies showed that, when compared to monolayer tracks, the natural pinning is stronger and the stochastic variation larger in SV nanotracks. Some of these differences may be attributable to the diversity of DW internal structure, which is usually poorly controlled in the mentioned SV studies, and to rougher track edges, due to limitations of the fabrication processes. Another important difference between monolayer and SV tracks is the efficiency of current induced DW propagation, which remains to be understood. It has also been shown that magnetostatic field produced by the reference layer has a large effect on the DWs in the free layer. All these aspects have limited the scope of the studies that could be performed on SV tracks, as well as limiting the type and complexity of logic structures to simpler traps and circuits than those studied in monolayer tracks. However, the electrical measurement of DW position possible in SV tracks would enable a finer study of many physical phenomena and of digital DW devices, as well as providing an electrical interface so important for technological applications.

# 1-7. References

# [2] Fabrication and measurement of spin-valve nanotracks

The fabrication and study of SV nanometric tracks, wherein DW propagation and pinning can be controlled, require the adaptation and development of several fabrication techniques. In this chapter we describe the main fabrication methods and measurement techniques used in this thesis.

All the fabrication steps were performed at Imperial College, with the exception of the SV stack deposition by ion beam-assisted sputtering, made by Drs Susana Freitas and Ricardo Ferreira at INESC-MN in Lisbon. All the low frequency measurements were performed again at the Nanofabrication Laboratory in Imperial College, and all the high frequency measurements were performed at the UMR Thales/CNRS in Paris with assistance of Drs Vincent Cros, Julie Grollier, Abdelmadjid Anane, Peter Metaxas, and colleagues.

**Table of contents**

## 2-1. Thin film deposition

There are many methods of deposition of thin films developed for a wide range of materials on a wide range of substrates. Thin film deposition is found in many different technological fields, e.g., architectural products, optical lenses, optical digital media, micro-electro-mechanic systems, and nanoelectronics [Waser 2005]. Due to the wide range of applications, thin film technology consists of a large group of very different techniques. Regarding lithography-based nanofabrication, most techniques involve deposition from a vapour phase [1], and can be broadly divided into *Physical Vapour Deposition* (PVD) and *Chemical Vapour Deposition* (CVD), depending on whether a chemical reaction was involved in creating the deposited species from a precursor in the vapour phase. Another important difference is the deposition isotropy, i.e. whether the deposited film conforms to the substrate shape. As a general rule, in PVD methods, the pressure is low and the molecules travel from a well-defined source to the substrate with few collisions, resulting in a highly anisotropic deposition. On the other hand, in CVD methods, the precursor gas fills the volume around the surface and the deposition is isotropic, uniformly covering the substrate features. The choice of deposition technique, along with the parameters used, change many other important characteristics of the deposited film, such as crystalline structure and grain size, film purity, film adhesion, roughness, thickness, and uniformity. The choice of deposition technique also depends on the material being deposited. A brief comparison between PVD and CVD techniques is shown in FIG. 2-1.

In this thesis, all metals were deposited using PVD techniques at low pressure: ion beam assisted sputtering deposition for the SV stack, magnetron sputtering and thermal evaporation for mask and electrical contact deposition.

---

[1] An important and widely used exception is elecrodeposition, which won't be covered here.

| | Physical vapour deposition | | Chemical vapour deposition |
| --- | --- | --- | --- |
| | Evaporation / MBE | Sputtering | |
| **Mechanism for production of depositing species** | Thermal energy | Momentum transfer | Chemical reaction |
| **Deposition rate** | High, up to 1.25 µm/s (low for MBE) | Low, except for pure metals | Moderate, up to 4.2 nm/s |
| **Deposition species** | Atoms and ions | Atoms and ions | Precursor molecules dissociate into atoms |
| **Energy of deposited species** | Low, 0.1 to 0.5 eV | Can be high, 1–100 eV | Low; can be high with plasma-aid |
| **Conformity to complex shaped substrate** | Poor, line of sight | Non-uniform thickness | Good |

**FIG. 2-1 Comparison of vapour deposition processes.** Adapted from [Waser 2005], in turn adapted from [Bunshah 1994].

## 2-1.1. Thermal evaporation

Thermal evaporation uses thermal energy to create the metal vapour, by boiling or sublimating the source material, which then travels with a linear path to the substrate where it condenses into a solid film.

The vapour pressures obtained are typically small (typically $10^{-5}$–$10^{-6}$ Torr; see FIG. 2-2 for vapour pressures of various elements) and so, to obtain a good purity film, this technique requires a low base pressure (typically $10^{-8}$ Torr). Another difficulty with this technique is the film uniformity and quality; as the particles travel linearly from a typically small source, strong shadowing effects occur around 3D features on the substrate. Evaporated particles also arrive with a very small energy (typically 0.1 eV, [2]), which can cause the film to be porous and have poor adhesion.

The vapour pressure depends strongly on the material being evaporated and partial pressures of different species often vary by several orders of magnitude, cf. FIG. 2-2.

---

[2] A naïve calculation, using the expected kinetic energy for an ideal gas at 1300 K, yields $K = \frac{3}{2} k_B T = 0.11$ eV .

This means that the stoichiometry of the vapour may differ from the melt, a problem when non-elemental substances are being deposited (see [Deshpandy & Bunshah 1991] and references within). Another important consideration with this method is the contamination from the crucible material and, for materials with low partial pressure, the concurrent sublimation of the crucible itself. This can be avoided by directly heating the source material when this is possible, i.e., by passing current. However, this is only possible when the material is electrically conductive and sufficient pressure is obtained below the melting point.

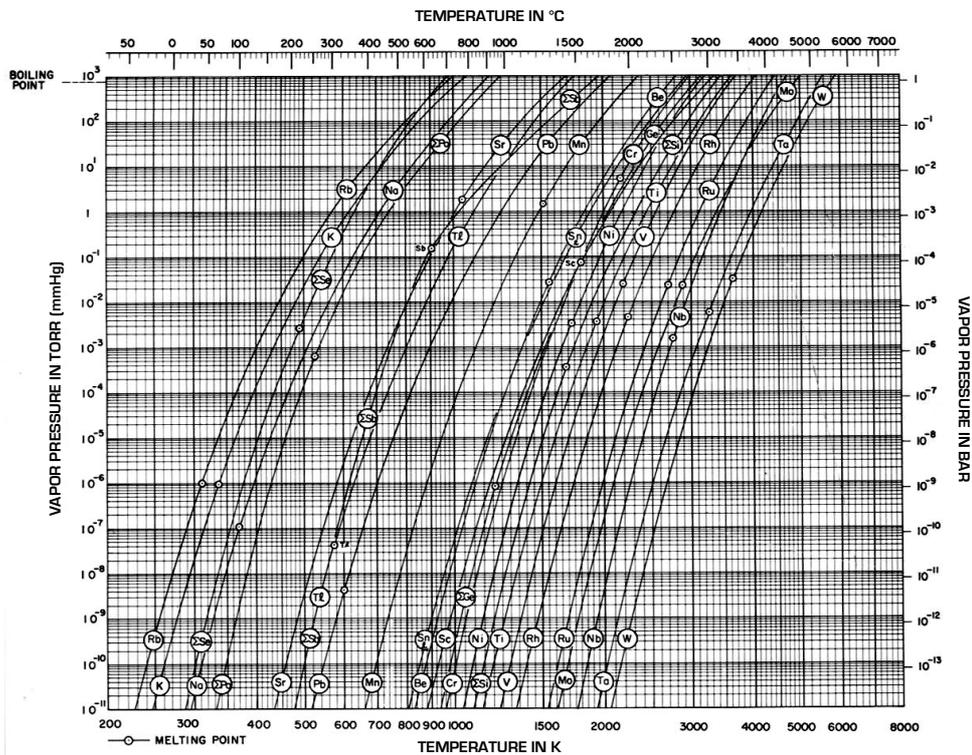

**FIG. 2-2 Vapour pressure vs temperature for various elements.** Adapted from [Honig & Kramer 1969].

The system used in this thesis is schematised in FIG. 2-3. There, the material is heated in one of three alumina crucibles containing an integrated tungsten heating wire [Megatech 2010]. The crucibles are held within a vacuum chamber at a base pressure of typically $10^{-8}$ Torr. The sample is held fixed at ~50 cm from the source, which is still much smaller than the mean free path [3]. The rate of evaporation is monitored with a quartz crystal monitor [4] and, by limiting the deposition time with a mechanical shutter, the evaporated thickness can be controlled.

---

[3] The mean free path is of the order of a few to 100s of meters, depending on the chamber pressure during evaporation.

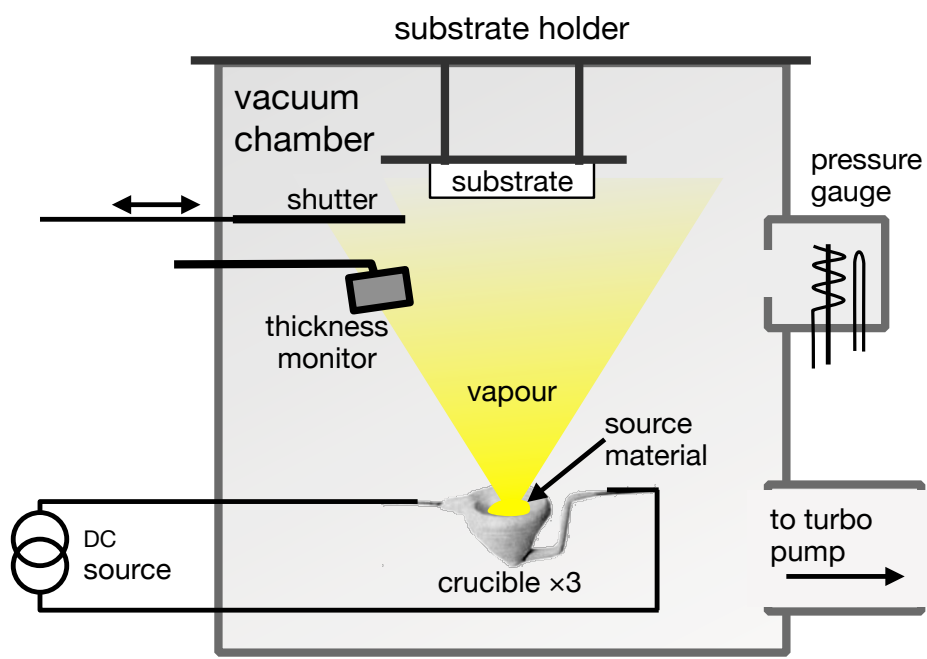

**FIG. 2-3 Thermal evaporation setup (schematic).**

Thermal evaporation was used to deposit titanium films for hard mask etching and Ti/Au [5] electrical contacts.

## 2-1.2. Sputtering deposition

Sputtering occurs when a free atom or ion enters a surface and collides with multiple surface atoms, depositing its kinetic energy, and causing some of the atoms to overcome their mutual binding energy and backscatter out of the surface. This is schematised in FIG. 2-4A. The yield of the sputtering process, i.e. the number of sputtered surface atoms per impinging atom, can be estimated by a simple *billiard-ball* model for a wide variety of target and impinging species. A finer model, given in [Bunshah 1994], gives the following expression for the sputtering yield, $S$:

$$S \propto e \frac{E}{U} a \left(\frac{M_T}{M_I}\right) ,$$

where $M_T$ and $M_I$ are the target and impinging atom masses, $e$ is the fraction of the transmitted kinetic energy (billiard ball-like),

$$e = \frac{4 M_I M_T}{(M_I + M_T)^2} ,$$

---

[4] The quartz crystal monitor is composed of a quartz crystal integrated in an oscillatory circuit. As more and more material condenses on the exposed crystal, the circuit's resonant frequency shifts proportionally to the thickness of the film, with a proportionality constant that depends on the material.

[5] The composition of multi-layer films or materials will be written in this thesis with the components separated by '/' , with the first (or bottom) component on the left and the last (or top) component on the right.

$E$ is the impinging atom energy, $U$ the (target) sublimation heat, and $a(M_T/M_I)$ an almost linear function of the mass ratio. This model adjusts well to the observed sputtering yields. In FIG. 2-4B, the yields for different target species as a function of incident ion energy are shown, revealing an almost linear variation with ion energy and target atom mass (tabled in FIG. 2-4C), as predicted by this model.

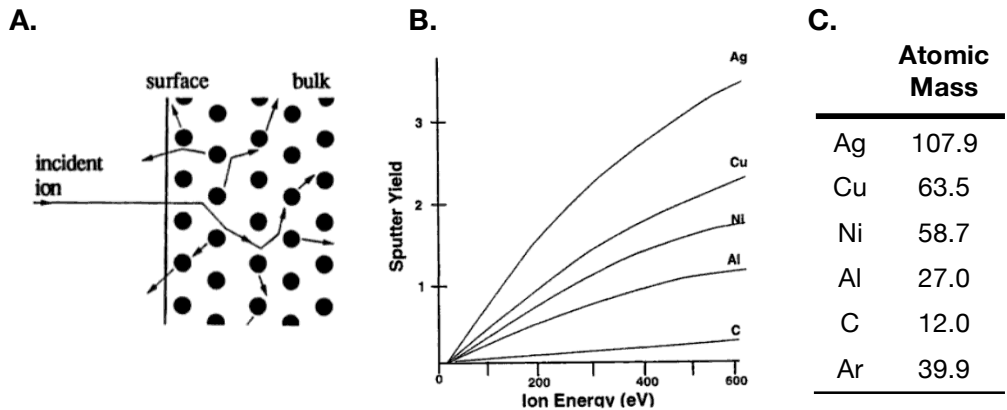

FIG. 2-4 **Sputtering. A.** Sputtering of surface atoms by incident ion. The arrows indicate the paths of the incident ion and of the collided target atoms. Notice how only some of the affected atoms leave the surface. **B.** Sputtering yield versus incident Ar+ energy for different target species. **C.** Some atomic masses. From [Auciello & Engemann 1993; Connelly et al. 2005].

We can see from this expression that the only material dependent factors are the atomic mass and the sublimation heat [6], and that the yield has an inverse first order dependence on the latter. This is very different from thermal evaporation, which depends exponentially on an activation energy [7]. The sputter process is comparatively *material agnostic*, with a yield that varies only moderately from material to material (as was seen in FIG. 2-4B), instead of the orders-of-magnitude differences found in thermal evaporation. Moreover, even with targets of mixed composition, the sputtering process shows a self-regulating behaviour, and the deposited film has the same stoichiometry as the target [8]. This happens because the sputtered atoms come from the material closest to the target surface. As such, after an initial phase of unbalanced stoichiometry, the higher yield atoms will have been partly depleted from the surface, to the point that all the partial sputtering fluxes match the target composition.

---

[6] The energy necessary for removing molecules from the solid target.

[7] Either the sublimation or evaporation heat.

Sputtering was used in this thesis for both film deposition (sputtering deposition) and for film etching (known as ion milling or etching).

### From DC to magnetron sputtering

The simplest way to deposit films by sputtering is to place the substrate and target on the anode and cathode of a planar diode plasma discharge, as schematised in FIG. 2-5. This is called **DC sputtering**. The plasma, usually Ar at a pressure of $10^{-4}$–$10^{-2}$ Torr, produces Ar+ ions that are accelerated towards the cathode (the $\oplus$ symbols in the figure). As the plasma is electrically conductive, and the ion and electron mobilities are very different, the voltage gradients occur at the borders of the plasma, with the largest voltage drop occurring close to the cathode (the voltage across the diode is represented by the dotted line). This causes the Ar+ ions to accelerate to almost the full-applied voltage without losing energy to collisions with elements in the plasma. They then sputter the target material, which is transported diffusively across the plasma and condenses on the substrate.

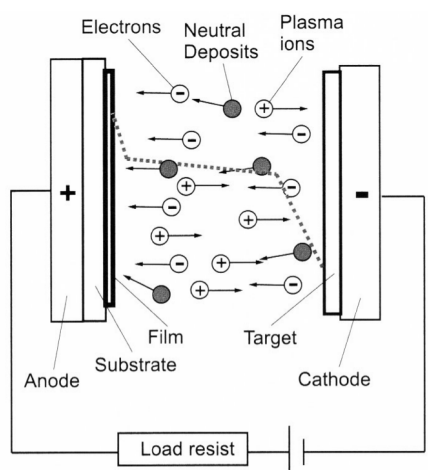

**FIG. 2-5 Schematic of a DC sputter system**. The dotted line represents the electric potential across the diode gap. Note that the velocity of the neutrals should be isotropic. From [Waser 2005].

---

[8] as long as diffusion in the target can be ignored and the different materials have approximately the same gas phase transport properties and substrate sticking coefficients.

DC sputtering has a very low deposition rate due to the low ionisation degree of the plasma. The **magnetron sputtering** source, schematised in FIG. 2-6, uses permanent magnets to contain the electrons in helical orbits close to the cathode. This increases the collision probability and the plasma ionisation, and consequently the sputtering rate, even at lower pressures. One consequence of magnetron sputtering is that the sputtering occurs unevenly on the target surface, leading to low target usage (typically 10–50% usage of target volume), and non-uniform film deposition. To increase uniformity of deposition, the substrate is positioned sufficiently far from the magnetron source (e.g. a few magnetron diameters away), and is typically rotated.

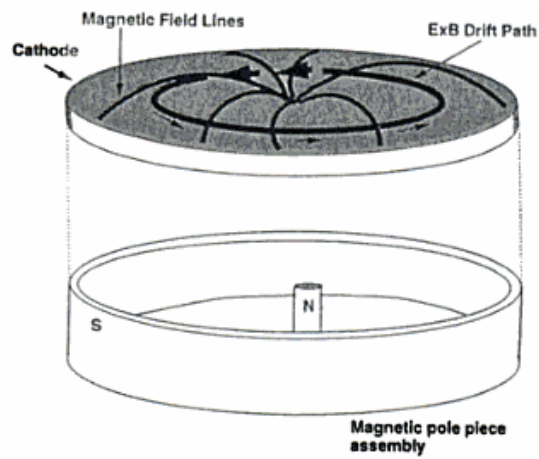

**FIG. 2-6** An expanded view of a magnetron source, showing the permanent magnet assembly, the magnetic field lines, and the electron orbits (E×B drift path). From [Auciello & Engemann 1993].

In this thesis, magnetron sputtering was used to deposit tantalum for electrical contact fabrication. The system used was manufactured by Kurt J. Lesker (schematised in FIG. 2-7) and was equipped with four magnetron sources, distributed radially and separated by ≈40 cm from the vertically mounted and rotating substrate holder. Deposition rates were previously calibrated for specific Ar flow and magnetron power settings (typical rates for Ta and other metals: 1 Å/s), and the deposited thickness controlled with a mechanical shutter.

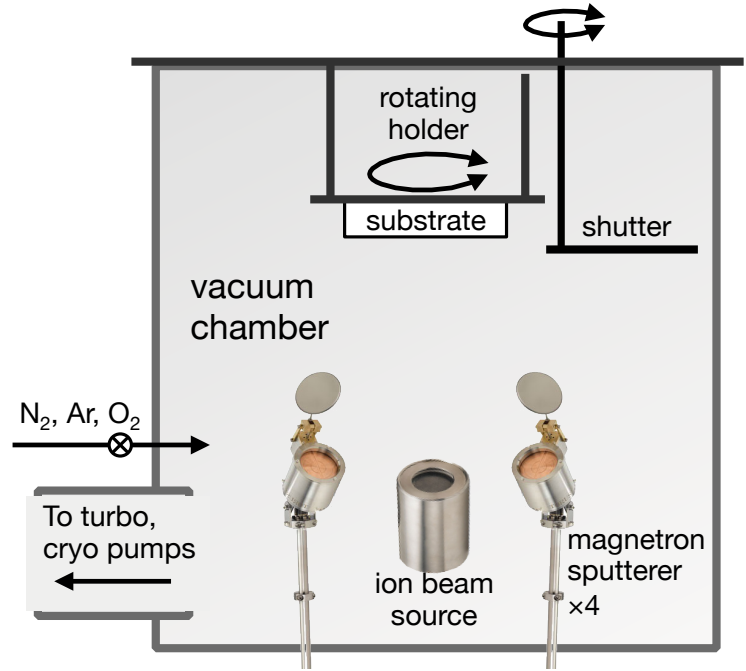

**FIG. 2-7 Sputtering and ion milling setup,** with the main used components represented.

## Ion beam and dual ion beam-assisted sputtering deposition

Ion beam assisted sputtering (IBS), schematised in FIG. 2-8, works by shining an ionic or atomic beam onto the target material, causing the sputtering of target atoms, which then condense onto the substrate. The ion source is usually some variation on a Kaufman source [Kaufman 1982] [9]. This source contains an RF-excited argon or xenon plasma, from which ions are extracted and accelerated using a series of charged grids, creating a broad, uni-directional ion beam of adjustable ion energy. An electron gun may be used outside the ion beam source to neutralise the ions, creating an atomic beam. Compared to DC magnetron sputtering, this technique has some advantages [Cuomo et al. 1989]: more flexibility in the choice of target materials (especially dielectrics), higher vacuum conditions (typically $10^{-5}$–$10^{-4}$ Torr), better film adhesion and surface roughness, and more efficient target usage.

---

[9] A bit of trivia: this ion source was developed by Harold Kaufman in the 1960s, while researching ionic space propulsion at NASA. Several related designs are used in space for this use.

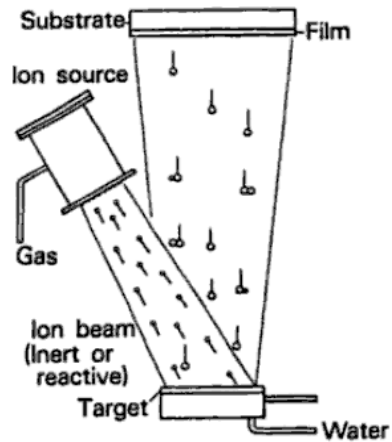

**FIG. 2-8 Ion beam sputtering** (schematic). From [Cuomo et al. 1989].

In this thesis, IBS was used for deposition of SV stacks. FIG. 2-9 shows a schematic of the main chamber of the machine used, a Nordiko 3000, installed at INESC-MN, Lisbon [Ferreira 2008]. The ion source (*deposition gun* in the schematic) is a variation of a Kaufman source (10 cm beam radius, using xenon; typical beam parameters: 0.6–1.4 keV, 5–35 mA). This beam is neutralised using a plasma electron source (*deposition neutralizer* in the schematic). This neutral beam then collides at an angle with a rectangular target (dimensions 15×10 cm²) causing it to sputter, as described above, and deposit onto the substrate, which is mounted vertically on a 6″ rotating table. The table has a permanent magnet that creates a 40 Oe magnetic field over the sample region for magnetically biased growth, and for defining the exchange pinning direction of the SV reference layer. The target is integrated in a shuttered wheel, and six different targets can be exposed separately. A second ion source (the *assist gun*) is present but not used in this thesis.

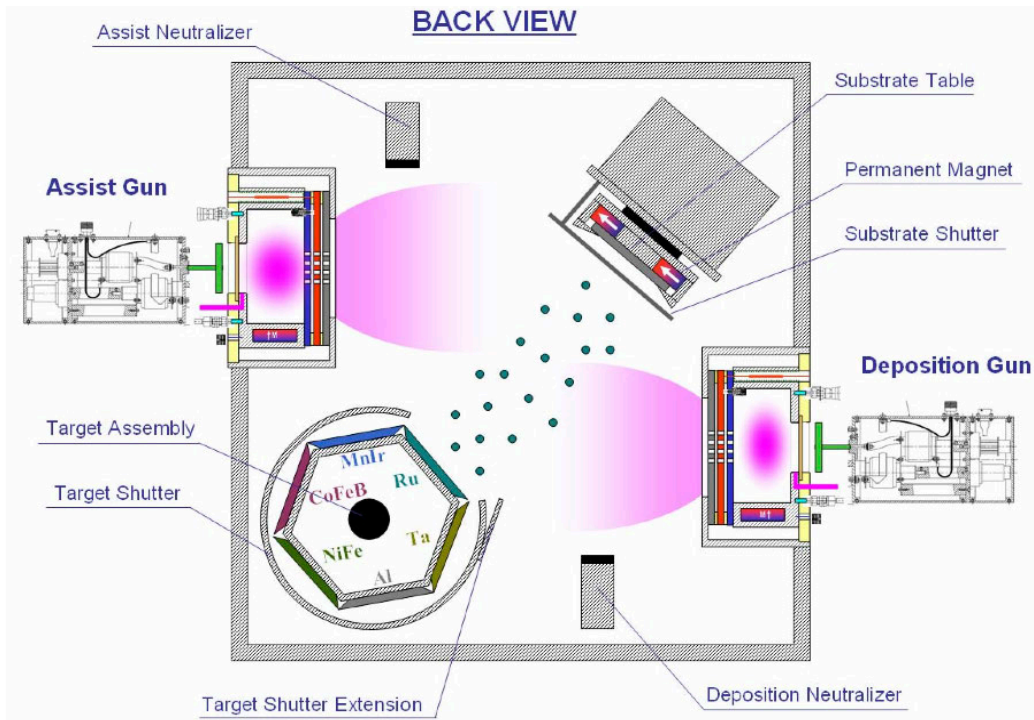

**FIG. 2-9 Schematic of the Nordiko-3000 beam-assisted deposition system.** The two Kaufman ion beam sources (assist and deposition gun) are also schematically represented. From [Ferreira 2008].

# 2-2. Patterning and pattern transfer

## 2-2.1. Electron beam lithography

Electron beam lithography (EBL) is a direct-write lithography technique that uses a highly-focused electron beam to sensitise a resist layer, capable of producing nanometre-scaled patterns. The process starts by creating a thin layer (typically hundreds of nm) of electron beam sensitive resist, on the sample by spin-coating from liquid solution. The sample is then exposed to an electron beam in a vacuum chamber. The EBL system is schematised in FIG. 2-10. The components used for creating, focusing, and deflecting the beam are similar to those in a scanning electron microscope (SEM), and indeed many EBL machines are adapted SEM systems. As with the SEM, the electron beam, generated by an electron gun, is focused into a small radius (2–10 nm [Rai-Choudhury 1997]) by a column of electromagnetic lenses. A computer controlled electrostatic deflector rasters the beam across a *write field* of typically tens to few hundred μm. The time each point is exposed to the beam is chosen so that the resist is exposed to a pre-determined dose (typically measured in charge of incident electrons per area unit, e.g. C/cm²). To generate the wanted patterns, the beam is blocked when it is shining on areas which are not to be exposed, usually by an electrostatic blanker installed in the column. The electron beam, of an energy between 1–100 keV and a current from pA to nA, induces chemical reactions in the resist that change the solubility of the exposed film. Afterwards, the sample is immersed in a solvent that selectively washes away either the exposed or unexposed areas, and the exposed pattern is revealed. If the exposed area is less soluble, and so is left on the surface after development, the resist is a *negative resist*, otherwise it is a *positive resist*.

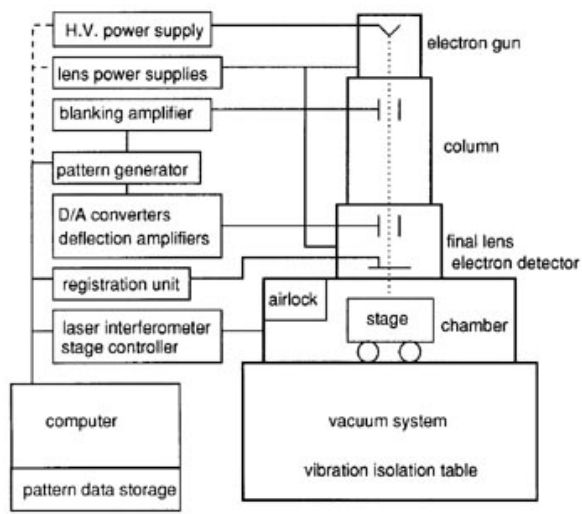

**FIG. 2-10 EBL machine** (schematic), showing the main components of a typical machine. From [Rai-Choudhury 1997].

Although the beam can be tightly focused, other phenomena lower the obtained resolution. These arise from the interaction of the electrons with the exposed sample, and thus are very dependent on the material properties and beam parameters. As the energetic electrons penetrate the resist, they slowly lose energy through multiple inelastic low angle scattering collisions, *forward scattering*, which broaden the beam profile (usually by few nms), and create multiple lower energy secondary electrons. These secondary electrons further stray from the beam profile (up to typically ~10 nm), and are responsible for most of the chemical changes to the resist. FIG. 2-11 shows an illustrative simulation of the electron trajectories in Si/PMMA substrates at two different beam energies, showing the importance of the beam parameters to the resolution. Another source of lowered resolution comes from the backscattered electrons. These occur when the incident electron suffers a large angle collision event and backscatters. As this can occur deep inside the sample, i.e. several μm, either in the resist or in lower layers, the backscattered electron can stray several μm away from the beam spot. The dose from the backscattered electrons, diluted over several μm², may not be enough to fully expose the resist, but it leads to what is called *proximity effect*, whereby the shape of exposed features is affected by the presence of neighbouring exposed features.

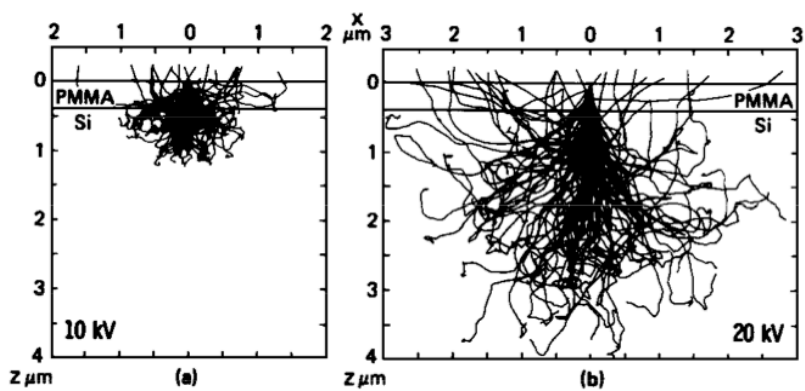

**FIG. 2-11 Monte Carlo simulated trajectories for 100 electrons in PMMA film on Si.**
Beam energy of 10kV (a), and 20kV (b). From [Kyser & Viswanathan 1975].

Another aspect to consider when optimising mask resolution is the raster process itself, i.e. the order through which the beam will expose each of the points in the write-field. The EBL system discretises each field into millions of points. By exposing points of the same structure sequentially, instead of e.g. scanning through the whole write-field line by line, the time interval between points of the same structure will be reduced. This is important because the sample stage is subject to mechanical vibrations, which introduce small but time-increasing positional errors to the mask.

The EBL machine used in this study was a Raith e-Line, with a beam energy of 20 kV and current from pA to several nA. The resist used in this thesis was polymethyl methacrylate (PMMA), from MicroChem [MicroChem 2001], in an anisole solution and with an average molecular weight of 950 10³ u. PMMA is a popular high resolution resist, that is a positive resist for the range of doses used (from 160 to 250 μC/cm²). The resist was spin-coated on previously ashed samples (spin-coating parameters: 2 min at 3000 RPM, giving a thickness of 250 nm). By ashing we mean exposing the sample to a RF-excited low power (10 W) $O_2$ plasma to remove organic residues from the surface. The resist was then heated on a hot plate to cure it (at 130–170 °C for 2 min). After exposure, PMMA was developed in a 1:3 solution of methyl-isobutyl ketone (MIBK) and isopropyl alcohol (IPA) for typically 1–2 min. Afterwards, the resist mask was again ashed (same parameters as above) to remove resist residues from the cleared areas.

## 2-2.2. Lift-off

Lift-off is a technique to transfer a pattern from a mask to a deposited film. In this thesis, we use an EBL patterned PMMA mask, onto which we deposited the film (by evaporation or sputtering). Part of the film will be deposited on the PMMA and part directly on the substrate. Immersing then the sample in acetone washes away the PMMA, aided by ultra-sonic excitation, and removes the areas of the film that were deposited on top of the PMMA, leaving a negative image of the pattern.

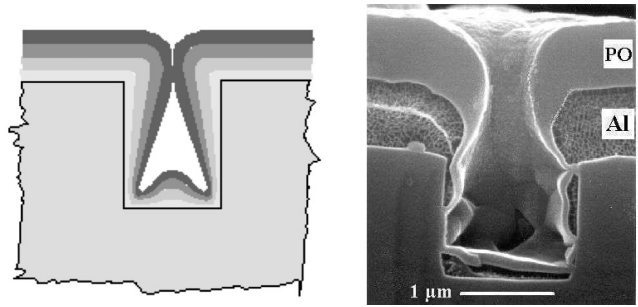

FIG. 2-12 **Isotropic deposition in trenches.** Schematic and TEM image of sputtered Al on Si. Isotropic deposition into high aspect ratio (≤1) trenches can lead to non-uniform film thickness, or even to a discontinuous film. From [Föll 2010].

Though conceptually simple, there are several aspects to consider when doing a lift-off process. Firstly, the borders of the pattern. Depending on the deposition process, particularly on its isotropy, the vertical walls of the PMMA mask may be covered with the deposited film (see FIG. 2-12), connecting the areas to be removed to those deposited on the substrate. One consequence of this can be that the edge profile shows an increased lateral roughness, caused by the breaking of the continuous film. It can also show what can be described as *rabbit ears*, left-over free-standing vertical walls of deposited film. In some cases, it could even mean that the lift-off is not possible, and that the film deposited on top of the PMMA is fully supported by these vertical walls. A solution for this problem is the engineering of over-hanging mask profiles, whereby the bottom of the resist mask is recessed and films deposited on the substrate and on the resist are no longer connected.

Another difficulty when designing a lift-off process is ensuring a uniform film thickness when using isotropic deposition techniques, or anisotropic deposition at an angle. This occurs as areas close to the PMMA walls will be shadowed by it, and will be thinner than areas farther away. In the extreme, if one of the feature dimensions is comparable to or smaller than the mask thickness, the material vertically deposited on the wall of the mask may even block the mask altogether. FIG. 2-12B shows such an example.

As it will be described below, the lift-off process was used to pattern a thermally evaporated titanium mask and to pattern electrical contacts.

## 2-2.3. Ion milling

Ion milling is an etch technique that uses the principle of sputtering (see §2-1.2) to remove unwanted material. It consists of directing a broad ionic or atomic beam onto a sample, causing the sputtering of the surface material. High beam currents are typically applied (0.01 to several amperes) to achieve high milling rates, at low beam energy (typically 800 eV or less) in order to minimise damage to the milled surface [Puckett et al. 1991]. The etched material will deposit on the vacuum chamber walls and back on the sample itself, like in the ion-beam assisted sputtering deposition described above. If a mask is fabricated on the sample, its pattern will be transferred, as only the uncovered areas of the film will be etched. Typical concerns with this technique are the erosion and other induced physical changes of the mask material (which may change its shape during the etching process and even disappear altogether), and redeposition of the etched material on the final surface or even on the mask (again changing its shape). The heating of the sample, and the kinetic energy deposited by the impinging atoms, can also induce inter-diffusion and chemical changes to the sample material.

This technique was used in this thesis for pattern transferring and surface preparation. It was performed in the setup schematised in FIG. 2-7.

## 2-3. Spin valve nanotrack fabrication

The spin valve stacks were deposited using an IBS setup (FIG. 2-9) at INESC-MN, onto a Si/SiO2 substrate. Several different stacks were used, typically containing an 8 nm thick Permalloy free layer, a Ta 5 nm capping layer, and a total thickness of about 25 nm.

Then, a thermally evaporated Ti mask was defined through EBL and lift-off. Afterwards, the sample was ion milled to remove the un-masked SV film and part of the mask thickness [10]. If a significant portion of the Ti thickness were to be left on the sample, it could shunt current from the SV track. Ti, however, has a high resistivity (420 nΩm versus 17 nΩm for Cu [Dieny 2004]), and the thickness of the Ti mask was calibrated and chosen so that the residual thickness was small (3–5 nm), decreasing the unwanted electrical shunting.

Electrical contacts were patterned by EBL, thermal evaporation of Ti/Au (Au thickness ~100 nm) and lift-off. Sample oxidation and resist residues left from the EBL process were found to cause the contact yield (i.e. percentage of low resistance contacts to SV structures) to lower. For this reason, before thermal evaporation of the Ti/Au film, the sample was subjected to a short ion milling step [11] and Ta sputtering (~20 nm) without vacuum break, in the setup of FIG. 2-7.

### 2-3.1. Samples

The results shown in this thesis were taken from four samples, labelled HM01, HM03, HM14, and HM18. These had three different compositions:

HM01:    (Si/SiO2)//Ta 2/Py 8/Cu 2/CoFe 2.2/Ru 0.8/CoFe 2.2/MnIr 6/Ta 5

HM03:    (Si/SiO2)//Ta 2/Py 8/Cu 2/CoFe 2.2/MnIr 6/Ta 5

HM14, 18:    (Si/SiO2)//Ta 3.5/Py 8/CoFe 2/Cu 2.4/CoFe 2/Ru 0.8/CoFe 2/MnIr 7/Ta 2

(thickness in nm, Py = $Ni_{81}Fe_{19}$, CoFe = $Co_{80}Fe_{20}$, MnIr = $Mn_{76}Ir_{24}$). Fabrication details can be consulted in Annex A.

---

[10] Calibration samples, consisting of the same SV stack and Ti film deposited on two different transparent glass substrates, were used to monitor when the SV had been fully etched and the residual Ti mask thickness.

[11] The ion milling step removed ~1–2 nm of the residual Ti mask and SV capping layer, without removing any active SV layers.

## 2-4. Measurement techniques

### Scanning electron microscopy

The previously mentioned Raith EBL system also allows imaging of the sample by SEM. An electron detector located right below the microscope column measures the flux of secondary electrons [12] produced by the interaction of electron beam probe with the substrate, which is rastered across the area being imaged. Secondary electrons are low energy (10s of eV) electrons that are ejected from the sample atoms by the incoming electron beam. Differences in material, topology, and electrical potential alter the flux of secondary electrons [Waser 2005], as schematically shown in FIG. 2-13. These differences are then represented in the image contrast. Image resolution is limited by the size of the beam probe and, more importantly, the dispersion of the electrons inside the material, as described before for the EBL process. This technique was used to measure the actual dimensions and shape of the fabricated structures, and to check the fabrication process itself, i.e. whether any patterning errors had occurred, or any large defects were present.

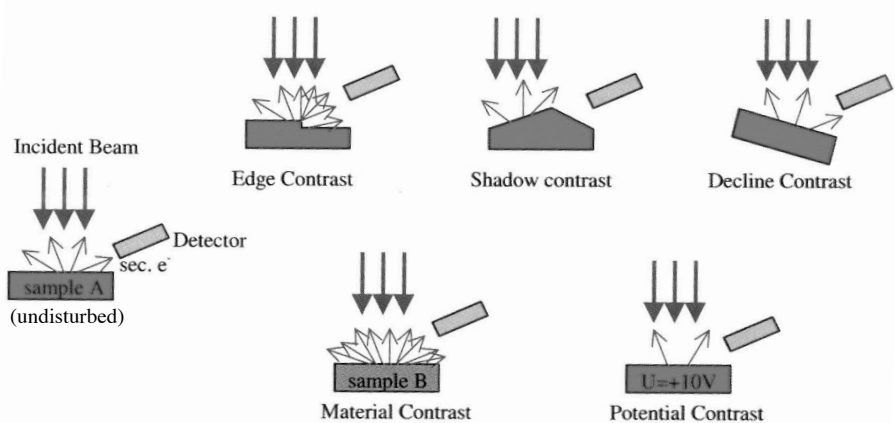

FIG. 2-13 **Mechanisms for SEM contrast** (schematic). From [Waser 2005].

### Low frequency electrical measurement

Resistance measurements were taken with a lock-in amplifier and a low injected current (typically 1–50 μA). Except where specified, measurements of the SV tracks were obtained as follows, also schematised in FIG. 2-14A. The sample was mounted onto a nanoMOKE system [Durham Magneto Optics 2011], which had a set of 8 micro-contact

---

[12] Other SEM systems detect different radiations, such has Auger electrons or X-rays

probes [13] that could be moved onto the sample. FIG. 2-14C, D show photographs of a mounted sample with the micro-contact probes. The micro-contact probes were placed on 80 μm sized gold pads patterned on the sample, which themselves were connected to the SV structures by 300 nm to 2 μm wide tracks (FIG. 2-14B shows a typical mask design of the contacts and contact pads). The total contact resistance was of a few 10s of Ω. The contacts on the samples were typically designed for a two-point resistance measurement with a common ground pad for every seven structures (cf. FIG. 2-14B). The contact probe set was connected to a contact box with a BNC connection for each of the probes, which allowed individual selection of the contacted structures. The chosen probe was then connected to both the input port (at high impedance, 1 MΩ) and the reference oscillator port of a lock-in amplifier [14]. The reference oscillator is frequency and voltage adjustable (maximum of 1 V at 50 Ω) and was set to a frequency between 1 and 40 kHz. In order to measure at constant current, a 100 kΩ resistor was placed immediately before the reference output (cf. FIG. 2-14A). The analogue output of the lock-in amplifier was connected then to the nanoMOKE computer system, which registered the signal and the Hall probe field measurement, and controlled the quadrupole electromagnet.

---

[13] model *Picoprobe* [GGB Industries 2011].

[14] model 7280 from Signal Recovery [Signal Recovery 2005].

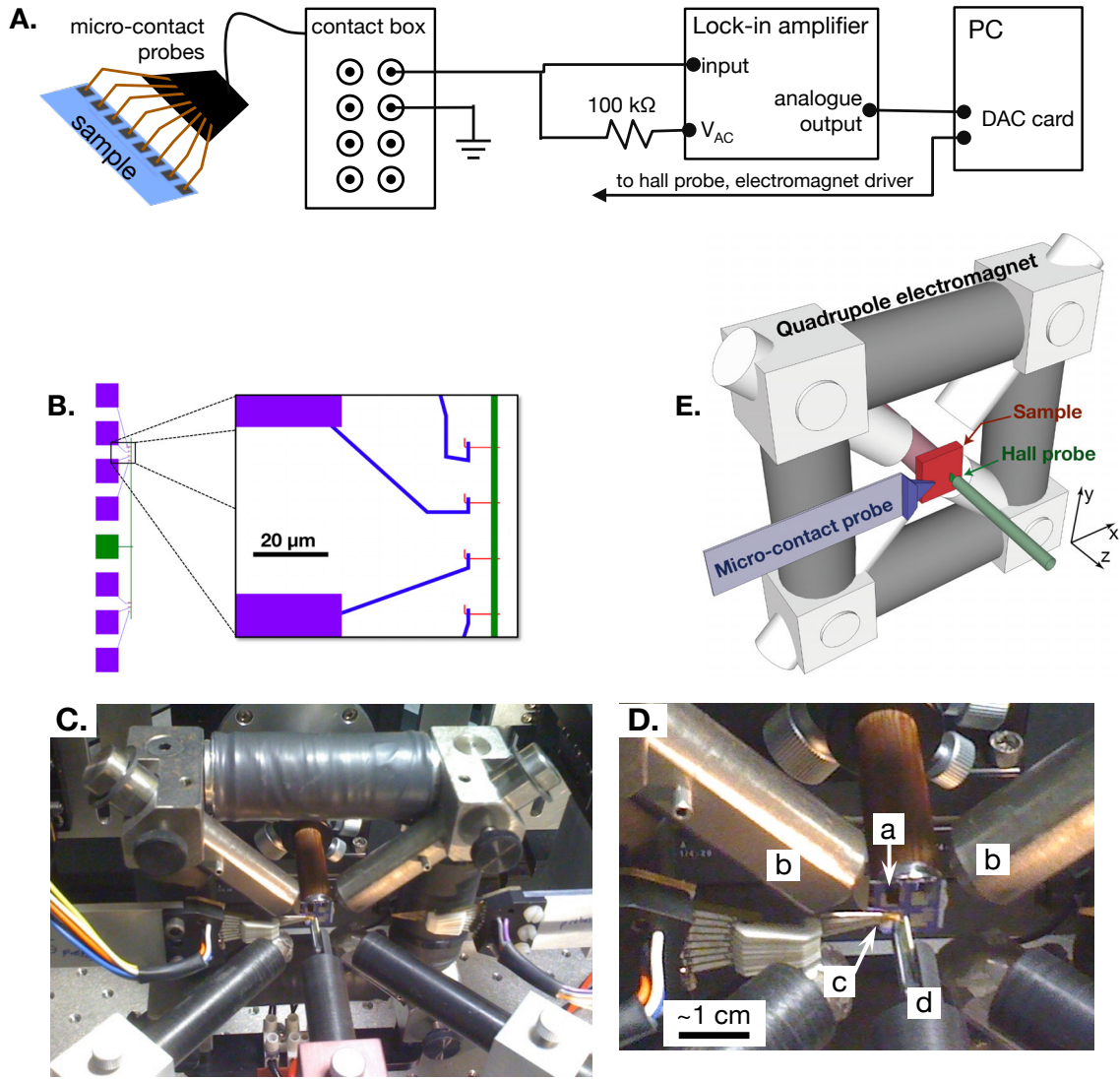

**FIG. 2-14 Electrical measurement setup**. **A.** Schematic showing the sample contact pads, the set of 8 micro-contact probes, the contact box, and electrical connections to the lock-in amplifier and controller PC. The connection lines consisted of coaxial cables. **B.** Typical lithography mask. The contact pads and tracks are in blue (individual structure contacts) and green (common contact), and the SV structures in red. **C, D.** Pictures of the measuring setup, showing the mounted sample (**a**), the quadrupole electromagnet (**b**), the micro-contact probes (**c**) and the Hall probe (**d**). **E.** Schematic of the setup (with the optical imaging system removed).

## High frequency electrical measurement

Resonance measurements were taken on a different setup at UMR CNRS/Thales, schematic in FIG. 2-15. The sample was wirebonded inside a brass box, with the ground contact connected to the grounded box and the signal contact connected to a coaxial SMA connector. It was then connected to a bias tee, with the DC port connected to a nano-voltmeter and a lock-in amplifier (with the same in-series resistor reference scheme described above). The AC port of the bias tee was connected to an RF source.

Magnetic fields could be applied with an electromagnet, and the sample could be rotated so to apply the magnetic field in different directions. The different sources and measurement devices were controlled by a computer using LabView scripts. This setup allowed for the same low frequency resistance versus field measurements described before, and for the measurement of DC voltages produced by injecting RF currents (see Chapter 5).

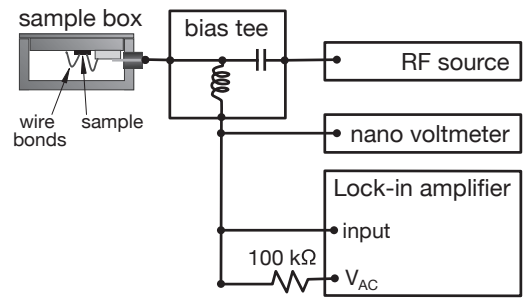

**FIG. 2-15 Schematic of high frequency measurements,** showing the sample wirebonded to a closed metal box, the bias tee, the RF power source, the nanovoltmeter and the lock-in amplifier. The wire segments consist of coaxial cables. The controller computer is not shown for clarity.

## 2-5. Micromagnetic simulations

Micromagnetic simulations were performed with the OOMMF package [Donahue & Porter 1999]. This code integrates the Landau-Lifshitz equations with a Runge-Kutta algorithm, using a uniform rectangular spatial discretisation and an adaptive step time discretisation. For Permalloy, the parameters used were 800 kA/m for the saturation magnetisation ($M_S$), $13 \times 10^{-12}$ J/m for the exchange stiffness (A), and either 0.01 or 0.5 for the damping constant ($\alpha$), the former for accurate time dependency and the latter for faster simulation of static equilibrium states. The spatial step size was chosen to be 5 nm, similar to the exchange length in Permalloy.

The OOMMF code allows for the adding of user-defined fields. This was used to eliminate non-physical demagnetisation fields at the simulation borders, and to add experiment specific fields, such as inter-layer magnetostatic fields or current induced Oersted fields.

# 2-6. References

# [3] Nucleation and domain wall propagation in spin valve tracks

In order to study the propagation of DWs in SV nanotracks and their interaction with different structures, and to study more complex DW-based logic devices, first a good control of DW creation and propagation is required. This control is the basis of the *DW conduit*, a nanotrack in which the external field needed to drive DWs is significantly lower than the field at which new DWs are injected. Making a SV DW conduit is the main goal of this chapter.

The track geometry—shape, cross-section, and roughness— affects crucially how DWs can be injected and how freely they can be driven in a nanotrack. Injection has usually been attained in nanotrack studies with local Oersted-induced magnetic reversing fields (e.g. [Himeno et al. 2003]), nucleation pads (e.g. [Briones et al. 2008]), curved track sections [Petit et al. 2008] (as done in this thesis), or just uncontrolled nucleation in a straight track (e.g. [Grollier et al. 2003]). All but the first technique are associated with a global applied field. The use of a curved track, which we do here, has the advantage over the other methods of separating DW injection from propagation, and of allowing some control over the injected DW structure, as it stabilises the transverse DW (TDW) even when it is only metastable [Kläui et al. 2004; Laufenberg et al. 2006].

The minimum propagation field in homogeneous tracks is limited by the lateral roughness, which creates pinning centres. Furthermore, it has been found that the lateral roughness also changes the propagation dynamics of DWs [Nakatani et al. 2003]. Earlier studies have found that, probably due to inter-layer magnetostatic interactions and to limitations of the patterning techniques, the pinning effect of lateral defects is very significant in SV tracks [Briones et al. 2008; Uhlir et al. 2010]. Though SV tracks have been used in several studies of DW propagation (see previous references), the high density of natural strong pinning centres is a strong limitation to the scope of possible experiments and to their sensitivity.

In this chapter we characterise SV tracks fabricated by the process described in Chapter 2, with the aim of obtaining tracks with good DW conduit characteristics. We study the track's magnetic properties, especially for injection and propagation of DWs, and its signal, i.e. the variation of resistance with the propagation of DWs through the GMR effect. We focus on how the track width and SV composition affect the measured signal, the DW propagation and the strength of pinning defects, the effect of the external field on the measured signal, and the signal of tracks not aligned to the reference layer direction.

**Table of contents**

# 3-1. A sv track

A set of L-shaped tracks of three different widths (80 nm – 260 nm) was fabricated with the titanium mask etch process (Chapter 2) from the following SV stack:

     (Si/SiO₂)// Ta 3.5/ **Py 8/ CoFe 2**/ Cu 2.4/ **CoFe 2/ Ru 0.8/ CoFe 2**/ MnIr 7/ Ta 2

(thickness in nms, schematic in FIG. 3-1B) [1]. The reference layer is in the SAF tri-layer CoFe/Ru/CoFe, which in turn is exchange pinned to the AFM layer (MnIr). The strength of the exchange pinning was beyond our setup maximum field ($H_{MAX}$ ~450 Oe); it was measured in similar stacks also deposited at INESC-MN to be ~600 Oe, although processing may have reduced it. The free layer, where the DWs will be created, is the double layer Py/CoFe, with an unpatterned coercivity Hc = 1.0 Oe and ferromagnetic shift HF = 10.0 Oe (arising from the coupling to the reference layer). The horizontal section of the track was 9 μm long, parallel to the direction of the magnetisation of the pinned reference layer. The contacts were placed 5.6 μm apart along the track, immediately after the curve, FIG. 3-1A. The curved end, a 90° arc, was used to inject DWs with a large 45° external field, as explained below.

---

[1] All fabrication parameters may be consulted in Annex A, sample HM18.

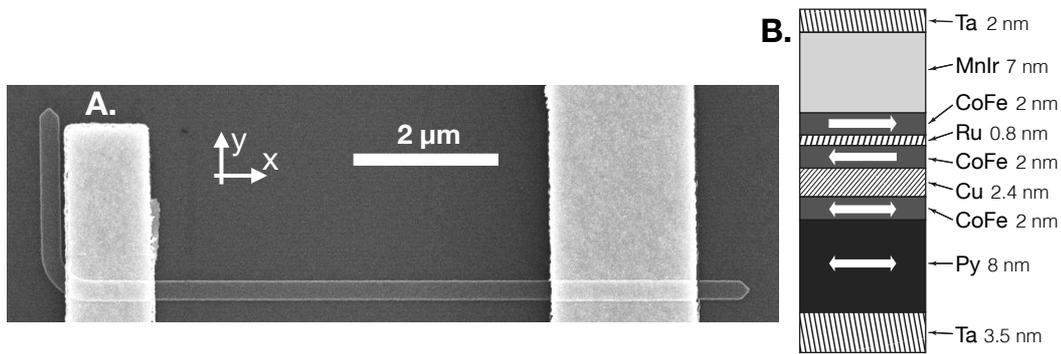

FIG. 3-1 L-shaped SV track. A. SEM image, showing the 260 nm wide SV track and the two gold electrodes. B. Schematic of the SV stack.

We measured the (two point) resistance between the contacts using micro-probes and a lock-in amplifier, as described in the previous chapter. The resistance between the contacts was 630 Ω for one 260 nm wide track. The top plot in FIG. 3-2 shows the resistance measured on 32 other structures in the same sample, of three different widths. The resistance followed closely the expected inverse function of width,

R = R_■ × length/width

where R_■ is a fitting parameter (called the geometrically normalised resistance or "r square"). The data fit shown in the plot yielded R_■ = 28.4 Ω, with a good fit quality (n = 32, adjusted R² was 0.98). The fact that the resistance varies so closely with the inverse of the width indicates a low contact resistance (10s of Ω). For some structures however the contact resistance was either very high or no electrical contact was possible at all. The most probable cause is the presence of some impurity on a specific region of the sample surface during fabrication, as the structures of high contact resistance were clustered together. These were few in number, and they were excluded from these plots.

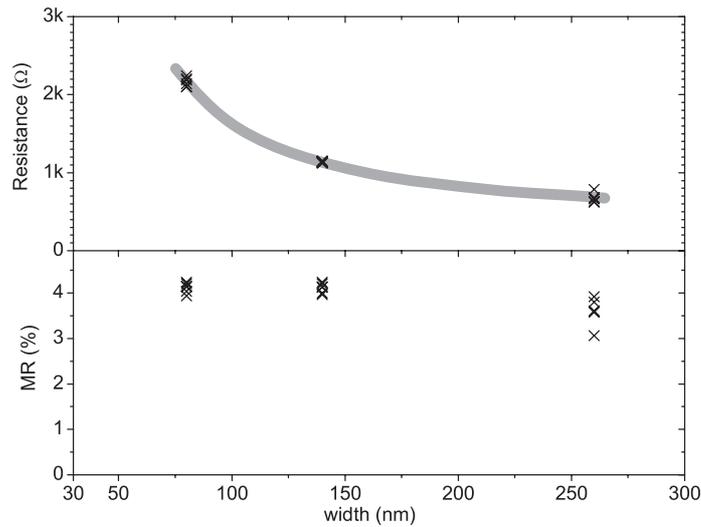

**FIG. 3-2 Resistance and MR vs track width.** Low-R state. The crosses are individual structures and the gray line is a hyperbolic fit (R=(R$_\blacksquare$ length)/width, R$_\blacksquare$=28.4 Ω, adjusted R$^2$=0.98).

The magneto-resistive ratio (MR) is calculated by MR $= \frac{R_{MAX} - R_0}{R_0}$, where R$_0$ is the resistance of the parallel magnetisations state (low-R state) and R$_{MAX}$ of the anti-parallel state (high-R state). It was measured by sweeping a horizontal field up to ±350 Oe, causing the reversal of the track magnetisation (see e.g. FIG. 3-4). It was 4% for the above-mentioned structure, and ranged 3–4.3% in the rest of the structures. This ratio is very similar to the one obtained for the unpatterned film (5.3%). The narrowest structures did not present a significantly lower MR, indicating there was no significant oxidation damage to the SV stack, a problem that often plagues the fabrication of sub-µm SVs [Chen et al. 1997]. From the variation of MR vs resistance and from the amplitude of the resistance measurements on nominally identical structures, we can indicate that the typical contact resistance is ~100 Ω.

## Current distribution

Due to the device geometry, the current is approximately parallel to the plane (this is known as current-in-plane geometry, CIP). Far from the contacts, the current density distribution is approximately constant along the plane dimensions, varying only across the depth of the SV stack. The different metallic layers have different resistivities and their thicknesses are comparable to the electron mean free path. Consequently, the current profile is between two extremes: a uniform profile (thin stack limit) and the parallel conductors profile (thick stack limit) [Dieny 2004], shown in FIG. 3-3 for $I = 1$ µA. The former assumes the current is the same for all layers, while the latter assumes that the current density is proportional to the layer conductivity. At $I = 1$ µA, j ≈ 0.4 $10^9$ A/m² in the free layer, and the effects of current on the magnetisation can be ignored (see Chapter 5).

It is also possible to see that approximately 20% of the current goes through the Py layer. This indicates that effective AMR signal should be ~1% [Li et al. 2001].

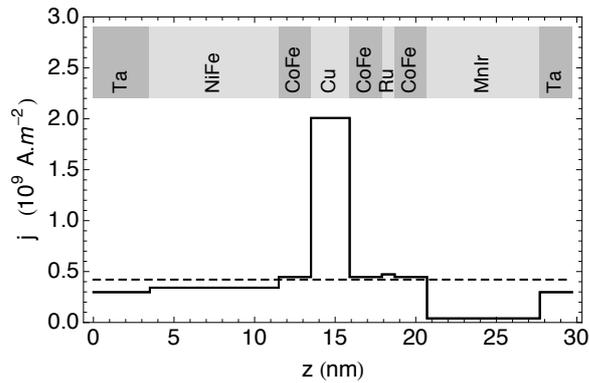

**FIG. 3-3 Current density profile.** The dashed line corresponds to the homogeneous current profile, and the solid line to the parallel conductors profile. For I = 1 µA, w = 80 nm, with resistivities taken from [Dieny 2004].

## Signal, noise and position resolution

We will now analyse the signal and noise of the measurement of SV tracks. This will serve as guidance to which structures and phenomena will be observable with the two-point measurement and lock-in technique we are using, described in detail in Chapter 2. This analysis is thus more a measure of our setup capability than of the intrinsic signal and noise of the structure.

The signal obtained with our setup allowed us to take single-shot measurements with typical experimental conditions: current of 1 µA, lock-in integration time 0.500 ms and field ramp rate of ~1 Oe/ms. A comparison between a single shot and a 10 averages measurement of a full MR transition is shown in FIG. 3-4. Defining the signal-to-noise ratio (SNR) as SNR = $\Delta\langle R\rangle/\sigma$ ($\sigma$ being the standard deviation) for periods of unchanging magnetic configuration, we obtain for this structure an SNR [2] of 47 ±11% (single-shot) and 98 ±6% (10 averages). If we assume a random noise, we would predict that SNR(10 avg.) = $\sqrt{10}$ SNR (single-shot), but 98 ±6% ≠ 149 ±11%. That the obtained value of the SNR for the average signal was lower than expected indicates that correlated noise between shots is present [3].

---

[2] We used the average $\sigma$ in the SNR formula. Error given by $\varepsilon_{SNR}^{rel} = \frac{\varepsilon_{R1}+\varepsilon_{R2}}{\Delta R} + \frac{\varepsilon_\sigma}{\langle\sigma\rangle}$, where $\varepsilon_\sigma$ was taken as half the difference between the two values of σ.

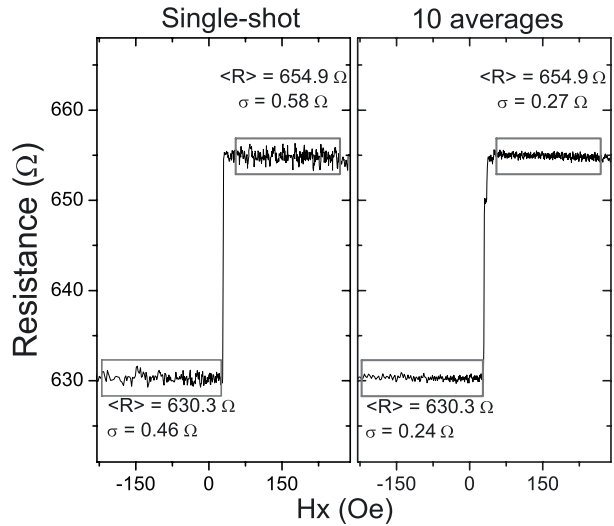

FIG. 3-4 **Single-shot and 10 averages of a full-MR transition.** The values of mean resistance and standard deviation (⟨R⟩ and σ) were taken from the intervals marked by the rectangles.

If a single DW is present in the system, then the resistance level can be linearly related to the DW position between the electrodes. The resolution of the DW position is given by the ratio of inter-electrode distance to the SNR. For the structure being described, this is 5.6 μm/98 = 57 nm for the 10 averages measurement and 119 nm for the single-shot. It is also interesting to consider what happens when the contacts are placed closer together. On one of the structures discussed in Chapter 4, the DW-DW interaction tracks, we can study such a case. This structure is also an L-shaped 260 nm wide track with an inter-contact distance of 1.7 μm (see FIG. 4–50). Taking one instance as an example, it presented a much lower MR of 2.3% due to the combination of a lower structure resistance (as it is over 3 times shorter in length), and thus a lower absolute MR change, and an increased contact resistance (caused by geometric limitations). Its SNR was measured to be 35 (single-shot) and 76 (10 averages), corresponding to a position resolution of 49 nm and 22 nm for single-shot and averaged measurements, respectively. This means that a net increase in resolution was obtained by placing the contacts closer together, even if the SNR was significantly degraded. Still, for both contact gaps, the resolution is better than the characteristic DW length (~track width).

---

[3] By correlated noise we mean non-random noise, most probably 50 Hz interference from the mains. In the averaged measurement, such correlated noise can generate higher or lower noise levels than in the single-shot measurement, depending on whether it interferes constructively or destructively.

Finally, it is important to point out that the real accuracy and precision of the DW position measurement might be adversely affected by other factors, such as MR variations unrelated to DW displacement (as described in §3-5 below) and setup limitations (e.g. quantisation errors and time integration). More importantly, the limitations in accuracy far exceed those of precision. A large source of inaccuracy is the finite width of the electrodes (typ. 500 nm), and contact position error (relative to the track) caused by imperfect lithographic registration (typically 100s of nm).

We also observed that over time periods of several minutes, much longer than those used for the field sequences (which are < 10 s), the resistance levels varied slightly; probably due to the influence of mechanical vibrations to the contact resistance between the contact probe needle and the gold pad on the sample. While for most measurements this effect has little influence, for some others it makes it necessary to normalise the curves before directly comparing them, with some inevitable loss of accuracy.

## Magnetic field precision

The field precision was limited by the lock-in integration time and field ramp rate. A transition, which occurs at timescales smaller than those treated here [4], suffers a *smear* due to the lock-in time integration of about 4 time constants (in the order of few ms). This measurement is in the quasi-static regime, i.e. no dynamic states can be observed, just the equilibrium static states. The time resolution also limits the field resolution; the results reported above, for example, had a 2 Oe resolution in the measurement of transition fields. Other experiments in this thesis, that required a higher precision, were performed at lower field ramp rates with higher field resolution. One aspect to note is that the field rate is typically not constant for the whole used field sequence: periods of interest, where the transitions under investigation are expected to occur, were done at a lower field rate (and thus higher resolution) than that of other periods, such as reset pulses. For example, in the field sequence of FIG. 3-6, the periods of interest (thick black line) were performed at 1 Oe/ms while the reset pulses (thin black line) at 4 Oe/ms. Field ramp rates can be lowered for higher precision, as mentioned above, but the setup equipment limits the duration of the whole field sequence to < 10 s, which in practice limits the ramp rate to >0.1 Oe/ms for the simplest used field sequence.

---

[4] The DW velocity is typically of the order of 100 m/s and the travelled distances ~10 μm, yielding a typical transition time of 100 ns, which is much smaller than the measurement interval (~1 ms).

Another limitation to the field precision is the noise of the field measurement. This noise is made of field noise, faithfully picked up by the transducer, and the unrelated electrical noise in the electrical circuit. For the single shot measurement of FIG. 3-4, the standard deviation of a static field measurement was 0.44 Oe. This value is much lower than most fields of interest (e.g. depinning fields in this thesis are typically of tens of Oe).

## 3-2. Nucleation of domains

The field necessary to nucleate domains in the track, the nucleation field ($H_{NUC}$), was determined as schematised in FIG. 3-5. A large (~450 Oe) reset field pulse is applied in the (-x,+y) direction, setting the magnetisation of the free layer in a single domain state, leftwards in the main arm (**i**). The field is then reduced to a small value of about ($H_X$, $H_Y$) = (0, 15) Oe. Such small vertical field has a limited effect on the measurement. A sweeping horizontal field is then applied in the opposite direction (**ii**). It eventually reaches $H_{NUC}$, when a new domain nucleates and reverses the wire (**iii**). The nucleation site is, most probably, at the track extremity where the demagnetisation field acts to increase the effective field. The same sequence is then repeated for the opposite field direction.

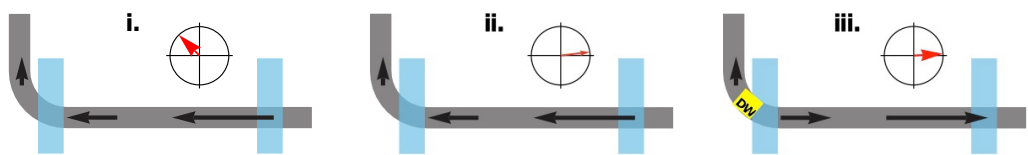

**FIG. 3-5 Magnetisation of the free layer during a nucleation sequence** (schematic). The track magnetisation is represented by black arrows, the DW by the yellow shapes, the contacts by the blue rectangles, and the direction of the external field by the red arrow. **i.** reset field. **ii.** sweeping horizontal field, with a positive y bias. **iii.** nucleation of a new domain and reversal of the main arm of the track.

An example of a measurement of $H_{NUC}$ in a 260 nm wide track is shown in FIG. 3-6, where $H_Y$ (top plot) and the resistance (bottom plot) are plotted versus $H_X$. The obtained hysteretic loop was highly repeatable, with no noticeable changes between measurements on the same structure, nor measurable variation in the value of $H_{NUC}$. There are two well-defined resistance levels, 630 $\Omega$ and 655 $\Omega$, corresponding to a $-x$ and $+x$ free layer magnetisation. It can be seen that the reversal of the track occurs in one single event at well-defined field values: $H^-_{NUC}$ =-159 Oe and $H^+_{NUC}$ =182 Oe.

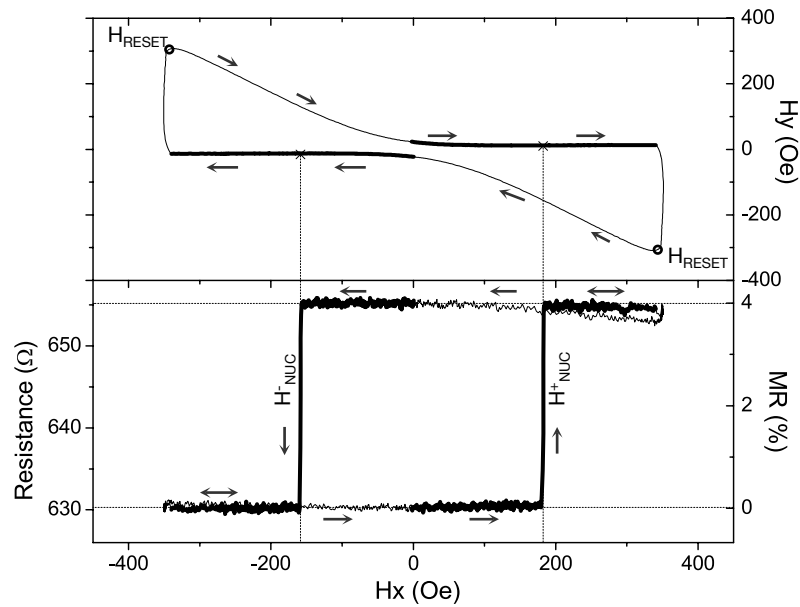

**FIG. 3-6 Nucleation sequence measured on a 260 nm wide SV track.** Averaged over 5 repetitions. The applied field sequence is shown on top, and the measured resistance on the bottom. The MR ratio is $(R-R_0)/R_0$, where $R_0$ is the low-R state. The crosses and vertical dashed lines indicate the magnetic reversal event ($H_{NUC}$), and the horizontal dotted lines indicate the resistance levels at $H_X = 0$. The thicker sections of the plots correspond to the periods of sweeping $H_X$ at $H_Y \approx \pm 15$ Oe.

The asymmetry in measured transition fields is a general feature of almost every SV measurement presented in this thesis. It arises from the interaction of the free layer with the constant magnetised reference layer. Its probable causes are RKKY coupling, Néel coupling, and magnetostatic interactions, the latter especially strong in the vertical edges of the track. The RKKY and Néel couplings are surface dependent, and as such can be measured on the unpatterned film. In the present sample, the unpatterned coupling shift was $H_{SH}$=10.0 Oe. In the above measurement, we can consider that $H_{NUC}$ is 170 Oe for both field directions if we accept a coupling shift $H_{SH}$ = +12 Oe is present, consistent with the coupling shift of the unpatterned film. Some variation (few Oe) around this value is seen from structure to structure on the same sample. It is also found that the stack composition is of great importance to this effect, as it will be discussed further below in this chapter.

# 3-3. DW propagation

Having measured the nucleation field, i.e. the field at which domains are injected, we consider now the propagation field (H$_{PR}$), i.e. the field at which DWs overcome the pinning of natural defects on the track and propagate through it.

The structure is the same as in the previous section, and the procedure, schematised in FIG. 3-7, is very similar to that used for measuring H$_{NUC}$. A large (~500 Oe) field pulse is applied in the (-x, -y) direction, setting the magnetisation of the free layer leftwards in the main arm and downwards in the vertical arm, creating a DW in the arc (**i**). The field is then reduced to a small value of about (0, -15) Oe. Such small vertical field has a limited effect on the measurement and prevents the DW from propagating upwards into the arc. The arc radius (0.5 μm) was chosen so its intrinsic pinning was negligible [Lewis et al. 2009]. A sweeping horizontal field is then applied in the opposite direction, eventually overcoming the pinning strength of the natural defects which pin the DW. The travelling DW then reverses the magnetisation between the contacts (**ii**). The same steps are then repeated for the opposite field direction.

The shape of the arc favours the injection of TDW with the central magnetisation pointing down [Lewis et al. 2009], as depicted in the figure. This has been observed with direct imaging techniques even in tracks with dimensions for which the TDW is only metastable [Kläui et al. 2004; Laufenberg et al. 2006; O'Shea 2010; Petit et al. 2010] and a vortex DW would otherwise be favoured [Nakatani et al. 2005].

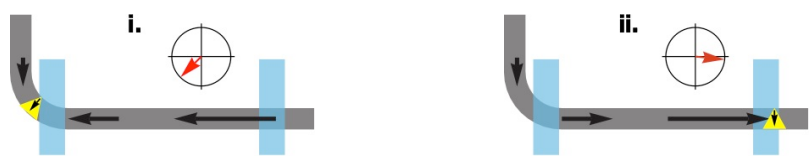

**FIG. 3-7 Magnetisation of the free layer during a propagation sequence** (schematic). The track magnetisation is represented by black arrows, the DW by the yellow shapes, the contacts by the blue rectangles, and the direction of the external field by the red arrow. **i.** Reset field creating a DW in the corner; **ii.** sweeping horizontal field, eventually reaching H$_{PR}$, when the DW propagates through the track reversing the magnetisation of the main arm.

The measurement done on the same structure as before is presented in FIG. 3-8, where $H_Y$ and the resistance are plotted versus $H_X$. We can see that again the magnetisation reversal occurs in one single event, at a well-defined field value. This is not always the case; some structures have strong pinning sites between the contacts, which generate a laddered switching curve. Others, having pinning sites before the contacts, show a single transition at an increased $H_{PR}$. In addition, the depinning fields of some of these natural defects show a measurement-to-measurement variation.

As with $H_{NUC}$ before, $H_{PR}$ is also shifted: $H_{PR} = 18$ Oe with $H_{SH} = +11$ Oe. This shift is again similar to the one measured in the unpatterned sample (10.0 Oe) and in the nucleation sequence (12 Oe).

$H_{PR}$ is much smaller than $H_{NUC}$ (18 Oe vs. 170 Oe), which indicates that the SV track is a good DW conduit: there is a large range of field under which DWs can propagate without nucleation of new domains.

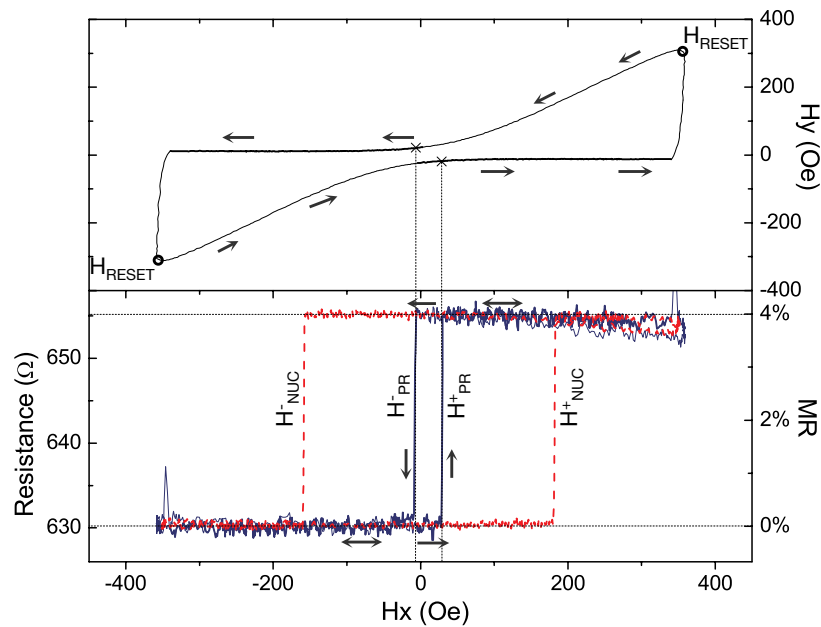

**FIG. 3-8 Propagation sequence measured on a 260 nm wide SV track.** Single shot. The applied field sequence is shown on top, and the measured resistance of on the bottom (blue line). The thicker sections of the plots correspond to the periods of sweeping $H_x$ and constant $H_Y \approx \pm 15$ Oe. The resistance measurement for the nucleation sequence (from Fig. 3-6) is also included in the lower plot for comparison (dashed red line). The crosses and vertical dashed lines indicate the magnetic reversal event.

# 3-4. Variation with width

FIG. 3-9 shows H_NUC and H_PR measured on 38 structures of three different widths (80, 140, and 260 nm). Due to the limits of the electromagnet driver, for the largest values of H_NUC, the transition occurred only during the reset field pulses (see FIG. 3-6) with a large vertical field component. Both H_NUC and H_PR are larger for narrower tracks, in an approximately 1/width relation. For H_NUC, this can be understood with a Stoner-Wolhfarth reversal mechanism, according to which the switching field is equal to $(\mathcal{N}_{YY} - \mathcal{N}_{XX})\,M_S \approx \mathcal{N}_{YY}\,M_S$ (where $\mathcal{N}$ is the demagnetisation tensor), which approximately follows a 1/width curve in tracks of length → ∞ (FIG. 3-10). Although the dependence with width is in qualitative accordance with this model, the measured nucleation field values are much lower than those predicted by the Stoner-Wolhfarth infinite track model: 216 Oe vs. 941 Oe for width = 140 nm. This indicates that reversal does not happen by uniform magnetisation rotation, but through nucleation and DW propagation. The nucleation should occur at the ends of the track, where the local shape anisotropy is much lower (using the expression above, $(\mathcal{N}_{YY} - \mathcal{N}_{XX})\,M_S$, near the ends the local $\mathcal{N}_{XX}$ is non-zero and the local $\mathcal{N}_{YY}$ is reduced).

**A.**

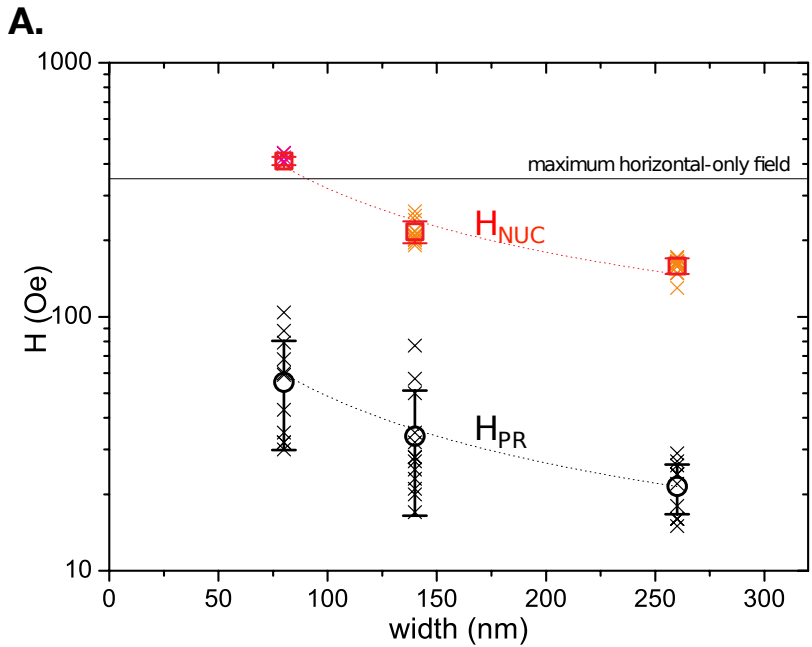

**B.**

| width (nm) | N | $\langle H_{NUC} \rangle \pm$ SD (Oe) | $\langle H_{PR} \rangle \pm$ SD (Oe) |
|---|---|---|---|
| 80 | 13 | *411 ± 16* ✱ | 55 ± 25 |
| 140 | 13 | 216 ± 21 | 34 ± 17 |
| 260 | 12 | 159 ± 11 | 21 ± 5 |

**FIG. 3-9 $H_{NUC}$ and $H_{PR}$ vs track width for 38 different structures of three different widths. A.** Semi-log plot. $H_{NUC}$ are the red crosses (×), and $H_{PR}$ the black (×). The red squares (□) and the black circles (O) are the sample averages for $H_{NUC}$ and $H_{PR}$, respectively. The error bars are the sample standard deviation. The horizontal line (at 350 Oe) marks the setup limit for horizontal-only fields, the data points above it having a significant $H_Y$ field (100s of Oe). The dashed lines are guides for the eye (hyperbolic fits). **B.** Data table. The $H_{NUC}$ value marked with a ✱ is beyond the setup limit for horizontal fields.

The $H_{PR}$ also shows an increase with narrowing tracks, of about the same relative magnitude. Though this variation is within the sample standard deviation, it is systematically observed in several samples. $H_{PR}$ is a measure of the pinning strength of the lateral shape defects on the track. Its variation with track width can be understood by assuming that the sizes of the lateral shape defects caused by the imperfect fabrication process are the same for all track widths. The pinning strength of those defects however is not: as the track narrows, the defects represent a much larger relative modulation of the track width. Furthermore, in narrower tracks the DWs are also narrower. Therefore, the same defect has a higher pinning strength in narrower tracks [Himeno et al. 2003; Faulkner 2004]. Thus, narrower tracks present higher $H_{PR}$.

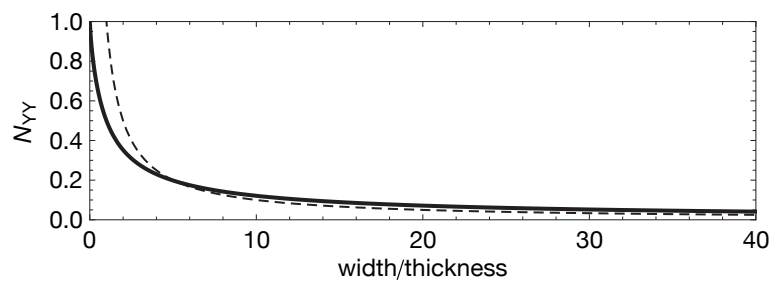

**FIG. 3-10 Demagnetisation factor vs cross-sectional ratio for an infinite prism.** $N_{YY}$ is the width-wise diagonal element of the demagnetisation tensor. Theoretical calculation (thick line; [Aharoni 1998]) and thickness-to-width ratio (dashed line).

# 3-5. Effect of a transverse field

In addition to the change in MR caused by the reversal of the track magnetisation, there was a smaller variation at high external fields that was not related to the propagation of DWs, as can be seen in FIG. 3-6 and FIG. 3-8. There, the application of large transverse ($H_Y$) fields during the reset period changed slightly and gradually the MR value, without any reversal of the track. To better observe this, the same propagation and nucleation field sequences were applied with larger reset field pulses ($H_{MAX}$ = 450 Oe), and the data plotted as MR versus $H_Y$, FIG. 3-11. During the reset pulse of the propagation sequence, the applied field goes from $(\pm H_{MAX}, 0)$ to $(\pm H_{MAX}, \pm H_{MAX})$, while during the reset of the nucleation sequence the applied field goes from $(\pm H_{MAX}, 0)$ to $(\pm H_{MAX}, \mp H_{MAX})$. We labelled the change in MR caused by these four different field directions as $\delta^{(\pm,\pm)}$ in the plot (the signs corresponding to the direction of the applied field in x and y, respectively).

The application of the 450 Oe Y-field increased the resistance of the low-R state and decreased that of the high-R state, though not by the same amount (about -0.5% for high-R state and +0.1% for the low-R state). This effect seems compatible with a rotation of the free layer with the external field. The sign is correct: a rotation of the magnetisation away from the horizontal direction would lead to a decrease of resistance in the high-R state and an increase in the low-R state. The asymmetry between the two magnetisation states might be caused by a small misalignment of the track with respect to the reference layer pinning direction, and by the AMR contribution.

This magnetisation rotation with $H_Y$ should depend on the track width: the wider the track, the smaller the transverse demagnetisation factor ($N_{YY}$) and the larger the rotation. To test this, we measured the $\delta^{(\pm,\pm)}$ values on 21 structures of three different widths, 80, 140, and 260 nm (7 structures of each). The results are shown in FIG. 3-12, where the averaged $\delta^{(\pm,\pm)}$ values are shown versus the track width. Firstly, we note that the resistance variation was as large as 0.7% at $H_Y$ = 450 Oe (in the widest tracks), equivalent to ~20% of the total MR change (for a typical MR of 3.5%). As before, we observe that, for almost every structure, the Y-field increased the resistance of the low-R state and decreased that of the high-R state, and that this effect was greater in the high-R state (the $\delta^{(+,\pm)}$ values) than in the low-R state (the $\delta^{(-,\pm)}$ values). Moreover, the wider tracks generally showed a greater change in resistance with $H_Y$ than narrow tracks, which is consistent with the hypothesis of a uniform magnetisation rotation.

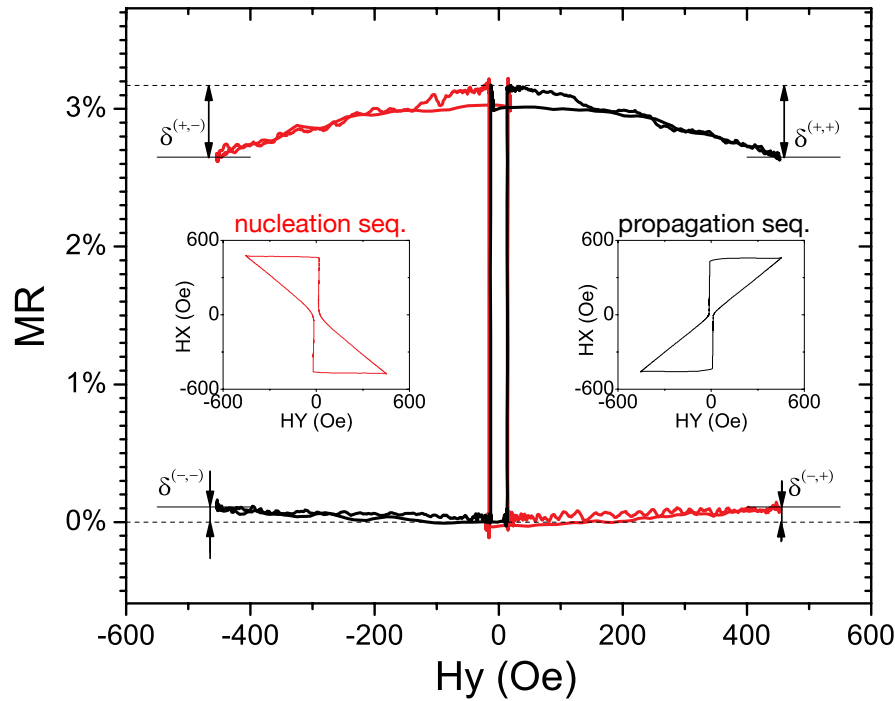

**FIG. 3-11 MR vs $H_Y$ during nucleation and propagation sequences,** on a 260 nm wide track. MR signal from the nucleation sequence in red, and from the propagation sequence in black. The thin horizontal markers indicate the MR level at the extremes, $\delta(\pm,\pm)$ labels defined in the text. The field sequences are plotted in the insets.

To analyse the magnitude of this rotation, micromagnetic simulations for the same values of field and track dimensions were performed, and the simulated horizontal magnetisation component ($m_X$) versus $H_Y$ is plotted in FIG. 3-13. In the simulated 260 nm wide track, at $H_Y = 450$ Oe, the magnetisation rotated about 27° and $m_X$ was reduced by 11%. This corresponds to a change in resistance of about 0.2% [5]. Likewise, the predicted change for the 140 nm wide track is 0.1% and, for the 80 nm wide track, 0.05%. Comparing to the experimental values shown in FIG. 3-12, we see that the values for the low-R state are generally consistent with these simulations, while the values for the high-R state are a few times larger. This difference may suggest a misalignment between the track and the reference layer pinning direction, or a significant AMR contribution.

---

[5] Note that note that $m_X$ goes from -1 to 1 between the low- and high-R states, which corresponds to the total MR variation, and so $2 \cdot \Delta MR = \Delta m_X$. Here we use a MR value of 3.5%.

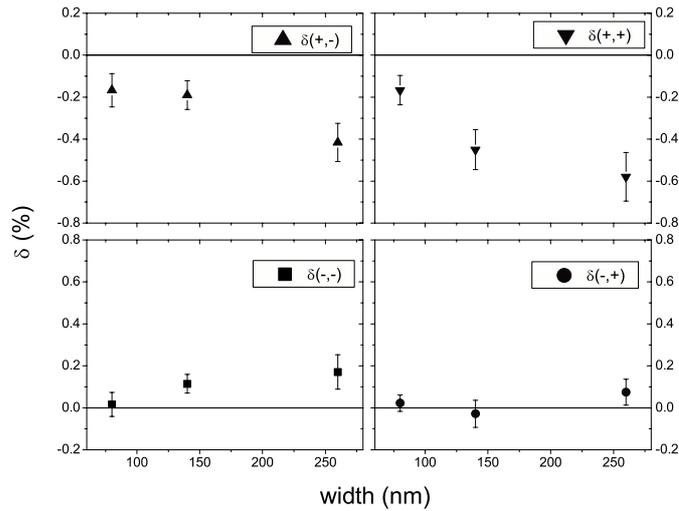

**FIG. 3-12 Variation of resistance with H$_Y$.** Average values (7 structures per width). Error bars are the sample standard deviation. δ(±,±) values as defined in the text.

We conclude then that fields applied transversely to the track direction induce a magnetisation rotation, which reflects on the measured MR signal. This effect can be quite large (we observed up to a fifth of the total MR variation at 450 Oe). This must be taken into account when analysing the measured signal, especially when measuring wide tracks.

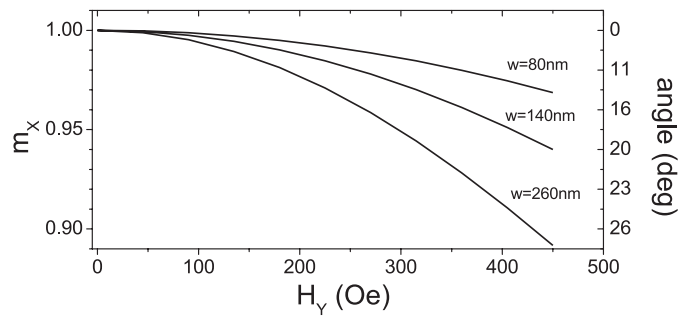

**FIG. 3-13 Magnetisation (horizontal component) of a single track vs vertical applied field** (numerical simulation). H$_X$ = 450 Oe (constant). Track dimensions: 2 μm×width×10 nm. Cell size = 5×5×10 nm$^3$, Ms = 800 kA/m. Simulated with OOMMF [Donahue & Porter 1999].

# 3-6. Signal of vertical tracks

So far, we have considered only the reversal of horizontal track segments, i.e. segments parallel to the direction of the (pinned) reference layer. To examine the MR signal produced by the reversal of vertical segments, we have used a four-turn rectangular spiral [6], FIG. 3-14A. The spiral track was 140 nm wide, with the outer loop measuring 15.6 by 10.5 μm, the inner 12 by 6.9 μm, with 0.6 μm loop separation. Two electrical contacts were placed, connecting a total of 7 spiral segments in parallel. For the purposes of this experiment, the spiral is similar to a series of L shaped tracks. Analogously to the propagation experiment in the L track, a 350 Oe $H_{RESET}$ was applied in the (−X,+Y) direction, injecting a DW in each of the spiral corners, 4 TT DWs in the bottom right corners and 4 HH DWs in the top left corners. The 8 vertical segments of the spiral were all magnetised upwards. Afterwards, a constant $H_X$ = +70 Oe was applied while $H_Y$ was cycled six times between +117 and -117 Oe at 30 Hz. This caused the DWs to propagate and reverse the vertical segments six times, while no DWs were propagated through the horizontal sections.

Due to the arrangement of the electrical contacts (cf. FIG. 3-14A), the current only traversed a part of the 8 vertical sections [7]. The registered signal, FIG. 3-14B, shows that the two vertical magnetisation states have distinct MR values. At $H_Y$=0, the *up* state has MR = 2.79% and the *down* state 2.63%. The signal shows the transitions between these states clearly, though not sharply, at $H_Y$ = -38±8 and +26±8 Oe. In addition to this hysteretic loop, there is also a linear variation of the MR signal of each state with $H_Y$, with opposing signs (+0.6 %/kOe and -1.2 %/kOe).

**A.**          **B.**

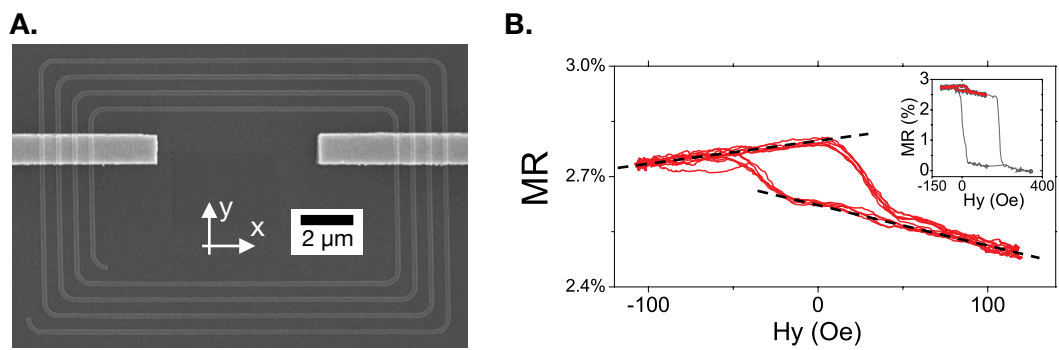

**FIG. 3-14 Reversal of the vertical track sections of a spiral track. A.** SEM image of the spiral track. **B.** MR vs cycling $H_Y$ (red). The black dotted lines indicate the linear variation of the MR with $H_Y$ (see text). The inset shows the signal during the reset pulse, with complete reversal of the horizontal track segments, for comparison.

---

[6] This structure will be further studied in Chapter 4.

[7] A more detailed analysis of the contributions of different spiral sections to the signal is presented in Chapter 4.

The observed difference in MR level between the up and down states might be unexpected, as the two states have the same horizontal magnetisation component, and so should have the same MR level. We offer the following explanation. The reference layer is never perfectly horizontal, due to inevitable errors aligning the sample during fabrication. As such, the magnetisation in the up and down states form slightly different angles with the reference layer, which generates the MR difference. As it is the up state (i.e. +y) that has the lowest resistance, the reference layer must have a slight positive vertical component. The signal transitions correspond thus to the track reversal, from up to down and vice-versa.

The measured transitions showed a field width of about 15 Oe. Part of this width is due to the lock-in signal filtering. The field ramp rate, at the moment of the transition, was $\approx 2.2$ Oe.ms$^{-1}$. With the lock-in time constant set to $\tau = 0.5$ ms, the typical lock-in time response is ~5 $\tau_0 \approx 2.5$ ms [8] or ~5.5 Oe. Another cause for the transition broadening is the fact that these transitions are generated by the propagation of 8 DWs, with some expected dispersal of $H_{PR}$ values. Indeed, the standard deviation of $H_{PR}$ measured before in L shaped tracks is 17 Oe (cf. FIG. 3-9).

As for the linear variation of MR signal with $H_Y$, recalling that the reference layer is oriented at 180° (-x), we can identify two origins for the observed signal. The first comes from the applied $H_X$ (+70 Oe), which causes the magnetisation to have a small +x component (cf. §3-5), antiparallel to the reference layer. At $H_Y = 0$, the resistance is higher than it would be at $H_X = 0$, although the increase is the same for both states. When $H_Y \neq 0$, however, this is no longer the case. If the (vertical) magnetisation and $H_Y$ are parallel, the +x component will be smaller, if they are antiparallel, the +x component will be larger, cf. FIG. 3-15. Thus, a rising $H_Y$ will have opposite effects in each state: in the up state, the signal will smoothly increase, in the down state, it will smoothly decrease, as we observed in FIG. 3-14A. Note that the magnetisation of the horizontal segments must also vary, albeit less than in the vertical segments and symmetrically in $H_Y$.

---

[8] Which is consistent with the transition (time) width of the large reset transitions (FIG. 3-14 inset), approx. 2.8 ms.

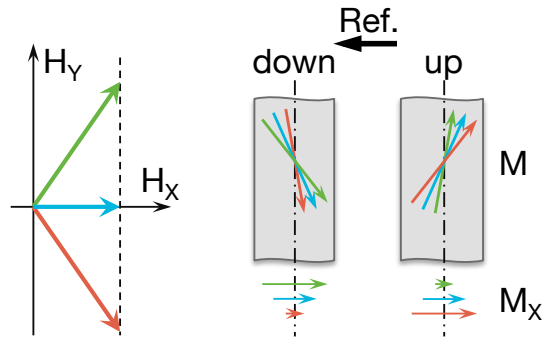

**FIG. 3-15 Free layer magnetisation of a vertical track under a sweeping Y field** (sketch). Applying a constant $H_X$ and a sweeping $H_Y$ yields a finite $M_X$ component, which varies smoothly with $H_Y$. The sign of the $M_X$ variation with $H_Y$ is opposite for the up and down states.

Concluding, we observed that the reversal of vertical track segments does generate an MR signal, though one small in comparison to the reversal horizontal segments. The origin of this signal can also be attributed to the GMR effect. It arises from the misalignment of the mask with the reference layer (during fabrication), which in fact means that vertical tracks are not perfectly perpendicular to the reference layer magnetisation.

## 3-7. Influence of SV stack composition

The composition of the free layer is extremely important to the DW propagation properties of the track, as it affects the nucleation field and the structure of the DW [Nakatani et al. 2005]. We have so far studied stacks with a Py 8 nm/ CoFe 2 nm free layer. In this section, we will compare this composition with single 8 nm-thick Py free layers, a composition for which there is extensive published research in monolayer tracks (§3-7.1). We will also compare these SV tracks with the monolayer Py tracks reported in literature (§3-7.2).

The other layers of the SV stack can affect the magnetic behaviour of the free layer. A sign of this can be seen in the shift of $H_{NUC}$ and $H_{PR}$ described earlier, which cannot be intrinsic to the free layer track but instead arises from its interaction with the pinned reference layer. Other SV studies have also reported that inter-layer interactions alter the pinning of lateral track defects [Briones et al. 2008] and influence the DW propagation dynamics [Ndjaka et al. 2009]. It is then important to investigate how different reference layer compositions affect the DW conduit properties of the track. So far, we performed measurements on structures fabricated from the stack described in §3-1 above. This stack included a SAF reference layer: CoFe 2/Ru 0.8/CoFe 2/MnIr 7 nm. We will here measure structures with a simple reference layer, CoFe 2/MnIr 7 nm, and present a comparison of the DW propagation properties (§3-7.3).

## 3-7.1. Free layer: Permalloy vs Permalloy-CoFe

Tracks of three different widths (33, 110, and 200 nm) were patterned with the titanium mask etch process from the following stack:

(Si/SiO₂)/Ta 2/**Py 8**/Cu 2/CoFe 2.2/Ru 0.8/CoFe 2.2/MnIr 6/Ta 5

(thickness in nms) [9]. The significant difference between this stack and the previous is the composition of the free layer (Py 8 nm vs Py 8/CoFe 2 nm). This composition should produce a smaller free layer magnetic moment due to the thickness decrease. It also generates tracks with a lower MR, now 1.5–2% (patterned), similar to what has been reported for NiFe/Cu SVs [Dieny 2004].

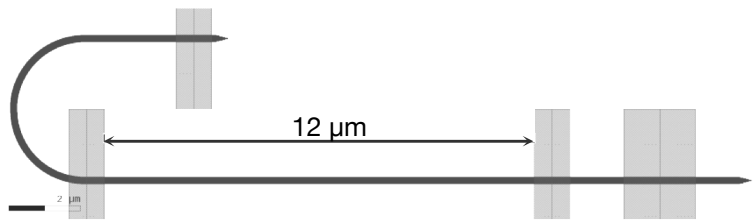

**FIG. 3-16 EBL mask for c-shaped wire,** showing the track (dark grey) and contacts (light grey). The contacts used for measurement were the inner ones, 12 µm apart.

Apart from the changes in width, these tracks also had some differences in the used masks, compare FIG. 3-1 to FIG. 3-16. The arc, used for DW injection, was a 180° and not 90° as before, though the DW was still injected in the same location. They are also longer (19 vs 9 µm). The contact separation was also different: 12 µm instead of 5.6 um. Some structures had large artificial traps placed far from the injection corner [10] and on these structures we did not measure H_{NUC}, which was lowered by the presence of the traps. All the listed differences should otherwise be irrelevant to the properties that we investigate here ($H_{PR}$ and $H_{NUC}$).

---

[9] All fabrication parameters may be consulted in Annex A, sample HM01.

[10] We shall discuss these artificial defects in the next chapter

We performed measurements of nucleation and propagation fields, and analysed them as a function of track width. We plotted the results in the semi-log plot of FIG. 3-17A, where we also included part of the results of the Permalloy-CoFe free layer structures (FIG. 3-9) for comparison. Although the track widths are not the same between the two samples, and there are only a few data points for $H_{NUC}$ in this sample, the propagation and nucleation fields seem to follow roughly the same variation with width.

We expect that a decrease in free layer thickness (from the previous 8+2 nm to 8 nm) to cause a decrease of $H_{NUC}$ for equivalent widths, due to a decrease in shape anisotropy (as analysed in §3-4 and FIG. 3-10). Using the previously discussed Stoner-Wohlfarth model of an infinite prism, for a 200 nm wide track, we predict that a 10 to 8 nm decrease in thickness leads to a 16% decrease in $H_{NUC}$. However, it is hard to compare this value to the experiment for experimental and theoretical reasons. Experimentally because there is no exact match between widths of the two samples, and there are only a few data points within the setup limits for horizontal fields. Theoretically because the $M_S$ of the double film, to which $H_{NUC}$ also linearly scales (as per our discussion in §3-4), is unknown.

It is worth of note that the 33 nm wide tracks presented here (FIG. 3-17A,C) are, to the best of our knowledge, the narrowest ever measured DW conduits in SV or in monolayer Py.

In summary, tracks fabricated with both these free layer compositions, Py 8 nm and Py 8/ CoFe 2 nm, showed a high contrast between $H_{PR}$ and $H_{NUC}$, with very similar quantitative variation with track width. We conclude thus that both compositions can be used to produce good DW conduits, i.e. tracks in which there is DW propagation without domain nucleation for a wide range of applied field. The tracks with the Py-only free layer have the advantage of being more directly comparable to studies on monolayer Py, a common composition in literature on magnetic tracks for DW studies. The tracks with the Py 8/ CoFe 2 nm composition, on the other hand, have the advantage of increased MR and thus higher SNR.

**A.**

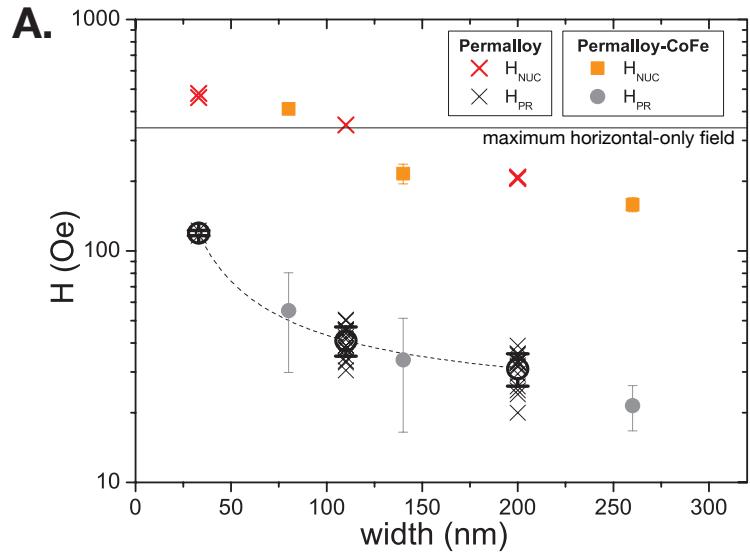

**B.**

| Width (nm) | N (H_NUC) | N (H_PR) | ⟨ H_NUC ⟩ (Oe) | ⟨ H_PR ⟩ ± SD (Oe) |
|---|---|---|---|---|
| 33 | 2 | 2 | *470* ✱ | 120 |
| 110 | 2 | 17 | *350* ✱ | 41 ± 6 |
| 200 | 2 | 17 | 207 | 31 ± 5 |

**C.**

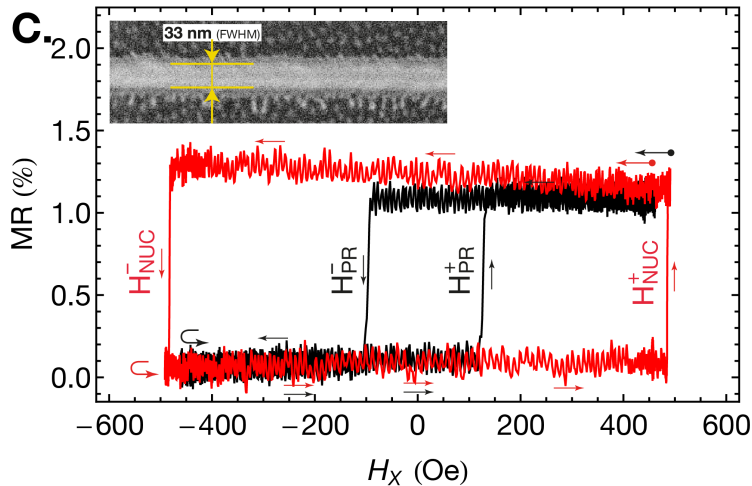

**FIG. 3-17** $H_{NUC}$ and $H_{PR}$ vs track width for structures of three different widths (sample with Py 8 nm free layer). **A.** Semi-log plot. $H_{NUC}$ and $H_{PR}$ of individual structures are the red and black crosses (×, ×), respectively. The black open circles (O) are the sample averages for $H_{PR}$ (error bars are the sample standard deviation). The data from the sample with the Py 8/ CoFe 2 nm free layer is included for easier comparison (grey circles ● and orange squares ■). The horizontal line (at 340 Oe) marks the setup limit for horizontal-only fields, the data points above it having a significant $H_Y$ field. The dashed line is a guide for the eye (hyperbolic fit of the ×). **B.** Data table. The $H_{NUC}$ values marked with a ✱ are beyond the setup limit for horizontal fields. **C.** Propagation and nucleation single shot measurements on a 33 nm wide structure (**inset:** SEM image of the track).

## 3-7.2. Comparison with monolayer tracks

FIG. 3-18 shows the $H_{NUC}$ and $H_{PR}$ values for monolayer Permalloy tracks fabricated with EBL and lift-off [11], reported in earlier studies.

| Width (nm) | Thickness (nm) | $H_{NUC}$ (Oe) | $H_{PR}$ (Oe) |
|---|---|---|---|
| 100 | 10 | 213 ±14 | 12 ±3 |
| 98 ±9 | 10 ±1 | 254 ±14 | 14 ±5 |
| 180 ±7 | 10 ±1 | 125 ±7 | 10 ±5 |

FIG. 3-18 **Propagation and nucleation in monolayer Permalloy tracks.** Data taken from [Lewis et al. 2009] (first row) and [Jausovec 2008] (second and third rows).

Overall, comparing with the SV tracks shown before, the monolayer tracks present lower switching fields, both $H_{PR}$ and $H_{NUC}$, and a much smaller variation of $H_{PR}$, indicating that they have smoother track edges. This is expectable, considering the fabrication process of both types of tracks. Py tracks and the Titanium etch mask used in SV tracks are patterned with the same process, and so should have similar lateral roughness. SV tracks are then subjected to an extra dry etching step, which may increase its lateral roughness, due to uneven etching and redeposition of magnetic material. However, as it was seen before, direct comparison of $H_{PR}$ and $H_{NUC}$ in samples differing both in thickness and width can be difficult, especially in such thin layers, where the error in thickness is very significant. Nonetheless, we can use the $H_{NUC}$ to $H_{PR}$ ratio as a figure of merit of the track as a DW conduit. This ratio is typically 7–10 in SV tracks (see e.g. FIG. 3-17) and 13–18 for the monolayer tracks.

While SV tracks do function as a DW conduit, they show a smaller propagation to nucleation gap than monolayer tracks patterned by the same lift-off method. We suggest that this is caused by a higher lateral roughness of the SV tracks, generated during the dry etch step of the process, which increases the $H_{PR}$.

---

[11] with a method completely identical to the Ti mask fabrication step used in the samples in this thesis.

### 3-7.3. Reference layer: simple CoFe vs SAF

To study the role of the reference layer on the conduit properties, structures made with a different stack were studied and compared to the previous results. The stack composition was

(Si/SiO₂)/Ta 2/**Py 8**/Cu 2/**CoFe 2.2**/MnIr 6/Ta 5

(thickness in nms). This stack had a simple Py free layer (as characterised in §3-7.1) and a simple pinned reference layer, CoFe, instead of the previous SAF triple layer. Tracks of three different widths (75, 210, and 330 nm) were patterned with the titanium mask etch process, using the same mask of FIG. 3-16 [12], and measurements of H_PR and H_NUC are shown in FIG. 3-19. In this sample, we observed large differences between HH and TT DWs, and for this reason we show the propagation values separately for both DW polarities. We observe the same overall relation with width as before, and similar values for nucleation and for propagation of TT DWs. Propagation field of HH DWs was however very large. On most structures, it was ≈H_NUC, and probably not truly corresponding to a DW propagation event. We believe this is the result of large stray fields induced by the reference layer. These fields may have impeded nucleation of HH DWs, or have pushed it towards the vertical arm.

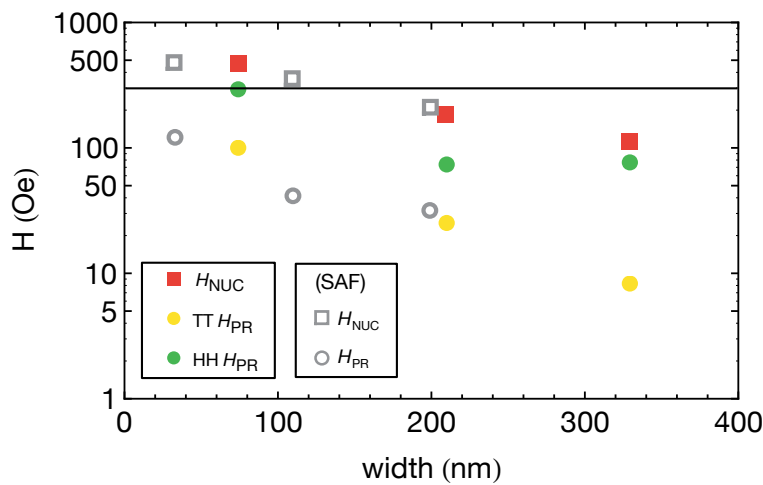

**FIG. 3-19 Propagation and nucleation in a sv with a simple reference layer.** Nucleation and propagation for HH and TT for the simple reference layer sample (full symbols). Average over 3 structures of each width. Data from the sample with the SAF reference layer and Py 8 nm free layer was also included from FIG. 3-17C (open symbols). The horizontal bar marks the setup maximum horizontal field (~ 350 Oe).

---

[12] The fabrication details can be consulted in Annex A, sample reference HM03.

Moreover, we found that in these tracks there were several strong pinning sites along the track, for both TT and HH DWs, which resulted in multiple steps in the propagation curve, FIG. 3-20. As the fabrication process was the same as the previous discussed samples, the lateral roughness should also be comparable. Thus, we suggest that the presence of strong pinning sites is not caused by this specific sample having larger edge defects but, instead, by the same defects having an enhanced pinning.

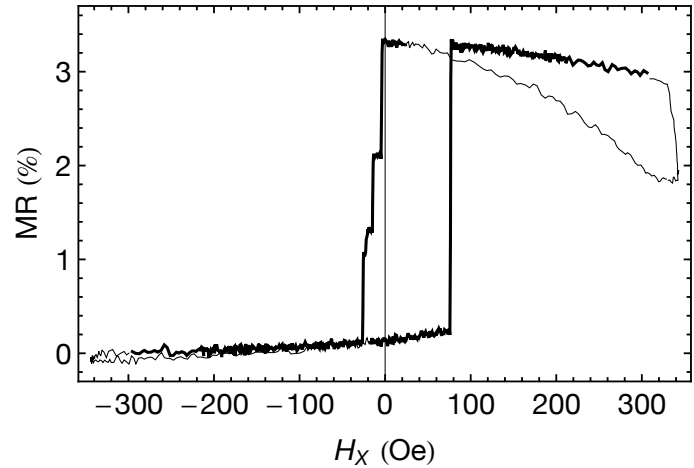

**FIG. 3-20 Pinning in simple reference tracks** (single shot MR measurement). Propagation sequence signal for a 330 nm wide structure. The thin sections correspond to the reset pulses with large applied $H_Y$.

The edge defects, in a SV with a simple reference, generate large local stray fields from the reference layer, which are absent in SV tracks with a balanced SAF reference layer. To test this hypothesis, two rough-edged tracks (FIG. 3-21A) were simulated separately, a monolayer Py and a SV track (consisting in a triple layer: Py 8 nm/vacuum 2 nm/CoFe 2nm). The CoFe layer was biased with a 350 Oe field to emulate the effect of the exchange pinning. FIG. 3-21B shows the stray field from the CoFe layer in the Py layer (blue is 0, red is ≥250 Oe), revealing several small regions of high field magnitude. Simulations of DW propagation in the monolayer and in the SV track show that this highly local and intense fields pin the DW at a defect more strongly than the simple defect would without the stray field, FIG. 3-21D. This is consistent with experimental and micromagnetic simulation results on large artificial notches reported by Briones et al. [Briones et al. 2008] where it was found that the local stray fields altered the pinning characteristics of an artificial notch.

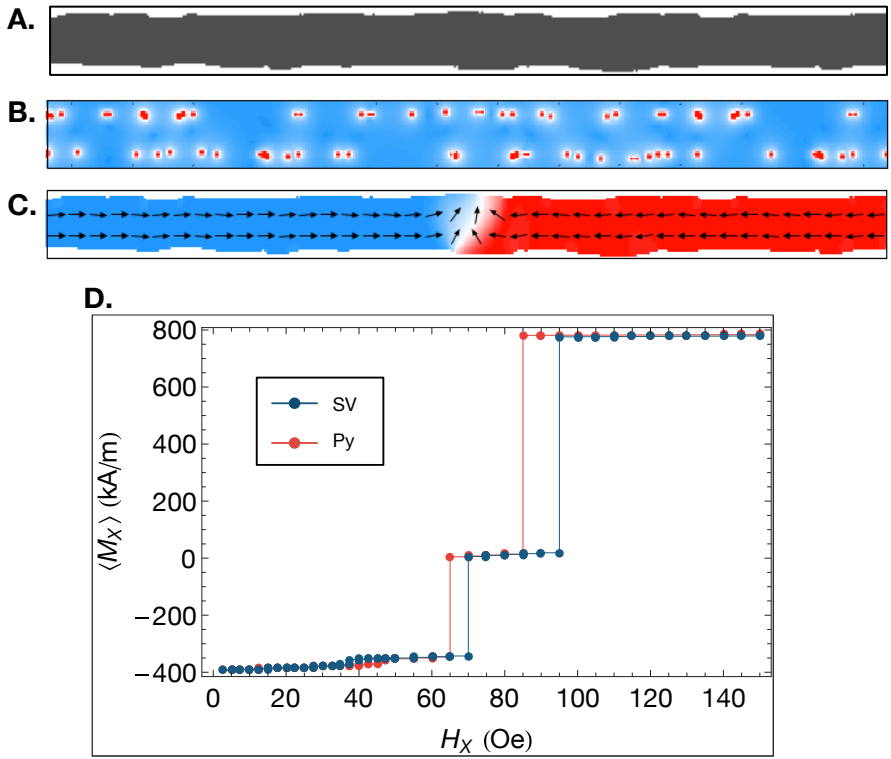

**FIG. 3-21 Enhanced defect pinning in sv tracks with simple reference layers** (micromagnetic simulation). A Py rough-edged track was simulated with and without the presence of a biased CoFe layer. **A.** Py track mask. **B.** Magnitude of stray field from the CoFe layer in the Py layer (blue is 0 Oe and red is ≥ 250 Oe). **C.** Magnetisation showing a DW pinned at one of the edge defects. **D.** Average magnetisation vs. applied field in the absence of the CoFe layer ("Py") and with the CoFe present ("SV"). Simulation parameters: $M_S$(Py)= 800 kA/m, $M_S$(CoFe)= 1010 kA/m, thickness of 8 nm (Py) and 2 nm (CoFe). The CoFe layer was 2 nm above the Py and was biased with a 350 Oe field. Cell size was 5×5×2 nm³, width ≈ 150 nm. Simulation performed with OOMMF [Donahue & Porter 1999].

This effect may also be responsible for the frequently reported high density of multiple natural pinning defects in sv tracks (see e.g. [Grollier et al. 2003; Himeno et al. 2003; Lacour et al. 2004; Uhlir et al. 2010]).

Another aspect of the use of simple reference layers in sv is that, having a magnetic moment, they are much more sensitive to external applied fields than the SAF reference. This is visible in the very large variation of the high-R state resistance with the reset pulse vertical field in FIG. 3-20. During the positive reset pulse, the MR goes from 3.2% to 1.8% (while $H_Y$ goes from 14 to 350 Oe and $H_X$ = 350 Oe), indicating that the mean angle between the free and reference magnetisations changed from ≈180° to ≈100° [13]. In the negative reset pulse however the MR variation is negligible, indicating the free-to-reference angle is maintained at 0°. Using the same simulation technique of FIG. 3-13, the free layer is estimated to rotate from 0 to 28° during the positive reset pulse (and from 180° to 203° during the negative) [14]. As the exchange bias of the reference layer is unknown, the reference layer angle cannot be estimated likewise, but it can be deduced from the already determined inter-layer angles, which yield a reference layer angle of ≈203° during the negative pulse and of ≈128° during the positive pulse, FIG. 3-22. These are much larger deviations than what was observed in the SAF samples (cf. FIG. 3-10). Such large deviations of the reference angle with external field hinder the measurement of the free layer magnetisation and particularly of the DW position, and induce additional stray fields which may alter the behaviour of the DW.

---

[13] As can be seen from the non-zero MR value at $H_X$= 0 for the low-R state (FIG. 3-20), there is a small misalignment of the reference layer, and therefore the remanence angle is not exactly 180°.

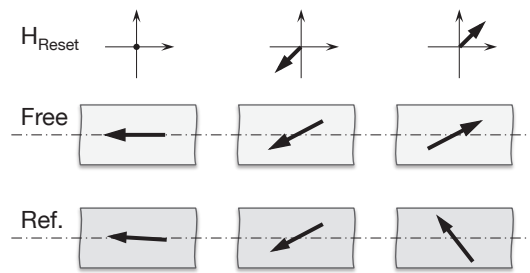

**FIG. 3-22 Magnetic layer angles in a SV with a simple reference layer during reset pulses.** Free layer angles were estimated by micromagnetic simulation (cf. FIG. 3-13) and reference layer angles deduced from the free layer angle and the measurement (FIG. 3-20).

We can conclude then that the use of simple reference layers is very prejudicial to the making of tracks with good DW conduit properties: it introduces asymmetries between DW polarities, causes the pinning of track defects to strengthen, increases the variability between structures, generates stray fields, and its finite-moment makes it sensitive to external fields which encumbers the measurement of the free layer magnetisation. We suggest that the use of simple reference layers is the strongest contributor to the high density of strong pinning centres that has been found in many earlier studies (see references above).

---

[14] In the estimation of the free layer angle the stray field from the also tilted reference layer was not taken into account; the free layer angle deviation is therefore overestimated.

# 3-8. Summary and conclusion

We have presented here L shaped SV tracks of small width (down to **33 nm**), which had a MR ratio similar to the unpatterned SV. Using resistance measurements to monitor the DW position, we demonstrated the **injection and propagation of DWs with field**, and measured the propagation and nucleation fields, $H_{PR}$ and $H_{NUC}$.

We started by characterising the measured electrical signal with applied fields, using our quasi-static setup. We showed that the signal noise was equivalent to a DW displacement of only a few tens of nms, and the field noise of about 1 Oe, well below the values of switching fields we study (10s of Oe).

We then studied the magnetic properties of these tracks. We observed that the switching fields presented a constant shift ($H_{SH}$), which we associated with the surface magnetic coupling between the free and reference layers, also observed (with similar magnitude) in the switching of the unpatterned SV. Several track widths were studied and compared, from 33 nm up to 260 nm. We found that both the $H_{NUC}$ and $H_{PR}$ fields decreased with increasing width, which we attributed to a lowering of the demagnetisation factor (for $H_{NUC}$) and to an increase of the pinning strength of lateral defects with reducing width (for $H_{PR}$). We also studied how a transverse field changed the MR signal of the track, and attributed this to a small rotation of the free layer magnetisation.

We have tested several SV compositions, two free layer compositions (Py and Py/CoFe) and two reference layer compositions (a SAF trilayer and a simple CoFe layer, both exchange pinned). The **two free layer compositions** resulted in tracks with comparable DW conduit properties, with the Py/CoFe showing a larger MR ratio. The two **reference layer compositions**, on the other hand, produced very different results, with the tracks fabricated from the simple CoFe reference layer showing asymmetries with DW polarity, increased pinning strength of track defects, higher variation between nominally identical structures. Using micromagnetic simulations, we showed how the strong magnetostatic stray fields produced by the simple reference layer might be responsible for all these differences. Finally, we compared the DW conduit properties of these SV tracks with the state of the art of Py monolayer tracks. We found that SV tracks presented smaller $H_{NUC}/H_{PR}$ ratios, and suggested that this is due to larger lateral defects in the SV tracks, caused by the necessarily more complex patterning process.

Previous studies on SV tracks were often hindered by a large density of pinning sites of stochastically varying strength. Due to their stochastic nature, these defects introduced uncertainty in the measurement of depinning fields. They also hindered the study of phenomena sensitive to DW deformation caused by strong pinning centres, such as current induced DW propagation. In addition, they limited the studied tracks to relatively wide and geometrically simple structures, compared to what has been achieved in monolayer Py tracks (see e.g. [Grollier et al. 2003; Himeno et al. 2003; Lacour et al. 2004; Uhlir et al. 2010]). Compared to these earlier studies, the tracks presented here showed much weaker pinning centres (as measured by $H_{PR}$), even at much narrower widths. We suggest that the use of SAF reference layers, along with a patterning process suitable for low lateral roughness such as the here-described hard-mask etching process, is critical to the fabrication of good SV DW conduits.

In conclusion, we have fabricated and optimised narrow SV tracks (down to 33 nm) without significant loss of MR signal (compared to the unpatterned film), in which we could inject and drive DWs. These tracks showed good DW conduit behaviour, with very low propagation fields compared to what has been reported in SV tracks. To the best of our knowledge, these are the narrowest DW conduits, in SV or Py, and the SV tracks with the widest propagation/nucleation gap, ever demonstrated to date.

# 3-9. References

# [4] Domain wall logic

In the previous chapter, we studied L shaped SV tracks that functioned as DW conduits, i.e. where DWs could be controllably injected and propagated with external fields. We use them in this chapter to study track systems with more complex shapes, and to demonstrate some complex DW logic devices.

**DW logic in magnetic nanotrack systems** are typically based on two ideas: that the bi-stable nature of a narrow magnetic nanotrack can be used to store binary information as a series of magnetic domains, and that the propagation of DWs and their interaction with tracks of different geometric shapes can be used to manipulate that information. Controlled DW injection, propagation, and pinning can be thought as the most fundamental operations in a DW logic device. Technologically relevant devices have been proposed based simply on these principles, such as the data-storing DW racetrack [Parkin et al. 2008] and the field turn-counter spiral [Mattheis et al. 2006]. By altering the shape of the track, many other logic functions have been demonstrated in **Py systems**, including various pinning traps (e.g. [Faulkner 2004; Petit et al. 2008a]), DW gates [Petit et al. 2008b; 2010], unary and binary logic operators (NOT, AND, OR, …) [Allwood et al. 2005], complex circuits that conjugated several of these operators, and even a data-storing shift-register based on a series of NOT gates [Allwood et al. 2005; O'Brien et al. 2009b; Zeng et al. 2010b].

The study of **DW logic in SV tracks** is motivated by its application to the electronic integration of future devices, as well as by the powerful measurements it allows. Single shot measurements and the precise determination of DW position, as the device is operating, allow for a detailed characterisation of the device operation, as well as the unveiling of new phenomena in the interaction of the DW with the shaped track.

However, many of these systems require complex track shapes and artificial geometric defects. These defects in a SV track with a simple reference layer, as we have seen in the previous chapter, create strong pinning centres that behave differently for different DW polarities. Along with poorer DW conduit performance, this has much limited the gamut of logic devices studied in SV tracks. Nonetheless, **earlier SV studies of digital devices** included the field turn counter spiral [Mattheis et al. 2006], simple pinning traps (e.g. [Briones et al. 2008]), and asymmetric pinning traps [Himeno et al. 2005; 2006].

In this chapter, we use the DW conduit developed in Chapter 3 to demonstrate and study some DW digital devices: a **spiral track turn counter** §4-1, a **T shaped DW gate** §4-2, a **NOT gate** §4-3, and a **DW-DW interaction structure** §4-4. For these structures, we characterise their operation and, when possible, compare it to their Py counterpart. We also use them as a tool to study some new phenomena related to the interaction of the DW with the track and with other DWs. These digital devices, whose complexity surpasses the devices so far demonstrated in SV, serve as a demonstration of the suitability of SV tracks to the use in future devices.

**Table of contents**

# 4-1. A spiral track

This simple device consists in a continuous unmodified spiralling SV track (see FIG. 4-1). As will be shown below, this shape allows the storage and propagation of multiple DWs without mutual annihilation via the application of a rotating (in-plane) field. It can also be used as a field turn counter, as was first proposed by Mattheis and colleagues [Diegel & Mattheis 2005; Mattheis et al. 2006]: as the field turns, and the DWs propagate sequentially through the spiral segments, the magnetisation of the spiral changes between a finite number of discrete states. Depending on the spiral shape, the placement of the electrodes, and the initial magnetic state, these discrete states produce different resistance values, which can be mapped to different values of the turning angle. Electrical current needs to be applied only when reading the turning angle, an advantage over other turn sensors. Mattheis et al. have studied GMR turn sensors consisting in different spiral designs coupled to a nucleation pad to inject DWs (see previous references, and citations therein). Here we characterise a pad-less spiral, demonstrate its turn counting function, and measure its operating margin (range of external field amplitude that produce the desired behaviour).

## Device design

For this experiment, we fabricated a 140 nm wide, rectangular, four-turn, clockwise (CW) spiral track [1], using the titanium etch process (cf. Chapter 2), FIG. 4-1.

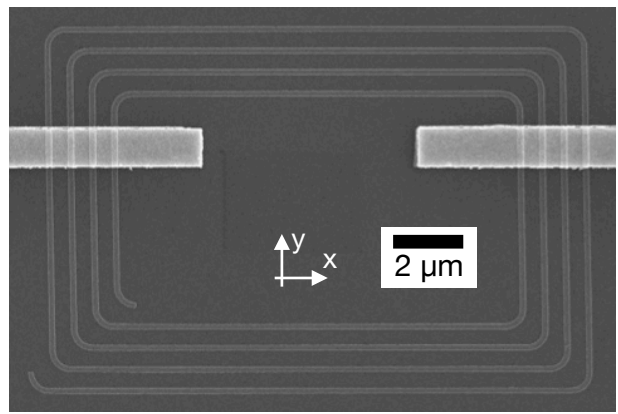

FIG. 4-1 Spiral track (SEM image), showing the track and gold contacts. The outer turn is 15.6 by 10.5 μm, the inner 12 by 6.9 μm, with 0.6 μm track separation.

Due to the position of the contacts, there are seven spiral segments electrically connected in parallel between the two contacts, four on top and three on the bottom. Notice that part of the inner and outer loops are unmeasured, as no current travels through them. Each segment contains vertical portions, where the track direction is perpendicular to the reference layer, and horizontal portions. Also the segments have increasing length and increasing resistance (4.8 μm difference between consecutive spiral turns, which is 2.4 μm between consecutive half turns, or 1.2 μm between consecutive sides). As we will see in §0, this causes the contribution to the total MR from the different segments to be different in amplitude. As part of the measured track is vertical, which contributes to the total resistance but has nearly zero MR change on reversal (as was seen in the previous chapter), the total MR [2] of this structure will be lower than the case of an L-shaped track. We measured the resistance with a lock-in amplifier as described in Chapter 2 [3].

---

[1] Sample fabrication details may be found in Annex A, under sample reference HM18.

## 4-1.1. Counting turns

First, we shall look into the different magnetic states and their MR values as we vary the number of DWs and their positions along the spiral. For this purpose, we performed the following sequence of fields, with the measurement shown in FIG. 4-2A:

– **Initialisation.** The track was initialised with a (-350, +325) Oe field, which created four HH and four TT DWs in the spiral, with one of the TT DWs being in an unmeasured track segment (diagram FIG. 4-2B-**i**). With all the horizontal segments in the low-R state (magnetisation pointing leftwards), this is the lowest measured resistance state (500.2 Ω).

– A **CW rotating field** was then applied, H$_{CYCLE}$, of 110 Oe, for 3.6 turns, until t = 0.93 s. During this time, seven sharp transitions were measured, which we shall label a–g. These transitions were alternately positive and negative and of decreasing amplitude, corresponding to the reversal of a different number of horizontal track segments.

– The **first transition (a)** occurs as the field goes from –x to +x direction and the eight injected DWs reverse all the eight track segments (FIG. 4-2B-**ii**). This is the largest observed transition (2.238%). Shortly after, the outer TT DW reaches the end of the spiral and is annihilated.

– The field continues its rotation and, as it progresses from +y to -y and completes half a turn, the seven DWs reverse all but one vertical segments (FIG. 4-2B-**iii**). As the segments are vertical, there is no clear MR transition [4]. There is though a clear continuous MR variation. This occurs during the passage of the DW through the rounded corners (of arc length ≈1.6 µm), which probably occurs in multiple steps as the DW follows the field.

– As the field progresses another quarter turn, and goes from +x to –x, the seven DWs reverse the seven outmost horizontal segments, leaving the inner horizontal segment pointing leftwards (FIG. 4-2B-**iv**). This causes a **sharp transition (b)**, which is smaller in amplitude than the previous (a). Again, the outmost DW, a HH now, is annihilated, leaving six DWs in the system.

– The field continues to rotate and a DW is ejected every half turn. As the number of DWs decreases, fewer horizontal segments are reversed each time (FIG. 4-2B-**v-vii**), producing sharp transitions of decreasing amplitude (**c–d**). As before, a small transition is observed as the field goes from ±y to ∓y, also of decreasing amplitude (from 0.10% before to 0.09%, 0.07%, 0.08%, ≈0%...), caused by the reversal of fewer and fewer vertical segments.

– **Counter-rotating field.** At t=0.93 s, only one HH DW remains, and only two domains are present. At this point, the H$_{CYCLE}$ chirality was reversed from CW to CCW (FIG. 4-2B-**vii**), for 3.6 turns, until t=1.86 s. Again seven transitions of alternating amplitude were produced, now corresponding to the single DW reversing each horizontal segment consecutively every half turn, starting with the second outmost one (**viii**) and ending with the inner one (**ix**).

---

[4] although possibly a transition of 0.10% occurs at t=0.141 s when the field is at θ=338°.

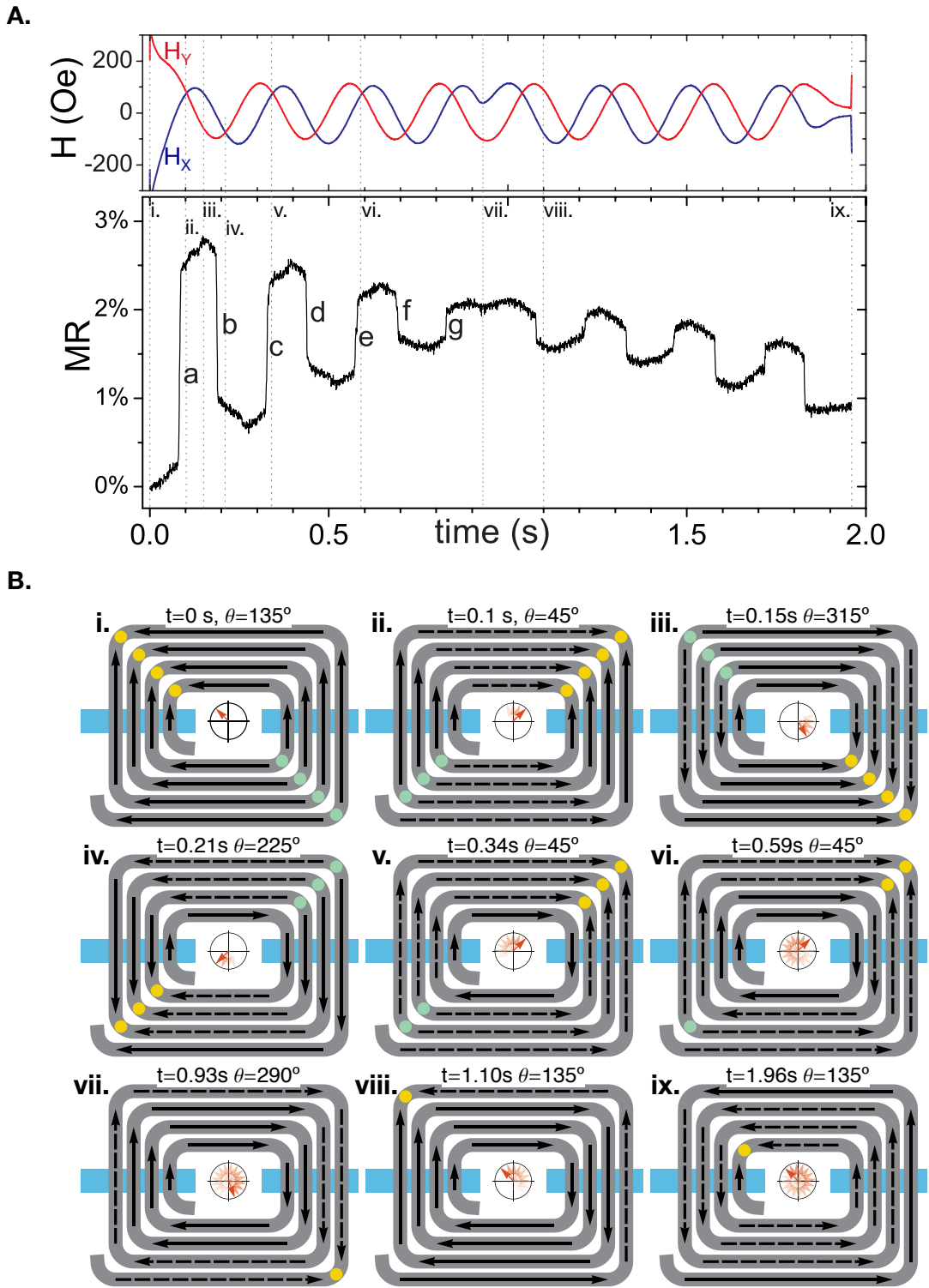

**FIG. 4-2 Winding and unwinding multiple DWs in a spiral track. A.** Applied field components and measured MR signal vs time. First seven transitions labelled **a–g**. Single shot measurement. **B.** Diagram of track magnetisation at nine points in time, **i–ix**, points also marked with the vertical dashed bars in A. Domain magnetisation is represented with black arrows, the DWs with circles (HH in yellow ●, TT in green ●), and the angle of the cycling field with a red arrow. The domains with the dashed arrows are those that reversed since the previous schematic. The angle of the field, θ, is 0° when rightward and increases CCW. The reference layer points leftwards.

## Angle and amplitude of the transitions

The transitions a–g are listed in FIG. 4-3C, along with before & after state schematics, list of reversed segments, amplitude, field angle and field X component. These transitions differ from the propagation experiments in Chapter 3 by the presence of a very significant transversal field (i.e. $H_Y$) at the instant of DW propagation (~105 Oe). Still, we see that they occur when $H_X$ is close to $H_{PR}$. We have determined the propagation field for L-shaped tracks of the same width (under negligible Y field): $H_{PR}$=34 ±17 Oe with typical $H_{SH}$=+11 Oe, or, equivalently, $H_{PR}^-$=-23 Oe and $H_{PR}^+$= 45 Oe (with the same error bars). These values are similar to those shown in FIG. 4-3C [5].

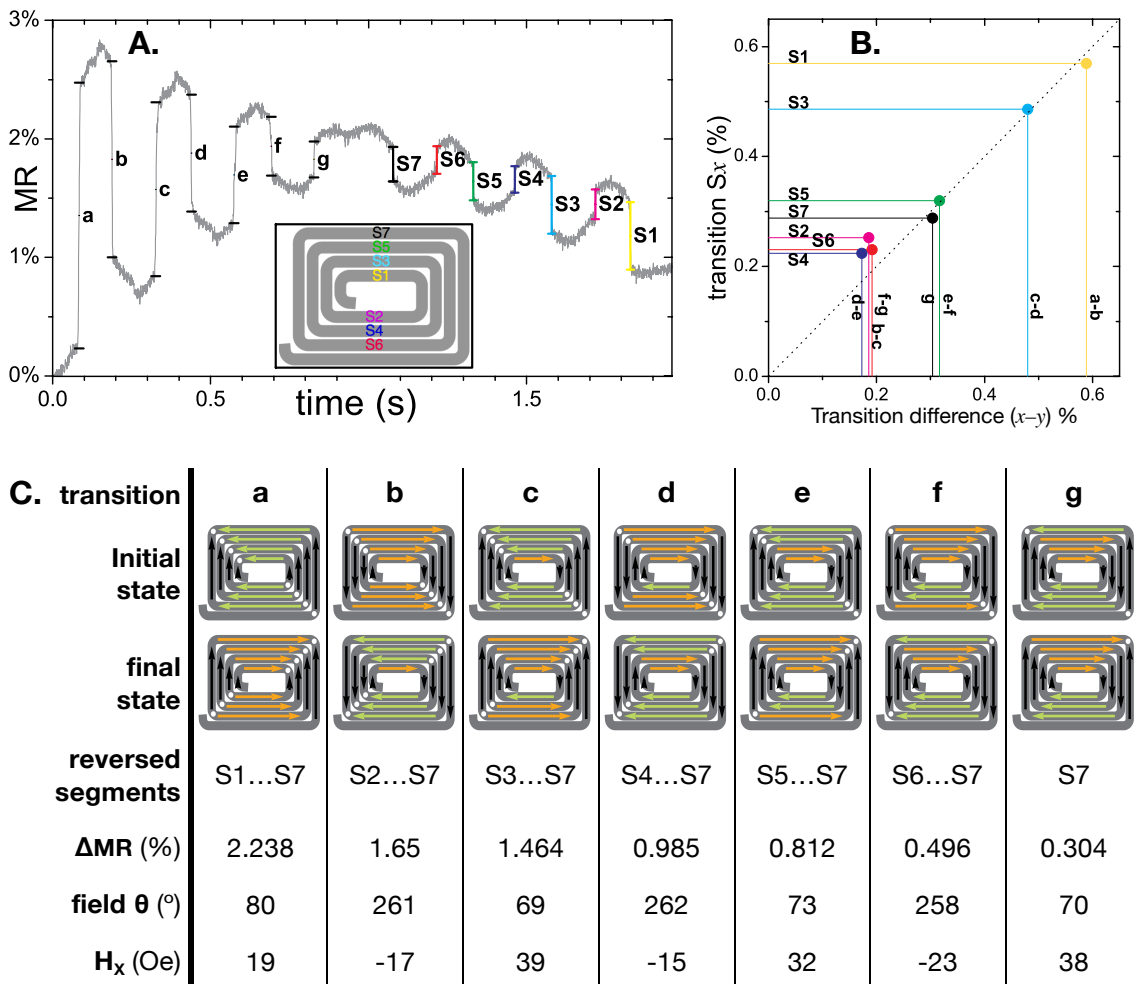

| C. transition | a | b | c | d | e | f | g |
|---|---|---|---|---|---|---|---|
| Initial state | | | | | | | |
| final state | | | | | | | |
| reversed segments | S1…S7 | S2…S7 | S3…S7 | S4…S7 | S5…S7 | S6…S7 | S7 |
| **ΔMR** (%) | 2.238 | 1.65 | 1.464 | 0.985 | 0.812 | 0.496 | 0.304 |
| field θ (º) | 80 | 261 | 69 | 262 | 73 | 258 | 70 |
| $H_X$ (Oe) | 19 | -17 | 39 | -15 | 32 | -23 | 38 |

**FIG. 4-3 MR transitions of a spiral containing eight DWs under a CCW field. A.** Replot of data of FIG. 4-2. Labels **a–g** are transitions involving multiple DWs and multiple spiral segments. Labels **S1…7** refer to single DW reversal of single horizontal segments. **Inset:** segment labels. **B.** Comparison of the values of the single DW reversal (S1…7) and the corresponding difference between multiple transitions (*x-y*). Dashed line is the identity function. **C.** Transitions a–g: magnetisation schemes, reversed segments, field

---

[5] Similar results were obtained with the repeated measurements below, e.g. FIG. 4-7.

angle, and variation of MR. In the schemes, the white dots are the DWs and the arrows are the domains (green for low-R state, orange for high-R, black for vertical segments).

We will now examine more closely the amplitudes of the observed transitions. For sake of clarity, we label the horizontal segments of the spiral S1 (the innermost) through S8 (the outermost), cf. inset of FIG. 4-3A. As we referred before, the first transition, (a), corresponded to all the seven contacted horizontal segments (S1…S7) reversing, and second transition, (b), to all but the innermost segment reversing (S2…S7), and so forth. Consequently, we can calculate the MR signal generated by the reversal of each segment by subtracting the amplitudes of consecutive transitions: if (a) corresponds to the reversing of S1…S7, and (b) to the reversing of S2…S7, (a)-(b) will yield the MR signal of reversing S1. We can then compare these subtracted values with the ones obtained when the single HH DW reversed the spiral one segment at the time (see labels in FIG. 4-3A). These two sets of values are plotted in FIG. 4-3B. There, we observe that agreement between the two obtained was very good, showing that the magnetisation is changing in the way we schematised in FIG. 4-2B and FIG. 4-3C.

We also observe that the MR of the four upper segments increases with decreasing loop length, i.e. S1 > S3 > S5 > S7, and that this decrease is not uniform. The case for the three lower segments is more unclear, as we get S2 > S6 > S4 (using the single DW transitions) but S6 > S2 > S4 (using the subtracted values). The signal of the lower segments is also much lower than that of the upper segments. This is a consequence of two factors: the individual segment resistance and its $\Delta R$ during horizontal reversal. The measured MR signal is deducible from the formula for parallel resistance:

$$R_{S1\|S2\|\cdots} = \left( \sum R_i^{-1} \right)^{-1}$$

$$\Rightarrow \frac{\partial R_{S1\|S2\|\cdots}}{\partial R_i} = \left( \frac{R_{S1\|S2\|\cdots}}{R_i} \right)^2; \qquad \text{(eq. 1)}$$

$$\Delta R_{S1\|S2\|\cdots} = \left( {R_{S1\|S2\|\cdots}} \big/ {R_i} \right)^2 \Delta R_i \ + o(\Delta R_i{}^2) \qquad \text{(eq. 2)}$$

$$\Rightarrow \frac{\Delta R_{S1\|S2\|\cdots}}{R_{S1\|S2\|\cdots}} = \frac{R_{S1\|S2\|\cdots}}{R_i} \frac{\Delta R_i}{R_i} \qquad \text{(eq. 3)}$$

where $R_{S1\|S2\|\cdots}$ is the total resistance, $R_i$ are the segment resistances, and $\Delta$ refers to the reversal of the horizontal tracks.

The first term in eq. 3 indicates that, with everything else remaining constant, the segments of smaller resistance (i.e. shorter length) will yield a larger measured signal, leading us to expect S1 > S3 > S5 ≳ S2 > S7 ≳ S4 > S6 (cf. FIG. 4-1). The amplitude difference between segments is predicted to be inversely proportional to their length [6] ratio (ignoring the contact-to-track resistance).

The second term of eq. 3, the loop effective MR, depends on the ratio of horizontal to vertical track, i.e. of active to dead track, signal-wise. For the geometry used, this has the order S1 > S3 > S5 > S7 > S2 ≳ S4 ≳ S6. The results of the application of both factors in eq. 3 to our mask (taking R∝length) are shown in FIG. 4-4A, plotted as a function of contact position deviation, and the measured amplitudes are shown again in FIG. 4-4B for comparison.

We can conclude that this model, of considering only the horizontal track segments with resistance proportional to length, predicts well the transition amplitudes, with only a small quantitative deviation due probably to differences in contact-to-track resistance.

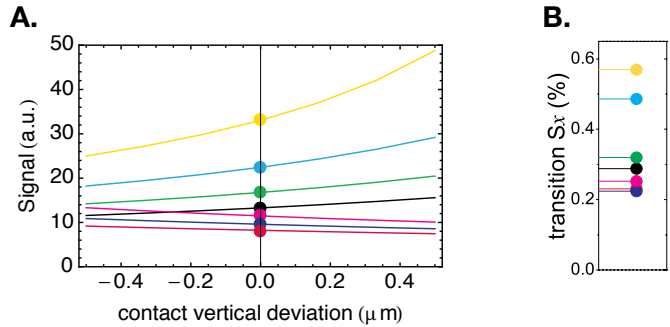

**FIG. 4-4 Calculation of segment contributions to the spiral signal. A.** Relative signal contribution of a horizontal segment reversal to the measured signal (see text and eq. 3), vs contact lithography deviation. The lines correspond, from top to bottom, to segments S1 ●, S3 ●, S5 ●, S7 ●, S2 ●, S4 ●, and S6 ●. **B.** Measured signal amplitude of the S1…S6 transitions for comparison (as in Fig. 4-3C).

## 4-1.2. Operating margin

So far, a cycling field of 110 Oe of amplitude has been used to measure this structure and detail its magnetic states. This value was chosen as it is clearly between $H_{PR}$ and $H_{NUC}$ (cf. Chapter 3), ensuring DW propagation and no nucleation, as was observed. We shall look now into what happens when a cycling field of different amplitude is applied. We expect at least three different behaviours:

---

[6] Here, length refers to the inter-contact length, and not to the horizontal track length.

- at low fields ($|H_{CYCLE}| < H_{PR}$), there should be no change of magnetic state;

- at intermediate fields ($H_{PR} < |H_{CYCLE}| < H_{NUC}$), there should be a change of magnetic state in phase with the field turning (correct operation as studied before);

- and at higher fields ($H_{NUC} < |H_{CYCLE}|$) new domains can nucleate, and the spiral segments should reverse (twice) every field cycle.

Mapping the values of $H_{CYCLE}$ to each of these regions will create an *operating margin map*, which has particular importance to the field of DW logic: the larger the region of desirable operation of a DW logic device, the easier will be to apply an adequate field value. A large operating margin is also important when operating a large number of logic devices, either connected or in parallel, when the field amplitude must simultaneously satisfy all operating margins.

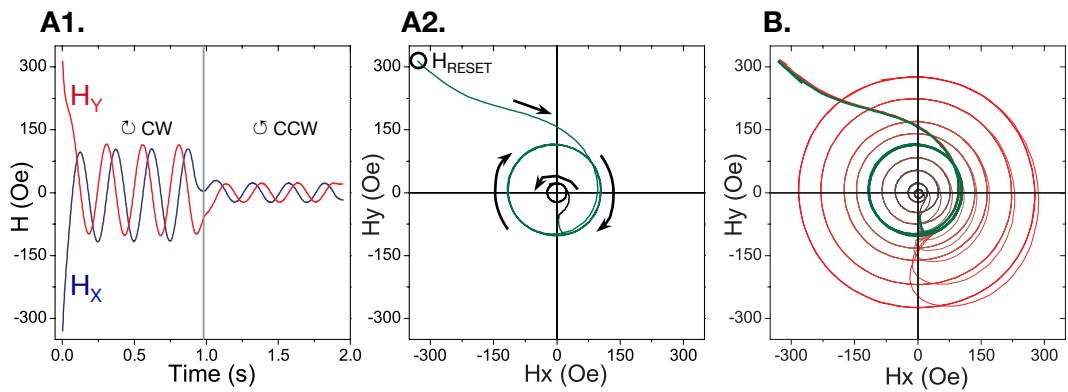

FIG. 4-5 **Field sequences for determining the operating margin of a spiral track.** **A1.** Applied field vs time. Until t = 0.98 s (grey bar), $H_{CYCLE}$ = 110 Oe (CW), as in the previous measurements. From t = 0.98 to 1.96 s, $H_{CYCLE}$ = 23 Oe (CCW). **A2.** Replot of A1, $H_X$ vs $H_Y$. The green portion is the CW initialisation (0–0.98 s) and the black portion is the CCW period (0.98–1.96 s). **B.** The set of 9 sequences used to test the operating margin. The sequences are identical during the CW initialisation (green) but have different $H_{CYCLE}$ values during the CCW period (black to red).

To measure the operating margin, and to obtain detailed statistics of the structure performance under the different fields, the following experiment was performed. The spiral was initialised to the single HH DW state as before (FIG. 4-2-vii), using the same field amplitudes. A CCW $H_{CYCLE}$ was then applied of chosen amplitude. This sequence was then repeated with nine different values for the CCW $H_{CYCLE}$, and for each value multiple single-shot measurements were taken (typically 40). The plots of the applied field for these 9 different sequences can be seen in Fig. 4-5B, with one of them plotted separately as an example (Fig. 4-5A).

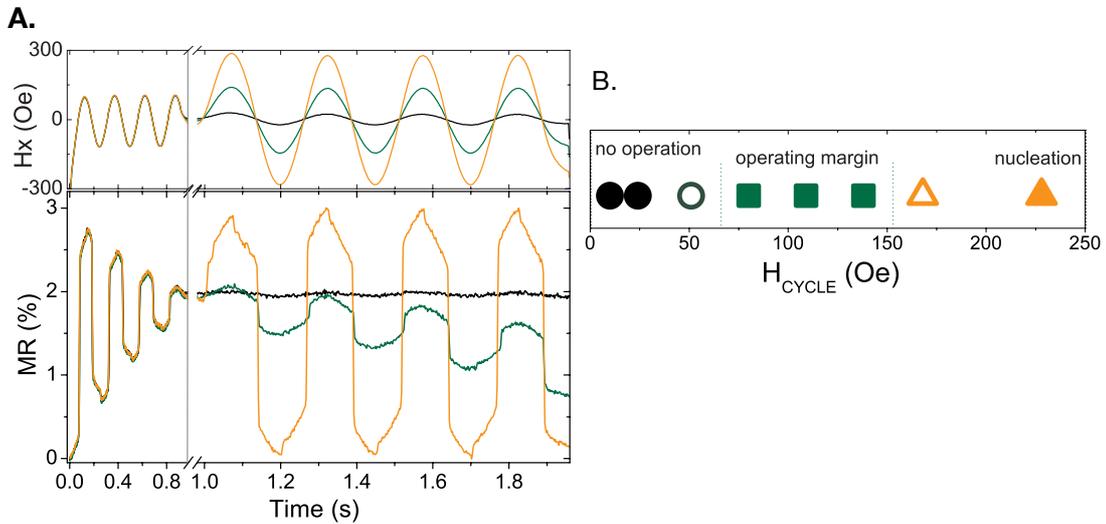

**Fig. 4-6 Operating margin of a spiral track. A.** Single shot measurements at $H_{CYCLE}$ of 24 Oe (black) showing no DW propagation, 139 Oe (green) showing in-phase DW propagation, and 280 Oe (orange) showing nucleation. **B.** Operating regime vs $H_{CYCLE}$, measured with 40 single shots per $H_{CYCLE}$ value. Black circles for when no transitions are observed, green squares for in-phase operation, and orange triangles for nucleation. The empty symbols represent intermediate behaviour. The best guess for the operating margin is the range marked with the dashed lines, 65–153 Oe (by the criterion in the text).

The three behaviours enumerated before were measured and the resulting curves are shown in see Fig. 4-6A: no state change at low field amplitude (black curve), in-phase state change at intermediate field amplitude (green curve), and nucleation at high amplitude (orange curve).

The different operating modes are plotted versus applied field amplitude in Fig. 4-6B. Below 24 Oe, no transitions are observed. At 51 Oe we observed what we call a **flickering behaviour**: there is at times DW propagation, but not in all single-shot measurements, nor even in all the field turns in a single measurement. An example of 40 single shot measurements in this regime can be seen in Fig. 4-7A. In this measurement, the stochastic behaviour mainly occurred when the DW traversed a particular region of the spiral (between the corners after S5 and S7, t > 1.7 s), probably where the strongest natural pinning defect is located. For measurements with $H_{CYCLE}$ of 80–139 Oe we obtained the in-phase propagation behaviour described before. At 168 Oe we observed again a flickering behaviour (Fig. 4-7B); in most of the acquired single-shots we measured the in-phase propagation behaviour, but in a few of them (2 out of 40) we observed nucleation of new domains.

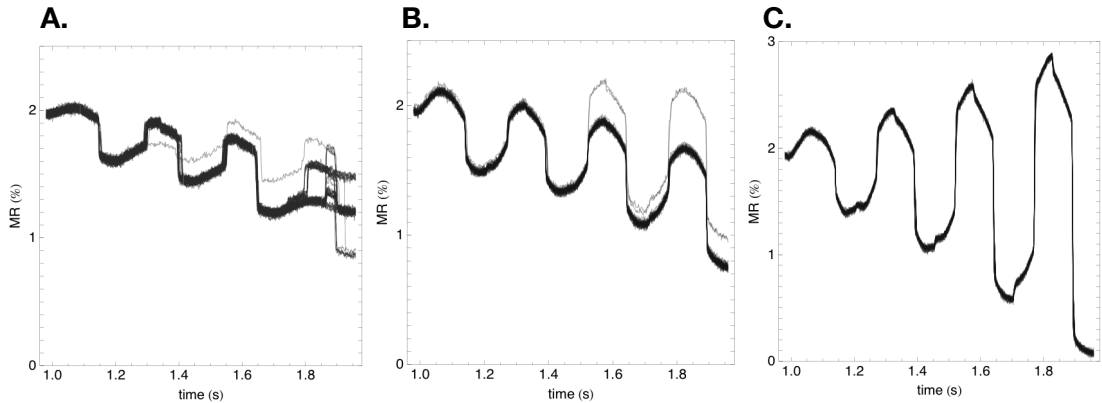

**FIG. 4-7 Intermediate behaviours.** Signal at different $H_{CYCLE}$ values, using the field sequences of Fig. 4-5, 40 single shots each. **A.** Intermediate behaviour at $H_{CYCLE}$ = 51 Oe: flickering propagation. **B.** Intermediate behaviour at $H_{CYCLE}$ = 168 Oe: flickering nucleation (2 events in 40). **C.** Intermediate behaviour at $H_{CYCLE}$ = 224 Oe: injection at end of the spiral.

We observed two regimes for the nucleation behaviour. At intermediate fields nucleation only occurred at the track outer end, once per semi-cycle (FIG. 4-7C). The seven observed transitions here do not correspond to the sequential single reversal of S7, S6, ⋯ S1, but to the reversal of S7, S7 + S6, S7 + S6 + S5, etc. Note that nucleation may also be occurring in the inner end of the spiral, but these DWs, however, cannot propagate into the spiral due to the handedness of the field. The second nucleation regime occurs at higher fields. Here all segments reverse twice per cycle, with domains being nucleated directly in the middle of the track (FIG. 4-6A).

Using the mid-point value as an estimate of the operating margin borders, we obtain an operating margin of 65–153 Oe (±15 Oe; see dashed lines in FIG. 4-6B). We have used here the criterion of less than 1 error per 40 single shot measurements to determine the operating margin. Different criterions, however, would lead to slightly different operating margins. Note also how this margin is significantly smaller than $H_{NUC}$–$H_{PROP}$ obtained for L-shaped tracks before (Chapter 3), with both a increased minimum propagation field and a decreased nucleation field. This occurs for a number of reasons. Firstly, the ends of this track are square-shaped, instead of being tapered, which decreases the nucleation field. Secondly, the $H_{NUC}$ measured for the L-shaped tracks was applied in one single direction (horizontally), while here we apply $H_{CYCLE}$ in every direction. As H_NUC depends on the field angle [Stoner & Wohlfarth 1948], this can explain how nucleation happens at a lower field. Lastly, here we propagate the DWs through a much larger distance (58 µm vs 6 µm) and so the probability of encountering large natural pinning defects is increased, leading to an increase of the lower edge of the operating margin.

### 4-1.3. Digital applications

As was referred before, spiral tracks have been investigated due to their use as digital *turn counters* [Mattheis et al. 2006; Diegel et al. 2007; 2009] and, indeed, the investigated structure exhibits this ability. Starting in the single HH DW state (FIG. 4-2-vii), this structure changed its resistance every half field turn with a total of 8 distinct states, FIG. 4-8. Repeats of CW and CCW field turns changed reliably the resistance state from one state to the next, as long as the total winding number did not surpass the spiral size (i.e. four turns), as is shown in FIG. 4-9.

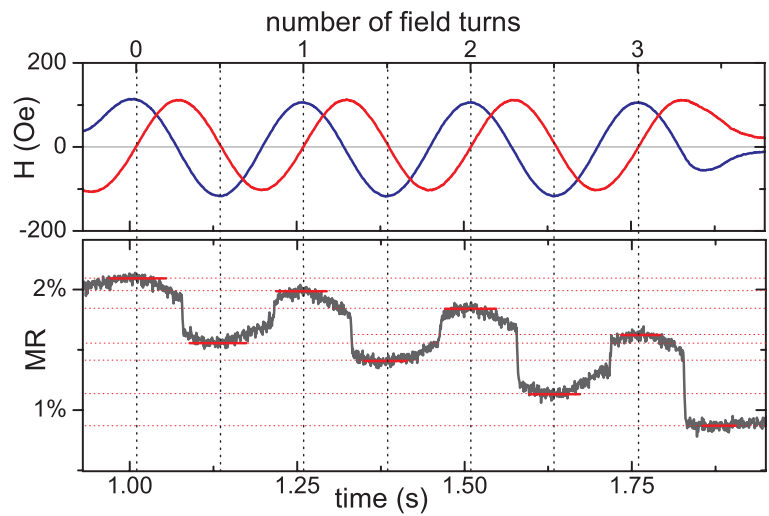

**FIG. 4-8 Resistance level vs number of field turns** (Replot of data from Fig. 4-2). Top: applied field (H_X in blue and H_Y in red). Bottom: **MR** signal. The red horizontal lines mark the **MR** levels at H_Y = 0.

As can be inferred from FIG. 4-8, as more turns are added to the spiral, which has a fixed MR ratio, the resistances of the discrete states get closer together. It is not then a scalable design. In order to improve the scalability, the contacts could be rearranged so to homogenise the resistance separation between states. Also, with a small addition to the fabrication complexity, more than two contacts could be patterned. The multiple resistance measurements could then be linearly processed to extract more accurately the magnetisation state (an example of such a scheme based on the Wheatstone bridge has been proposed in the above references). Eventually, the scaling limitation is one inherent to the spiral design: each added turn occupies a larger area (and adds a larger track length) than the previous one, i.e. the occupied area and the track length scale with $N^2$. Also, with increased length comes increased probability of higher strength pinning sites and narrower operating margin.

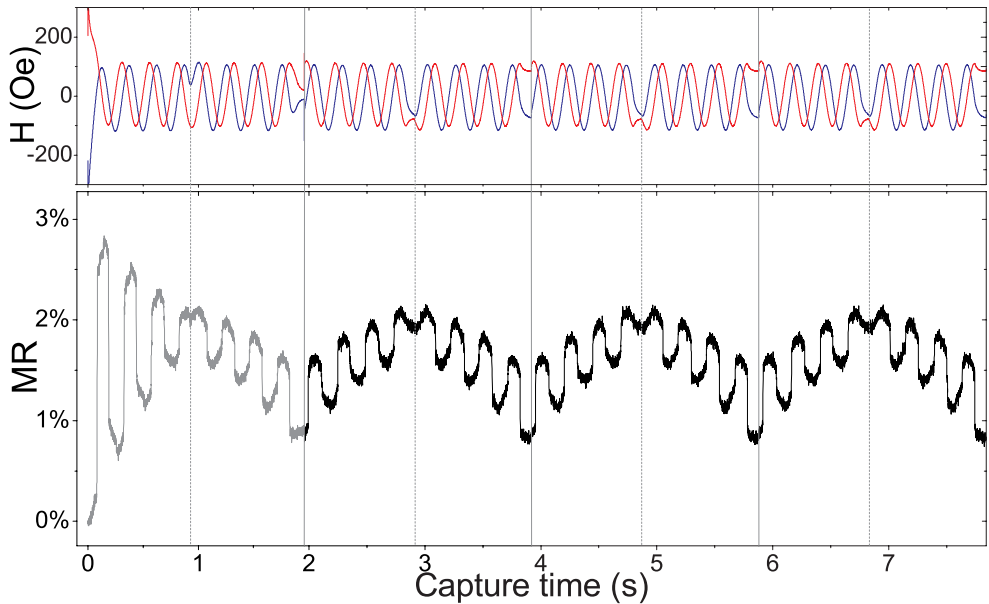

**FIG. 4-9 Repeated winding and unwinding of a DW in a spiral track.** Field vs time in the top plot ($H_x$ in blue, $H_y$ in red) and MR vs time in the bottom plot (previously shown data in grey). The spiral was initialised to a single HH DW in the inner loop state (t= 1.96 s), as described before. A (110 Oe) CW field was applied for 3.6 turns (t= 1.96–2.94 s), and was followed by a CW field for the same number of turns (t= 2.94–3.92 s). This CCW-CW sequence was repeated two more times. For setup specific reasons (see Chapter 3), this measurement was recorded in 4 separate periods (marked with the thick grey lines, each 1.96 s long).

Another possible use is the storage of data, where each of the spiral's half-loops is used to store one bit (which, for the studied spiral, yields n = $2^7$= 128 states). However, the type of magnetisation scheme used—of injecting 8 DWs, selectively annihilating all but one DW, and finally choosing the discrete position along the track to place that DW— limits the number of accessible states to n = 8 = 1 + 7 ([7]). A simple addition to this scheme, namely leaving a "train" of more than one DW in the spiral, would increase the number of accessible states to n = 29 = 1 + 7 + 6 + ⋯ + 1 ([8]). Reaching the total number of digital states, however, would require the ability to inject arbitrary numbers of non-consecutive DWs. This could be done by modulating H_CYCLE between a level in the operating margin and the level of FIG. 4-7C, at which DWs are injected in the ends of the spiral, akin to [Allwood et al. 2005], or electrical injection of DWs [Himeno et al. 2003; 2006].

---

[7] this corresponds to 1 state for the DW placed before the measured segments (similar to having 0 DWs), and seven states for the DW placed after each segment.

## 4-1.4. Summary

We have demonstrated here a simple DW logic device: a 4-turn rectangular SV track spiral with no nucleation pads. We have demonstrated the injection and propagation of multiple DWs with no mutual annihilation by the application of a rotating field. We analysed its operating margin, i.e. the field range producing the desired operation, and observed that it was smaller than that of the L shaped track, due to the design of the track ends and to inherent properties of the spiral, namely longer length and consequent higher number of natural defects, and application of fields at several angles. Comparing to the previous work done by Mattheis and colleagues [Mattheis et al. 2006], the device presented here shows a wider operating margin (65–153 Oe versus reported 120–170 Oe). There are several differences that could explain this increased margin: smaller device size and thus shorter track length and fewer natural defects (though both had 4 turns, and were of similar track width, the present device is almost 50 times smaller), absence of a nucleation pad, and finally fabrication differences.

We have also analysed the signal obtained by the reversal of different spiral segments. We showed that, properly initialised, it could change between discrete magnetisation states in phase with a cyclical external field, and that this is measurable as discrete changes in its (two-point) electrical resistance. These resistance states can be mapped to the binary magnetisation of the horizontal segments of the spiral, and subsequently to the number of elapsed field turns. We then showed how the spiral could be used as a digital turn counter or a data storage device, and discussed its limited scaling ability.

---

[8] As before, this corresponds to 1 state for 0 DWs (or 1 DW placed before the seven segments), 7 states for 1 DW placed after each segment, 6 states for 2 DWs placed after two consecutive segments, etc.

## 4-2. DW manipulation using the T gate

Having studied a logic system based on an unmodified SV track, we shall study here an example of a DW trap created by changing the shape of the track: the T trap, FIG. 4-10. It consists off a small track stub joined at 90° to the main track where DWs propagate. This structure was well characterised in Permalloy tracks [Petit et al. 2008b; Lewis et al. 2009; Petit et al. 2009]. In those studies it was found that the pinning strength of the trap depended greatly on the DW structure, stub magnetisation, and stub and DW relative orientation. It was also found that, in certain configurations, the reversal of the T gate magnetisation produced two drastically different pinning strengths, which constitutes the principle of a controllable DW gated valve.

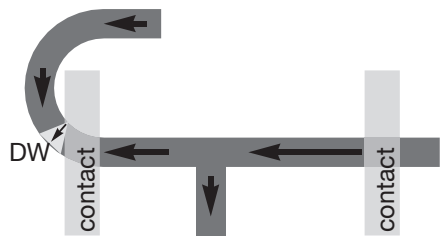

**FIG. 4-10 C shaped track with a T gate** (schematic). Example of a track with a T gate at the bottom of the track. Black arrows represent magnetisation, a TDW is shown in the corner, and electrical contacts are in light grey.

There are several degrees of freedom in the DW–trap interaction, even considering only transverse DWs [9], which will be labelled according to:

**i.**  DW central magnetisation (*up* or *down*),

**ii.**  DW polarity (HH or TT),

**iii.**  position of the stub (either at the *top* or at the *bottom* of the wire),

**iv.**  and finally trap magnetisation (*up* or *down*).

We shall see that in SV tracks, as in Py tracks, several of these permutations are symmetric, barring a small influence of the coupling to the reference layer. As such, we will categorise the interaction using the following two binary variables:

**i.**  whether the stub magnetisation and the DW central magnetisation are parallel or anti-parallel (*P/AP*),

---

[9] cf. discussion in Chapter 3.

**ii.** whether the stub is at the narrow or wide side of the transverse DW (*narrow/wide*).

In Py studies (referenced above), these 4 cases (*P/AP* and *narrow/wide*) produced very different pinning strengths and magnetic configurations. The two parallel cases produced either a very low [10] pinning strength in the *P-narrow* case (≈H_PR) or a medium pinning field (*P-wide* case). For this latter case, the pinning was found to be symmetric, i.e. identical for both (horizontal) directions of the applied field. The case was very different for the anti-parallel cases, however. Both showed very high depinning fields, comparable to H_NUC. Moreover, it was shown that depinning in those cases occurred with the nucleation of a new domain (and the injection of at least one new DW). Also, the pinning was highly asymmetric, with the depinning in one direction comparable to H_NUC while, in the opposite direction, comparable to H_PR.

Besides serving to demonstrate a complex DW logic device, the study of T gates in SV tracks also provide important information about the DW structure in these tracks. As we shall show in this section, the ability to measure the DW position in the SV track will also allow the study of some aspects of the complex behaviour of the T gate so far unobserved, including the generation and splitting of pinned 360° DWs and the magnetostatic interaction of the stub with the DW.

## Structures

The measurements were done on C-shaped tracks, width 110 nm, with a T trap placed ~4 μm from the arc where DWs are injected. These structures were fabricated using the titanium etch process [11], and were already used in Chapter 3 to study DW propagation. Contacts were placed 4 μm before the trap and 7.5 μm after (FIG. 4-12C). Two types of structure were fabricated corresponding to two possible positions for the T trap: either top or the bottom the track, FIG. 4-11A/B.

---

[10] even immeasurable in those studies.

[11] Fabrication details in annex A, sample reference HM01.

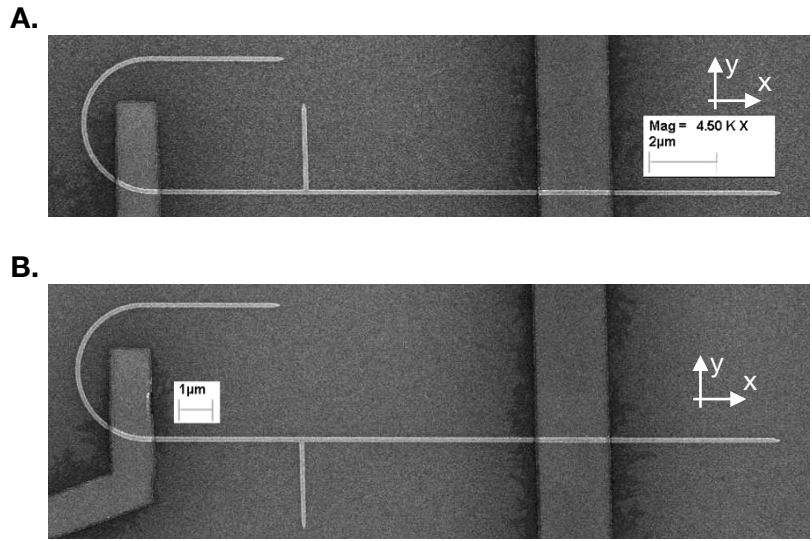

**FIG. 4-11 Track with T trap** (SEM images). **A.** Track with the T trap at the top of the track, and **B.** a track with the T trap at the bottom.

We used the field sequences described in Chapter 3 to measure the $H_{NUC}$ and $H_{PROP}$. This latter measurement also served to determine the transmission field ($H_{TR}$), i.e. the field at which the DW depins from the T trap and moves forward. The values of $H_{PR}$ can be found in Chapter 3.

We observed a significant lowering of $H_{NUC}$ from the value measured on C-shaped tracks without the T trap: from 350 [12] to 208 ± 8 Oe (averaged over 14 structures). This indicates that nucleation occurs in the region of the trap and not at the track ends.

## 4-2.1. Depinning from a T trap

We examine here the signal obtained with the propagation sequence when the DW, injected at the arc, is pushed through the region of the trap. We shall look into two structures (with the T trap on the top and bottom) and study the general features of the DW transmission through this kind of trap, and associate the measured behaviour to the four interaction cases described above. We present also results taken on several identical structures, and an analysis of the probability of each interaction case on a large number of measurements.

---

[12] this value was beyond the setup maximum reachable field; see Chapter 3 for details. $H_{NUC}$ for the simple C-shaped track is also presented in that same chapter.

### DW pinning at a T trap placed on the bottom of the track

In FIG. 4-12 we can see a typical single shot measurement of a structure with the T stub at the bottom. In contrast with the tracks without a (large) pinning site (cf. Chapter 3), there are now four well-defined MR states, with the magnetisation reversal happening in two steps at different fields, separated by a period of mainly unchanging MR signal.

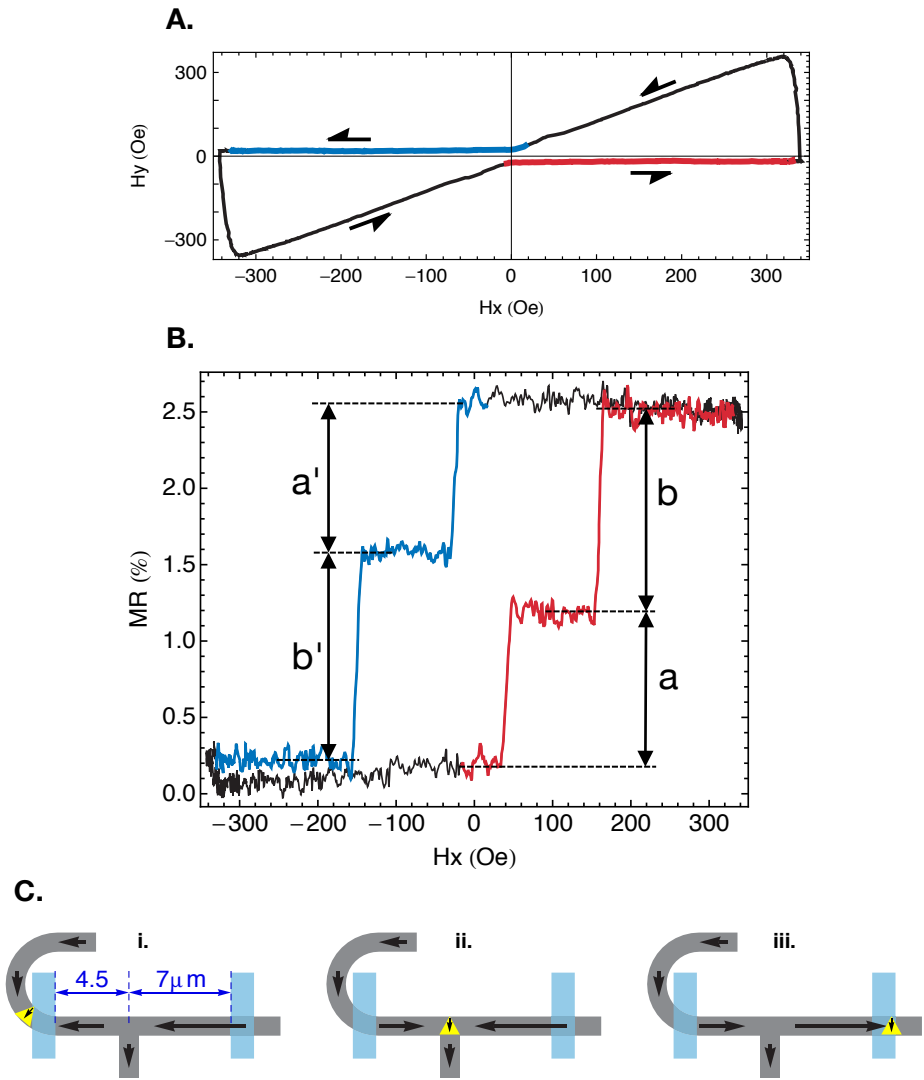

**FIG. 4-12 DW pinning at a T trap placed on the bottom of the track.** Track width is 110 nm, and the interaction is of type *P-wide*. **A.** Propagation field sequence, $H_x$ vs $H_y$. The periods of interest, when $H_x$ is sweeping up or down under a constant and small $H_y$, are coloured in red and blue, respectively. **B.** Single shot measurement, showing the four MR levels and the four transitions (labelled a,b and a',b'). **C.** Schematic of track magnetisation during HH DW propagation (i → ii) and depinning (ii → iii). Track in grey, contacts in light blue, and DW in yellow.

The double transition shows that the track is reversed in two separate halves: the first section, between the arc and the T-trap, reverses when the DW depins from the corner at $H_{PR}$; the second section, from the T-trap to the end, only reverses when the sweeping field reaches the characteristic pinning field, $H_{TR}$, and the DW is transmitted through the trap and propagates to the end (FIG. 4-12C).

The amplitude of the MR transitions should be proportional to the length of the reversed track between the contacts. As the sections before and after the trap have different lengths—4.5 and 7 μm respectively—we are able to test this relation. In particular, we expect $a = a'$ and $b = b'$, and ${}^{a}/_{b} = 4.5\,\mu m/7\,\mu m \approx 0.64$. For this particular structure, ${}^{a}/_{b} = 0.71 \pm 0.5$ and ${}^{a'}/_{b'} = 0.70 \pm 0.5$ [13], which is consistent with the expected value but could indicate a contact lithography deviation of a few hundred nms. Such magnitude of deviation is typical of long exposures on our EBL setup. Note that the difference in amplitude allows for the determination of which segment has reversed.

As before, we use the average value of the corresponding HH and TT transitions to determine the value of $H_{PR}$/$H_{TR}$/etc., indicating with $H_{SH}$ the shift between this average and the two individual values. For this particular structure and measurement, $H_{PR}$ = 31 Oe ($H_{SH}$ = +9 Oe), $H_{TR}$ = 153 Oe ($H_{SH}$ = +8 Oe). The value of $H_{SH}$ is consistent with the unpatterned sample shift (10 Oe). Nucleation was also measured, $H_{NUC}$ = 223 Oe, with a negative shift: $H_{SH}$ = -3 Oe. The difference between $H_{TR}$ and $H_{NUC}$ indicates that DW depinning occurs without nucleation (as opposed to re-nucleation and injection of a new DW at the trap). This is not always the case, as we shall now see.

## Multiple values of H_TR

We took 20 single shot measurements with the same field sequence as before [14], which are plotted in FIG. 4-13. The data show that $H_{TR}$ was not unique: two distinct values occurred, for both HH and TT DWs [15] (FIG. 4-13B). We can observe there that the lower transition field is the most probable (with 34 of the 40 observed transmission transitions). The $H_{SH}$ is consistent with the $H_{SH}$ of this SV stack ($H_{TR}$= 154 Oe with $H_{SH}$ = +5 Oe). The higher value transmission is rarer (occurring 6 times in 40), with an average value of $\langle H'_{TR} \rangle$ = 213 Oe and $\langle H_{SH} \rangle$ = -6 Oe. This higher value probably corresponds to a nucleation of a new domain, being close to $H_{NUC}$ in amplitude (223 Oe).

---

[13] Using the signal standard deviation as the error of the transition amplitudes and using the canonical rules for error propagation.

[14] the previous measurement (FIG. 4-12) is included in these 20.

[15] No correlation exists between observing either transition on the HH and on TT in the same measurement acquisition.

**A.**

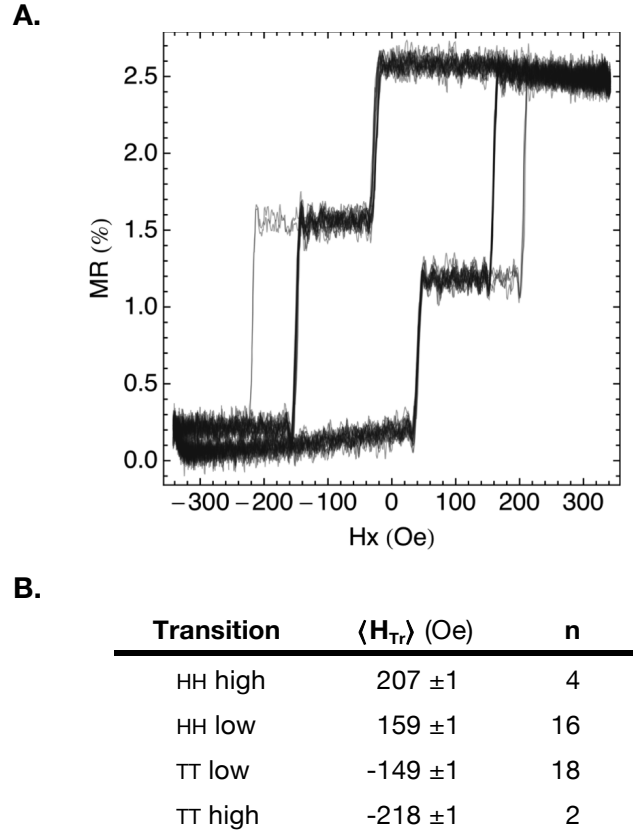

**B.**

| Transition | $\langle H_{Tr} \rangle$ (Oe) | n |
|---|---|---|
| HH high | 207 ±1 | 4 |
| HH low | 159 ±1 | 16 |
| TT low | -149 ±1 | 18 |
| TT high | -218 ±1 | 2 |

**FIG. 4-13 Multiple transmission measurements on a track with a T bottom trap.** Twenty single-shot measurements were taken, of which we show in **A.** the superimposed MR vs $H_X$ plots, and in **B.** the statistics for the transmission transitions. Errors are the sample standard deviation (n>6) or half amplitude otherwise. $\langle H_{PR} \rangle$ was 34 ±2 Oe.

These two values of transmission are consistent with what was found in Py systems [Petit et al. 2008b; Lewis et al. 2009]. They correspond to the two possible central magnetisation states of the transverse DW. At the moment of injection the central magnetisation of the DW is set by the reset field (pointing down for HH and up for TT) and is parallel to the arm magnetisation. However, during the 4.5 μm travel to the trap, the DW suffers a cyclical reversal of its central magnetisation, a phenomenon called Walker breakdown [Schryer & Walker 1974]. The number of reversals varies strongly with the applied field and is stochastic [Glathe et al. 2008], which randomises the observed in the state of the DW on arrival at the trap.

As the data shows, depinning from the T trap is a process highly dependent on the relative direction of the arm and DW central magnetisations. In the first case the original DW depins from the trap (schematic in FIG. 4-14A), while in the second case a new domain in nucleated (FIG. 4-14B). The fate of the newly injected DWs in the system in this latter case will be determined more accurately later. In the notation introduced earlier (p. 128), these two cases correspond to *P-wide* and *AP-narrow*, as in the first it is the wide side of the DW that interacts with the trap while, in the second case, it is the narrow side of the flipped DW that interacts with the trap.

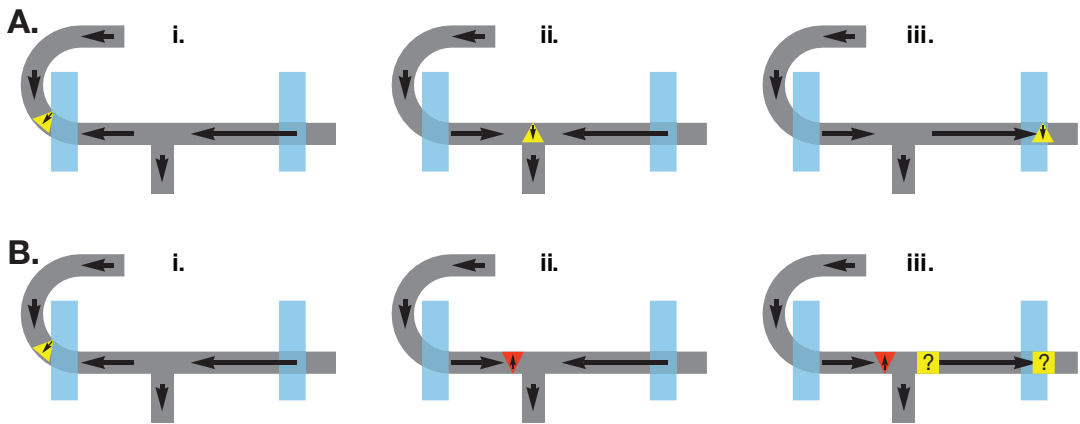

**FIG. 4-14. Pinning of a DW at a T trap on the bottom of the track** (schematic). **A.** *P-wide* case. The initial transition occurs at $H_{PR}$ as the DW travels to the trap (i→ii), where it is pinned. At $H_{TR}$, the DW depins and reverses the rest of the track (ii→iii). **B.** *AP-narrow* case. As the DW travels to the trap, its central magnetisation flips due to the Walker breakdown process (i→ii). Reversal of track occurs via nucleation of a new domain (and new DWs), at a higher $H_{TR} \sim H_{NUC}$ (iii).

## DW pinning at a T trap placed on the top of the track

In the previous paragraphs we have studied a structure with a T trap placed at the bottom of the track. The stochastic DW reversal allowed us to observe two of the four possible cases of DW–T trap interaction (for both HH and TT DWs). To study the other two cases, namely *P-narrow* and *AP-wide*, we examined a structure with the trap on the top (FIG. 4-11B). We took, as before, 20 single shot measurements with the field sequence of FIG. 4-12A, the results are plotted in FIG. 4-15 and FIG. 4-16. $H_{NUC}$ was also measured: $H_{NUC} = 201$ Oe ($H_{SH} = +12$ Oe).

**A.**

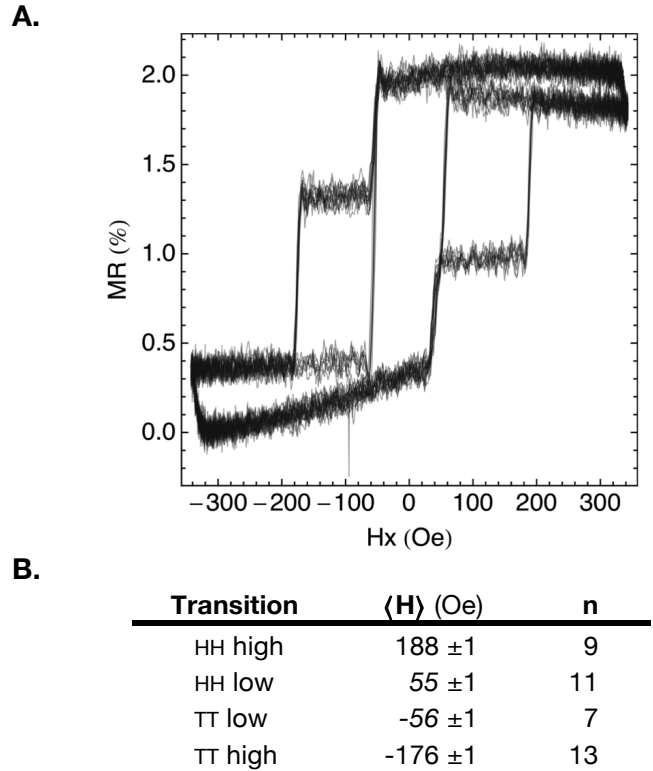

**B.**

| Transition | ⟨H⟩ (Oe) | n |
|---|---|---|
| ʜʜ high | 188 ±1 | 9 |
| ʜʜ low | *55* ±1 | 11 |
| ᴛᴛ low | *-56* ±1 | 7 |
| ᴛᴛ high | -176 ±1 | 13 |

**Fɪɢ. 4-15 Multiple transmission measurements on a track with a ᴛ top trap.** Twenty single shot measurements. **A.** MR vs $H_X$ plots. **B.** Statistics for the transmission transitions. Errors are the sample standard deviation. ⟨$H_{PR}$⟩ was 47 ±2 Oe ($H_{SH}$ = -8 Oe). The low transitions are limited by $H_{PR}$.

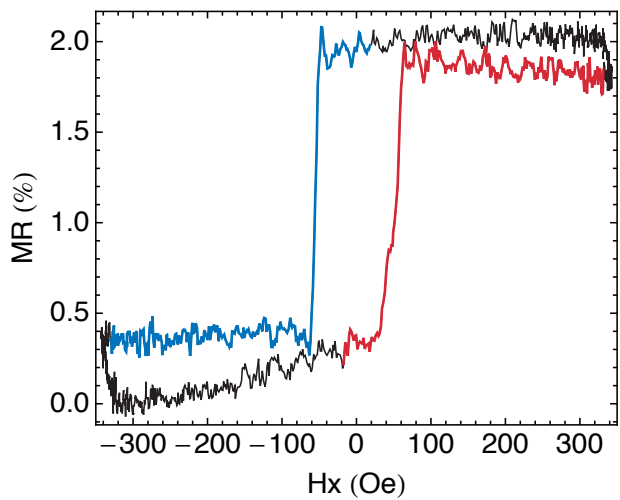

**Fɪɢ. 4-16 Transmission measurement on a track with a ᴛ top trap.** MR signal vs Hx, single shot. Same colour code as in Fig. 4-12. In this particular measurement, both the ᴛᴛ and the ʜʜ ᴅᴡs were transmitted at the lowest $H_{TR}$.

As with the structure with the trap on the bottom, we also observed two transmissions fields: $H_{TR}$ = 56 Oe ($H_{SH}$ ≈ 0) and $H'_{TR}$ = 182 Oe ($H_{SH}$ = +6 Oe). The highest value, corresponding most probably to a nucleation event, corresponds to the case *AP-wide*, schematised in Fɪɢ. 4-17ʙ. The lowest value, which corresponds to the case *P-narrow*,

FIG. 4-17A, strikingly, is very close to HPR. Examining a single measurement, FIG. 4-16, we have trouble finding any step in the reversal transition [16]. Indeed, we can say that this measurement is limited by HPR, meaning that the true depinning field of a DW in a т trap in the *P-narrow* configuration might be lower than the value measured. This findings are also in good agreement to what was found in Py systems (referenced above).

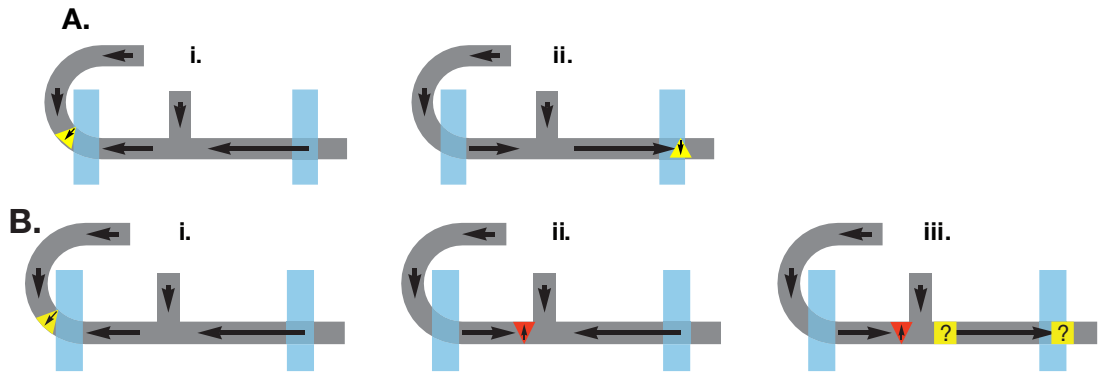

**FIG. 4-17 Pinning at a т trap on the top of the track** (schematic). **A.** The low HTR case (*P-narrow*), on which the pinned state is not detected. **B.** The flipped DW, high HTR case (*AP-wide*).

## Pinning configuration: statistical analysis

As was written before, the central magnetisation of the TDW arriving at the defect is altered by the Walker breakdown process, with some degree of randomness. This can be detected by the double depinning field of each т trap. To study this, we present here a statistical study of the incidence of the different pinning configurations.

Several identical structures were measured as above, 6 with the т trap on the top of the track and 7 on the bottom. In every structure, we observed the transition pattern described above. For each structure, 20 single shot measurements were taken, and the frequency and average value of each of transition were recorded. The sample standard deviation of HTR (for the same structure) was typically 0.5–2 Oe. The averages across all structures are shown in FIG. 4-18.

---

[16] And the reason we can present a value for HTR at all is that we define HTR here as the field at which the second half of the track reverses. Specifically, we used the point at which the MR curve crossed the mid-level of the expected transitions to determine HPR and HTR.

| Geometry | N | Case | $\langle H_{Tr} \rangle$ (Oe) | $H_{SH}$ |
|---|---|---|---|---|
| 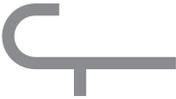 | 7 | *P-wide* | 150 ± 5 | +6 |
| | | *AP-narrow* | 211 ±13 | +10 |
| 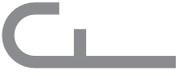 | 6 | *P-narrow* | *49 ± 8* | +6 |
| | | *AP-wide* | 171 ± 8 | +7 |

**FIG. 4-18 Transmission transitions in several structures.** Average over *N* structures (20 single shot measurements per structure) of the transition fields for each interaction case. $H_y$ bias of 20 Oe. The *P-narrow* case is limited by $H_{PR}$. The errors are the (structure to structure) sample standard deviation.

For the two structures described in the previous sections, the frequency of the P and AP cases were not the same (FIG. 4-13B and FIG. 4-15B). To study this, the relative frequency of the parallel case in 20 single shot measurements (*f*) was measured and is shown for all 13 structures in FIG. 4-19. We observe that the frequency of the parallel DW is

i) not always 0 or 1, i.e. there can be P and AP cases measured on the same structure;

ii) nor always ½, i.e. the number of P and AP cases are different in some structures;

iii) it varies from structure to structure, and, moreover,

iv) it is not the same for HH and TT DWs.

This is consistent with the hypothesis that the DW inverts by Walker breakdown, as this phenomenon is dependent on the field at propagation, which, as we saw, had shot-to-shot variation (i, ii), was different from structure to structure (iii), and was different for HH and TT (iv). That the DW magnetisation is stochastic, i.e. is not always P or AP, we can be sure by looking at FIG. 4-19. The other observations above (ii–iv) could be a product of chance, and require further statistical examination before we take any conclusions.

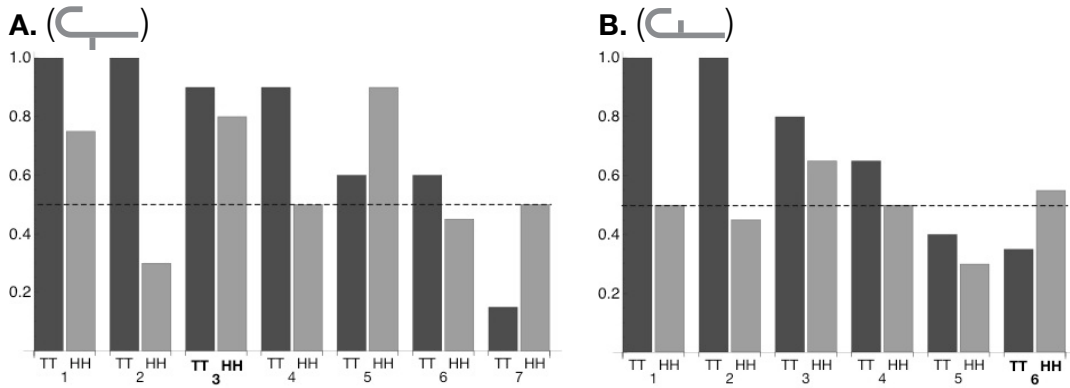

**FIG. 4-19 Relative frequency of the parallel case for each structure. A.** Frequency of the *P-wide* case in the 7 structures with the T on the bottom. **B.** Frequency of the *P-narrow* case in the 6 structures with the T on the top. Counted on 20 single shot measurements per structure. The structure numbers are arbitrary; #3 on A and #6 on B are the structures studied in the previous sections. The dashed line marks *f*=0.5. Dark grey bars correspond to TT DWs, light grey to HH.

For this, we shall now test whether the observed frequency values ($f_i$) indicate, with statistical significance, a non-½ probability of P to AP DWs ($p_i$) in any structure. We shall use the Binomial Test, with a double tail p-value. This means, for a measured frequency $f_i$ in one specific structure and DW polarity, we calculate the probability of measuring a frequency as or more extreme than $f_i$, assuming the null hypothesis ($H_0$: $p_i = 0.5$):

$$\text{p-value} = p\left(|F - \tfrac{1}{2}| \geq |f_i - \tfrac{1}{2}|\right)$$

If this probability (the p-value) is smaller than a threshold, 5% [17], we consider the measurement statistically significant. The Binomial Test p-values for a n = 20 sample are shown in FIG. 4-20A. With a p-value < 5%, the observed frequencies that are ≤ 0.25 or ≥ 0.75 are statistically significant: that is the case in 11 of the 26 measurements (8 TT vs. 3 HH, 8 trap up vs. 3 down) [18]. This indicates that indeed there are structures for which $p \neq \tfrac{1}{2}$, confirming the observation (**ii**) above.

---

[17] A p-value < 5% is equivalent to say that we expect this test to fail 5% of the time, where failing means incorrectly rejecting $H_0$. Thus, with our 26 samples, we expect 1.3 errors. Likewise, with 1%, we would expect 0.26 errors.

[18] For a p-value < 1%, that would reduce to 8 cases, 7 TT vs. 1 HH, 6 with the defect down vs 2 up.

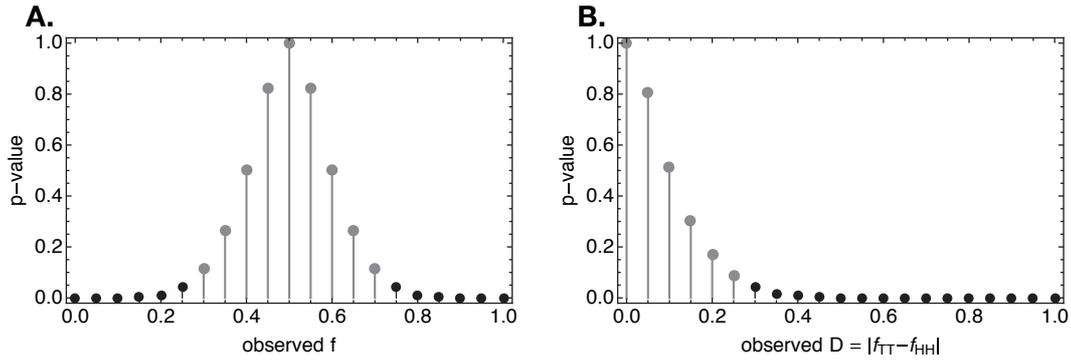

**FIG. 4-20 P-value for statistical tests.** The values with p-value < 5% are black, > 5% are grey. **A.** Binomial test (n=20, p=0.5). Double tail p-value vs observed frequency. **B.** Binomial difference test (n=20). Single tail p-value vs observed D (frequency difference).

We are then left with confirming observations (**iii**) and (**iv**), namely that there are structures with different P case probability $(p)$, and that there are structures for which the P case probability of HH and TT DWs $(p_{HH}$ and $p_{TT})$ is different. We shall use the single-tail p-value test, now on the variable $d = \left| f_i - f_j \right|$ (the observed difference in frequency), for a null hypothesis $H_0$: $p_i = p_j$. The p-values for this test were calculated and are shown in FIG. 4-20B. This indicates that differences $\geq 0.3$ are significant.[19] As it is possible to pick out pairs of structures with $f_i$ differing more than 0.3, in both trap positions and in both DW polarities, observation (**iii**) is confirmed. Also, several structures show significant difference in frequency between HH and TT DWs (4 structures with the T on the bottom, and 2 with T on the top) [20], confirming the observation (**iv**).

Overall, in the 13 structures, we observed that the p case is more frequent $(f = 0.63)$, indicating that there is a strong tendency for the maintenance of the original injected DW configuration. This is consistent with statistical results on Py structures [Jausovec 2008].

## Walker breakdown and fidelity length

The measurements above have shown that the central magnetisation of the DW at the trap is stochastic, which we have attributed to Walker breakdown coupled to stochastic depinning from the corner. This effect is cyclic, i.e. the DW reverses every spatial period, with a period which is a function of applied field, track dimensions, and material parameters. However, natural defects in the track make the reversal period highly irregular [Glathe et al. 2008], even supressing it altogether is some cases [Nakatani et al. 2003]. This also means that within some distance of the starting point the DW should have the original configuration. Thus, by placing the trap close enough to the DW injection point, the as-injected DW configuration can be tested. This experiment was conceived by Lewis et al. who, in Py tracks, did observe that for small travel lengths (shorter than a *fidelity length*) the DW magnetisation was always maintained (i.e. $p = 1$, in the notation above) [Jausovec 2008; Lewis et al. 2009; Lewis 2010], though not the reverse: regardless of trap distance, there were some structures that always showed the parallel configuration.

---

[19] For a p-value < 1%, only differences > 0.5 are considered significant. Incidentally, note that this test is more stringent than the previous, as now both $p_i$ and $p_j$ are unknown.

[20] For a p-value < 1%, these would reduce to 1 case with the T on the bottom, and to 2 cases with the T on the top.

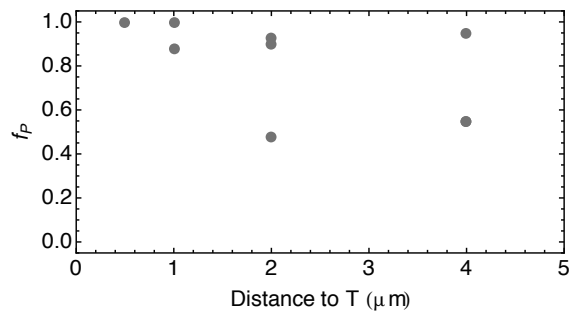

**FIG. 4-21 Frequency of the P case vs. travel length to T trap.** These frequencies were measured in 8 structures, with 20 single shot measurements.

To confirm that also in SV tracks the DW is initially in the P state, a study was performed in SV tracks with the T trap at four different distances from the injection point (0.5–4 µm). These structures were fabricated with a narrower track width (50 nm) and in a different sample from the one used in this section [21]. Apart from an increased $H_{PR}$, $H_{TR}$, and $H_{NUC}$ (cf. width dependency studied in chapter 3), the pinning behaviour of the T trap was essentially the same as described above. Eight structures were measured, and the results are shown in FIG. 4-21. Though the number of structures is limited, the observed result is consistent with what was found in the above-cited studies in Py: up to a characteristic length, the probability of P DWs ($p$) is 1 and, above this threshold, there are structures for which $p \neq 1$, though overall $p > 0.5$. Our measurements are not fine enough (nor do we have enough statistics) to safely determine this characteristic fidelity length. It does support, however, the hypothesis

---

[21] sample reference HM14, see Annex A.

that the injected DW is a transverse DW of central magnetisation parallel to the reset field.

**Summary**

So far, we have seen that:

– it is possible to measure H$_{PR}$ and H$_{TR}$ through the trap, given H$_{TR}$ > H$_{PR}$;

– a T trap has two possible H$_{TR}$ values, corresponding to the two possible DW central magnetisation directions (P and AP);

– a T trap on the top or on the bottom yields different H$_{TR}$ pairs, that can be attributed to the four possible interaction cases (*P/AP-narrow/wide*);

– the H$_{TR}$ for AP cases is ≈ H$_{NUC}$, possibly corresponding to a new nucleation;

– the H$_{TR}$ values are similar between DW polarity (HH and TT);

– the DW central magnetisation (P/AP) is stochastic, with different probability on each structure and DW polarity;

– though, overall, the P case is more probable.

## 4-2.2. Nature of the pinning: push-pull experiments

We have so far shown transmission measurements, in which the DW is injected, pushed to the pinning trap (at H$_{PR}$), and finally depinned *through* the trap. This measurement determines the depinning field, but it tells us nothing on the nature of the pinning: whether the DW is attracted to the trap and pinning lowers the system energy, i.e. the trap is an *energy well*, or whether the DW is repelled from the trap, and the trap is an *energy barrier*. In fact, a pinning trap may present a more complex energy landscape than a simple well or barrier, even without the added complexity of multiple magnetic states that the T stub presents [Petit et al. 2008a]. One way to study the pinning nature is to measure the backward depinning field, where the DW is pushed to the trap (at H$_{PUSH}$ < H$_{TR}$) and then pulled back in the opposite direction. By comparing the field necessary to depin the DW through the trap (H$_{TR}$) and back from the trap (H$_{PULL}$), it is possible to determine whether the DW is pinned at a well (H$_{PULL}$ ≈ H$_{TR}$) or at a barrier (H$_{PULL}$ ≪ H$_{TR}$).

**Push-pull field sequences**

The field sequences used in the push-pull experiments are shown in FIG. 4-22. For experimental convenience, HH and TT DWs are tested separately. The figure shows the field sequences for HH DWs; the sequences for TT are the same but reversed. In these sequences, the reset pulse injects a DW in the lower corner of the arc (as before). Then, a sweeping $H_X$ field is applied under a constant $H_Y$ bias (≈-40 Oe), pushing the DW to the trap (the *push* phase). When $H_X$ reaches a certain value ($H_{PUSH}$), it is swept back to ≈-330 Oe, pulling the DW from the trap (the *pull* phase). The experiment is then repeated with different $H_{PUSH}$ values (10 single shot measurements per $H_{PUSH}$ value).

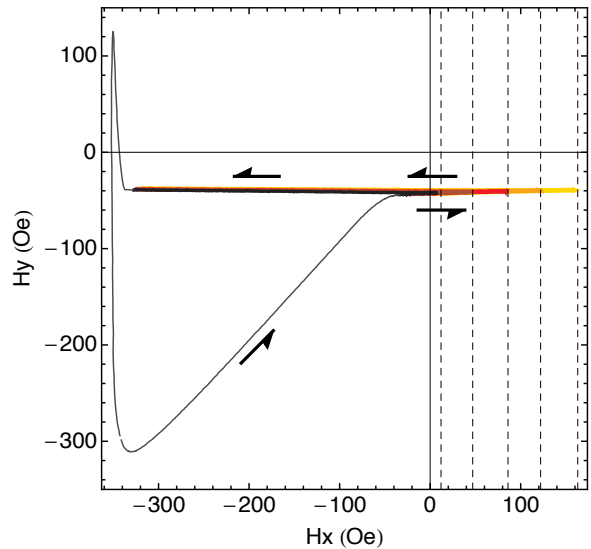

**FIG. 4-22 Field sequences for push-pull measurements.** $H_Y$ vs $H_X$ for five field sequences that inject a HH DW and push it at five different $H_{PUSH}$ values (12, 47, 86, 122, and 163 Oe for curves in black to yellow). The thin line corresponds to the reset period of the field sequence (identical for all five), and the thick lines to the measurement period. The sequence duration was 1 s.

## Results for structures with the T on the bottom

The results for a structure with the T stub on the bottom of the track are shown in FIG. 4-23. We observe that the number of transitions, their amplitudes, and the field at which they occur, depend on the value for $H_{PUSH}$. For most values of $H_{PUSH}$, we observe that the single shot measurements show two distinct patterns, which are coloured in red and blue in the figure. We shall see that these patterns correspond to the two possible interaction cases in this structure: *P-wide* and *AP-narrow*.

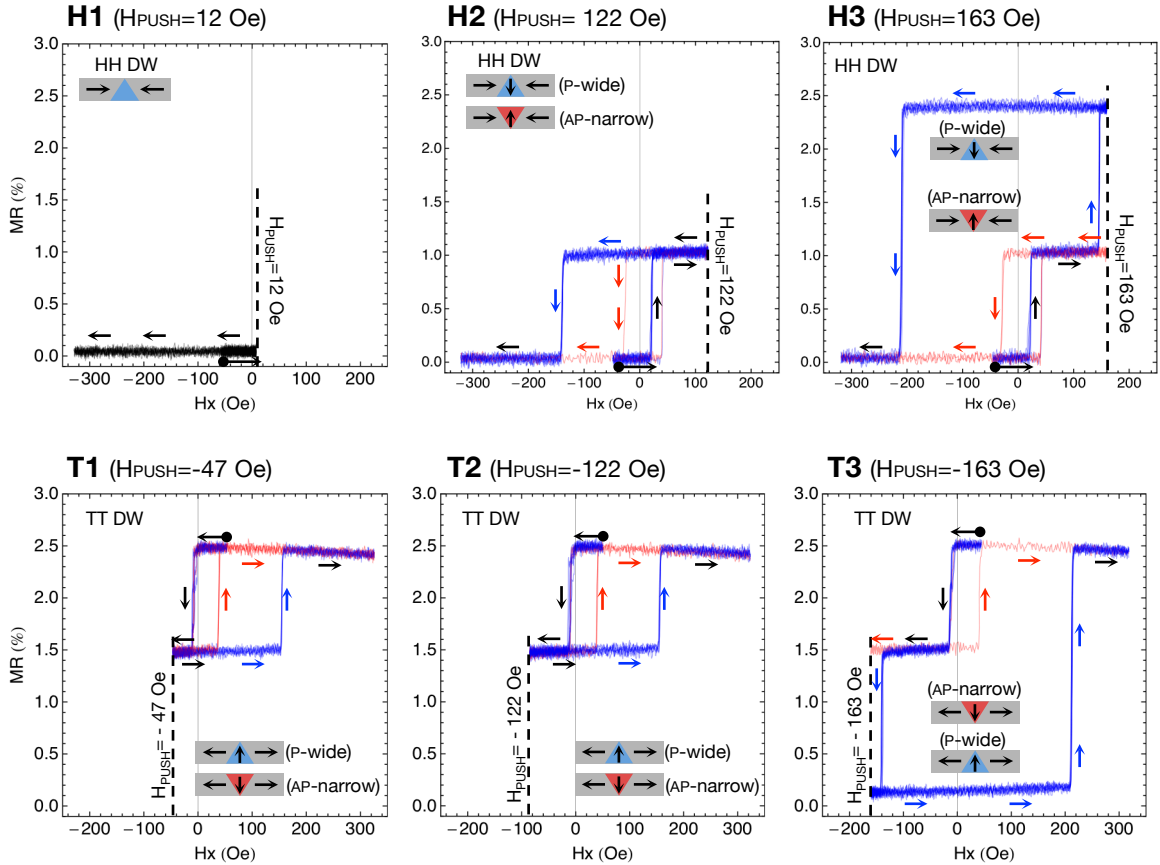

**FIG. 4-23 Push-pull measurements on a track with a T trap on the bottom.** Each plot contains 10 single shot measurements. MR was calculated with the same $R_0$. When the two transition patterns are distinguishable, the measurements are coloured (red and blue for the *AP-narrow* and *P-wide*, resp.). For clarity, the points from the reset period (cf. FIG. 4-22) are excluded. Plots **H1–3** correspond to the injection of a HH DW, **T1–3** to a TT DW. The start of the measurement is marked by the circle.

Using the same reasoning as before, we can associate each transition to a particular track segment (before and after the trap). As such,

– when $H_{PUSH} < H_{PR}$, no transitions are observed (plot H1).

– After this threshold, and with $H_{PUSH} < H_{TR}$, each loop has two transitions: one at $H_{PR}$ during the push phase, when the DW travels from the arc to the trap, and one during the pull phase at a characteristic field, $H_{PULL}$, when the DW depins from the trap and propagates back to the arc. (plots H2, T1, and T2). We can see that there are two distinct values for $H_{PULL}$.

– For $H_{TR}^P < H_{PUSH} < H_{TR}^{AP}$ (resp. 149 and 207 Oe for HHs in this structure), we observe three transitions: at $H_{PR}$, when the DW travels to the trap; at $\approx H_{TR}$, still in the push phase, when the DW depins through the trap; and finally during the pull phase, at $H_{NUC}$, when the track reverses in the absence of the original DW. We observe that some loops show these three transitions ($H_{PR} - H_{TR} - H_{NUC}$), while others show the previous two ($H_{PR}$ − $H_{PULL}$). This is due to the two different $H_{TR}$ values. Only one value of $H_{PULL}$ is now observed.

In plots H1–3 we see that there are two possible values for $H_{PR}$ for HH DWs: 41 Oe (similar to what was found before, 43 Oe) and 21 Oe. There is no correlation between the value of $H_{PR}$ and the value of other transitions. As $H_{PR}$ is in reality the depinning field from a natural trap in the arc, we may dismiss this phenomenon as some stochastic pinning to a natural trap, that is particularly visible with these specific field sequences.

| Case | $H_{PR}$ (Oe) | $H_{TR}$ (Oe) | $H_{PULL}$ (Oe) |
|---|---|---|---|
| HH *P-wide* | 31 | 149 | -140 |
| HH *AP-narrow* |  | 207 | -28 |
| TT *P-wide* | -8 | -140 | 156 |
| TT *AP-narrow* |  | -200 | 20 |

FIG. 4-24 **Push-pull measurements on a track with a T trap on the bottom** (summary table).

We have seen that there are two values of $H_{TR}$ (§4-2.1) and now two values of $H_{PULL}$. By analysing cases with different $H_{PUSH}$, we can associate each $H_{PULL}$ to each $H_{TR}$, and finally to the two possible interaction cases: *P-wide* and *AP-narrow*. For example, in H2 we observe two $H_{PULL}$ values: -28 (n=1) and -140 Oe (n=9). With a greater $H_{PUSH}$ (H3), the most frequent DW is now transmitted through the trap (at 149 Oe; n=8), while the lowest $H_{PULL}$ value remains: -28 Oe (n=2). We can then infer that $H_{TR}$ = 149 Oe corresponds to the same interaction case as $H_{PULL}$ = -140 Oe (*P-wide*), while the second DW type has $H_{PULL}$ = -28 Oe and $H_{TR}$ = 207 Oe (*AP-narrow*; §4-2.1). These results are listed in FIG. 4-24.

## Results for structures with the T on the top

The results for the structure with the T stub on the top of the track[22] are shown in FIG. 4-25. Similar to the previous structure, this one has two possible interaction cases: *P-narrow* and *AP-wide*. However, unlike before, we can only measure $H_{PULL}$ for one of the two cases, as in the other it is impossible to reliably put the DW at the trap, $H_{TR}$ being ≲ $H_{PR}$.

---

[22] This is also the same structure of FIG. 4-15.

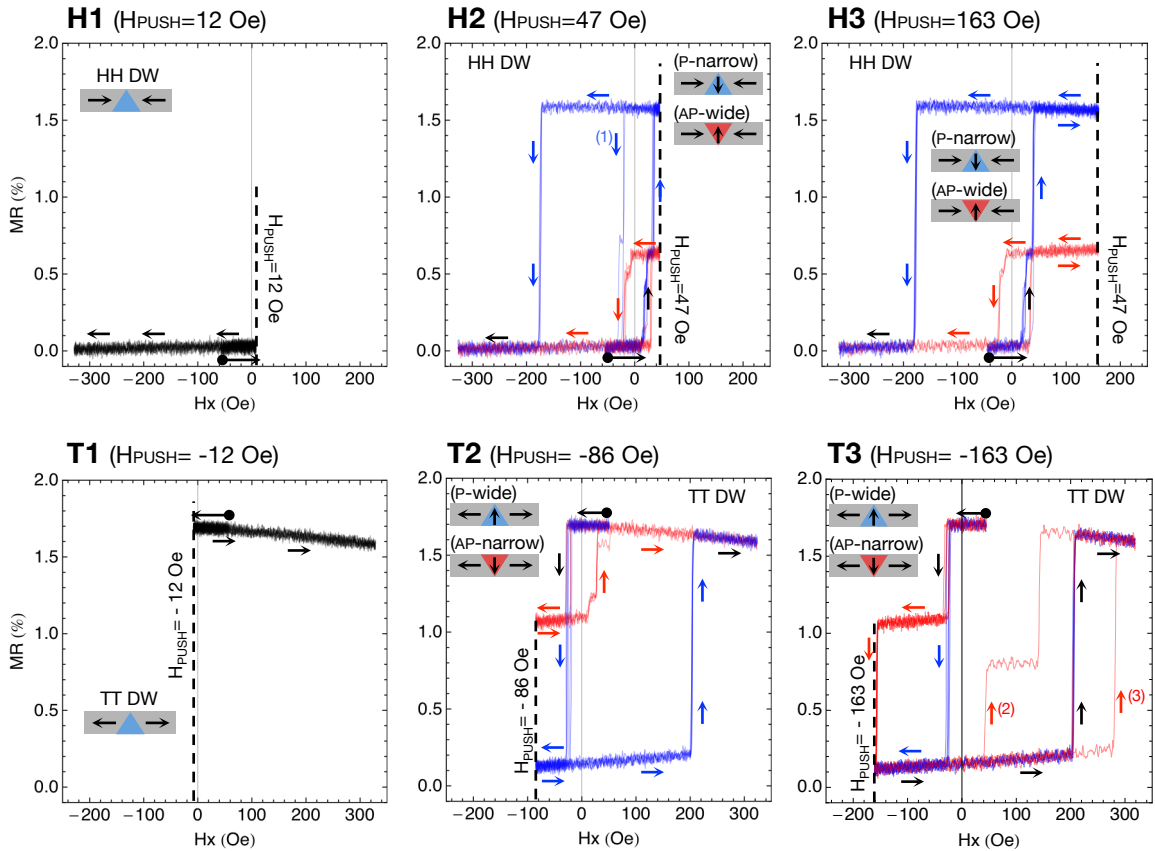

**FIG. 4-25 Push-pull measurements on a track with a T trap on the top.** Each plot contains 10 single shot measurements. MR was calculated with the same $R_0$. When transition patterns are distinguishable, the measurements are coloured (red and blue for the *AP-narrow* and *P-wide*, resp.). For clarity, the points from the reset period are not plotted. Plots **H1–3** correspond to the injection of a HH DW, **T1–3** to a TT DW. The start of the measurement is marked by the circle.

As before, for $\mathbf{H_{PUSH}} < \mathbf{H_{PR}}$ (H1, T1), we see no transition. For $\mathbf{H_{PUSH}} > \mathbf{H_{PR}}$ (H2–3, T2), we observe that, during the push phase, either the DW propagates through the entire track or stays pinned at the trap. The first case corresponds to the *P-narrow* interaction, the second to *AP-wide*. Applying the pull field, we observe that the in the first case the track reverses at $H_{NUC}$ (there being no longer a DW in the track) while in the second the pinned DW is pulled at $H_{PULL} \approx H_{PR}$. A rare *P-narrow* case is also observed (n=2), marked as **(1)** in the figure, where the transmitted DW does not annihilate at the track's end, and instead travels backwards when $H_{PULL}$ is applied. Finally, for $\mathbf{H_{PUSH}} > \mathbf{H_{TR}}$ [23] (T3), we observe that for both interaction cases the track reverses at $H_{NUC}$ during the pull phase, as no DW is left in the systsm. There are two other reversal patterns events for the *AP-wide* case, marked in the figure as **(2)** and **(3)**, which will be addressed later on page 148.

---

[23] $H_{TR}$ of the *AP-wide* case, that is.

These measurements show a larger spread of values $H_{PR}$ presents than in the transmission measurements. No correlation was found between $H_{PR}$ and any of the other fields.

## Measurements on several identical structures

These measurements were taken in a total of 8 structures, 5 with a trap on the top and 3 on the bottom. These were a subset of the 13 structures on which the transmission measurements were previously made. The average results are shown below, FIG. 4-26.

In the *P-wide* case $H_{Pull} \approx H_{TR}$, while in both *AP* cases $H_{PR} \approx H_{PR} \ll H_{TR}$. We can thus say that in the former case the trap is a (symmetric) *energy well* and, in the latter, an *energy barrier*.

This is in agreement to the findings on Permalloy structures, micromagnetic simulations, and TEM studies [Petit et al. 2008b; 2009; O'Shea 2010].

| Geometry | N | Case | $\langle H_{Pull} \rangle$ (Oe) | $H_{Sh}$ (Oe) | $\langle H^{*}_{Pull} \rangle$ (Oe) | $H_{Sh}$ (Oe) | $\langle H_{Tr} \rangle$ (Oe) |
|---|---|---|---|---|---|---|---|
| 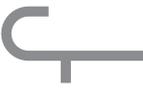 | 3 | *P-wide* | 145 ±3 | +11 | 200 ±13 | +15 | 150 |
| | | *AP-narrow* | 40 ±5 | +3 | – | – | 211 |
| 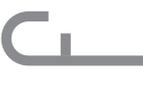 | 5 | *P-narrow* | – | – | 192 ±13 | +11 | 49 |
| | | *AP-wide* | 25 ±4 | +6 | – | – | 171 |

**FIG. 4-26 H_PULL measured on several structures.** The errors are the half sample amplitude. The presented values are the mean of the HH and TT values, and $H_{SH}$ half their difference. $H^{*}_{Pull}$ is the reversal field when the DW has been transmitted over the trap. $\langle H_{TR} \rangle$ is taken from Fig. 4-18, shown for comparison.

## Magnetostatic interaction between the DW and the trap

It is also clear that there is a difference between $H_{Pull}$ values of the *AP-narrow* and *AP-wide* cases (40 ±5 Oe versus 25 ±4 Oe). This is due to the magnetostatic interaction of the T stub and the DW. Consider a HH DW as an example, schematised in FIG. 4-27. The DW has a positive magnetostatic charge of +2Q, where $Q = \mu_0 M_S S$ is the track characteristic charge, $M_S$ the saturation magnetisation, and S the active layer cross-section. The pinning region is surrounded by 3 tracks, so it can have charge -3Q, -1Q,

+1Q, or +3Q. In the *AP-wide* case, with two inward and one outward magnetised tracks, the charge is +Q while in the *AP-narrow* case it is -Q. Thus, in the former case there is a repulsive interaction while in the latter there is an attractive interaction, causing the difference in the observed H_Pull values.

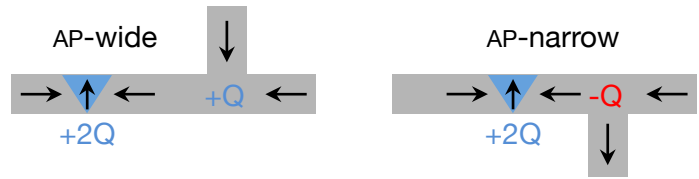

**FIG. 4-27 Magnetostatic interaction between the T trap and the DW** (schematic), for the two AP configurations.

Furthermore, during the push-pull measurements, this repulsion in the *AP-wide* case causes the DW to move away from the trap at fields lower than H_Pull / H_PR, to a point in the track where the natural pinning balances the magnetostatic repulsion. As can be seen in FIG. 4-28, of the five structures with the T on top, three show that the DW moved back before H_PR while none of the three with the T on the bottom did so [24]. Note that in two of the structures the DW moves back against the applied field (at +19 and +17 Oe, structures W1 and W3), an indication to the strength of the magnetostatic interaction. In these three structures, the DW is pushed back 400–800 nm from its pushed position before depinning.

---

[24] In the parallel cases, or better, in the *P-wide* case, also no repulsion was observed.

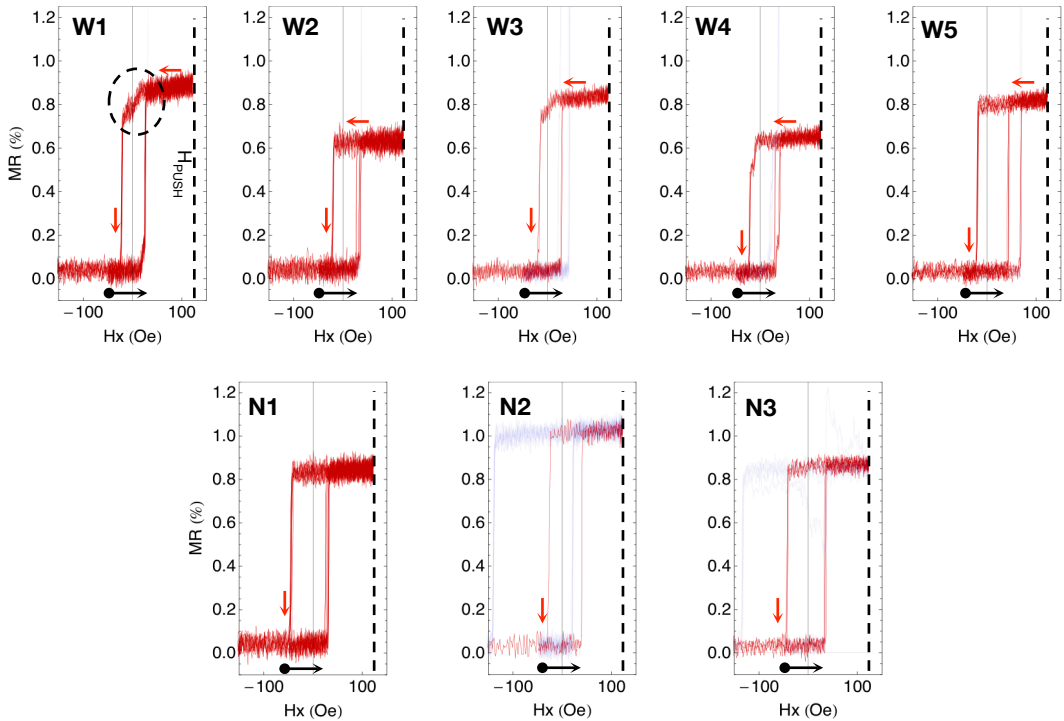

**FIG. 4-28 Comparison of push-pull measurements of *AP-wide* and *AP-narrow* cases** (HH only). All 8 studied structures are shown. The cases W1–W5 are *AP-wide* (red), from structures with the T on the bottom, and N1–N3 are the *AP-narrow* (red) from structures with the T on top. The parallel cases are also shown in light blue. $H_{Push}$ was 122 Oe.

## 4-2.3. 360° DWs left over in the transmission of *AP* cases

In the case of anti-parallel DW and arm, $H_{TR}$ is close to $H_{NUC}$, which indicates that the reversal of the second arm occurs via nucleation of a separate domain. This opens the possibility of the original DW still existing after the track has reversed, and that 360° DWs may be formed.

360° DWs can be thought as a coupled pair of a HH and a TT (180°) DWs of anti-parallel central magnetisation. Its stability arises from the balance of the magnetostatic attraction between them and the exchange repulsion caused by their opposed winding directions [Muratov & Osipov 2008]. We shall see that the push-pull measurements presented above, and others shown below, show that the formation of a 360° DW is one of the possible cases, albeit not the only one.

**Possible combinations of original and nucleated DWs**

In the transmission of AP configurations, the DW has its central magnetisation anti-parallel to the arm, and it is pinned at the left of the trap as it experiences an energy barrier (FIG. 4-29A-ii). When $H_X$ reaches ~$H_{NUC}$, the right segment reverses by nucleation. The nucleation can occur at the track's right end, middle, or at some location near the T trap. The significant lowering of $H_{NUC}$ in structures with the T trap indicates that it occurs at the trap. In any case, the right segment will reverse by one or two DWs, one of which will be annihilated at the track end and the other will be left at the trap, FIG. 4-29A-iii. There, we have the original anti-parallel HH DW on the left, and the new TT DW on the right. The actual position of the new DW, as well as its central magnetisation, is unknown. In a track without any traps, two adjacent TDW would collapse or form a 360° DW, depending on their central magnetisations being parallel or anti-parallel, respectively [Kunz 2009]. Here, however, T trap may change the interaction between the two TDW.

There are then five cases to consider: two trap configurations, two possible central magnetisations of the new DW, and the possible case that no DW is present in the system at all. These cases are shown in FIG. 4-29B, diagrams #1 and #2 for structures with the trap on the top, and diagrams #3 and #4 for the trap on the bottom. Diagram #5 represents the case of no DW (identical for both trap positions). Of these, cases #2 and #4 are symmetric [25], and thus should present the same behaviour, bar some small influence from the coupling to the reference layer. In the schematics, the new DW is always drawn to the right of the trap. If nucleation happens at the T, it is also possible that the injected DW is at the left side of the trap. In that case, if the central magnetisations of the injected and the original DWs are parallel, they will annihilate [Kunz 2009] (case #5), otherwise it would lead to cases #1 and #3.

---

[25] The cases   FIG. 4-29B-#2 and #4 can be obtained from each other by symmetry operations which preserve the micromagnetic behaviour. For example, starting from case #2 to reach #4, perform a time reversal (reversing all magnetisation arrows), then a vertical mirror operation, and finally a horizontal mirror operation.

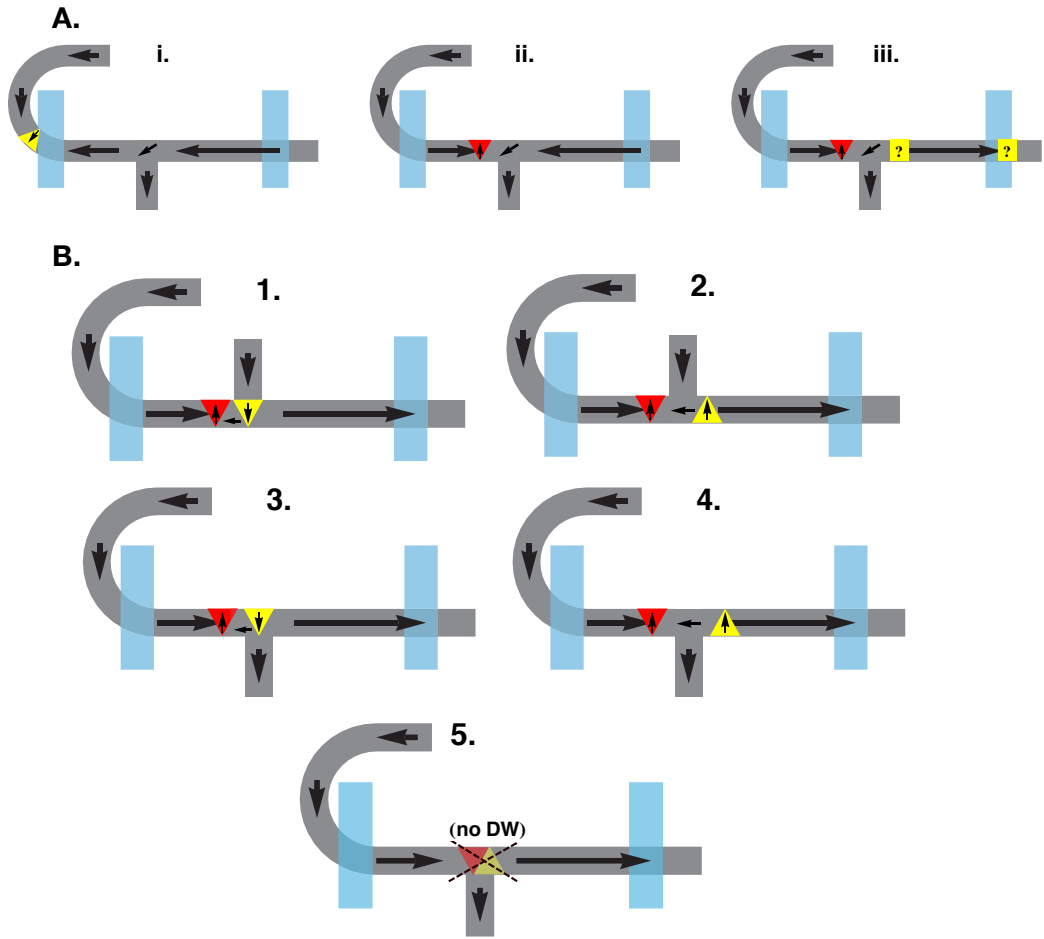

**FIG. 4-29 Multiple DWs at the T trap after nucleation in the *AP-narrow* or *AP-wide* case** (schematic). **A.** The *AP-narrow* transmission. A structure with the trap on the bottom is initialised (**i.**), the DW is pushed to the trap and, as it propagates, its central magnetisation flips (**ii.**). The DW is then pinned at the right of the trap. When $H_X$ reaches $H_{TR} \approx H_{NUC}$, the right segment reverses by nucleation and DW propagation (**iii.**). While one DW is annihilated at the track end, the second DW stays at the trap. **B.** The five possible DW pairs (for an original HH DW).

## Push-pull measurements after transmission

The used field sequences for the push-pull measurements had a maximum $H_{Push}$ lower than the $H_{TR}$ of the *AP-narrow* case (163 vs. ~211 Oe), and just about over the $H_{TR}$ of the *AP-wide* case (158 Oe [26]). As such, only cases #1, #2 and #5 could be tested.

We observed that the reversal of the track by the pull field, after the *AP-wide* transmission/nucleation event, presented 3 different reversal patterns. An example of a structure showing all these three cases can be seen in FIG. 4-30 (same data included in FIG. 4-25-T3). There, 10 single shot push-pull measurements of a TT DW are juxtaposed, with the *P-narrow* cases in blue and the *AP-wide* cases in red. During the push phase, the *AP-wide* are transmitted at 158 Oe, while the *P-narrow* cases are transmitted at ~$H_{PR}$ (as analysed earlier). During the pull phase, there are three different reversal patterns: **a double transition** (37 and 141 Oe), **a single transition at ~$H_{NUC}$** (190 Oe), and **a single transition at a field higher than $H_{NUC}$** (268 Oe).

---

[26] These values are lower than those reported in FIG. 4-18. This is due to the higher $H_Y$ bias of the push-pull sequences (40 versus 20 Oe), which lowers slightly the depinning fields.

The single transition at ~$H_{NUC}$ corresponds to case #5, reversal with no DW present. This is the same field value observed in the *P-narrow* case, where we know that no DWs were indeed present.

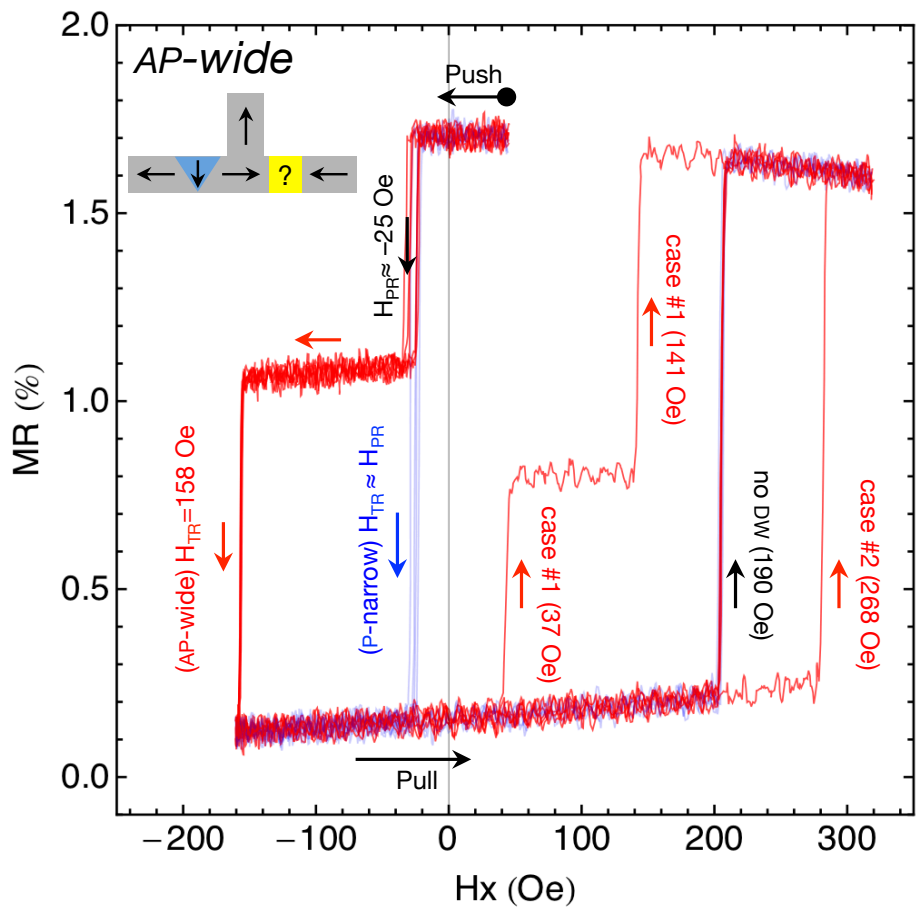

**FIG. 4-30 Push pull measurement (TT DW) on a structure with the T on the bottom** (10 single shot measurements; same data as Fig. 4-25). The *P-narrow* case is in blue, and the *AP-wide* case in red. The three possible cases for the pull from an *AP-wide* transmission are labelled.

The double transition corresponds to case #1, anti-parallel DWs. There we observe first the reversal of the shorter left segment at a field comparable to $H_{PR}$, corresponding to the left DW propagating leftwards. This is similar to the single DW *AP-wide* pull case studied before [27] (cf. FIG. 4-25-T2, FIG. 4-26). The second reversal, of the longer right segment, corresponds to the depinning of the injected DW that was left behind. This DW is in a *P-wide* pinning state (see FIG. 4-29B-#1), and indeed the reversal field is similar to the $H_{TR}$ of *P-wide* (141 vs. 150 Oe).

The higher transition (at 268 Oe) must then correspond to case #2, parallel DWs. The mechanism responsible for the increased reversal field will be clearer with the micromagnetic simulations below.

We repeated this experiment on 5 structures with the trap on the top. The $H_{TR}$ of the *AP-wide* case was reached in every structure and for both DW polarities but once. In a total of 90 single shot measurements (50 TT and 40 HH, over the 5 structures), we observed the AP case 39 times, always with similar results to the case shown before. The **average transition field values** were:

– **case #1**, anti-parallel DWs, was observed 19 times. The left segment reversed at 37 Oe, and right segment reversed at 141 Oe ($\approx H_{TR}$ of the *P-wide* case, 135 Oe [26]);

– **case #2**, parallel DWs, was observed 12 times. The track reversed in a single transition at 268 Oe (about 41% greater than $H_{NUC}$).

– **case #5,** the reversal with no DW present, was observed 9 times. The track reversed with a single transition at 190 Oe ($H_{NUC}$ [26]).

**Micromagnetic simulations**

To better understand the DW behaviour in each of these cases, micromagnetic simulations were performed. A track with a T trap was initialised in each of the four independent cases (#1, #2, #3, and #5). To emulate a push-pull measurement, a horizontal field was ramped up to +340 Oe and then back to -500 Oe. The magnetisation at key values of field is shown in FIG. 4-31. These simulations predict that not all these DW couples are stable: the parallel DW pair (case #2 and, by symmetry, #4) collapses with injection of a DW in the T arm.

---

[27] This transition occurs at a slightly different field value than the single DW *AP-wide* pull case. We will analyse this difference below.

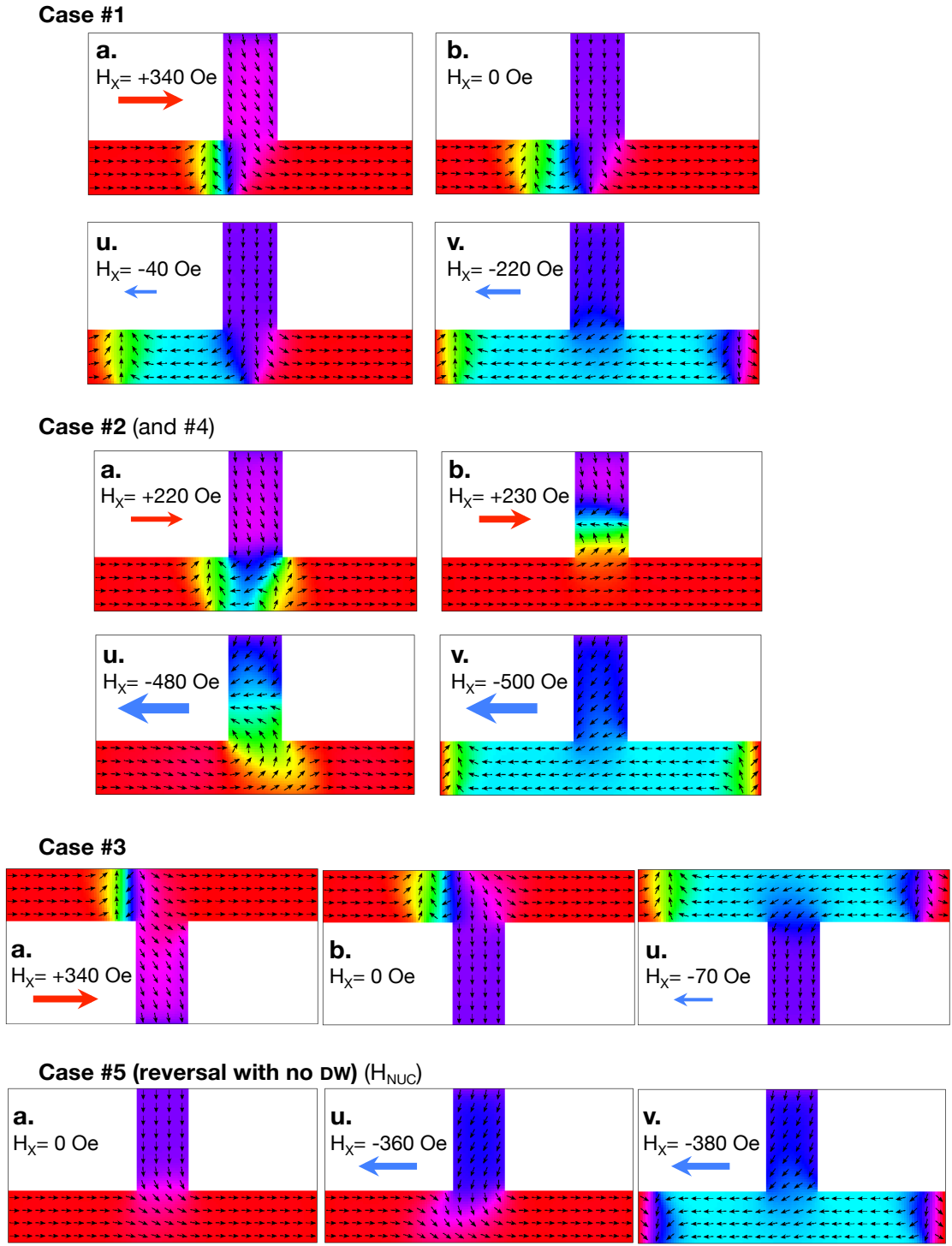

**FIG. 4-31 Pull after nucleation in the *AP-narrow* or *AP-wide* case** (micromagnetic simulation). Colour and arrows are mapped to the magnetisation. The track was initialised with the cases of Fig. 4-29B (case #2 and #4 are identical). $H_X$ was ramped to +340 Oe and then back to -500 Oe. $H_Y$= -40 Oe. **Case #1** showed two transitions during pull (at -35±5 and -210±10 Oe). In **case #2** the DW pair collapsed (at +225±5 Oe), and a single transition occurred during pull (at -490±10 Oe). **Case #3** showed a single transition during pull (at -65±5 Oe). **Case #5** ($H_{NUC}$) occurred at -370±10 Oe. Cell size=5×5×8 nm³, simulation volume= 600×300×8 nm³, width=100 nm. Simulations performed using OOMMF [Donahue & Porter 1999].

The simulations also predict that these different cases yield different push-pull measurements. Note that field values from these micromagnetic simulations cannot be directly compared to experiment as the simulation is ran at 0 K. During the pull phase, the simulations show:

- **In case #1**, two distinct transitions during the pull phase: the original DW reverses the left segment at a finite field, $H_{Separation}$ = -35 ±5 Oe, and the nucleated DW, pinned in *P-wide* configuration, reverses the right segment at the respective $H_{TR}$, -210 ±10 Oe.

- **In case #2 (and #4),** a DW was injected in the T stub during the push field. This DW *increases* the nucleation field, and the track reverses at a field about 32% greater than $H_{NUC}$ (-490 ±10 Oe, vs. $H_{NUC}$ = 370 ±10 Oe).

- **In case #3**, the original DW is being pulled from an energy barrier (*AP-narrow*) and the nucleated DW is weakly pinned (*P-narrow*). Both DWs propagate at a finite field, $H_{Separation}$ = -65 ±5 Oe.

- **In case #5** ($H_{NUC}$), the track reverses with a single transition, at 370 ±10 Oe.

The simulations of cases #1, #2, and #5 are in excellent agreement with what we found in experiments, both in the order of track reversal (e.g. in case #1, the reversal of the left segment before the right segment), and in relative magnitude of the transition fields [28]. The measurements shown before and these simulations strongly support the proposition that we have observed the creation and the splitting of 360° DWs.

## Splitting of 360° DWs

It is interesting to notice the micromagnetic simulations presented in FIG. 4-31 predict that the field needed to split the 360° DWs is non-zero and different depending on the pinning configuration: it is 35 ±5 Oe for case #1, and 65 ±5 Oe for case #3. A simulation of a 360° DW isolated in a straight track, presented in FIG. 4-32, shows that the unpinned DW splits at field higher than both those cases, 75 ±5 Oe.

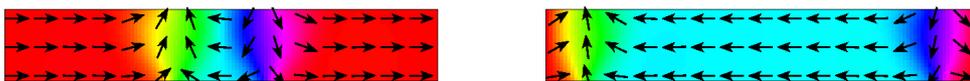

**FIG. 4-32 Splitting of a 360° DW** (micromagnetic simulation). The track was initialised with a 360° DW (left image) and a horizontal, leftwards field was applied. The DW was

---

[28] As explained before, direct comparison of experimental and simulated transition fields is not possible as the simulated system is at 0 K.

split at H_Separation = -75 ±5 Oe (right image). . Cell size=5×5×8 nm³, simulation volume= 600×100×8 nm³. Simulations performed using OOMMF [Donahue & Porter 1999].

It is also interesting to compare the splitting of the pinned 360° DW, cf. FIG. 4-31, with the pull of an *AP-wide* or *AP-narrow* DW. We will analyse them, as before, in terms of the magnetostatic interaction between the T gate and the DW. In order to do this, the *AP-narrow/wide* cases were simulated in the same conditions as FIG. 4-31. The results of all these simulations, along with the experimental values where available, are summarised in FIG. 4-33. Also included in the image are the schematics of the magnetostatic charges. As before, $Q = \mu_0 M_S S$ is the characteristic magnetic charge of the track.

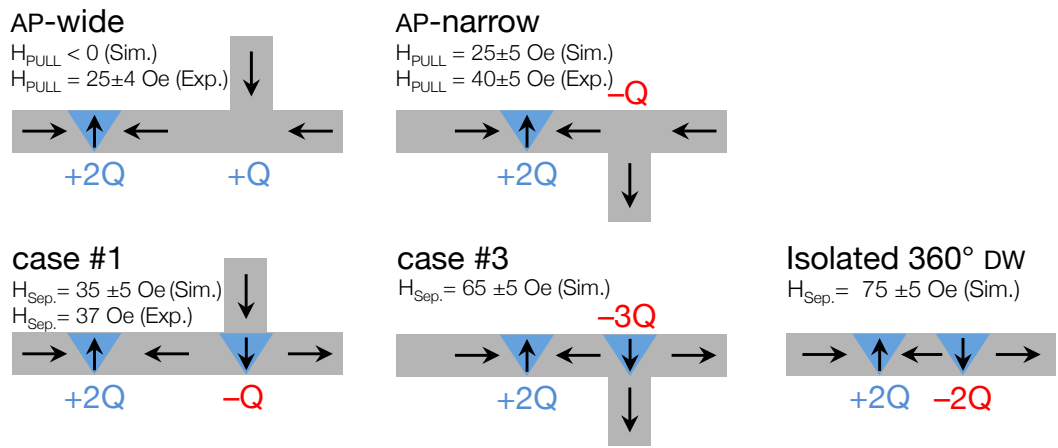

**FIG. 4-33 Magnetic charges during pull of single HH DWS and splitting of pinned and isolated 360° DWS.** The simulated ("Sim.") and measured ("Exp.") H_PULL / H_Separation fields are also included (cf. FIG. 4-31, FIG. 4-32).

We observe there that the first transition of case #1 and H_PULL of the *AP-narrow* case both have a +2Q/-Q magnetostatic interaction, and indeed their transitions occur at similar fields (37 Oe vs. 40 ±5 Oe). In simulation, the transition of case #1 occurs at a lower field than the *AP-narrow* pull (35 vs 25 Oe). It is possible that such difference is masked by the finite H_PR in experiment.

For the other configurations we have only the simulated results. Overall, the simple monopolar charge model is consistent with the simulated H_PULL/H_Separation values. In the only case with two charges of the same polarity (*AP-wide*) the DW is repelled without application of a field. The two cases with +2Q/-Q show a similar H_PULL (25 and 35 Oe), much smaller than the +2Q/-2Q case (75 Oe) or the +2Q/-3Q case (65 Oe).

There are however some unexplained differences. Considering the cases with the 360° DW, the charge model suggests that the isolated 360° DW would separate at a field smaller than case #3, which is not the case. We suggest this occurs because of two reasons. **(1)** The monopole model presented before does not explain fully the magnetostatic interaction. This model ignores the large dipolar moment of the 180° DWs, which contributes to their attraction [Muratov & Osipov 2008]. This dipolar moment is fully present in the isolated 360° DW but is partly quenched by the presence of the T trap. **(2)** The T trap distorts the magnetisation of the pinned DW, and changes the exchange energy of the joined and split DW. A larger decrease in exchange energy from the joined to the split state would cause a lowering of $H_{Separation}$.

The presence of 360° DWs can alter significantly the reversal of magnetic elements used in sensor and DW logic devices, affecting adversely their sensitivity and reliability [Schafer et al. 1993]. Furthermore, they are stable under even large applied fields, and cannot be field propagated away as its 180° counterparts. That same stability, on the other hand, coupled its small size, suggests its application to data storage devices. For all these reasons, it is interesting to note the manipulation capabilities demonstrated by T trap, specially in the configuration of case #1: it can be used to inject, pin, and split at a significantly reduced separation field a 360° DW.

## 4-2.4. Summary

We have measured and analysed here the pinning behaviour of a T trap. We observed that, analogously to its monolayer Permalloy counterparts [Petit et al. 2008b; Lewis et al. 2009; Petit et al. 2009], the **T trap has 4 possible pinning configurations**, corresponding to two different positions of the trap (*up/down*) and to two possible DW central magnetisations (*P/AP*). Each of these cases presented **different pinning fields**, which, in two of the four cases, depended on the depinning direction. Two cases with special interest to the field of DW logic are produced by the structure with the T stub on the top. We observed there that a reversal of the gate magnetisation yields two drastically different pinning strengths. This demonstrates the principle of a **gated DW valve**, able to effectively open or close the main track to the passage of DWs. Again these findings are consistent with experiments in Permalloy structures.

The sensitivity of the T trap to the DW structure allowed us to study the **reversal of the TDW central magnetisation** by the Walker breakdown process. We observed that this reversal is **stochastic**, **varies between structures**, and between DW polarities, with the parallel case (the assumed as-injected configuration) being overall more probable. These findings are consistent with what was reported before for Permalloy [Lewis et al. 2009] and SV tracks [Glathe et al. 2008]. Analogous to what was found in Permalloy, we also observed that below a certain length threshold, **the fidelity length**, the DW structure was always the same (parallel), allowing us to confirm the structure of the injected DW.

By using the measurement of DW position, we were also able to observe the repulsive magnetostatic interaction between the DW and the trap under certain configurations.

We showed that the **transmission of a DW in the AP cases occurs via nucleation and injection of new DWs**. Furthermore, we showed that the original and newly-injected DWs could either annihilate, produce a DW in the T stub, or form a 360° DW. We then showed that, with the two possible T gate orientations, these formed four independent configurations. Using push-pull measurements, we observed three of these configurations, and showed that they could occur in the same structure. We analysed these configurations with a micromagnetic study, which showed an excellent agreement with the experimental observations.

These measurements and simulations revealed that the **T stub acted as an injection point and a pinning site for 360° DWs**. Furthermore, it was also shown that it **reduced significantly the splitting field of the 360° DW**, when compared to an isolated DW. We interpreted this in terms of a change of magnetostatic interaction and deformation of the DW. These findings suggest that the T gate can be used to create and manipulate 360° DWs, a DW type particularly difficult to manipulate.

It is important to notice the role in these studies of the single shot measurements of DW position possible in SV tracks, without which many of these findings would be impossible or ambiguous.

As for **DW logic applications**, we demonstrated that the T gate has several possible uses. It can serve as an artificial DW trap; it can serve a controllable DW gate; and, finally, it can be used to inject, pin, and split at a significantly reduced separation field a 360° DW.

# 4-3. The NOT gate [29]

We shall now study the DW NOT gate. This kind of structure, first introduced by Cowburn and colleagues [Allwood et al. 2002], transforms TT DWs into HHs (and vice-versa) under a rotating field. As binary information can be mapped onto domain or DW polarities, this structure effectively performs the NOT logic operation. The structure consists in a cusp, schematised in FIG. 4-34, which stabilises the magnetisation of both its arms in a symmetric configuration (FIG. 4-34A). In the presence of a CCW rotating field, DWs present in the left arm (the *input arm*) will be pushed into the cusp. As the DW reverses the apex of the cusp (FIG. 4-34B-ii), it is annihilated and a new DW, of opposite polarity, is created (FIG. 4-34B-iii). As the field continues its rotation, the new DW propagates outwards through the right arm (FIG. 4-34B-v). Two identifying characteristics of the NOT gate are polarity reversal of DWs (the NOT operation itself) and a delay in the propagation of the DW (as the field had to rotate 180° before the right portion of the arm could reverse).

---

[29] The used NOT gate designs were developed by Dr E Lewis and Dr H Zeng (cf. FIG. 4-35). Sample fabrication and measurements on the asymmetric NOT gate were done in close collaboration with Dr Zeng, and related results are partially included Dr Zeng's PhD thesis [Zeng 2010].

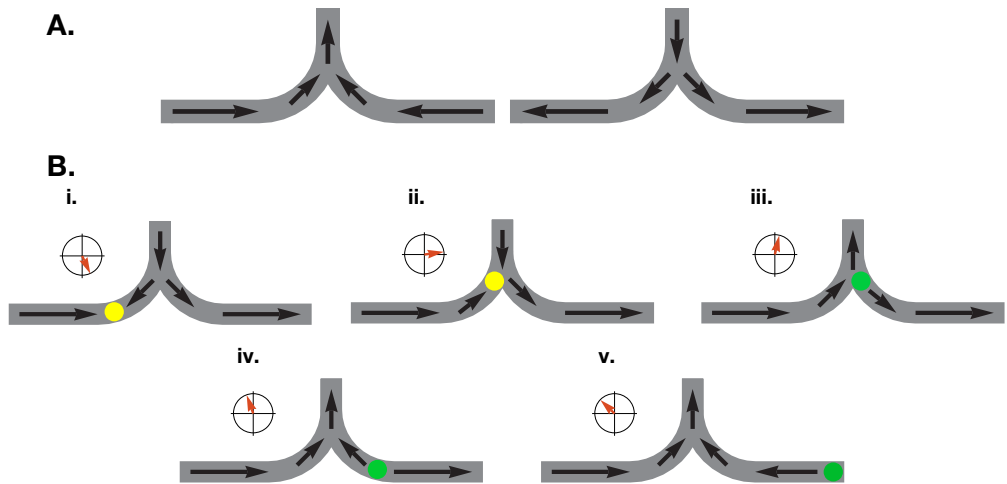

**FIG. 4-34 NOT gate operation** (schematic). **A.** The two stable states of the gate (*empty gate*). **B.** Gate operation under a rotating field (red arrow). A HH DW enters from the left (**i**) and travels up into the cusp (**ii**) annihilating and creating a TT DW (**iii**). As the field progresses, the TT DW travels out into the right arm (**iv**, **v**).

The NOT gate also provides a way to propagate multiple DWs in a clocked fashion. If several NOT gates are strung together they form a shift-register: a data-storing structure that allows the simultaneous containment and propagation of multiple, non-annihilating DWs [Allwood et al. 2002; O'Brien et al. 2009b].

The configuration, shape of the structure, and field directions indicated in FIG. 4-34 are merely for illustration. These are properties that depend on the design of the structure, design which has known several iterations in the search for an optimised and dense gate (FIG. 4-35 and references therein).

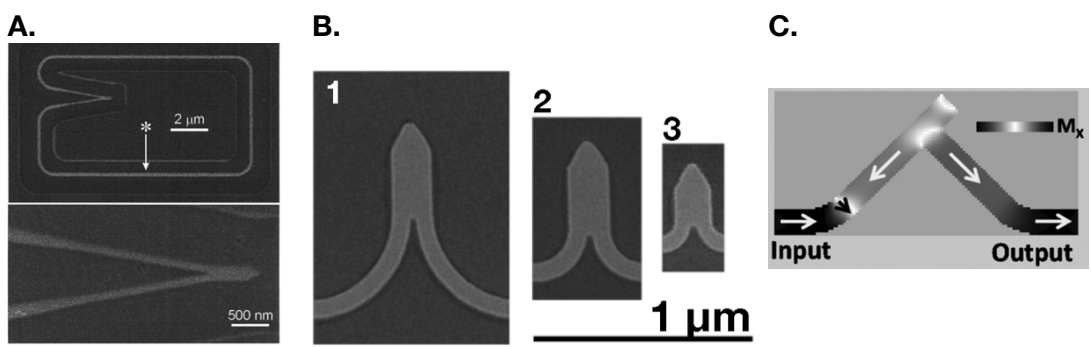

**FIG. 4-35 Different designs of the DW NOT gate. A.** Classic design, focused ion beam fabricated gate (SEM image), adapted from [Allwood et al. 2004]. **B.** Three different symmetric compact designs of EBL fabricated gates (SEM images), adapted from [Lewis 2010]. **C.** Asymmetric compact gate for dense and rectangular tessellation (simulation), adapted from [Zeng et al. 2010b].

Our interest in fabricating NOT gates in SV tracks is two-fold. First, being a complex structure with a complex behaviour, it provides a strong test to demonstrate the suitability of SV tracks to DW logic applications. Secondly, it provides a way to probe the internal magnetisation state of the gate during operation, which should aid the study and improvement of gate design (studies so far have relied on measurement of the input and output arms magnetisation).

## 4-3.1. Measuring a SV NOT gate

Several ring-shaped tracks containing one NOT gate were fabricated from a SV using the titanium etch process [30]. An SEM image of the structure can be seen in FIG. 4-36. The ring dimensions were 20 by 10 µm, width of 160 nm (ring track) tapering down to 60 nm in the cusp. Contacts were placed on the upper side of the ring, 4.8 µm apart. Though there are two current paths, the large difference in length (4.8 vs 57 µm), and the higher resistance of the narrow sections of the gate, make the signal correspond almost completely to segment between the contacts.

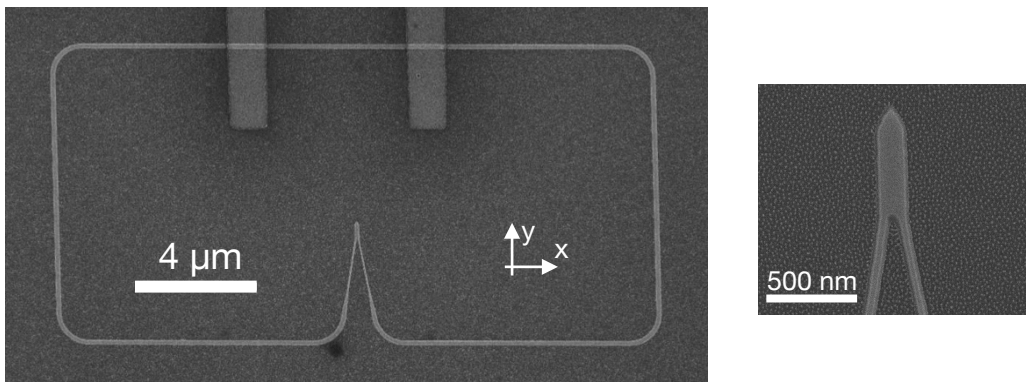

**FIG. 4-36 A SV ring with a NOT gate** (SEM image).

An elliptical CCW rotating field, $H_{CYCLE}$, was then applied. The horizontal and vertical amplitudes ($H_X$, $H_Y$) were varied, and some measurements at different field amplitudes are plotted in FIG. 4-37. Different behaviours were observed for different values of $H_{CYCLE}$, which could be grouped in three patterns:

   – at low $H_{CYCLE}$, no transitions or only intermittent transitions (at $H_X = H_{PR}$) were observed (**A**);

---

[30] Fabrication details can be found in Annex A, under sample reference HM01.

- at medium H<sub>CYCLE</sub>, transitions occurred every 1.5 field cycles (period of 3 field cycles; **B**);

- and at large H<sub>CYCLE</sub>, transitions occurred every field cycle (**C**).

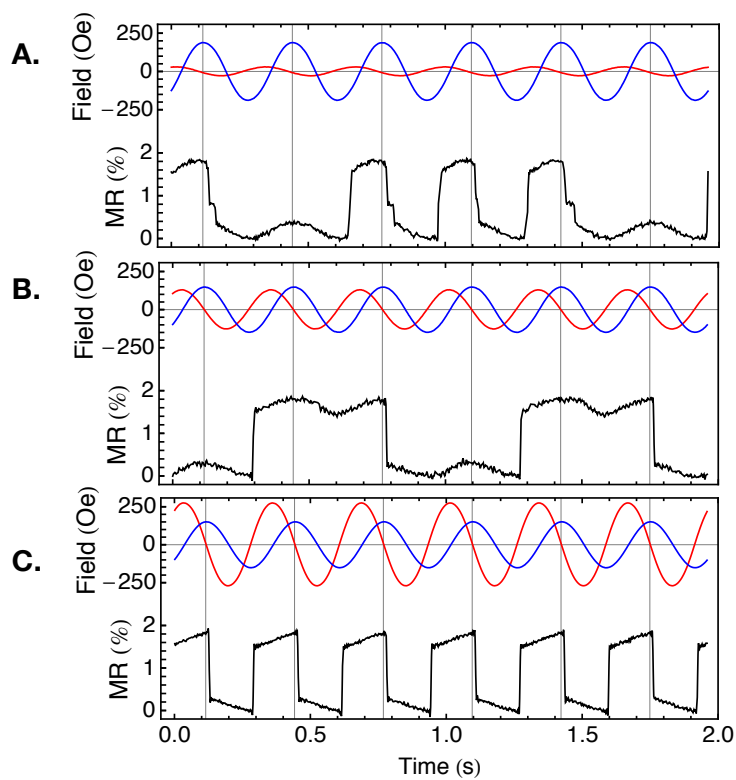

**FIG. 4-37 Ring with a NOT gate at three different field amplitudes** (MR measurement). The x and y field components are plotted in red and blue, respectively. The vertical grid lines mark field angle = 0. The field amplitudes for the three plots are **A.** (30, 189) Oe, **B.** (129, 149) Oe, and **C.** (273, 151) Oe. A and B are single shot measurements, C is averaged (n=5). Sinusoidal variation is caused by H<sub>Y</sub> sensitivity, cf. §3–5.

The first case (A) correspond to the case where the DW cannot overcome either the natural pinning of the track or the pinning of the NOT gate. As such, the only observed transitions are due to a DW going back and forth as H<sub>X</sub> switches, a DW that cannot complete a full ring turn [31].

The second case (B) corresponds to the correct NOT gate operation with a single DW in the ring, with its characteristic 3 cycle period. As mentioned above, the gate delays the DW by half a field turn and, as such, it takes 1+½ field turns for the DW to complete a full round. As it takes 2 trips for the DW to reverse back to the original polarity, the total period is then of 3 cycles. The sequence of magnetic states in this operation is schematised in Fɪɢ. 4-38A, with the applied field angle represented by the red arrow and the DWs by the green (HH) or yellow (TT) disks. The other stable mode of operation, with three DWs (Fɪɢ. 4-38B) was not observed. This was due to the way we tested different values of Hᴄʏᴄʟᴇ: we started always from low to high amplitudes. As the triple DW state is unstable at low fields, if ever it was present, it would have been annihilated during the measurement set.

---

[31] In the particular case of FIG. 4-37A, it is the H<sub>Y</sub> that is below H<sub>PR</sub>, leaving the DW corralled in the top segment of the ring.

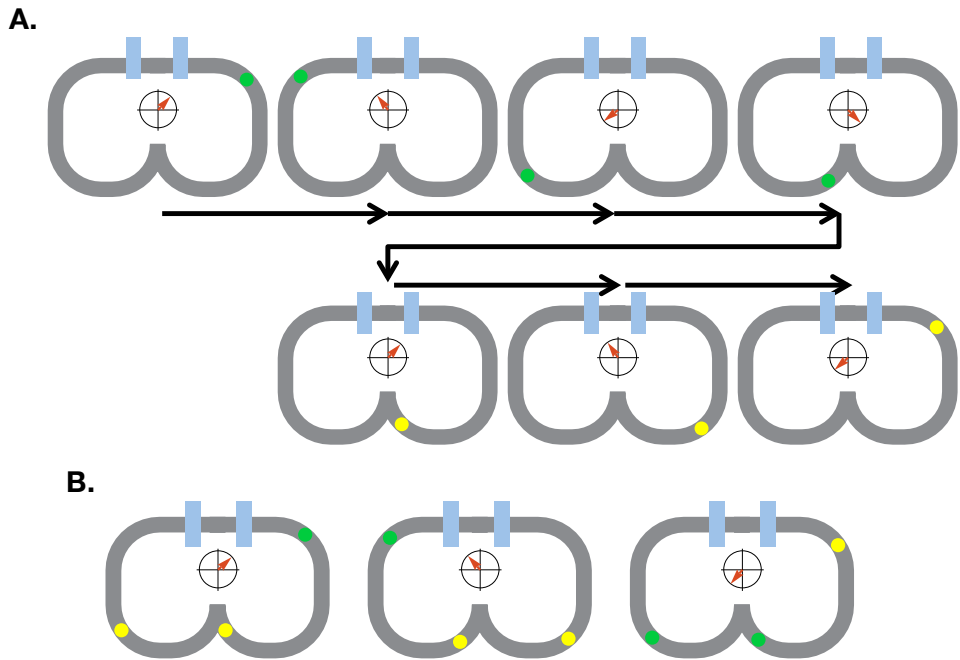

**Fɪɢ. 4-38 Operation of a ɴᴏᴛ gate in a ring track** (schematic). Half a period is represented for **A.** single DW mode and **B.** triple DW mode. The red arrow represents the applied field direction, the disks the DWs (green for HH and yellow for TT), and the blue rectangles the electrical contacts.

In the third case (C) the field is strong enough to nucleate new DWs (most probably at the gate) and the single DW state is no longer stable. In this structure, this regime is similar to the triple DW behaviour (Fɪɢ. 4-38B).

In all cases, due to the position of the contacts, the field at which the transitions occur are characteristic of the depinning from the corner and not of the ɴᴏᴛ gate itself.

## Operating margin

The three operation patterns we have identified above depend on the Hᴄʏᴄʟᴇ amplitudes in an asymmetrical and non-simple way. As before for the spiral device (cf. Fɪɢ. 4-6), the applied field amplitudes that yield the desired operation can be mapped, Fɪɢ. 4-39, and the *operating margin* (or *region*) can be found. This plot shows that the operating margin is limited at high fields by nucleation and at low fields by propagation, as was the case for simpler devices. Worthy of note is the vertical to horizontal asymmetry: while the nucleation limit is similar for both $H_X$ and $H_Y$ ($\approx$275 Oe), the propagation limit is very dissimilar ($\approx$40 vs $\approx$140 Oe). This is caused by the DW having to overcome a pinning step before entering the head of the cusp, something that happens when the field is oriented vertically [32].

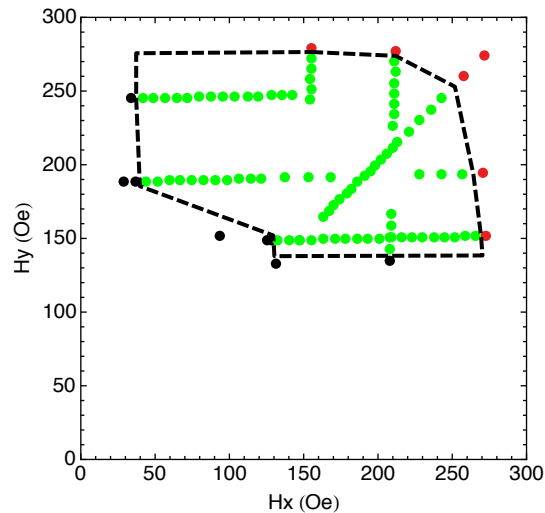

FIG. 4-39 **Operating margin of a NOT gate.** Each point represents a measurement like the ones of Fig. 4-37. The black points ● correspond to no operation (i.e. below gate transmission), the green points ● to correct operation, and the red points ● to nucleation. The dashed line marks the operating margin, interpolated between measurements or extended where unequivocally possible.

### Correct operation threshold criterion

As previously discussed, at low field values we observe intermittent transition behaviour, caused by DW propagating back and forth in the ring. As the field reaches the start of the operating area, the structure may intermittently show correct behaviour, interspersed with cycles during which the DW was not able to overcome the pinning field of the gate. It is then necessary to define the *threshold criterion*: in how many operation cycles we search for non-correct operation. For example, in the data of FIG. 4-37 and FIG. 4-38, this number would be 4 (4 processed DWs per measurement). While a limited number of observations increases the sampling error, for most structures and gate designs this is rarely significant to the determination of the operating margin, as the contrast of intermittent and correct operation is high. It is important to notice that in industrial devices, such as data-storage media, the criteria for correct operation are usually far stricter, with error rates measured in errors per $10^{15}$ operations [Mielke et al. 2008].

---

[32] We shall examine more closely the internal pinning fields below.

**Gate orientation**

An identical structure rotated by 90° was also fabricated, and its operating margin is shown in FIG. 4-40. In this structure, the gate is parallel to the reference layer. This rotation caused the observed x/y asymmetry of the operating margin to reverse, confirming its origin in the gate. Also, more interestingly, the margin for both structures is apparently of comparable size. To compare operating margins from different structures more quantitatively, we shall use the *normalised operation area* [Allwood et al. 2004]: the ratio of the margin area to the product of its centroid coordinates, describing loosely how much relative variation of the applied field is allowed while still obtaining the correct gate behaviour. The normalised operation area for the original (vertical) gate is 0.81 and for the horizontal gate is 0.82, indicating that the influence of the coupling field from the reference layer upon the gate operation is limited.

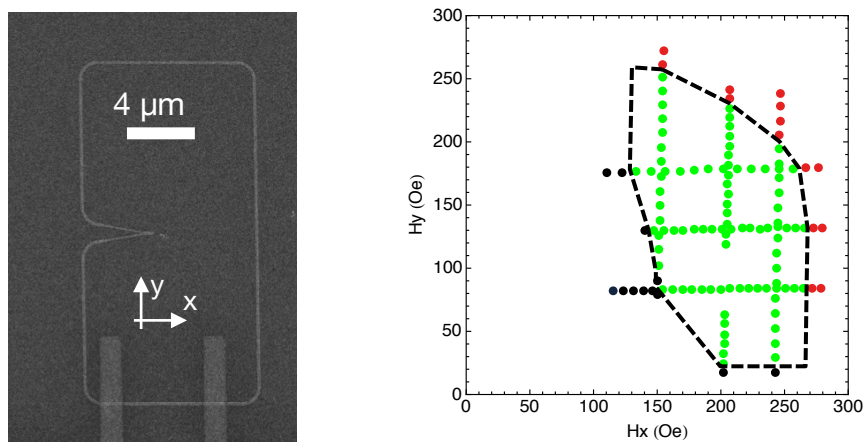

**FIG. 4-40 Ring with a horizontal NOT gate. Left.** SEM image. Electrical contacts are visible on the bottom. **Right.** Operating margin (same colour code as Fig. 4-39).

## 4-3.2. Different gate designs

We fabricated and measured the operating margins of several structures of three designs: the previous symmetric design (FIG. 4-36), a symmetric compact design (FIG. 4-41A), and an asymmetric compact design (FIG. 4-40B).

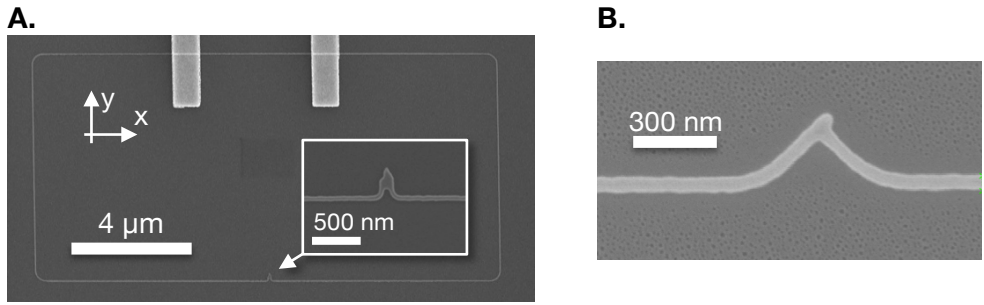

**FIG. 4-41 Compact NOT gate designs in sv tracks** (SEM images). **A.** Symmetric compact NOT gate (track width = 65 nm). **B.** Asymmetric compact NOT gate (track width 50 nm).

Both compact designs worked, i.e. showed a finite operating margin. The operation plots of both types of designs are shown in FIG. 4-42. The normalised operation areas for the shown structures are very different: 0.64 (symmetric design) and 0.072 (asymmetric). We shall compare them further below. The blue points in FIG. 4-42 correspond to the intermittent regime, to which we will return later.

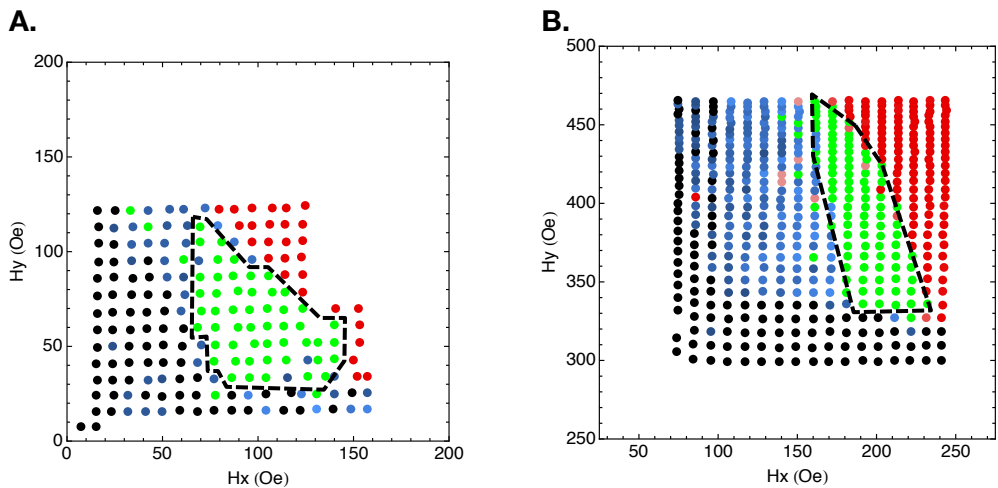

**FIG. 4-42 Operating margins of compact not gates. A.** Symmetric compact gate. **B.** Asymmetric compact gate (note the plot origin). The dashed line indicates the operating margin. Red, green, and black points correspond to nucleation ●, correct operation ●, and below propagation ● measurements, as before. The blue points ● correspond to intermittent operation, with dark/light blue indicating few/many observed transitions (i.e. close to and above the propagation threshold, resp.). Some noise is present caused by the transition counting algorithm. The operation threshold criteria were 4 (A.) and 8 (B.).

### Comparison to similar gates on single-layer Permalloy

A few more NOT gates were measured of each type, and their normalised operation areas are shown in FIG. 4-43. For comparison, the normalised operation areas of similar design gates fabricated on single-layer Py are also included (from the indicated sources). Gates for which no operation point was found were not included. Though the number of tested structures is low, as a general rule, the normalised operating margin of the SV gates is smaller than their Py counterparts (by about 12 to 75%). We suggest that this is due to a larger lateral roughness of the SV structures, particularly visible in the gates of reduced size (see related discussion of narrow tracks in Chapter 3).

| Design | N | Spin-valve Normalised operation area | Permalloy (literature) Normalised operation area |
|---|---|---|---|
| Classic | 3 | 0.82, 0.81, 0.81 | ~ 0.93 |
| Sym. compact | 3 | 0.64, 0.19, 0.17 | ~ 0.42–0.95 |
| Asym. compact | 4 | 0.10, 0.09, 0.08, 0.07 | ~ 0.39 |

**FIG. 4-43 Comparison of SV and Py NOT gates.** The Permalloy data was measured on plots in [Lewis 2010] for the classic and symmetric designs, and in [Zeng et al. 2010b] for the asymmetric compact design.

**Intermittent regime**

The asymmetric gate showed a very extensive intermittent operation region (in blue in FIG. 4-42B) which, as we described before, consists in two regimes: back and forth DW propagation without transmission through the gate, and intermittent correct operation. The former occurs for lower fields, and the latter for fields closer to the operation threshold. To study how gradually the intermittent correct operation regime fades into the full operation regime, 100 single shot measurements were taken at four values of applied field near the operation threshold of an asymmetric compact gate, FIG. 4-43. In the shown operating margin plot, the points where the DW propagates but does not transmit through the gate are shown in black (what would be blue in the previous plots). We observe the intermittent regime in a wide interval (at least 143–159 Oe), where the probability of correct operation stays at ~50%. A small step of the applied $H_x$—step smaller than 10 Oe, corresponding to 6% of the value of $H_x$—was sufficient to go from the intermittent to the full operation regime of 99% correct operation (a threshold criterion of ~100). This shows that the operating margin is sharply defined.

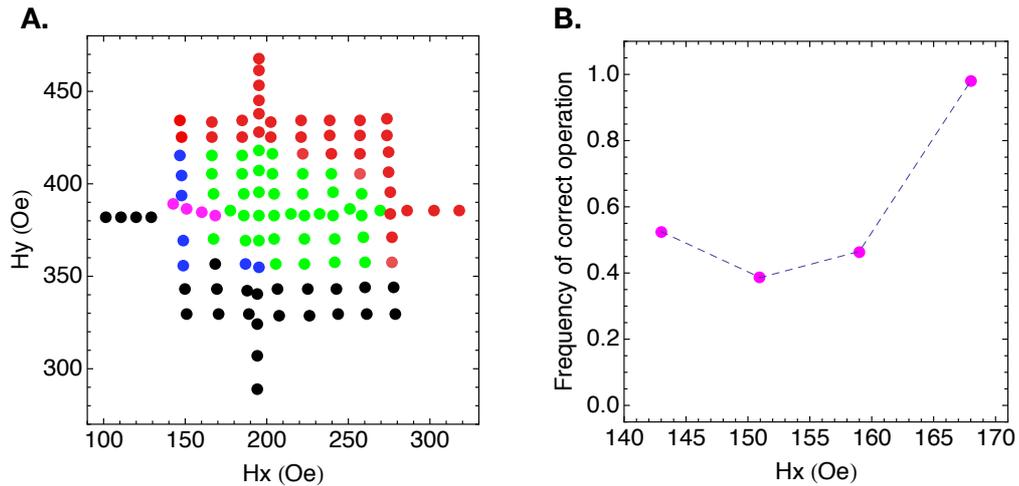

**FIG. 4-44 Intermittent operation in asymmetric compact NOT gates. A.** Operating margin. The green points ● mark correct operation; the blue points ● intermittent correct operation, but not cases where there was DW propagation but no transmission through the gate (as was the case on previous plots). Red and black points correspond to ● nucleation and ● no DW transmission. Purple points ● correspond to the measurements in plot B. **B.** Relative frequency of correct operation (in 100 tests per data point) vs applied $H_x$.

## 4-3.3. Probing the internal magnetisation

Structures were fabricated where the NOT gate was situated between the contacts, allowing for the measurement of magnetisation changes while the DW passed through.

**The classic design gate**

The measurement of a L-shaped wire with a classic NOT gate at three different applied field amplitudes is shown in FIG. 4-45. As with the ring structure, the track width is 160 nm (outside the gate) and 60 nm (in the cusp arms). The applied field sequence is composed of a reset field pulse, at +(200, 350) Oe, and 2 to 3 CCW field rotations at the chosen amplitude. The measurements are single shot acquisitions of a repeating field sequence. Due to setup limitations, the nucleation pulse is divided between the beginning and the end of the acquisition time (at 0 and ≈1.6–2.0 s). The quantitative determination of the MR contributions of each of the structure sections is unfortunately encumbered by the variety of track width and of track angle, and by the high $H_Y$ fields applied (which influence the MR signal, cf. Chapter 3).

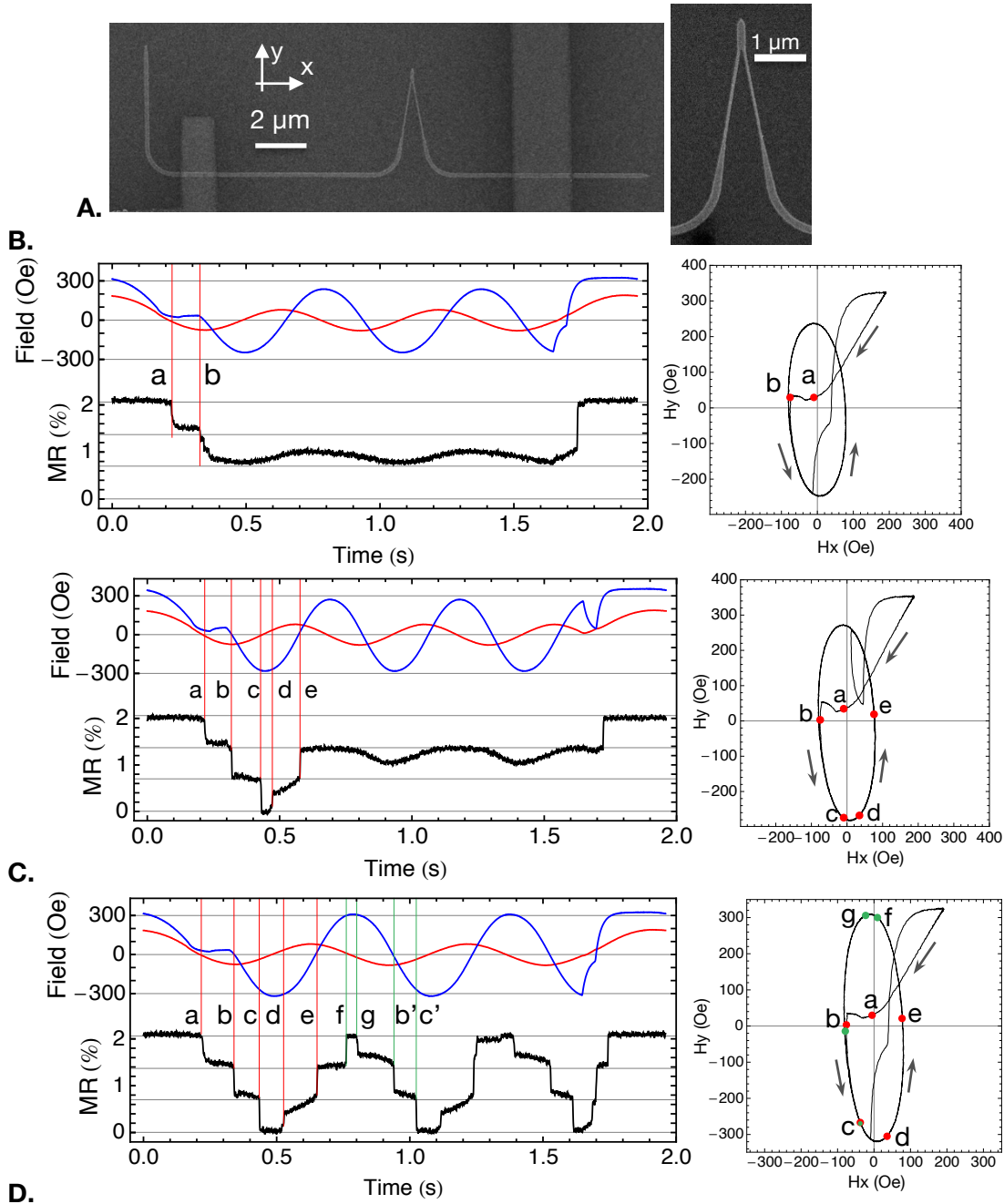

**FIG. 4-45 MR signal across a classic gate. A.** SEM image of a classic NOT gate on a L-shaped wire. **B, C & D.** MR signal across the gate, respectively: below the propagation threshold, correct operation, and nucleation. The red and blue curves are the applied $H_X$ and $H_Y$ (respectively). The field is replotted at the right to show the field angle at each transition.

On the first plot, two transitions are observed (*a* and *b*) plus a reset transition (at $t \approx 1.7$ s). Transitions *a* and *b* occur at negative $H_X$ (-11 and -74 Oe), and correspond to two injected TT DWs reversing the horizontal sections of the track (as schematised in FIG. 4-46). The amplitude of the transitions is 0.6 and 0.5% (respectively), indicating, albeit without certainty, that transition *a* corresponds to the DW to left of the gate (the left contact is separated by ≈6.5 μm from the gate, vs. ≈3 μm for the right contact). The very disparate field values indicate that one of the corners has a high pinning field.

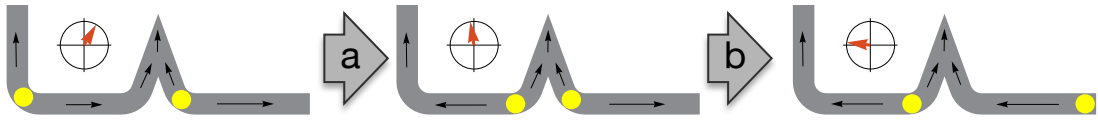

**Fɪɢ. 4-46 Magnetisation below the propagation threshold** (schematic). The ᴛᴛ ᴅᴡs are represented by yellow points. The red arrow shows the applied field direction.

In the measurement of the signal during correct operation (Fɪɢ. 4-45ᴄ), we observe the two transitions of before and three new (*c*, *d*, and *e*), which we shall map to the magnetisation changes of Fɪɢ. 4-47. All these transitions occur in the first half field turn after the reset pulse, and no transitions are observed afterwards, as is expected of the transit of one ᴅᴡ through the ɴᴏᴛ gate. The transitions *a–c* have the same signal (negative), as they all switch the magnetisation to the left (by propagation of a ᴛᴛ ᴅᴡ). Transitions *d* and *e* on the other hand are positive, as they correspond to magnetisation changing to the right (by propagation of a ʜʜ ᴅᴡ). Between *c* and *d*, the magnetisation of all segments has a –x component, yielding the lowest observed resistance state. The entrance and exit of the ᴅᴡ to and from the gate (transitions *c* and *d*) occur at almost vertical applied field (angle of -92° and -83°, resp.) and high field amplitude (-275 Oe), a sign of the high pinning caused by the interface between the cusp and the arms (which effectively is a width step of 120 to 60 nm). After the ᴅᴡ leaves the gate and leaves the track (transition *e*), the applied field continues rotating but no further transitions occur.

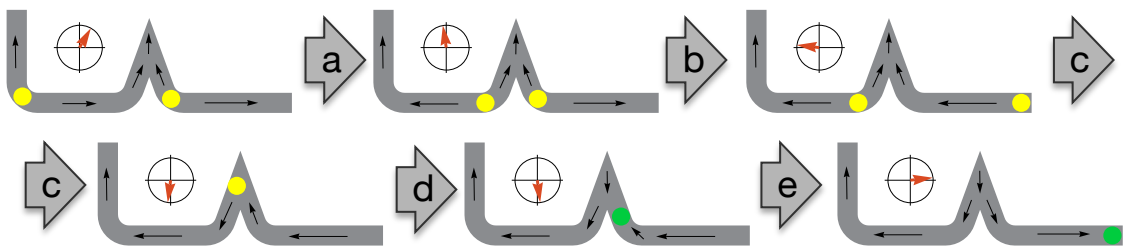

**Fɪɢ. 4-47 Magnetisation during correct operation** (schematic). The ᴅᴡs are represented by yellow and green points (for ᴛᴛ and ʜʜ, respectively).

Fɪɢ. 4-45ᴅ is a measurement above the nucleation threshold. The transitions observed before are present here again, showing that the ᴅᴡ still describes the same path as during correct operation. However, even after the original DWs have left the track, every time the field rotates new DWs are generated at the cusp (transitions $f$ and subsequent).

**The symmetric compact gate**

The measurement of the MR signal across a symmetric compact gate in a ring track is shown in FIG. 4-48. The electrodes are placed 5 μm apart with the gate in the middle. Due to the compact gate's much smaller size, and because it is of the same track width as the rest of the track, the gate resistance, and therefore its signal, is very small; as such, no transitions corresponding to internal magnetisation changes were observed.

As before, we show in the figure measurements corresponding to applied field amplitudes below the propagation threshold, in the operation area, and above the nucleation threshold.

In the operation area (FIG. 4-48C), we observe pairs of opposite signal transitions ($a$ & $b$, $c$ & $d$) separated by half a field turn. These are caused by the incoming and outgoing DWs of opposite polarity; eg. HH for $a$ and and TT $b$, respectively. The separation between the transitions (one half and then one full field turn) is characteristic of a looping single DW. Both transitions for the incoming of the DW and the outgoing DW occur at roughly opposing field angles. This simply corresponds to $H_X$ reaching the depinning field of the corners, which is similar to $H_{PR}$. The operation limiting depinning field, that from the cusp itself, is not observable here, due to the small amount of reversed magnetisation it accompanies.

The signal measured above nucleation (FIG. 4-48D) shows a full reversal transition every half field turn. This corresponds to the three DW mode: a DW enters the gate at roughly the same time as another DW is leaving, and their transitions are compounded. In some transitions, a stochastic variation of the depinning field allows the distinction of the two transitions, and a small step is visible.

The signal measured below the propagation threshold (FIG. 4-48B) is typical of the intermittent operation regime: stochastic variations of pinning fields make the correct operation of the gate possible but infrequent, interspersed by failed operations. The correct gate operation is seen in transitions a & b, but failure to transmit is also observed in c & d. The field angle at transition also shows a spread.

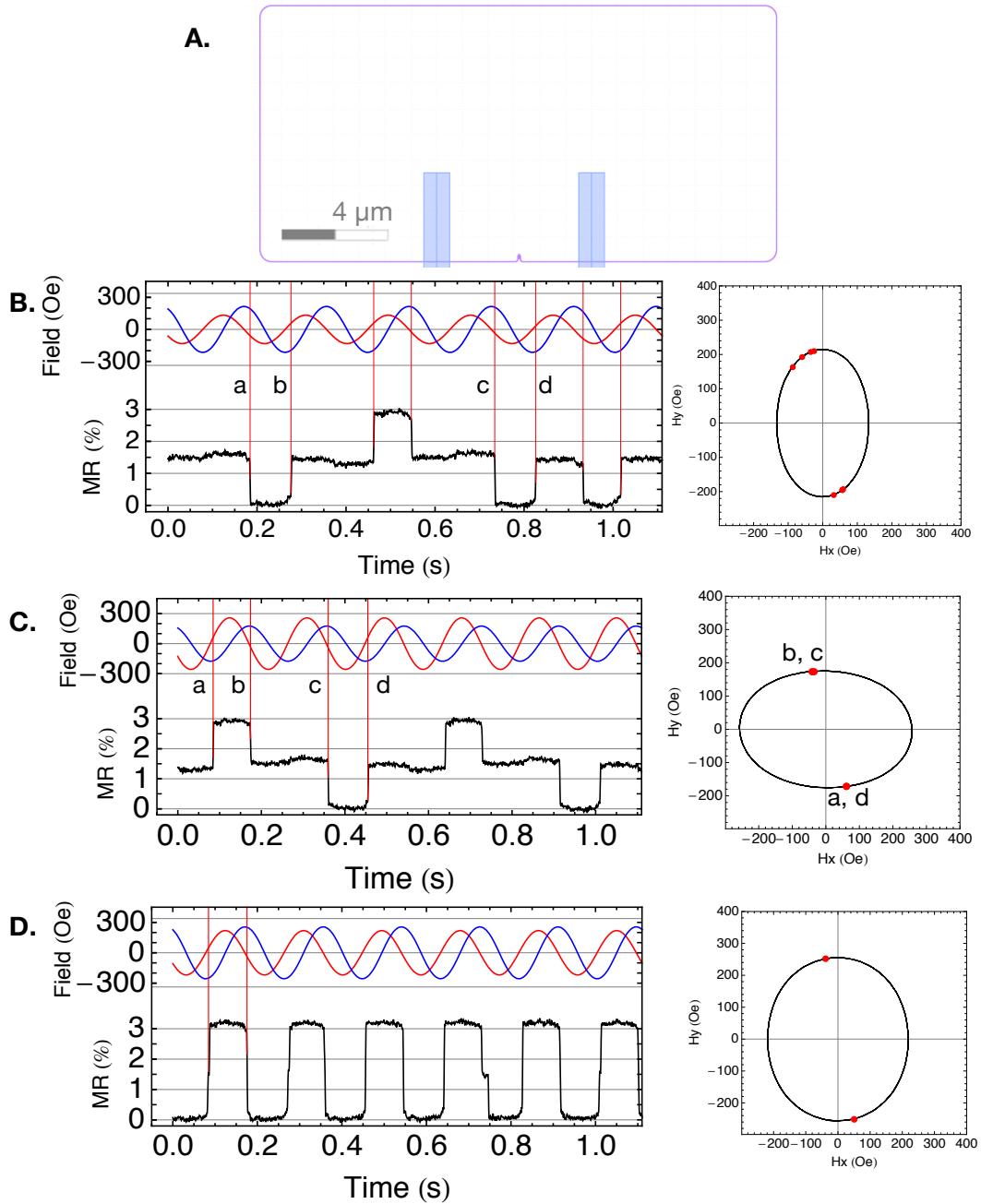

**FIG. 4-48 MR signal across a symmetric compact gate. A.** Mask of the measured structure. **B, C & E.** MR signal across the gate, respectively: below the propagation threshold, intermittent operation, correct operation, and nucleation. The red and blue curves are the applied $H_X$ and $H_Y$ (respectively).

## 4-3.4. Multiple NOT gates: a DW shift register

As mentioned before, the coupling of several NOT gates is very desirable for digital applications, as it allows the formation of data storing shift registers, frequency dividers, and module turn counters [Diegel et al. 2009]. In Permalloy, shift-registers consisting of multiple gates have been demonstrated for all three designs studied in this section [Allwood et al. 2002; O'Brien et al. 2009b; Zeng et al. 2010b]. There are two main challenges to a dense, multiple gate shift register: compounding operating margins, i.e. the operation points of the group have to lie inside the operating margins of all the gates, and the possible inter-gate interactions.

We have fabricated and measured working shift registers in SV tracks, of both symmetric and asymmetric (not presented) gate design. FIG. 4-49 shows measurements on a seven gate shift register of the symmetric compact design embedded in a ring track. The MR measurement was done with two electrodes 5 μm apart with the gates in between (see FIG. 4-49A). As before, due to the small gate size, internal magnetisation changes in the gates have a very small effect in the MR signal. The signal above the nucleation threshold is shown in FIG. 4-49D with the typical two transitions per field turn.

The signal at correct operation is shown in FIG. 4-49C, with the expected half-period of 4.5 field turns of the single DW mode. This half-period is the sum of 1 field turn for the DW to go around the ring, plus 7 x ½ field turns to go through the seven gates. Being an odd number of gates, the DW polarity is reversed after passing through the gates, and as such the magnetisation period is actually 9 field turns. This is easily seen in the recorded signal. At the (negative) transition marked in red, a TT DW leaves the shift register through the right. After one field turn, it has circled the ring and appears again at the left of the gates, causing another negative transition (in green). Then, after 3.5 field turns, the DW emerges again in the right, this time being a HH DW, and thus causing a positive transition (in purple).

Two structures of this type were measured, with normalised operation areas of 0.19 and 0.33 [33], which is comparable to that of single gate structures (0.17–0.64, cf. FIG. 4-43).

---

[33] This value, 0.33, was measured on the plot of FIG. 4-49B, where the operation area was not completely determined; as a consequence, the real value is possibly greater than 0.33.

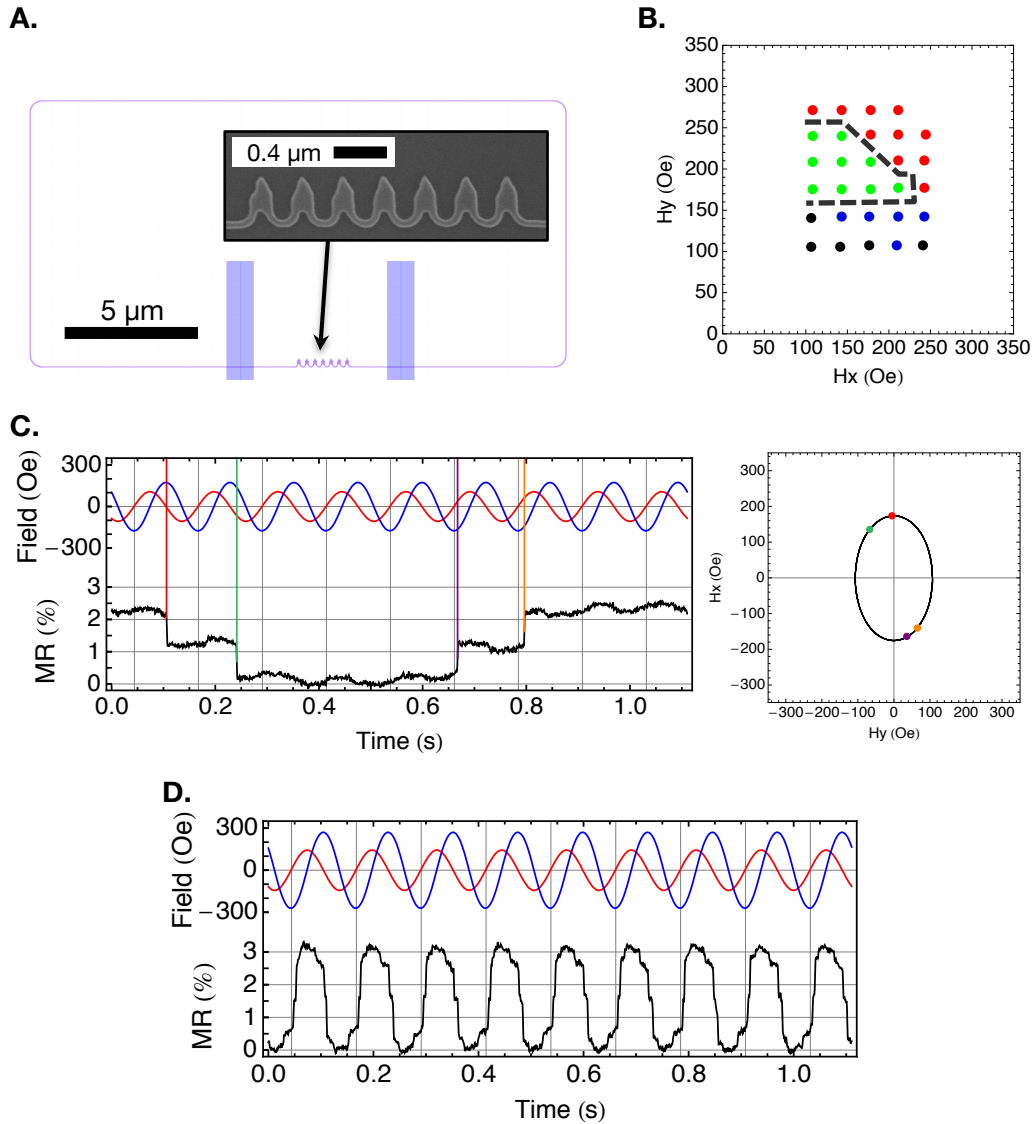

**FIG. 4-49 Shift-register with seven symmetric compact NOT gates. A.** Mask design and SEM image. **B.** Operating margin (● – correct operation, ● – nucleation, ● – no propagation; ● – intermittent operation). **C.** Measurement (single shot) of a correct operation single DW mode. **D.** Measurement (single shot) of the nucleation mode.

## 4-3.5. Summary

We measured DW NOT gates in a SV track, the first electrical measurement of such gates to this date. We have demonstrated the operation **three types of DW NOT gates,** including highly compact gates with narrow tracks (down to 50 nm width). For all these, we have measured the **operating margin**, and compared it to reported experiments on Permalloy structures. We found that the margins in SV tracks were smaller, though of comparable size (a reduction in the margin of 12 to 75%). We also observed that **gates in different orientations** had operating margins of identical size, which suggests that the pinning caused by the reference layer is small.

Using single shot measurements, we observed that the asymmetric compact NOT gate showed an **intermittent operating behaviour** for a wide region of applied field amplitudes. We characterised the border between the intermittent and correct regimes, and found that this border was relatively narrow, with a change of 50% to 99% correct operation occurring in just 10 Oe change of the applied field.

By placing electrical contacts in both sides of the track, we characterised the magnetic transitions of the gate during operation.

We then demonstrated the operation of a **shift register** composed of seven adjacent NOT gates of the symmetric compact design. We measured its operating margin, and found it to be of comparable size to the margin of single gates. To this date and to our knowledge, this is arguably the most complex DW logic circuit to be demonstrated in SV tracks.

## 4-4. Inter-track DW interaction

We have seen that to control DW propagation and pinning in SV tracks it is necessary to limit the action of magnetostatic coupling to the reference layer; the reasons for which we have concluded for the use of SAF reference layers in Chapter 3. If tracks are placed very close together, as is needed for many logic applications, magnetostatic coupling effects may occur between adjacent tracks which affect the propagation of DWs.

These coupling effects arise wherever there is magnetic charge, such as track ends, or junctions, or DWs. Here, magnetic charge is defined as $q_M = -\mu_0 \int \nabla \cdot M \, dV$, leading to a calculation of the magnetostatic energy between two magnetic charges that is similar in form similar to the electrostatic attraction ([34]):

$$U = \frac{q_m \, q'_m}{4\pi\mu_0 \, r}$$

Inter-track DW interactions can be quite significant; previous work has shown that DWs of opposite polarity in closely placed Permalloy tracks attractively pin one another, with decoupling fields well above propagation (up to ~5x) [O'Brien et al. 2009a]. It was also found that this interaction also influences the stability of different DW structures [Laufenberg et al. 2006]. It was shown that the magnetostatic interaction locks the DWs together with minimal structure deformation in a virtually parabolic potential well [35]. This effect is extremely sensitive on track separation and on charge distribution within the DWs. Transverse DWs, having most of their charge concentrated at the track edge on their wide side [Zeng et al. 2010a], show the strongest coupling when interacting wide with wide side, while other configurations (wide with narrow transverse DWs and vortex DWs) show coupling strengths orders of magnitude smaller. The interaction of adjacent DWs has possible technological applications, such as DW oscillators [O'Brien 2010], while the possibility of distortion free pinning is useful for fundamental studies of DWs in nanotracks.

---

[34] As further described in Chapter 1.

In this section, we study the interaction of DWs in two adjacent SV tracks. The ability to monitor simultaneously the positions of the two DWs, and to take single-shot measurements, will allow us to observe directly the inter-DW pinning, and to study new phenomena arising from the interaction. It also serves to demonstrate that DW interactions occur equally in realisable SV structures, an important step towards practical applications to future devices.

## Device design

The structures consisted of two tracks, one L-shaped and the other U-shaped, placed in close proximity, FIG. 4-50, fabricated using the usual titanium etch process [36]. The track gap varied down to ~50 nm (two track widths were used: 140 and 260 nm). Three electrodes were placed in a common ground arrangement (the ground contact being the rightmost one). The inter-electrode distance is 1.7 μm on the L-track and 2.4 μm on the U-track. The DW positions will be described by their distance to the edge of the common electrode, *d*. During the contact lithography, the sample was semi-automatically re-aligned on each structure exposure for better contact position accuracy (typical deviation <100 nm).

---

[35] in the limit of small DW separation and for transverse DWs. In the limit of large separations, the interaction follows a -1/x profile typical of a point-charge interaction.

[36] Fabrication details may be consulted in Annex A, under sample reference HM18.

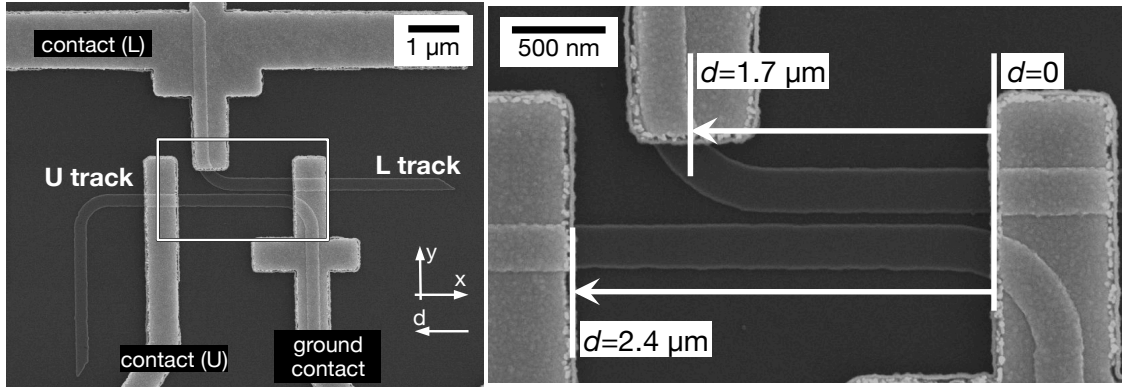

**FIG. 4-50 Parallel tracks for DW interaction** (SEM images). The L-shaped and U-shaped tracks can be seen on the top and bottom of the structure, respectively. The distances between the common (ground) electrode and each of the track electrodes is marked on the image on the right. The DW positions, $d_L$ and $d_U$, are defined as the distance to the common electrode. Both tracks have a 4.7 μm horizontal segment, with ~2.2 μm of it in close proximity, width = 260 nm.

The configuration of the tracks was chosen so two transverse DWs could be injected by a (-H$_{Reset}$, -H$_{Reset}$) field (FIG. 4-51A-i) and made to propagate with their wide sides facing each other when driven by a sweeping horizontal field (FIG. 4-51A-ii, iii). This measurement we label as the *interaction measurement* (which is actually identical to the propagation sequence used in Chapter 3). The use of an L- and U-shaped tracks allows also the testing of an important control: propagation of a single DW in the U-track without interaction (FIG. 4-51B). This is done by using a (-H$_{Reset}$, +H$_{Reset}$) pulse, which creates a single DW in the U-track. This measurement we label as the *control measurement* (which is identical to the nucleation sequence of Chapter 3).

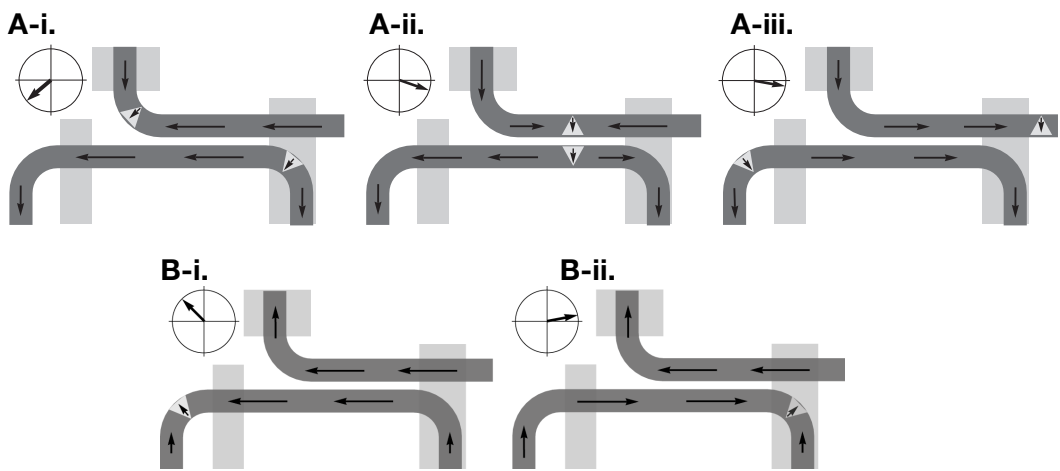

**FIG. 4-51 Inter DW coupling** (schematic). The red arrow indicates the applied field direction and the blue rectangles the electrode positions. **A.** Interaction measurement. Two DWs of opposing polarity are injected by a (-x, -y) reset field (**i**), followed by a sweeping horizontal field. At $H_x \approx H_{PR}$, the DWs propagate from the corners, becoming coupled at some point in the wire (**ii**); as $H_x$ reaches the decoupling field, they depin (**iii**).

The reverse case is done analogously, by reversing all applied field arrows. **B.** Control measurement. The reset field in the (-x, +y) direction injects a single DW in the U- track (**i**); the sweeping $H_X$ then causes the DW to propagate without any DW-DW interaction (**ii**). As $H_X$ reaches $H_{NUC}$, the L-track reverses by nucleation (not shown).

## Detecting DW-DW interaction events

The progress of the DWs in the interaction measurement should be similar to the case where the DW pins at a track defect (e.g. FIG. 4-12): reversal of a portion of the track at $H_{PR}$, up to the point of interaction, followed by the reversal of the rest of the track at some $H_{Pin}$ caused by the DW-DW interaction. If just one of the tracks were to be measured, the DW-DW coupling would be hard to distinguish from pinning at one of the many natural defects that unfortunately are common in these closely placed tracks. The distinguishing feature of a DW pair interaction is coincidence in time and position of DW displacements in both tracks, leading to the need to measure the resistance of both tracks simultaneously. Though the non-simultaneous measurement of both tracks could also detect DW-DW interactions by singling out those transitions with the same depinning field (analogous to [O'Brien et al. 2009a]), in the presence of significant pinning centres with stochastic depinning fields, this technique is far less accurate and sensitive than the simultaneous alternative.

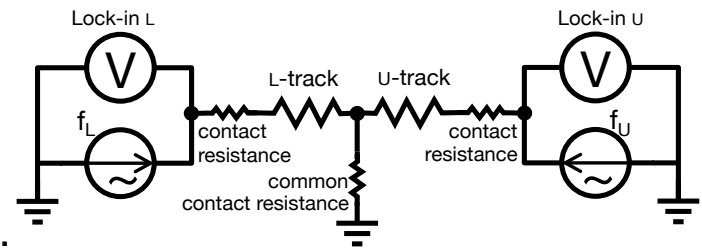

FIG. 4-52 **Electrical setup for simultaneous measurements.** Two currents modulated at different frequencies ($f_L$ and $f_U$, from 4 to 40 kHz) are injected in the two individual electrodes, while the common electrode is grounded. As the output resistances of the lock-in amplifiers and current sources are larger than the track resistance (1 MΩ and 100 kΩ vs ~500 Ω), the currents each only flow through the respective track. Cross-talk between leads is further avoided by the frequency discrimination of the lock-in amplifiers.

*Signal cross-talk* between the resistance measurement needs to be avoided: if a transition in one track would produce voltage steps in both signals, it would be indistinguishable from a DW interaction event. With a less careful measuring arrangement, cross-talk can easily occur in a system with a shared electrode and significant electrode resistance. The setup used to carry this experiment is schematised in FIG. 4-52. Two independent lock-in amplifiers, and two reference sinusoidal current sources, were connected to the two track electrodes, and the common electrode was grounded. As the output resistances of the amplifiers and the signal sources are much greater than the track resistance (1 MΩ and 100 kΩ vs ~500 Ω), the current flows from each signal source, through the respective track, to the common ground electrode, without flowing through the opposite track. The two reference signal sources were also set to different frequencies, so that voltage signal caused by the resistance of the common electrode could not induce cross-talk. The control experiment (FIG. 4-51B), which creates large amplitude and non-coincident MR transitions, can be used to confirm that no cross-talk exists, and is shown in FIG. 4-53. There, it can be seen that the reversal transitions of each track do not in fact produce any visible transitions in the other track's signal, confirming the isolation of the two simultaneous resistance measurements. The difference in the noise level between the two measurements, visible in FIG. 4-53 and in other data in this section, is both a result of the use of two different lock-in amplifiers (at different modulation frequencies and with different noise filters) and of different intrinsic noise levels (due to different resistance levels, contact quality, and other measurement limitations).

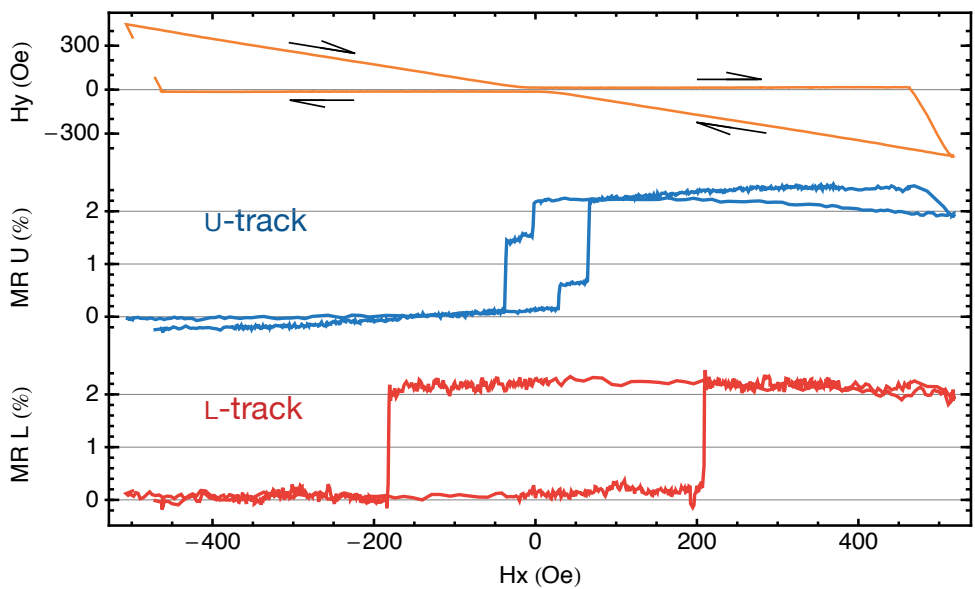

**FIG. 4-53 Signal of the L- and U-tracks during a control measurement** (simultaneous, single-shot measurement). The L-track reversed at $H_{NUC}$ (~200 Oe) and the U-track at $H_{PR}$ (~16 Oe). The step on the top curve is caused by a (natural) pinning centre ($H_{Pin}$ = 50 Oe). Track width = 260 nm. Note that there are no coincident transitions.

## 4-4.1. Interaction measurement: DW–DW pinning

An example of an interaction measurement is shown in FIG. 4-54A, where MR vs. field is shown for both the U- and L-tracks. For clarity, only the positive sweep, 0 to +150 Oe, is shown. The two simultaneously measured MR signals show several transitions in the 10–50 Oe region, one of them corresponding to inter-DW pinning (as we shall see), the others to pinning at multiple natural defects. In both tracks, the transitions are positive, corresponding to a HH DW travelling rightwards in the L-track and a TT DW travelling leftwards in the U-track (cf. FIG. 4-51A).

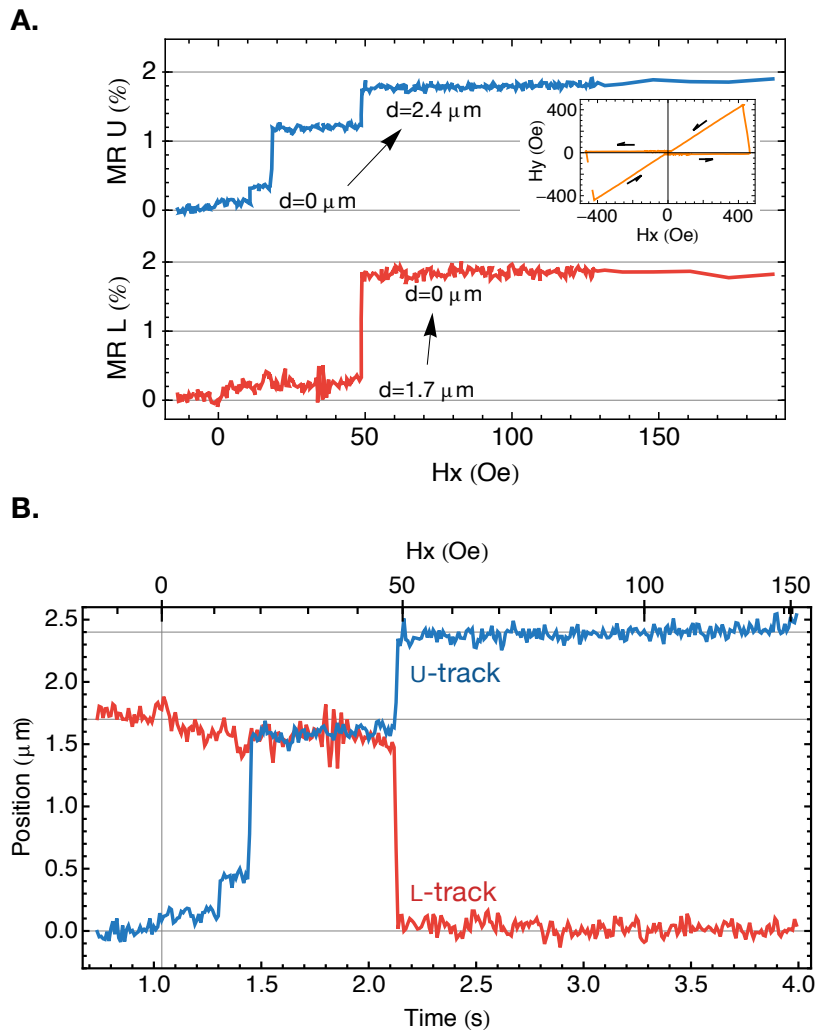

FIG. 4-54 **Inter-DW pinning and depinning in adjacent tracks** (simultaneous, single-shot measurement). Track width is 260 nm, separation 70 nm. The red and blue curves correspond to the DWs in the L- and U-tracks, respectively. **A.** Signal vs. sweeping $H_x$ during an interaction sequence (field shown in the **inset**). For clarity, only the data during the positive $H_x$ sweep is shown. **B.** DW position ($d$, cf. FIG. 4-50) versus time and $H_x$ (same data). The blue curve is the position of the DW in the U-track, the red of the DW in the L-track. The transition pair at 49 Oe occurred simultaneously (within the error described in the text).

The measured MR signal can be linearly transformed to DW position $d$, FIG. 4-54B. Note that in the extremes, $d$=0 or 1.7 μm for the L-track and $d$=0 or 2.4 μm for the U-track, the DWs are not in between the contacts and can actually be far away from the measured segments (this is particularly true for the final states). In this plot, we see that the DW in the U-track first depins from its corner at 12 Oe, is pinned at some natural defect at $d$=0.43 μm, until at 19 Oe it travels to $d$=1.58 μm where it meets the DW in the L-track, which is still at its initial corner. The two DWs are coupled together, until they finally decouple at 49 Oe. While coupled, the positions of the DWs are virtually identical, differing by only $12 \pm 20$ nm (averaged in the interval [1.45, 2] s), which is far within the accuracy of this measurement (see section below). This measurement was repeated on the same structure, with similar observations, though with some variation of the field at which the pair is decoupled (ranging 36–49 Oe in 10 measurements). This variation will be analysed further below.

In order to confirm that these observations do indeed correspond to a DW-DW interaction event, we will examine the likeness of the *null hypothesis*: that the two DW depinning events are independent and their simultaneity is a random coincidence. For this, we will analyse the accuracy of the field and position measurements.

**Simultaneity of the measurement**

It is conceivable that two uncoupled DWs depin from different defects with very similar pinning strengths. In order to avoid erroneously counting such a case as a DW coupling event, we shall estimate how close in depinning field two independent depinning events would have to be in order to be mistakenly counted. The parameters of the measurement in FIG. 4-54 are a typical example of the values used throughout this section:

    – Field rate [37]                    23 ms/Oe

    – Digitalisation step [38]         9.1 ms

    – Measured transition duration [39]    1–2 digitalisation steps.

---

[37] In the region of interest. In other measurements, this parameter ranges 23–46ms/Oe.

[38] This value ranges 3–9.1 ms in other measurements.

[39] The real transition is virtually instantaneous in the timescales being considered. This duration is generated by the lock-in noise filters and time integration.

The transitions in the two tracks were simultaneous within one digitalisation step. In values of field, that means they occurred very closely within 0.4 Oe of each other at 49 Oe. This difference is very small, and is orders-of-magnitude greater than the shot-to-shot variation of the simultaneous depinning field (which ranged 36–49 Oe in ten single shot measurements of the same structure; see analysis further below). Furthermore, in structures with a larger track separation, no simultaneous transitions were observed, another indication against the hypothesis of uncorrelated coincidence. It is thus extremely improbable that the measured simultaneous depinning events are uncorrelated.

**Position accuracy**

As described in Chapter 3, the determination of DW position from the MR signal suffers mainly from a problem of accuracy: determining accurately the zero of the distance scale while in the presence of contact position error ($\lesssim 0.1$ μm) and finite contact width (0.5 μm). Furthermore, part of the track corners is between the contacts. Unlike the reversal of a straight track section, the progress of a DW in a corner consists of numerous small steps, with a varying MR contribution that is hard to quantify. This makes the calibration of position $d$ versus MR level less accurate and more prone to measurement-to-measurement variation.

The noise level also limits the precision of the measurement. Though the noise varied greatly from structure to structure, we can take as an illustration the measurement of FIG. 4-54, where the noise level was equivalent to ±60 nm [40]. As the DWs are usually static for a large number of measurement points that can be averaged (reducing the position error by $1/\sqrt{n}$, with $n$ typically greater than 10), this is less of a limitation [41].

We can conclude then that we have observed an interaction event, where the two DWs were coupled by the magnetostatic interaction, and decoupled simultaneously. This conclusion is based on three factors: the close correspondence of the positions of both DWs, the simultaneity of their depinning, and the fact that, though the depinning field varied for different measurements, it was always the same for the two DWs.

---

[40] this value is the standard deviation of the L signal in the interval [2.2, 3.5] s (FIG. 4-54).

[41] Noise filters have been applied in the making of several of the shown plots for clarity sake. All the measurements, though, refer to the unfiltered signal.

## Stochastic variation of pinning field and position

The interaction measurement on the same structure of FIG. 4-54 was repeated 10 times and, in every measurement, the DWs pinned closely together and depinned simultaneously, though at different field values and positions. The depinning field and the DW positions before depinning ($d_L$ and $d_U$) are plotted in FIG. 4-55. The observed difference between $d_L$ and $d_U$ ranges 0.0–0.2 μm, which is within the expected accuracy for $d_L = d_U$, as discussed above.

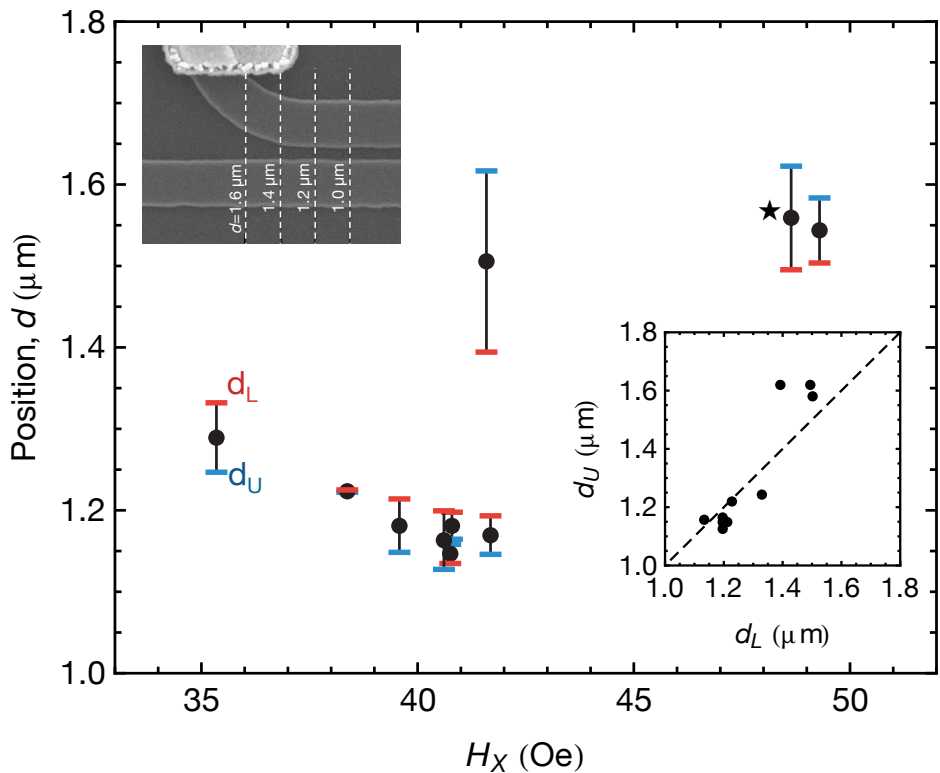

**FIG. 4-55 Stochastic variation of DW-DW depinning.** Plot showing pinned position vs. depinning field in several measurements on the same structure as FIG. 4-54, width=260 nm. The red and blue dashes mark the positions of the DWs in the L- and U-tracks, respectively, and the black circles mark the average of the two. The star (★) marks the measurement of FIG. 4-54. **Inset (top):** SEM image of part of the structure, showing some values of the position $d$ (same data as in FIG. 4-50). **Inset (bottom):** Plot of $d_U$ vs. $d_L$.

These single-shot simultaneous measurements reveal that there is more than one position where the DWs are able to pin together, at d≈1.2 and d≈1.6 μm, and that their depinning field also varied significantly, 35–49 Oe. Both variations, of position and field, are greater than the experimental error (the field measurement noise is ±1 Oe, and the position error between measurements may be estimated from the position difference of the coupled DW pairs).

Similarly to the depinning from defects, a small stochastic variation in the decoupling field is to be expected. The data in FIG. 4-55, though, shows a variation much larger than what was observed in the structures presented before. In reported studies of SV tracks, the stochastic variation of depinning fields was found to be quite large, sometimes of the order of the depinning field itself [Himeno et al. 2003; Lacour et al. 2004; Briones et al. 2008; Jiang et al. 2010; Mihai et al. 2011]. The source of this variation is not however completely understood, with some authors suggesting the existence of multiple adjacent pinning sites or small differences in the coupling of the free and reference layers to explain this variation.

A possible explanation for the variation seen here is then that the DW pair is sitting at different natural defects; as these DWs are located very near the L-track arc (see inset image in FIG. 4-55), different positions lead to different inter-DW distances, changing the DW coupling strength. As the inter-DW distance increases with increasing d, this hypothesis predicts stronger pinning for lower values d, which is not observed. A second hypothesis is that the strength of the pair coupling is modulated by the geometry of the particular natural defect whereat the DW pair sits. As the pair is pinned at different adjacent defects, its coupling strength changes. The defects are after all geometrical indents or protrusions on the track edge. These change the edge magnetic charge distribution and alter the coupling strength, as the coupling is caused almost completely by those edge charges [O'Brien et al. 2009a].

On other structures, it is not always the case that the DWs pin in every measurement. This can be understood to result from the stochastic variation of the $H_{Pin}$, which can cause $H_{PR}$ to be greater than the depinning field in some measurements and lower in other measurements. In some other structures, it is also not always the case that pinned DWs depin together. If one of the DWs is sitting at a natural defect stronger than the inter-DW pinning, the DWs will separate before both leave the initial position. Simultaneous depinning is then only observed when $H_{PR} < H_{Pin}$ (pair) < $H_{Pin}$ (defect).

Concluding, we observed DW-DW interaction, and these measurements show that these interactions are more complex in tracks with significant pinning sites. In particular, we observed two phenomena in the DW-DW interaction in SV tracks that were undetected in earlier studies of Permalloy tracks [O'Brien et al. 2009a]. First, that the DW pair can sit at different positions along the tracks. Variations of pair pinning position in Permalloy tracks, should they have existed, would not be detected by the referenced study. Secondly, that depinning field values can be multiple. For the latter, we suggested that it is caused by the DW pair being pinned at different natural defects, defects which modulate the DW-DW interaction. No variation of depinning field was observed Permalloy possibly due to smoother edges of those single-layer tracks.

## 4-4.2. Distant attraction

It was observed in a large portion of the structures that two unpaired DWs, at some initial (horizontal) distance, moved simultaneously towards each other. FIG. 4-56 shows two examples of such measurements, where we observe DWs initially separated by 1.1 and 1.4 μm moving simultaneously to a coupled position. In different measurements and structures, this simultaneous movement occurred at varied values of applied field, and of initial and final positions.

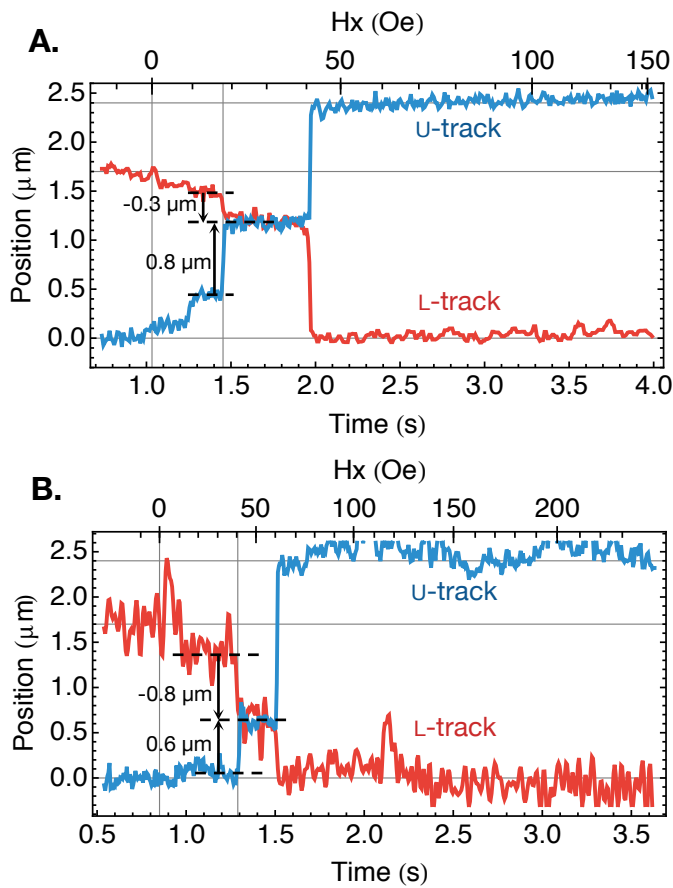

FIG. 4-56 **Attraction between two DWs.** In these measurements, the two DWs moved simultaneously to the coupled position. The red and blue curves correspond to the DWs in the L- and U-tracks. **A.** Measurement on the same structure as FIG. 4-54, track width 260 nm, separation 70 nm. The DWs move together at $H_x$=18 Oe, moving +0.8 and -0.3 μm. At $H_x$=61 Oe the DWs depin simultaneously. **B.** Measurement on a 140 nm wide track. The DWs moved +0.6 and -0.8 μm, at $H_x$=41 Oe.

Analogously to the argument shown before, the hypothesis that these movements are independent and just coincidental can be rejected. It would imply that the two independent pinning sites would have $H_1 \approx H_2$ within the measurement resolution of ~0.4 Oe, which is improbable. Also, the shot-to-shot variation of the field at which the DWs moved together (~2 Oe) was larger than the maximum difference between the DW movements (~0.4 Oe).

Though they are correlated, they are not simultaneous. They appear so in the measurement because our setup is orders-of-magnitude slower than the DW movements (few ms vs 100s of ns). In FIG. 4-57, the DW movements are schematised. The two DWs are pinned at two sites separated by $x_i$ (**i**), with different depinning fields ($H_1 < H_2$), under a sweeping horizontal field $H_X$, and mutually attracted by a magnetostatic interaction $H_{Interaction}(x_i)$. As $H_X$ reaches $H_1 - H_{Interaction}(x_i)$, one DWs moves and pins at the final position (**ii**). With a smaller inter-DW separation, the magnetostatic attraction is strengthened, and $H_X + H_{Interaction}(x_{ii}) > H_2$, causing the second DW to move and couple to the first (**iii**).

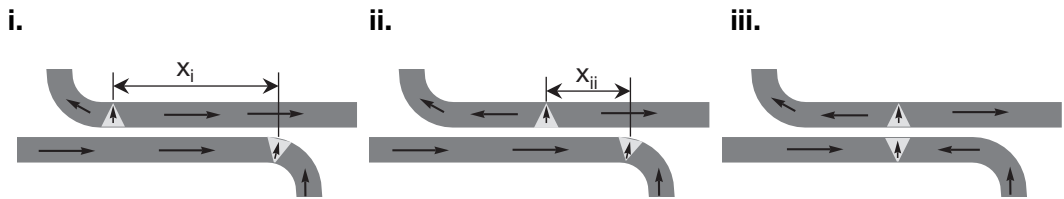

**FIG. 4-57 Mechanism of simultaneous DW encounter. i.** The DWs are initially pinned at two natural defects with pinning fields $H_1$ and $H_2$. **ii.** As $H_X$ reaches $H_1$, one of the DWs moves to a defect of higher pinning field, $H_3$. The magnetostatic interaction field, $H_{Magnetostatic}$, which is approximately inversely proportional to the DW distance, is now stronger. **iii.** The applied $H_X$ plus the increased of $H_{Magnetostatic}$ are now larger than $H_2$, causing the second DW to depin, move, and pin at the first DW.

For this mechanism (FIG. 4-57) to be likely, it is necessary that the change in the magnetostatic attraction, $H_{Interaction}(x_{ii}) - H_{Interaction}(x_i)$, be significantly larger than the field resolution (0.4 Oe) [42]. To test this, the following simulation was performed.

---

[42] Otherwise it would be as frequent as the hypothesis, rejected before, of having two independent pinning sites with $H_1 \approx H_2$.

**Strength of the interaction** [43]

A micromagnetic simulation of the average stray field created by one DW on the other, $\langle H_{\text{Magnetostatic}} \rangle$, is plotted versus DW separation in FIG. 4-58. When the DWs are farther away than the characteristic dimension of its charge distribution, such as is the case in FIG. 4-56, this is a good approximation of $H_{\text{Interaction}}$, i.e. the actual non-uniform magnetostatic interaction field. As the DWs come close together, however, the presented curve underestimates $H_{\text{Interaction}}$.

For the case of FIG. 4-56B, where the DWs were initially at 1.4 μm from each other, the simulation predicts that $\langle H_{\text{Magnetostatic}} \rangle$ was 1.2 Oe. The first DW dislocation was either 0.6 or 0.8 μm, depending on which DW moved first, and the simulation predicts that $\langle H_{\text{Magnetostatic}} \rangle$ increased to 3.2 or 5.6 Oe, respectively. Both these values are significantly larger than the field resolution (~0.4 Oe).

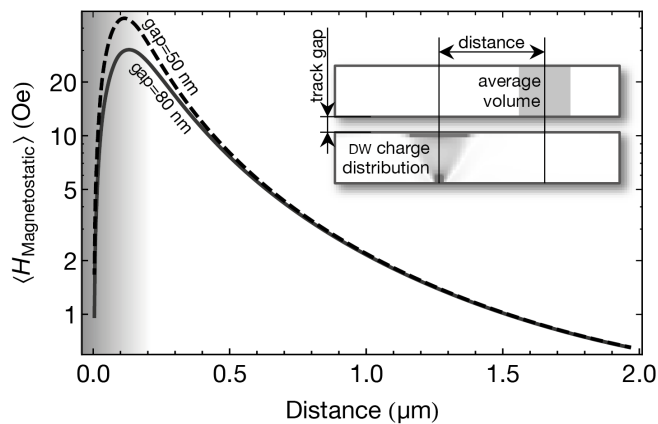

**FIG. 4-58 Interaction strength vs distance** (micromagnetic simulation). Horizontal component of stray field of one DW, averaged over a rectangular volume, vs. horizontal distance (log scale; see inset for geometry). The track cross section was 140×10 nm², and the track gap 50 nm (blue curve) or 80 nm (red curve). The average volume is 140×140×10 nm³. The grey gradient marks d ≲ width, when the spatial charge distribution would have to be considered. $M_S = 8 \cdot 10^5$ A/m. The simulation was performed with OOMMF [Donahue & Porter 1999].

We can then conclude that we observed the mechanism described before (FIG. 4-57), whereby the movement of one DW closer to a second pinned DW induces the depinning of the latter, via their magnetostatic interaction. In our measurements, the range of this interaction was up to 0.6 μm (FIG. 4-56B), equivalent to ~4 track widths. The observation of this phenomenon, so-far unseen, shows once more the usefulness of the DW position measurement made possible by the use of SV tracks.

---

[43] A detailed analysis of the interaction potential of DWs in adjacent tracks, including depinning field from the coupled state, can be found in [O'Brien et al. 2009a].

Finally, it is worth to compare this long-range magnetostatic interaction to the case of DW pinning at T defect in the *AP-wide* configuration, where a long-range repulsion caused the DW to move away from the defect (FIG. 4-28). In both cases there is an interaction between two magnetic charge distributions of roughly of the same spatial extension, though the interaction in the DW-pair is attractive (equal charges of opposite sign) while in the DW at the T defect it is repulsive (with charges of the same sign and different magnitude).

### 'Drive-by' depinning

It was also observed that the DWs influenced each other even when they did not form coupled DW pairs. Frequently, a travelling DW was observed to cause the other DW, pinned at some natural defect, to depin without the formation of a stable DW pair; two examples are shown in FIG. 4-59. There one can observe the simultaneous movement of DWs that are initially at a significant distance of each other (0.94 and 0.67 μm for FIG. 4-59A and B, respectively). These two examples were chosen because their large initial distances rule out the possibility that the DWs are actually paired at the same position, and that the measured positions are mistaken due to a contact lithography misalignment. At the moment of the transition, one of the DWs depinned and travelled towards the end of the track, passing by the other DW while not being pinned by it. The passing of the first DW induced the depinning of the second one, which then travelled to the opposite end. The two DW movements, though not completely simultaneous, appear so due to measurement time resolution being much lower than the DW travel time (9 ms versus tens of ns). As before, which of the two DWs moved first cannot be determined by this measurement.

This can be interpreted as a similar case to the previous, but one where the applied $H_x$ happens to be greater than the decoupling field of the DW pair at the moment of the DW crossing. To further understand this, we conducted a micromagnetic simulation of this depinning phenomenon.

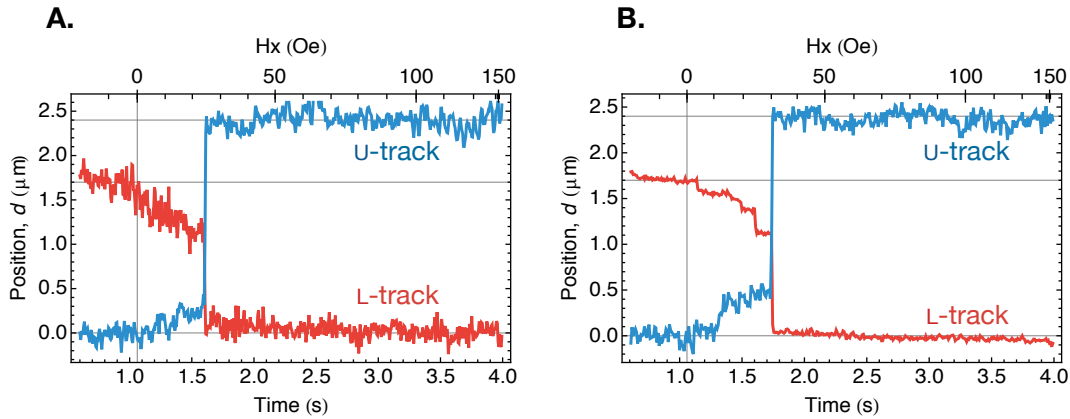

**FIG. 4-59 Drive-by depinning.** The two DWs moved simultaneously and did not form a stable coupled pair. The red and blue curves correspond to the DW in the L- and U-tracks. The two plots correspond to two different structures, both 260 nm wide. The initial DW-DW separations were 0.94 μm (A) and 0.67 μm (B), and the final separation was 2.4 μm in both cases. The value of the field at the transition was 25 Oe (A) and 30 Oe (B).

**Micromagnetic simulation**

FIG. 4-60 shows a micromagnetic simulation of drive-by depinning. The dimensions were smaller than the experiment for computational limitations. The simulation comprised of two DWs in parallel tracks pinned at notches of different depth (depinning fields, simulated separately, were 179 and 216 Oe), initially 0.4 μm apart (magnetisation shown in FIG. 4-60B, DW positions in FIG. 4-60A). A slowly ramping field [44] was then applied, which causes the DW in the shallowest notch to depin at $H_x = -179$ Oe. With the parameters used, the travelling DW showed the typical Walker breakdown behaviour with precessing central magnetisation and oscillating velocity [45] (FIG. 4-60C-i).

As the travelling DW approached the pinned DW, the mutual attraction caused the latter to depin towards the incoming DW (FIG. 4-60C-ii), even though the applied field (179 Oe) was 37 Oe smaller than the unaided depinning field of the notch. This occurred when the two DWs were ~100 nm apart (roughly equivalent to 2 DW widths). With both DWs free from the pinning notches, and with $H_x$ larger than their coupling field (115 Oe, simulated separately), the DWs then propagated separately towards opposite ends of the tracks (FIG. 4-60C-iii).

---

[44] Ramp rate was 0.03 Oe/ns, and thus the field was approximately constant for the duration of the DW movement.

[45] Separate simulations were also performed with a higher $\alpha$ for which there was no Walker breakdown. Still, the inter-DW depinning effect was essentially unchanged.

In summary, the simulation confirms a nearby travelling DW induces depinning, even at applied fields much smaller than the $H_{Pin}$ of the defect, and is a very likely explanation for the observed measurements.

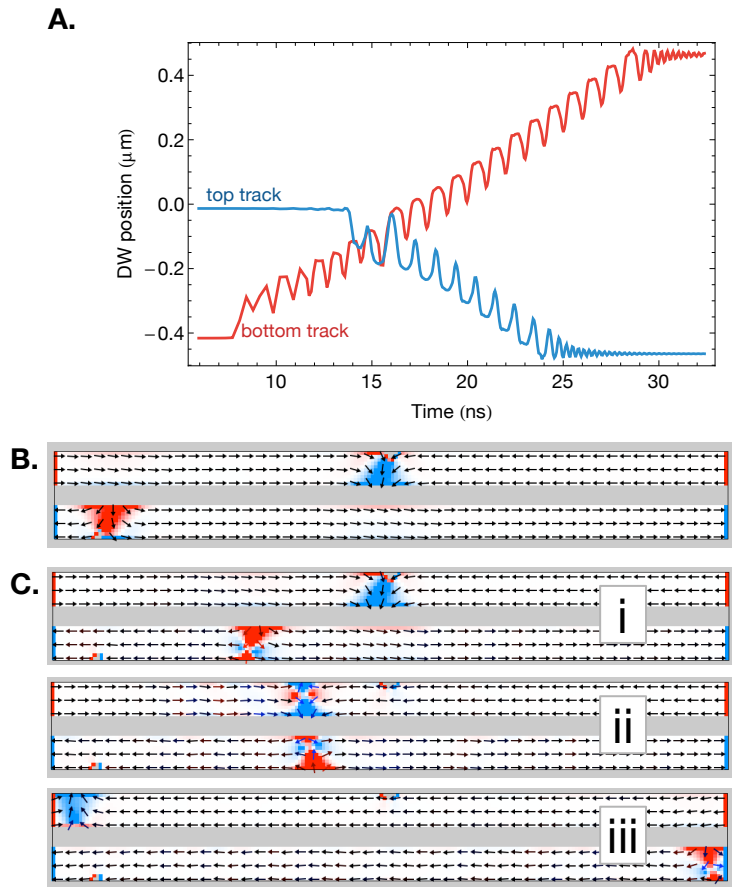

**FIG. 4-60 Drive-by depinning** (simulation). **A.** Horizontal position of DWs vs simulation time (**red** and **blue** for the bottom and top DWs, resp.). **B, C.** Snapshots of the magnetisation (arrows) and magnetic charge density (colours; **red** for negative and **blue** for positive charge). Track geometry: 50×8 nm² cross-section, 1 μm long, 30 nm gap. The notches were 0.5 μm apart, 10 and 5 nm wide, with $H_{Depin}$ of 179 and 216 Oe. Simulation parameters were $M_s$=8·10⁵ A/m, α=0.01, cell size 5×5×8 nm³. The simulation was performed with OOMMF [Donahue & Porter 1999].

### Possible application to racetrack devices

This depinning effect of the DW-DW interaction also suggests the possibility of using a travelling DW to serially depin multiple DWs in a parallel main track, which could be applied to planar racetrack devices.

One instance of these data storing devices consists in a track with multiple DWs, alternately TT and HH, pinned at periodically placed notches [Parkin et al. 2008]. Current injected in the track depins and moves all the DWs in the same direction by the spin-torque effect, maintaining the DWs separated while they hop from one notch to the next. Some of the problems in implementing these devices are the large currents needed to induce depinning, the variation in depinning current of the notches, and the variation DW velocity under the spin-torque effect. Such effects can cause the DWs to come in contact with each other and annihilate, destroying the encoded information. A travelling DW in a parallel un-patterned track could serially depin the DWs in the notched wire, which would then travel, one at a time, in the same direction under the action of the spin-torque current (schematised in FIG. 4-60). The notch size and design would be chosen so that, at the applied currents and fields, depinning could only occur with the passage of the travelling DW. This scheme could potentially require a lower injected current, would be more robust towards notch variability, and would synchronise the DW depinning, avoiding DW annihilation.

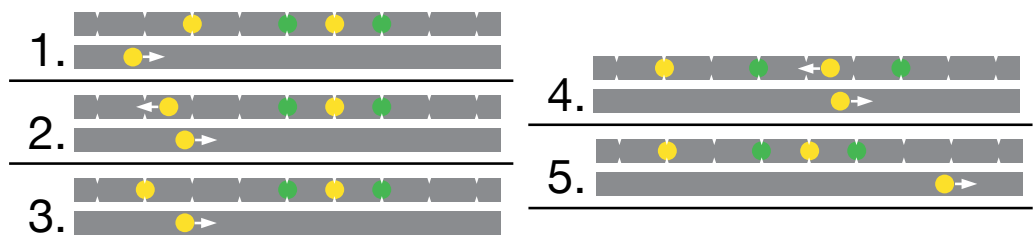

**FIG. 4-61 Application of drive-by depinning to the DW racetrack** (schematic). The device consists in two parallel tracks, a smooth track, and a notched track containing the encoded information in a series of pinned DWs alternately HH (●) and TT (●). Current is injected in opposite direction in the two tracks, pushing the DWs on the notched track leftwards, and on the smooth track rightwards. A travelling DW in bottom track serially depins the DWs on the notch track (**1–5**).

One aspect assumed by the scheme of FIG. 4-61 is that drive-by depinning works the same way for same polarity DW pairs as it does for opposite polarity DWs. However, the interaction is very different in the two cases: attractive for opposite polarities but repulsive for the identical polarity. The simulation of the same polarity case, for field driven DWs, can be seen in FIG. 4-62. In the simulation, analogously to the previous case, the repulsive DW-DW interaction does aid the depinning of the pinned DW when both DWs are pushed in the same direction. When the two DWs are pushed in opposite directions, however, the repulsive magnetostatic field would work against the desired depinning direction.

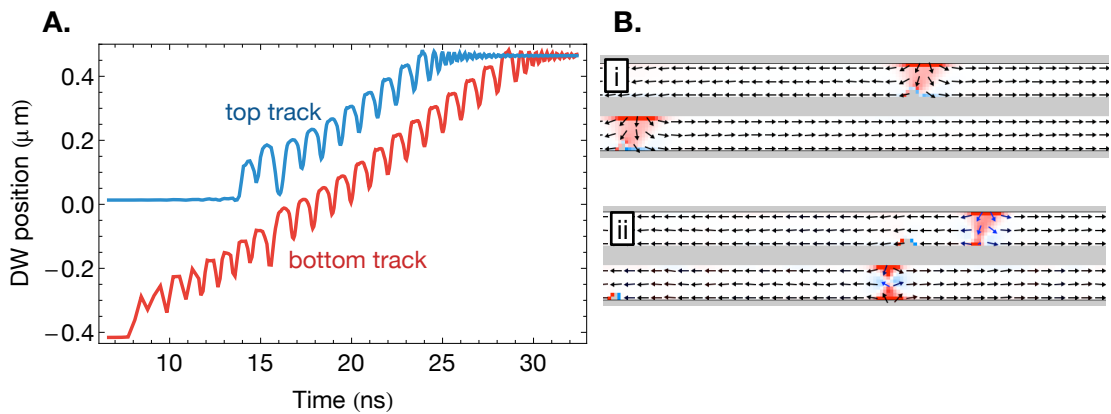

**Fig. 4-62 Drive-by depinning on DWs of the same polarity** (simulation). The travelling DW (in the bottom track) depins the top DW, under $H_x$= 187 Oe. **A.** Horizontal position of DWs vs simulation time (**red** and **blue** for the bottom and top DWs, resp.). **B.** Snapshots of the magnetisation (arrows) and magnetic charge density (colours; **red** for negative and **blue** for positive charge), at the initial state, and during depinning. Same geometry, notch, field and simulation parameters as Fig. 4-60. The depinning field for the bottom DW was slightly higher (187 vs 179 Oe) due to the added repulsive DW-DW interaction.

While it is not the scope of this section to design a working racetrack device based on the DW-DW interaction, we have investigated briefly an alternative scheme that could solve this problematic asymmetry. As was referred before, there are several aspects that can be modified in this scheme, one of them the notch shape. With a larger and stronger notch such as that of Fig. 4-63C, we observed in simulation that a (field driven) travelling DW could depin the trapped DW if it was of opposite polarity, and would passed unperturbed if it was of the same polarity (Fig. 4-63).

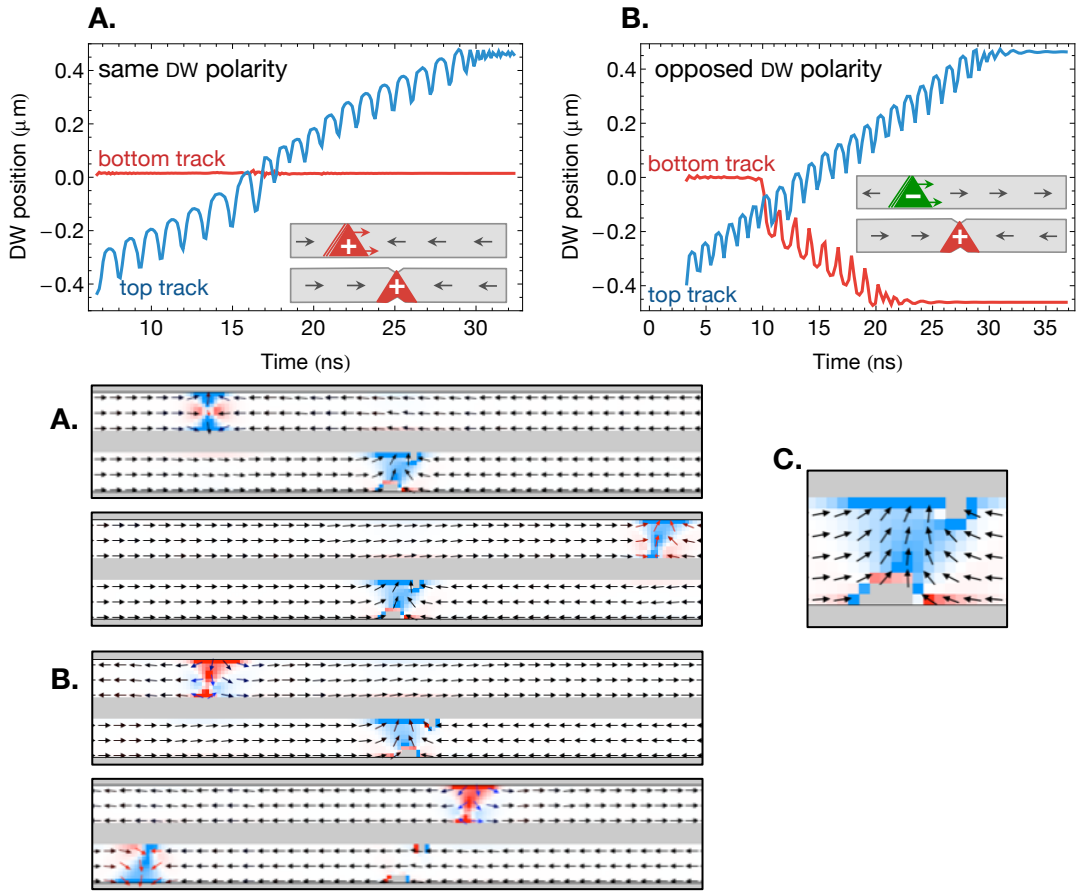

**FIG. 4-63 Drive-by depinning on DWs of the same polarity — strong pinning site** (simulation). The (field-driven) travelling DW is in the top track, the pinned DW in the bottom. **A.** A HH DW passing by a pinned HH DW under $H_x$ =-150 Oe. The plot shows the horizontal position of DWs vs simulation time (**red** and **blue** for the bottom and top DWs, resp.) and the images show two snapshots of the magnetisation (arrows) and magnetic charge density (colours; **red** for negative and **blue** for positive charge). **B.** A TT DW passing by and depinning a HH DW DW, under $H_x$ = 150 Oe. Same colour code. **C.** Image of the notch. The notch was a potential well with depinning field 255 Oe (for both left and rightwards depinning). Same geometry and simulation parameters as FIG. 4-60.

This leads to the revised racetrack scheme of FIG. 4-64 using such notches. At the passage of a travelling HH DW in the smooth track, all the pinned TT DWs would hop one notch (as before), but the HH would hold their positions (FIG. 4-64, 1–4). Then, at the passage of a travelling TT DW, the pinned HH DWs would finally hop one notch also (FIG. 4-64, 5–7). This requires that the DWs are spaced one or more notches, and the passage of two synchronising DWs per shift. Additionally, this scheme could be field driven, as DW of opposite charge are being driven always in opposite directions. The DWs in the smooth track could be generated by a nucleation pad.

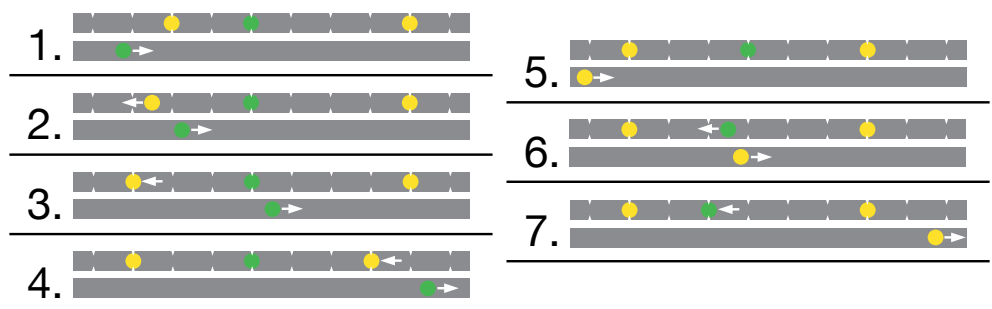

**FIG. 4-64 Revised scheme for using drive-by depinning in a DW racetrack.** The device consists in a smooth track and a notched track containing the encoded information in a series of pinned DWs alternately HH (🟡) and TT (🟢). Current is injected in opposite direction in the two tracks, pushing the DWs on the notched track leftwards, and on the smooth track rightwards. A travelling TT DW in bottom track serially depins the HH DWs on the notch track while not affecting the TTs (**1–4**) and then a travelling HH DW does the reverse (**5–7**).

## 4-4.3. Summary

We successfully fabricated closely spaced SV tracks (separation down to ~50 nm) and demonstrated **single-shot, simultaneous measurements of interacting DWs**. We were able to calculate the positions of the DWs with a resolution better than ~100 nm. This study confirmed the **formation and separation of coupled DW pairs in SV tracks**, and showed that there is a **stochastic variation of the decoupling field and of the position** of the DW pair, and suggested that differences in the position of DW pair were associated with differences in decoupling field. It also revealed new phenomena caused by the DW-DW interaction, viz. **pair formation by distant attraction** and '**drive-by' depinning**. These phenomena were repeatable and were observed in several structures. The spatial and field accuracy, along with the ability to take several single-shot measurements, allowed us to show that the null-hypothesis, i.e. that these simultaneous DW displacements were produced by independent DW depinning events, was highly improbable.

The results were compared to a magnetostatic model of the DW-DW interaction and to micromagnetic simulations, which reproduced successfully the main experimental findings.

Finally, we used micromagnetic studies to study how drive-by depinning, i.e. depinning induced by a travelling DW in the adjacent track, could be used to synchronise the displacement of a series of pinned DW, with potential applications to *racetrack*-like [Parkin et al. 2008] data storage devices.

These structures may also be useful in future experiments to study DW-DW interactions. For example, it has been predicted that coupled DW pair has a non-linear resonation mode [O'Brien 2010; O'Brien et al. 2011], which could be electrically measured in a structure similar to this one.

## 4-5. Conclusion

Based on the SV DW conduit developed in Chapter 3, we have successfully demonstrated four families of device structures in SV tracks: the spiral conduit, the T gate, the NOT gate, and the DW interaction double track, all of which have potential applications in digital logic field. The MR measurements have enabled us to study these devices, validate micromagnetic simulations, compare them to single-layered Permalloy structures when possible, perform statistical characterisations of their operation, and discover new aspects of their operation.

We started by analysing a 4-turn **spiral track** with no nucleation pads or other modifications to the track geometry. There, we demonstrated the injection and propagation of multiple DWs while avoiding their mutual annihilation. We measured its operating margin, which we found to be smaller than that of L shaped tracks (Chapter 3), but significantly larger than that of SV spirals with nucleation pads reported in literature. We also measured and determined the change to the MR measurement caused by the reversal of each of the spiral segments, and shown how this could be used to implement a digital turn counter (previously shown in literature for similar spirals) or a data storing device.

We then studied an example of a DW artificial trap, the **T gate**. We observed four pinning configurations associated to the gate magnetisation and TDW configuration, and we measured the different depinning fields in both directions. We found similar results to those reported for Permalloy systems, which showed that the extensive characterisation of T gate in Permalloy could be applied to SV systems. Importantly for DW logic applications, we could reproduce the DW valve effect: the pinning field of the trap could be changed from ~$H_{PR}$ to ~$H_{NUC}$ by simply reversing the trap magnetisation. We then used the properties of the T gate to investigate the reversal of the TDW magnetisation during propagation due to the Walker breakdown process. We found that the TDW reversal was stochastic, with a probability that varied between structures and DW polarities (these results were consistent with reported findings in Permalloy tracks). By using the measurement of DW position, we were also able to observe the effects of the repulsive magnetostatic interaction between the DW and the trap under certain configurations.

We were also able to show that the T stub acted as an injection point and a pinning site for **360° DWs**, and that it reduced significantly the DW splitting field. These results suggest that this trap can be used as a tool to study 360° DWs, a magnetic structure of importance to DW logic and sensor devices, but one often complex to inject or manipulate.

We then demonstrated the operation of the **NOT gate**. We studied several designs, including highly compact gates with narrow tracks (down to 50 nm width). For all these, we measured their operating margins and compared them to the reported values for Permalloy gates. We found that the margins in SV tracks were smaller, though of comparable size. We also measured the DW position during operation of the gate. The single shot measurements allowed us to characterise the flickering behaviour, which occurs at low applied field amplitudes.

We then demonstrated a **shift register** composed of seven adjacent NOT gates, in a narrow track (65 nm wide). To this date, this is arguably the most complex DW logic circuit to be demonstrated in SV tracks.

Finally, we studied the magnetostatic **interaction of DWs in adjacent SV tracks**. Using simultaneous, single-shot measurements of the DW position in both tracks, we demonstrated the formation and separation of coupled DWs pairs, and showed that there was a stochastic variation of pinning field and position. We also discovered some new phenomena caused by the DW interaction, namely the pair formation by distant attraction and the depinning induced by a travelling DW. We used micromagnetic simulations to understand these effects, and to suggest a potential application to storage devices (the *DW racetrack*).

We believe that the structures studied in this chapter considerably support the idea that technological relevant DW logic circuits can be implemented in SV tracks, allowing the electronic integration of future devices. Apart from the technological relevance the electric measurement of SV tracks may present, the single-shot measurement of DW position in very narrow tracks, demonstrated here, allowed the study of new phenomena in these systems.

# 4-6. References

# [5] Effects of electric current

The application of current in a SV track, essential to measuring its resistance, influences the magnetisation of the free layer via several physical effects, such as heating and spin transfer torque. In SV tracks, as in monolayer tracks, researchers have studied these effects motivated by fundamental scientific interest, and also by the potential technological use of current induced effects in controlling DW propagation, depinning, and structure [Kläui et al. 2005; Parkin et al. 2008].

Likewise, our interest in the effects of the current is two-fold. On the one hand, we are interested in determining how the current we apply to measure the resistance perturbs the magnetic behaviour of the structures we study. On the other hand, we are also interested in finding how the interaction of current and magnetisation can be used to further characterise our structures, and to manipulate their behaviour.

To this goal, we study the effect of current in two different situations: firstly, the effect of a DC current on DW depinning from an artificial trap (§5-1). Secondly, the effect of an RF current on the magnetisation of the track (§5-2).

**Table of contents**

# 5-1. Domain wall depinning with current bias

In this section, we examine the change of depinning field with the direction and amplitude of the applied DC bias current, $I_{DC}$. The pinning site used was a T trap as studied in the previous chapter. This well characterised trap will allow us to study the effects of current on several different pinning configurations.

As many symmetric parameters will be compared, it is important to carefully define each variable used. The spatial axes and the current sign can be seen in FIG. 5-1. The depinning fields, $H_{TR}$ or $H_{Pull}$, are the depinning field amplitudes and are always positive (though the actual applied field may be negative). The $\Delta H_{TR}$ (or $\Delta H_{Pull}$) is the difference of the depinning field between positive and negative currents, and is positive if the depinning field is greater with positive currents.

As we will see below, the several phenomena of interaction of current and magnetisation have different symmetries, in respect to current direction, depinning direction, and magnetisation. Measuring several symmetric cases (positive and negative current, HH and TT DWs, forward and backward pinning, and different pinning configurations), allows us to distinguish between these different physical phenomena affecting DW depinning and propagation under current.

## Structures

The structure examined was a C-shaped track with a T trap, identical to the structures studied in Chapter 4, FIG. 5-1. The sample was fabricated with the previously described titanium hard mask process (Chapter 2) from a SV stack with a free layer of Py 8 nm [1]. The track width was 110 nm. The forward and backward depinning fields were measured by applying an external field sequence, as described in Chapter 4 (see §§ 4-2.1 and 4-2.2). The forward depinning, $H_{TR}$, is measured by initialising a DW in the arc (with a diagonal field), pushing the DW towards the trap, and finally pushing the DW through the trap. Correspondingly, the backward depinning, $H_{Pull}$, is measured by propagating the DW towards the trap but depinning it back towards the arc.

---

[1] See Annex A for fabrication details, sample reference HM01.

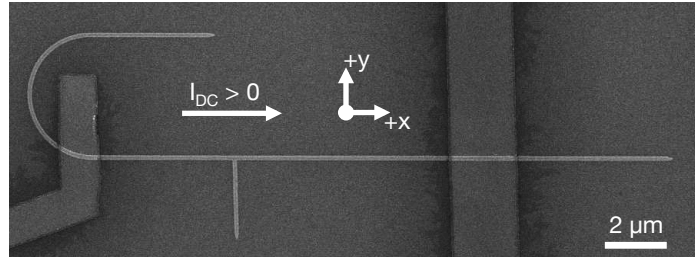

**FIG. 5-1 Structure used in injected current experiments** (SEM image). The spatial coordinates and positive current direction are also shown.

As described in Chapter 4 (and in [Petit et al. 2009]), the T trap at the bottom of the track shows two possible pinning potentials, corresponding to the two possible TDW central magnetisation states: the *P-wide*, a potential well, and *AP-narrow*, a potential barrier. The former is symmetric, with H$_{Pull}$ ≈ H$_{PR}$, and the latter asymmetric, with H$_{TR}$ ≈ H$_{NUC}$ and H$_{Pull}$ ≈ H$_{PR}$. The *P-wide* configuration occurs when the DW reaches the trap with its as-injected central magnetisation, and is generally the most common case. Still, in most structures, there is a finite probability for the DW magnetisation to reverse during its travel to the trap, resulting in *AP-wide* configuration. If the trap is placed at the top of the track, two other configurations are produced, *P-narrow* and *AP-wide*, where the latter generates a high potential barrier and the former a very small pinning potential [2].

Three structures were measured, one with the trap on top and two on the bottom, though not all possible pinning configurations were observed in every structure.

## 5-1.1. Measurements

The measurements of depinning field as a function of I$_{DC}$ are shown here for three different pinning configurations: *P-wide*, *AP-narrow*, and *AP-wide*. The first two were measured on tracks with T trap placed on the bottom of the main track, and were measured on two (nominally identical) structures. The third configuration was measured on a single structure with a t trap on the top. The applied current (I$_{DC}$) is limited to ≤1.5 mA by the destruction of the track through heating and electromigration.

The discussion of these observations is left for the next section.

---

[2] undetectable in these measurements.

**Depinning in the *P-wide* configuration**

The measurement of H$_{\text{TR}}$ in the *P-wide* case versus I$_{\text{DC}}$ is shown in F$_{\text{IG}}$. 5-2A and B. The curves are separated by current direction and DW polarity. In all cases, there is a lowering of H$_{\text{TR}}$ with current of 7–29 Oe at I$_{\text{DC}}$ = 1.5 mA. There is also a significant change of H$_{\text{TR}}$ with the current direction, quantified by ΔH$_{\text{TR}}$:

$$\Delta \text{H}_{\text{TR}} = \text{H}_{\text{TR}}|_{I = +I_{DC}} - \text{H}_{\text{TR}}|_{I = -I_{DC}}$$

For TT DWs, ΔH$_{\text{TR}}$ is negative, i.e. H$_{\text{TR}}$ is lower with positive currents than with negative. In contrast, for HH DWs, ΔH$_{\text{TR}}$ is positive. In both cases, ΔH$_{\text{TR}}$ is approximately linear with I$_{\text{DC}}$, and a linear fit model was applied,

$$\Delta \text{H}_{\text{TR}} = m \cdot \text{I}_{\text{DC}}$$

with the proportionality factors (*m*) -6.9 and +14 kOe·A$^{-1}$ for TT and HH DWs, respectively (F$_{\text{IG}}$. 5-2E).

The same depinning configuration was also measured for backward depinning H$_{\text{PULL}}$, F$_{\text{IG}}$. 5-2C and D. Again, the pinning field of the TT DW lowered with positive current (ΔH$_{\text{Pull}}$ < 0) while the pinning of the HH increased (ΔH$_{\text{Pull}}$ > 0). The proportionality factors (*m*) were -17 and +9.4 kOe·A$^{-1}$ for TT and HH DWs, respectively (F$_{\text{IG}}$. 5-2E). The backward depinning field was also measured on a second nominally identical structure, with similar results (proportionality factors -14 and +14 kOe·A$^{-1}$), F$_{\text{IG}}$. 5-3.

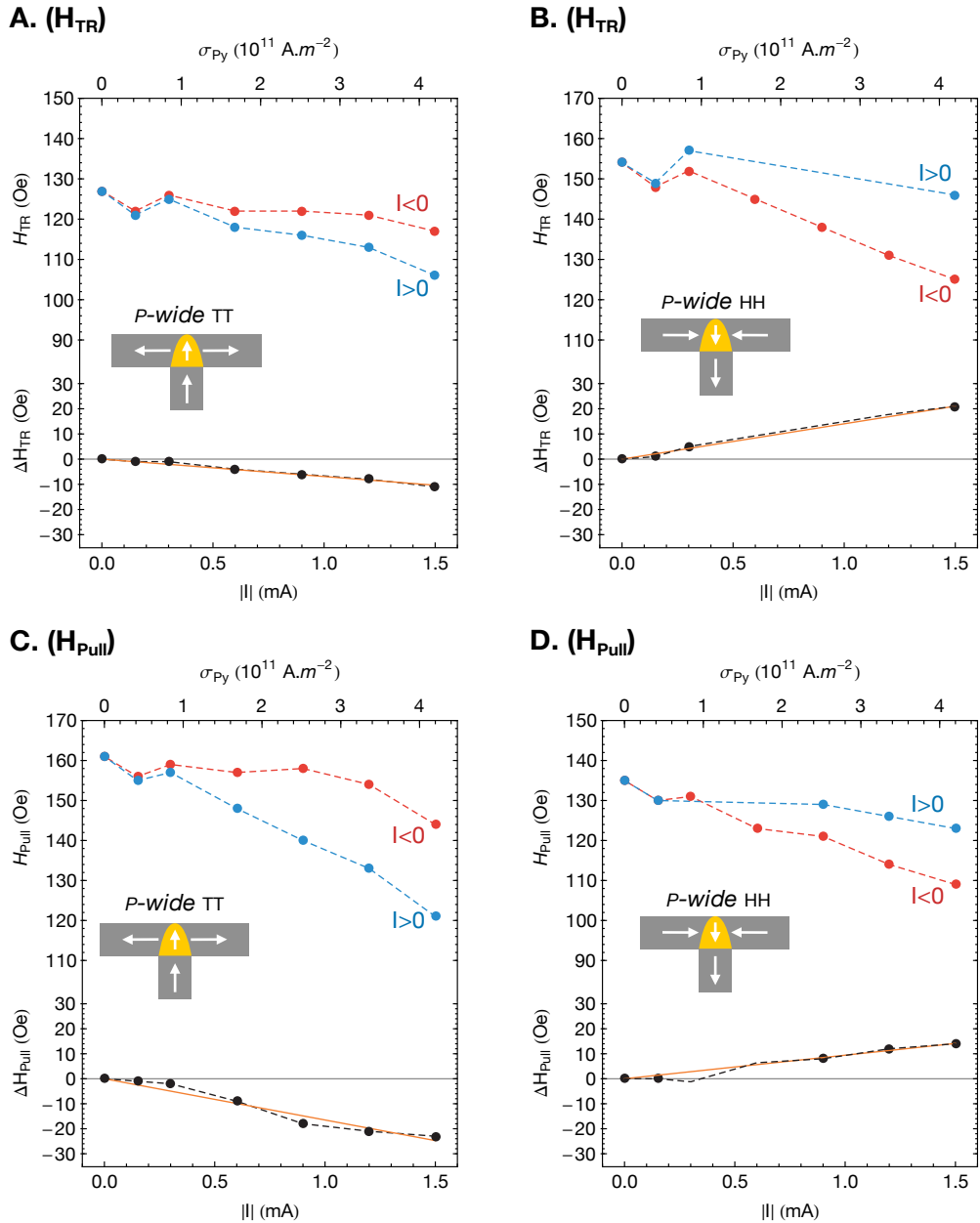

**A. (H_TR)**

**B. (H_TR)**

**C. (H_Pull)**

**D. (H_Pull)**

**E.**

| transition | configuration | m (kOe·A⁻¹) | adjusted R² |
|---|---|---|---|
| H_TR | P-wide TT | -6.9 ±0.3 | 0.99 |
| | P-wide HH | 14.0 ±0.6 | 0.99 |
| H_Pull | P-wide TT | -17 ±1 | 0.98 |
| | P-wide HH | 9.4 ±0.4 | 0.99 |

**FIG. 5-2 Variation of H_TR and H_Pull with current (*P-wide* case).** The upper plots show H_TR (**A, B**) and H_Pull (**C, D**) vs. current amplitude for negative and positive currents (red and blue points) and for TT and HH DWs. The lower plots show ΔH_TR or ΔH_Pull (the difference between depinning with positive or negative current; black points). The top axis is the estimated current density in the Py layer. The solid orange line and the table (**E**) are a linear fit with model $\Delta H_{TR/Pull} = m \cdot |I|$. (The *adjusted R²* is the coefficient of determination adjusted per degree of freedom).

| Transition | configuration | m (kOe·A⁻¹) | adjusted R² | N_Points |
|---|---|---|---|---|
| | *P-wide* TT | -14 ±1 | 0.99 | 2 |
| H_Pull | *P-wide* HH | 13 ±1 | 0.96 | 3 |
| | *AP-narrow* TT | -3 ±2 | 0.25 | 3 |
| | *AP-narrow* HH | 1.4 ±0.6 | 0.63 | 3 |

**FIG. 5-3 Variation with current of H_Pull on a second, nominally identical structure.** Parameters of a linear regression with model $\Delta H\_ = m \cdot |I|$ .

### Depinning in the *AP-narrow* configuration

The *AP-narrow* pinning configuration is obtained with the same field sequence as for *P-wide*. The stochastic reversal of the DW central magnetisation causes that in some measurements we observe the *AP-narrow* configuration, and the *P-wide* in others. They are distinguished by their very different depinning fields (see Chapter 4 for details).

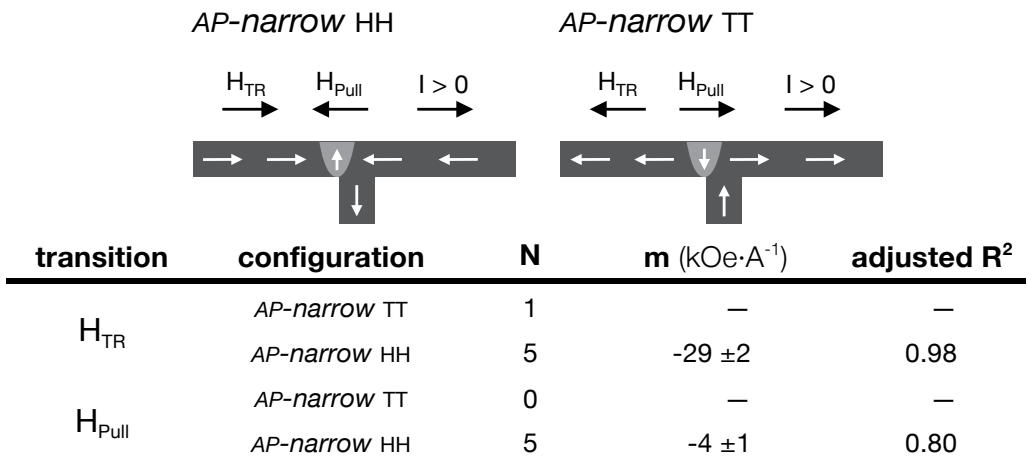

| transition | configuration | N | m (kOe·A⁻¹) | adjusted R² |
|---|---|---|---|---|
| H_TR | *AP-narrow* TT | 1 | — | — |
| | *AP-narrow* HH | 5 | -29 ±2 | 0.98 |
| H_Pull | *AP-narrow* TT | 0 | — | — |
| | *AP-narrow* HH | 5 | -4 ±1 | 0.80 |

**FIG. 5-4 Variation with current of H_TR and H_Pull (*AP-narrow* case). Top.** Schematic of the pinning configuration and field directions. **Bottom.** Fit parameters of the model $\Delta H\_ = m \cdot |I|$ to $\Delta H_{TR}$ and $\Delta H_{Pull}$. *N* refers to the number of $\Delta H_{TR}$ or $\Delta H_{Pull}$ measurement points (not counting I=0). The data plots may be consulted in FIG. B-1 in Annex B.

The measurements for the *AP-narrow* configuration are shown on FIG. 5-4 (for brevity, the H_TR/Pull vs. I_DC and $\Delta$H_TR/Pull vs. I_DC plots were included in Annex B, FIG. B-1). As this configuration requires that the central DW magnetisation reverse, which was very uncommon for TT DWs in this structure, resulting in few data points for TT DWs. For HHs, both H_TR and H_Pull lowered with positive current (i.e. $\Delta$H_TR < 0 and $\Delta$H_Pull < 0), though the variation for the backward depinning was much smaller (*m* factors -29 for $\Delta$H_TR vs. -4 kOe·A⁻¹ for $\Delta$H_Pull). In a separate structure, H_Pull of the *AP-narrow* configuration was measured for both TT and HH DWs, FIG. 5-3, and a very small variation with current was also observed (-2 ±2 and 1.4 ±0.6 kOe·A$^{-1}$ for TT and HH, respectively).

**Depinning in the *AP-wide* configuration**

This configuration can be obtained in structures with the T trap placed on the top of the track when the DW reverses from its injected configuration. It was measured on a different type of structure than the one studied before, as the T-trap is placed on the opposite side of the track. The results for forward and backward depinning (H$_{TR}$ and H$_{Pull}$) are shown in FIG. 5-5 (and FIG. B-2, in Annex B). Forward depinning showed a large variation with current, with $\Delta$H$_{TR}$ < 0 for TT and > 0 for HH (proportionality factors *m* -13 and 19 kOe·A$^{-1}$, respectively). No measurable change was found for backward depinning in TT DWs (*m* ≈ 0), and only a very small change with HH (*m* = -4.5 ±0.6 kOe·A$^{-1}$).

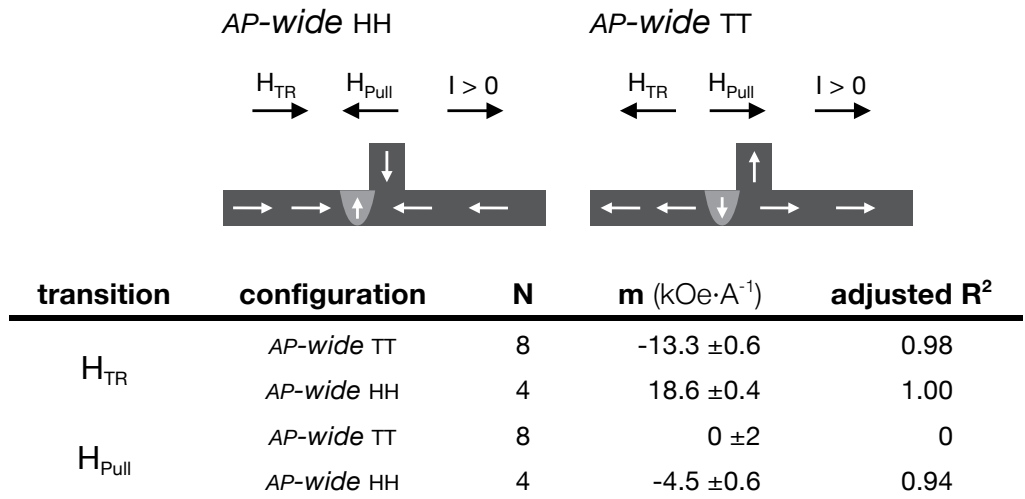

| transition | configuration | N | m (kOe·A$^{-1}$) | adjusted R$^2$ |
|---|---|---|---|---|
| H$_{TR}$ | *AP-wide* TT | 8 | -13.3 ±0.6 | 0.98 |
| | *AP-wide* HH | 4 | 18.6 ±0.4 | 1.00 |
| H$_{Pull}$ | *AP-wide* TT | 8 | 0 ±2 | 0 |
| | *AP-wide* HH | 4 | -4.5 ±0.6 | 0.94 |

**FIG. 5-5 Variation with current of H$_{TR}$ and H$_{Pull}$ (*AP-wide* case). Top.** Schematic of the pinning configuration and field directions. **Bottom.** Fit parameters of the model $\Delta H\_ = m \cdot |I|$ to $\Delta$H$_{TR}$ and $\Delta$H$_{Pull}$. *N* refers to the number of $\Delta$H$_{TR}$ or $\Delta$H$_{Pull}$ measurement points (not counting I=0). The data plots may be consulted in FIG. B-2 in Annex B.

All the proportionality parameters *m* (from the fitting model $\Delta H_{TR/Pull} = m \cdot I_{DC}$) measured before are summarised in FIG. 5-6 for ease of comparison.

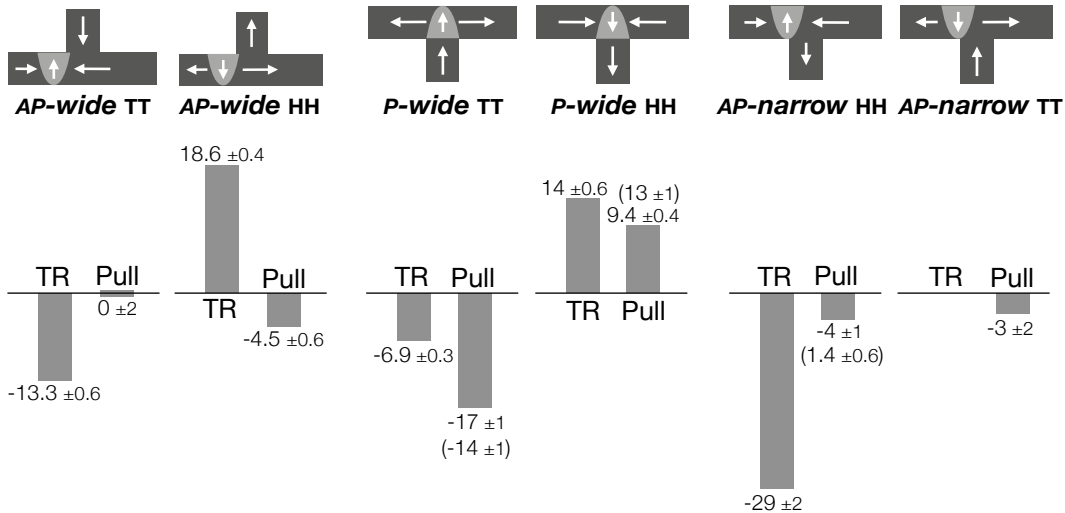

**FIG. 5-6 Measurement summary.** Data from FIG. 5-2, FIG. 5-3, FIG. 5-4, and FIG. 5-5. The bars (and labels) correspond to the proportionality factor $m$ from the fitting model $\Delta H_{TR/Pull} = m \cdot I_{DC}$. Some $m$ values were measured in two structures; for these, the second measurement is included in parenthesis.

## 5-1.2. Analysis

The variation of depinning fields with $I_{DC}$ differed with DW polarity, depinning configuration, and depinning direction (transmission vs. pull). We analyse here the contributions to the change in depinning field of three mechanisms — Joule heating, Oersted field, and spin-transfer torque — to the change of depinning field, taking advantage of their differing symmetries to distinguish their individual contributions.

### Joule heating

The electrical current injected in the track deposits large amounts of power (about 150–200 μW per μm of track at 1 mA). This power heats the track and induces a quadratic increase in the resistance (FIG. 5-7). The temperature increase it generates can be estimated from the variation of resistance, which, close to room temperature, is linear with temperature: $\Delta R/R_0 = \alpha \, \Delta T$ (where $\alpha$ is the material's temperature coefficient). Using the temperature coefficient of copper for a rough approximation [3] (0.004 °C$^{-1}$ [Nave 2010]), it is found that the temperature increases about 25 °C at $I_{DC}$ = 1 mA (see also the right axis in FIG. 5-7).

---

[3] The coefficient for Py is similar: 0.003–0.004 °C$^{-1}$. [Belous et al. 1967; Higashi & Johnson 1988]. Most of the current is concentrated in the Cu and Py layers.

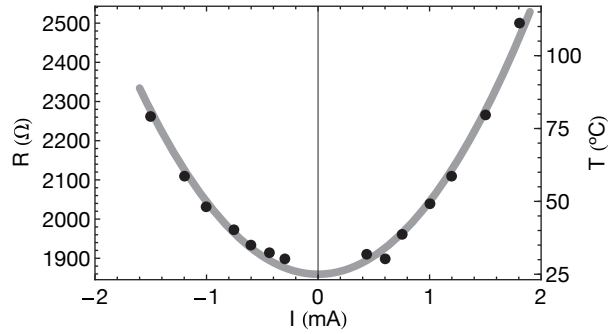

**FIG. 5-7 Variation of resistance with $I_{DC}$.** The grey line is a linear fit with model $R = R_0 + A \cdot I^2$, $R_0 = 1859 \pm 8\ \Omega$ and $A = 186 \pm 6\ M\Omega/A^2$. The temperature is estimated as described in the text.

Almost all curves of depinning field vs. $I_{DC}$ presented a clear negative slope (cf. FIG. 5-2, FIG. C-1, and FIG. C-2). This can be attributed to the effect of heating on depinning [Attané et al. 2006]. Depinning is a thermally activated phenomenon, meaning that depinning occurs when the pinning barrier is comparable in height to the thermal energy available to the free layer spins; increasing the temperature universally decreases the pinning field. The actual relation between temperature and depinning field, though, is complex.

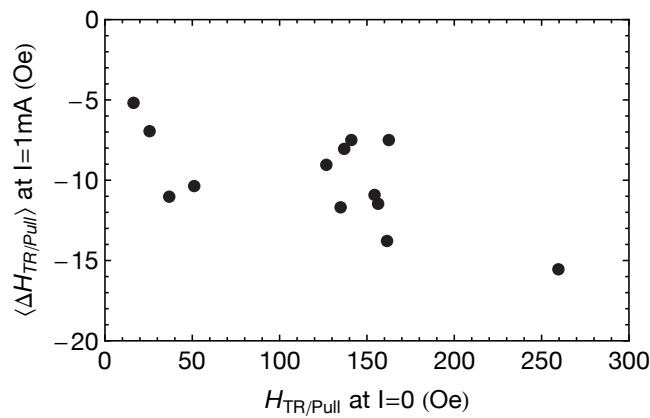

**FIG. 5-8 Thermal reduction of depinning field with current.** The horizontal axis corresponds to the depinning field (either $H_{TR}$ or $H_{Pull}$) at negligible current. The vertical axis corresponds to the change of depinning field at $|I_{DC}| = 1$ mA, averaged for positive and negative $I_{DC}$.

This effect should only depend on the absolute value of the current and the pinning potential. As the other two significant effects (the Oersted field and the spin-transfer effect) both change sign with opposite current directions, the average between the values taken at positive and negative $I_{DC}$ should indicate the strength of the thermal effect. This average is shown for 1 mA ($\Delta T \approx 25°C$) in FIG. 5-8. There we observe that the average depinning field lowered by -5 to -16 Oe depending on the measurement, with no strong correlation to the magnitude of the depinning field.

## The Oersted field

The current in a SV generates a significant Oersted field on the free layer, $H_{Oe}$. For cross-sectional aspect ratio of the studied tracks, this field is most significant in the Y direction [4] and, with this particular SV stack composition, it is positive (+Y) inside the free (Py) layer. To determine what is the role of the Oersted field in the observed results, we shall calculate its magnitude and the expected sign and amplitude of its effect on the studied depinning fields. In order to eliminate the strong effect of heating, $\Delta H_{TR}$ ($\Delta H_{Pull}$) will be analysed instead of the depinning fields directly.

### Magnitude of $H_{Oe}$

To calculate the magnitude of $H_{Oe}$, it is necessary to estimate the distribution of current within the stack, which varies along its depth. The different stack layers have different resistivities and their thicknesses are comparable to the electron mean free path. Consequently, the current profile lies between two extremes: a uniform profile (the thin stack limit) and the parallel conductors profile (thick stack limit) [Dieny 2004]. The former assumes the current is the same for all layers, while the latter assumes that the current density is proportional to the layer conductivity. The resulting $H_{Oe}$ profile is shown in FIG. 5-9. The mean $H_{Oe}$ in the Py layer is similar for both theoretical profiles: 28 and 34 Oe at $I_{DC} = 1$ mA.

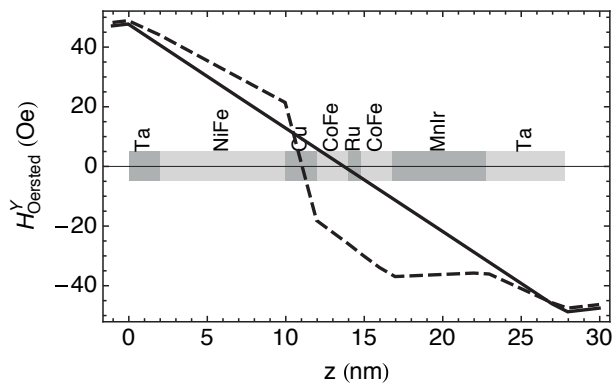

**FIG. 5-9 Oersted field in a SV track**. Depth profile of the Y component of $H_{Oe}$ induced by a 1 mA current in a 110 nm wide track with the composition of the studied structure. Two models for the current profile were used: uniform current density (solid line) and

---

[4] As opposed to the smaller Z component. The spatial axis is defined in FIG. 5-1.

parallel conductors (dashed line). The real current density lies between these extreme cases. Material resistivities taken from [Dieny 2004].

## Parity and symmetry

One distinguishing characteristic of the influence of $H_{Oe}$ on the depinning field is its parity: the effect of $H_{Oe}$ produced by a positive current on a HH DW should be the same as the one produced by a negative current on a TT DW in the same pinned configuration, and vice-versa. In other terms, the Oersted field induces a $\Delta H\_$ for HH DWs of the same magnitude but opposing sign than it does for TT DWs. In all the measurements, the observed $\Delta H$ changes sign for opposing DW polarities, and, in many of the measurements, the magnitude is approximately the same (see FIG. 5-10). This indicates that $H_{Oe}$ is responsible for a large part of the reported results.

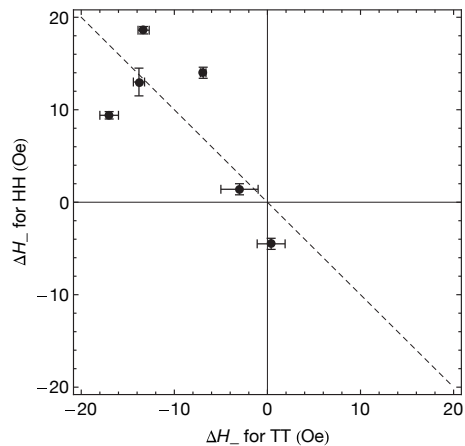

**FIG. 5-10 $\Delta H$ for HH vs. TT DWs.** The points represent either $\Delta H_{TR}$ or $\Delta H_{Pull}$ for the measurements where both HH and TT DWs were tested, at $I_{DC} = 1$ mA (from the fit data shown before). The dashed line marks $\Delta H(HH) = -\Delta H(TT)$.

The *P-wide* configuration is symmetric in X, which causes $H_{TR} \approx H_{Pull}$. As $H_{Oe}$ has no X component, its effect on these two depinning fields should be the same, i.e. $\Delta H_{TR} = \Delta H_{Pull}$ for the same DW polarity. The observed $\Delta H_{Pull}$ and $\Delta H_{TR}$ do agree in sign, though its magnitudes differ (see FIG. 5-6 for a comparison of $\Delta H_{Pull}$ and $\Delta H_{TR}$). This difference can be partly attributed to the effects of spin-transfer torque analysed below.

## Depinning and transverse field bias

We shall now compare the variation of depinning field obtained with the injection of current to the variation obtained by the application of a homogenous external field in the transverse (Y) direction, $H_Y$. The effect of $H_{Oe}$ should be approximately the same as that of $H_Y$ of equivalent magnitude [5]. To test this hypothesis, first the variation of $H_{TR}$ with $H_Y$ will be quantified, and compared to the observed effect of $H_{Oe}$.

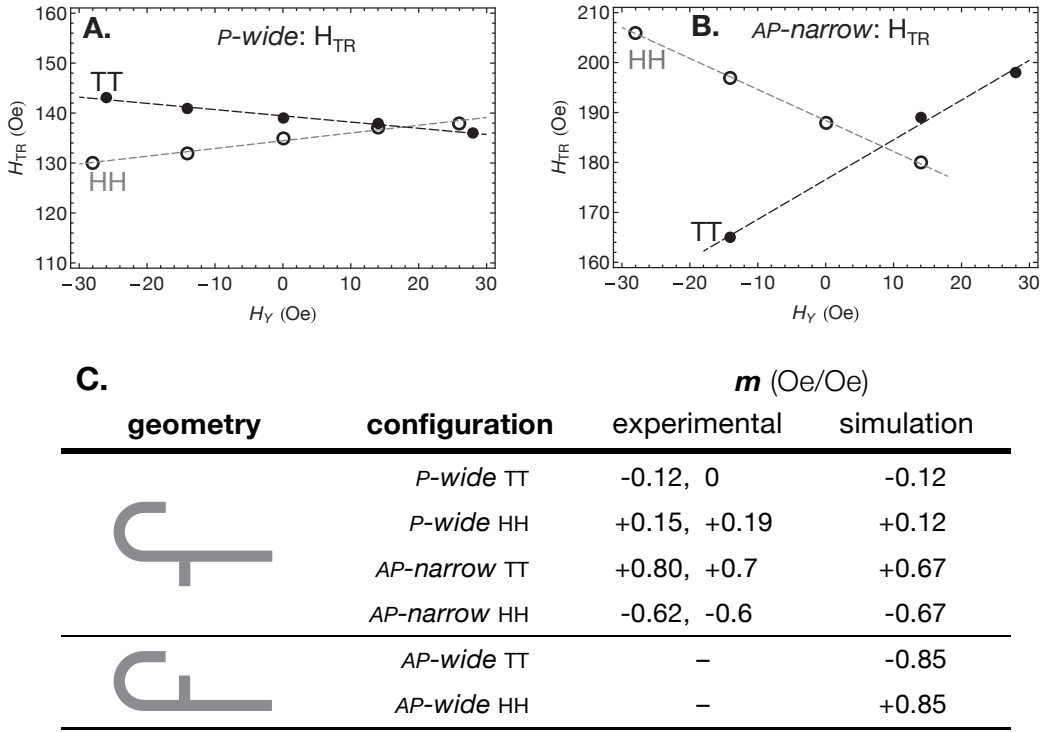

| geometry | configuration | experimental | simulation |
|---|---|---|---|
| | *P-wide* TT | -0.12, 0 | -0.12 |
| | *P-wide* HH | +0.15, +0.19 | +0.12 |
| | *AP-narrow* TT | +0.80, +0.7 | +0.67 |
| | *AP-narrow* HH | -0.62, -0.6 | -0.67 |
| | *AP-wide* TT | – | -0.85 |
| | *AP-wide* HH | – | +0.85 |

**Fig. 5-11 Variation of depinning field with a transverse bias field.** The $H_{TR}$ was measured with a constant transverse bias field ($H_Y$) for the *P-wide* (**A.**) and *AP-narrow* (**B.**) configurations. The dashed lines are linear fits. **C.** Parameter *m* of linear fits with model $H_{TR} = H_0 + m \cdot H_Y$, in two structures and in simulation. Simulations performed with OOMMF [Donahue & Porter 1999].

The variation of $H_{TR}$ with an applied $H_Y$ was measured on two structures and was micromagnetically simulated. The observed variation was linear in the range of $H_Y$ studied (-30 to 30 Oe), and was quantified by the slope of the curve $H_{TR}$ vs. $H_Y$, *m*, Fig. 5-11A and B. The agreement between the measurements and the simulation of $H_{TR}$ vs. $H_Y$ is very good (Fig. 5-11C). Note that only the transmission fields were measured or simulated. As the *P-wide* configuration is symmetric for horizontal inversions, $H_{Pull}$ varies with $H_Y$ similarly to $H_{TR}$, and as such presented values can be used for both cases. As for *AP-wide* and *AP-narrow*, $H_{Pull}$ is actually the depinning of a random natural defect at the left of the T-trap. As such, any variation with $H_Y$ should be specific to that defect, limiting the usefulness of a comparison to a different structure [6].

---

[5] We ignore here the fact that $H_{Oe}$ is not spatially homogenous, and that it has a small z component.

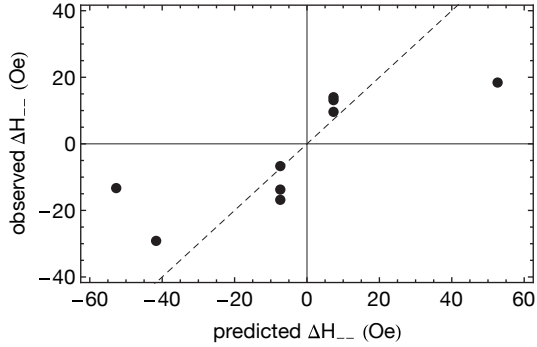

**FIG. 5-12 Current induced vs. predicted Oersted field induced depinning field change.** The Y-axis is the observed current induced $\Delta H_{TR}$ (or $\Delta H_{Pull}$) at 1 mA (from the fit data in FIG. 5-2 through FIG. 5-5). The x-axis is the $\Delta H_{TR}$ induced by a homogenous field of the intensity of the Oersted field at 1 mA (see text for formula). The dashed line is the identity function.

The previous calculation of $H_{Oe}$ ([7]), +31 Oe at $I_{DC}$ = 1 mA, and the parameter $m$ of FIG. 5-11 can be used to determine the $\Delta H_{TR}$ (and $\Delta H_{Pull}$) generated by a $H_Y$ = $H_{Oe}$, by the following relations:

$$H_{TR} = H_0 + m \, H_{Oe}$$

$$H_{Oe} = I_{DC} \cdot 31 \text{ Oe/mA}$$

$$\Delta H_{TR} = H_{TR}|_{I=+I_{DC}} - H_{TR}|_{I=-I_{DC}}$$

$$\Rightarrow \Delta H_{TR} = m\big(H_{Oe}|_{I=+I_{DC}} - H_{Oe}|_{I=-I_{DC}}\big) = 2 \, m \, I_{DC} \cdot 31 \text{ Oe/mA}$$

The observed current induced $\Delta H_{TR}$ is plotted against this predicted Oersted induced $\Delta H_{TR}$ in FIG. 5-12. The predicted values agree in sign and in order-of-magnitude with the observed values. The predicted effect of $H_{Oe}$ in the AP configurations is however overestimated.

**Summary**

The observed values of $\Delta H_{TR}$ and $\Delta H_{Pull}$, which quantify the variation of pinning strength with the direction of applied current, agree in symmetry, sign, and variably in magnitude to the action of the current induced Oersted field, which was independently determined.

---

[6] In simulation, it was either impossible to determine m or m ≈ 0.

[7] using the mean between the two current distribution models discussed previously.

**Spin-transfer effect**

The spin-transfer effect has been reported to induce field-like pressures on the DW of significant intensity in monolayer tracks of similar dimensions to those studied in this thesis: about 0.2–0.5 Oe per $10^{11}$ A·m$^{-2}$ [Kläui et al. 2003; Vernier et al. 2004; Parkin et al. 2008]. The same effect has been measured in SV tracks. In some SV studies, it was observed that the spin-transfer effect was orders-of-magnitude more efficient than in monolayer tracks: few Oe at $10^{10}$ A·m$^{-2}$ [Grollier et al. 2004; Ravelosona et al. 2007; Pizzini et al. 2009]. Other studies, however, have found that the efficiency of the effect in SV to be similar to monolayer tracks [Jiang et al. 2011; Mihai et al. 2011].

The pressure on the DW caused by the spin-transfer effect is in the direction of the electron movement and independent of DW polarity. With $I_{DC} > 0$, the electrons move leftwards and the spin-transfer effect should lower $H_{Pull}$ and increase $H_{TR}$, for both TT and HH DWs. The effect is field-like, implying that it can be linearly added with the external field (as long as no significant DW distortion occurs). The spin-transfer effect is not, however, independent of the depinning configuration [Beach 2008], and its efficiency decreases for higher pinning fields [Parkin et al. 2008]. The backward depinning ($H_{Pull}$) in the AP configurations, which occurs at the lowest observed depinning field ($\approx H_{PR}$, 20–40 Oe), is then a particularly good situation to measure this effect.

In order to determine the magnitude of the spin-transfer, the effects of heating and of the Oersted field have to be separated by using their different symmetries. The heating, being independent of current direction, is eliminating by considering $\Delta H_{TR}$ (or $\Delta H_{Pull}$), as described before. The effect of the Oersted field is eliminated by comparing the $\Delta H_{TR}$ of TT and HH DWs: the Oersted field effect reverses for opposite DW polarity while the spin-transfer is the same. As such, calculating the average of the $\Delta H_{TR}$ for TT and HH DWs the effect of the Oersted field is eliminated, leaving only the spin-transfer effect contribution: $\Delta H_{ST, TR} = \frac{1}{2} \cdot (\Delta H_{TR,TT} + \Delta H_{TR,HH})$. Though this definition may seem obscure, this variable quantifies the difference in depinning field with currents of opposite signs caused by the spin-transfer effect; if there were no change in temperature nor an Oersted field, this parameter would be the same as $H_{TR}|_{I=+I_{DC}} - H_{TR}|_{I=-I_{DC}}$.

| Configuration | $\Delta H_{ST, TR}$ | $\Delta H_{ST, Pull}$ |
|---|---|---|
| | ½·($\Delta H_{TR,\textbf{TT}}$ + $\Delta H_{TR,\textbf{HH}}$)  (Oe) | ½·($\Delta H_{Pull,\textbf{TT}}$ + $\Delta H_{Pull,\textbf{HH}}$)  (Oe) |
| 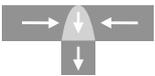 *P-wide* | +3.6 ± 0.5 | -3.8 ± 0.7 |
| *P-wide †* | — | -0.4 ± 1.1 |
| 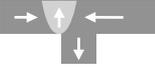 *AP-narrow †* | — | -0.8 ± 1.3 |
| 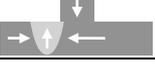 *AP-wide* | +2.7 ± 0.5 | -2 ± 1 |

**FIG. 5-13 Contribution of the spin-transfer effect on the depinning field.** The spin-transfer contribution to the depinning field ($\Delta H_{ST}$) was calculated at $I_{DC}$ = 1 mA, using the fit data shown in FIG. 5-2 through FIG. 5-5. Each line is a different structure, except lines marked †, which correspond to the same structure.

The calculated spin transfer contribution is shown in FIG. 5-13. Overall, the values are close to the error level of our observations. Still, it can be seen that $\Delta H_{ST, TR}$ is positive and $\Delta H_{ST, Pull}$ is negative. This is consistent with the spin-transfer effect: a positive current, with its left-travelling electrons, pushes the DW leftwards, increasing the forward depinning field ($H_{TR}$) and decreasing the backwards depinning field ($H_{Pull}$). Also, for the structures where $H_{TR}$ and $H_{Pull}$ were both measured, the spin-transfer contributions for the two depinning fields are close in absolute value, which is the expected behaviour.

Though the error band associated to these results is high, it is still possible to take some information from the magnitude of the effect. The observed magnitude ranged 0.4–3.8 Oe at $I_{DC}$ = 1 mA, with an estimated current density in the free layer of $2.8×10^{11}$ A·m⁻². This magnitude is consistent with the reported values in experiments with monolayer tracks (~0.2 Oe per $10^{11}$ A·m⁻²), and with the studies that observed a low efficiency spin-transfer effect in SV tracks [Jiang et al. 2011; Mihai et al. 2011]. The much enhanced spin-transfer effect reported in other SV studies [Grollier et al. 2004; Ravelosona et al. 2007; Pizzini et al. 2009] was not observed.

The origin of the enhanced effect, as well as the cause for the different results observed in different studies, is still to be determined. Pizzini and colleagues (referenced above) have suggested that the enhanced effect is caused by the spin-torque effect of perpendicular-to-plane spin currents near the DW. Such effect would depend strongly on the stack composition. We suggest that perhaps the different observed results are caused by the different stack compositions used in the referenced studies, specially since the copper separation layer is of a very different thickness in the structures with the enhanced effect (10 nm [Grollier et al. 2004] and 8 nm [Pizzini et al. 2009] versus 2.4 nm [Jiang et al. 2011] and 2 nm in this thesis [8]).

## 5-1.3. Summary

The influence of a DC current on the depinning fields in several pinning configurations was measured and analysed. The roles of the Joule heating, Oersted field, and spin-transfer were quantified. The **Joule heating** was responsible for a lowering of depinning field in all pinning configurations in the range of -5 to -16 Oe at 1 mA (temperature increase ≈ 25°C).

We attributed to the effect of the **Oersted field** a large observed difference in pinning strength between TT and HH DWs, dependent on pinning configuration and current direction. This difference ranged from <6 to 29 Oe at currents of 1 mA ($H_{Oe} \approx 31$ Oe). The variation of pinning fields with an external field of the same direction and magnitude as the expected Oersted field was consistent with the observed results.

The influence of the **spin-transfer torque** was determined to be in the range of a few Oe (0.4–4 Oe, with error bars ranging ± 0.5–1.3) at 1 mA ($j_{Py} \approx 2.8 \times 10^{11}$ A·m$^{-2}$). The magnitude of the observed effect is consistent with reported experiments on monolayer tracks of similar dimensions with similar strength pinning sites, as well as with some of reported studies in SV tracks. It is, however, in disagreement with reported experiments on SV tracks that have found a much increased spin-transfer efficiency.

It might be necessary in some future experiment to tailor these effects, namely to suppress the Oersted field, so that the spin-transfer effect might be observed with increased precision. This could be achieved by balancing the current profile at the expense of the MR signal: adding shunting layers to the SV stack so that the 0-crossing of the field profile are centred with the Py layer.

---

[8] Not included in this comparison are the two other cited studies, which analyse perpendicular-to-plane magnetised films. In these, Ravelosona et al. observed the enhanced effect in a SV with a separator layer of 6 nm of Cu, while Mihai et al. observed the effect of similar efficiency to monolayer tracks, in a SV with a separator layer of 2.4 nm of Pt.

# 5-2. Current induced ferromagnetic resonance

The effects of the current that have been characterised before can be used to influence the magnetisation of the track beyond the change of static depinning characteristics. In this section, we study how high-frequency (~GHz) currents can induce the ferromagnetic resonance (FMR) of the free layer. We also show how SV structures can be used to simultaneous excite and measure FMR excitations, and in this way use the study of FMR to characterise single, nano-scaled structures.

These resonant excitations are also interesting for their effects on pinned DW resonance [Bedau et al. 2007], DW pinning (particularly in depinning assisted by DW resonance) [Thomas et al. 2007; Parkin et al. 2008; Metaxas et al. 2010], and inter-DW resonance [O'Brien 2010].

## 5-2.1. Experiment

The experiment consisted in measuring the spectrum of the FMR response to an injected RF-modulated current, of a track with a T-trap.

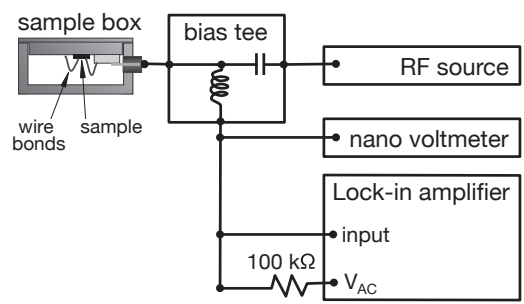

**FIG. 5-14 Setup for high frequency measurements** (schematic), showing the sample wire-bonded to a closed metal box, the bias tee, the RF power source, the nano-voltmeter and the lock-in amplifier. The lines represent coaxial cables. The controller computer is not shown for clarity.

The measurement setup, described in Chapter 2, is schematised in FIG. 5-14. The sample was put inside a metallic box, and the contact pads were wire-bonded to the box (ground pad) and to a through-connector (structure pad). The connector was then connected to an RF power source via a bias tee, so an RF current could be injected in the structure. To the DC arm of the bias tee, a nano voltmeter and a lock-in amplifier (along with an AC source) were connected. The nano voltmeter was used to measure the self-homodyne effect (explained below), while the lock-in amplifier was used to measure the track resistance (similarly to the previous setup). Static fields were applied to initialise the structures with an external electromagnet.

These experiments were performed on three C-shaped tracks with a T-trap, identical to those studied in the previous section (FIG. 5-1) [9]. Using the external field sequences described before (in the previous section as well as in Chapter 4), the structure was initialised in a mono-domain state, or in a bi-domain state with a DW pinned at the T-trap (in either a *P-wide*, *AP-narrow*, or *AP-wide* configuration). During the measurement, an RF current with a ramping frequency was injected, while static fields were applied and the dc voltage recorded (as will be further explained below). Two structure widths were tested, 110 and 200 nm.

There was an impedance mismatch between the measurement setup (which works at 50 Ω) and the structures (900 – 2000 Ω). Due to this mismatch, most of the applied RF power was reflected and not injected in the structures, and less so at higher frequencies. This impedes the precise determination of the injected power and current. We could however estimate the injected power by comparing the heat-induced resistance increase of a 0 dBm (at 50 Ω) RF signal with the same effect by a DC current. Measurements on one structure (R = 2 kΩ) showed that an applied power of 0 dBm produced about 0.3 mA (-11 dBm) at 0.5 GHz, and 0.2 mA (≈ -13 dBm) at 8 GHz.

## Self-homodyne measurement of a resonance peak

The resonance effects were measured with a self-homodyne scheme. This consisted in injecting the RF-current, $I_{RF}$, and measuring the generated DC voltage, $V_{DC}$. This voltage arises as both the current and the resistance (via the GMR effect) vary at the same frequency:

$$I_{RF} = I_0 \sin(\omega t) \, , \qquad R = R_0 + R_0 \, \text{MR} \, A_{MR} \sin(\omega t + \phi_{MR}) \qquad \text{(eq. 5-1)}$$

$$V = R \, I_{RF} = \tfrac{1}{2} I_0 R_0 \big( \text{MR} \, A_{MR} \cos(\phi_f) + 2\sin(\omega t) - \text{MR} \, A_{MR} \cos(2\omega t + \phi_{MR}) \big)$$

$$V = V_{DC} + V_\omega(t) + V_{2\omega}(t) \, ,$$

$$V_{DC} = \frac{I_0 R_0}{2} \text{MR} \, A_{MR} \cos(\phi_{MR}) \qquad \text{(eq. 5-2)}$$

Where $A_{MR}$ and $\phi_{MR}$ are the (frequency dependent) amplitude and phase lag of the mean variation of the horizontal component of the free layer magnetisation ($m_x$), and

---

[9] For fabrication and sample details see Annex A, sample reference HM01.

MR is the magneto-resistive ratio. As the frequency is swept through a resonant frequency $\omega_{FMR}$, the amplitude of the $m_X$ variation peaks and the phase increases by $\pi$. $V_{DC}$ is proportional not to $A_{MR}$ but to $A_{MR}\cos(\phi_{MR})$, and therefore no directly comparison is possible between $V_{DC}$ and $A_{MR}$. This transforms the peak shape, as shown in FIG. 5-15A. Depending on the oscillation modes present in the magnetic system, there might be a static phase delay: instead of $\phi_{MR}$ changing from 0 to $\pi$ (as in FIG. 5-15A), it may change from some initial value $\alpha$ to $\alpha + \pi$, producing the different peak shapes shown in FIG. 5-15B.

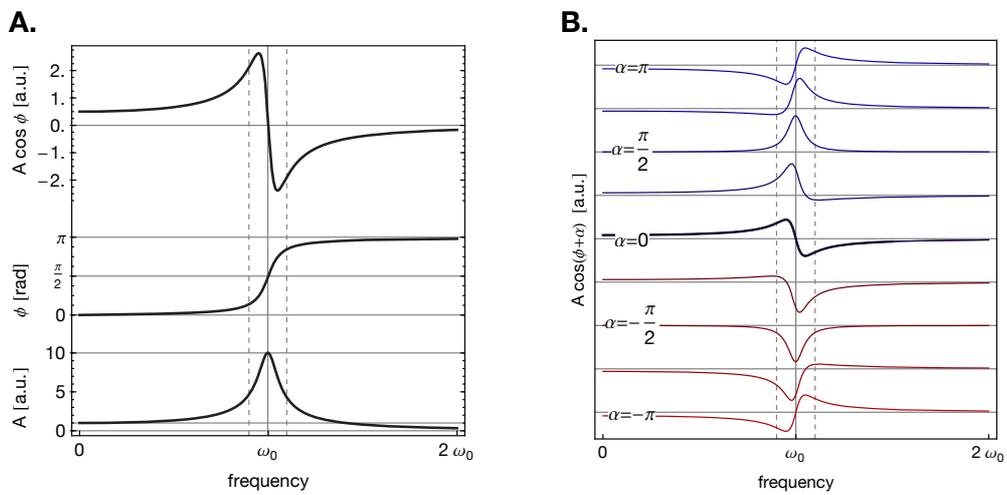

**FIG. 5-15 Shape of a harmonic resonance peak in a homodyne measurement. A.** Homodyne shape of the resonant response of a harmonic oscillator ($\zeta = 0.05$, $\omega_0 = 1$) vs. frequency. The bottom plot the amplitude $A$, the middle plot the phase $\phi$, and the top plot is the homodyne peak shape, $A\cos(\phi)$. **B.** Homodyne peak shape for varied fixed added phase, $A\cos(\phi+\alpha)$. In both plots the vertical lines mark $\omega_0$ (natural frequency) and the FWHM band.

### Comparison to other methods

The method described here uses current to excite the FMR, and uses the GMR effect to measure the resonance amplitude. An analogous method was reported before in the study of DW resonance in monolayer structures [Bedau et al. 2007; 2008], using the spin-transfer torque on the DW to drive the excitation, and the AMR effect to measure it. Comparatively, the presented method has some advantages: **1** the GMR greater signal, **2** the possibility of measuring thinner active layers [10], and **3** the ability to drive the resonance with an Oersted field. This allows more precise measurements, of smaller structures, and of magnetic configurations that are not subject to significant spin-transfer torques.

---

[10] As free layer thicknesses as low as 2 nm can be used in SVs with large GMR signal [Dieny 2004].

Other methods used to measure FMR in nanoscaled magnets involve the use of one or more external current lines to drive the excitation and measure the change in magnetic flux caused by the resonance phenomenon. In order to obtain a detectable change in magnetic flux, these techniques are usually applied to a larger magnetic volume. This typically requires the use of several adjacent copies of the tested structure, or the study of larger (or thicker) structures than the ones tested here (see e.g. [Silva et al. 1999; Buchanan et al. 2005; Giesen et al. 2005; Costache et al. 2006]).

Another important method to measure FMR in nanoscaled magnets is the use of magnetic tunnel junctions excited by spin-transfer torque. Such structures have been widely researched, motivated by its application to microwave oscillator devices (see e.g. [Russek et al. 2010] for a review). The very large MR ratio of such devices, as well as their more efficient spin-transfer torque effect, enable the study of FMR with greater precision in smaller magnets. Still, the method shown here is advantageous in some situations, as magnetic tunnel junctions have a more complex fabrication process, and are typically limited in area (due the degradation of MR ratio with increasing area).

## Uniform mode resonance

In the so-called uniform mode of FMR, the magnetisation oscillates with the same amplitude and angle in the whole magnetic volume. In the equations above, $A_{MR}$ and $\phi_{MR}$ are the amplitude and phase of the MR variation, and not the amplitude and phase of the magnetisation angle. $A_{MR}$ and $\phi_{MR}$ can be related to the magnetisation angle variation amplitude and phase ($A_\theta$ and $\phi_\theta$) by the following equations:

$$R = R_0 + R_0 \, \mathrm{MR} \, m_X(t) \qquad \text{(eq. 5-3)}$$

$$\begin{cases} m_X = \cos(\theta) \\ m_\perp = \sin(\theta) \end{cases}, \qquad \theta = \theta_0 + A_\theta \sin\left(\omega t + \phi_\theta\right)$$

$$\Rightarrow m_X = \cos(\theta_0)\cos\left(A_\theta \sin(\omega t + \phi_\theta)\right) - \sin(\theta_0)\sin\left(A_\theta \sin(\omega t + \phi_\theta)\right)$$

Expanding about $A_\theta$=0,

$$m_X = \cos(\theta_0) - A_\theta \sin(\theta_0)\sin\left(\omega t + \phi_\theta\right) - \frac{1}{2} A_\theta^2 \cos(\theta_0)\sin^2(\omega t + \phi_\theta) + O\!\left(A_\theta^3\right)$$

$$m_X \approx \cos(\theta_0) - A_\theta \sin(\theta_0)\sin\left(\omega t + \phi_\theta\right) - \frac{1}{4} A_\theta^2 \cos(\theta_0)\left(1 - \cos(2\omega t + 2\phi_\theta)\right)$$

Ignoring the constant terms and terms that vary with $2\omega$ (as they do not contribute to $V_{DC}$):

$$\Delta m_X = -A_\theta \sin\left(\theta_0\right) \sin\left(\omega\, t + \phi_\theta\right) \quad .$$

Remembering eqs. 5-1 and 5-3, we reach finally:

$$A_{MR} = -A_\theta \sin\left(\theta_0\right), \qquad \phi_{MR} = \phi_\theta$$

Note that this predicts the oscillation around the horizontal direction ($\theta_0 = 0$) produces no GMR signal [11]. As such, a track in a single domain state, under moderate fields, should produce no signal [12]. However, as was seen in the previous chapters, there are always small deviations of the reference layer direction, which generate a finite signal. Also, Costache et al. in a study of FMR resonance of Permalloy tracks subject to a similar symmetry signal constraints [Costache et al. 2006], were able to detect a signal, which they attributed to small deviations of the resonant mode from perfect symmetry.

**Other resonant modes**

Finally, it is important to realise that there can be resonant modes that are not described by a single magnetisation vector oscillating in angle. In a multi-domain state, when a DW is present, the (total) magnetisation magnitude can also oscillate. Also, the different regions — domains and DWs — can have different orientations and different demagnetisation fields and, as such, different resonant frequencies and modes.

## Sources of resistance variation

The model described above considers only the GMR contribution to the variation of resistance. Other contributions, such as AMR, might also be significant. The determination of the actual weight of each contribution would require the precise measurement of the resistance amplitude of the observed resonant peaks, and the knowledge of the injected current magnitude and distribution — quantities that are not known with sufficient precision. However, whether the resistance variation is produced by GMR or AMR (or a combination of the two), still the same above-described principles apply.

---

[11] Intuitively, this can be understood as the magnetisation vector describing a cone around the horizontal axis, causing its horizontal projection to be constant.

[12] This is true also for the AMR contribution as, at $\theta_0 = 0$, $\partial R_{AMR} / \partial\theta = 0$.

**Fit model**

The voltage peaks will be fitted to an adapted damped harmonic oscillator model:

$$V_{DC}(\omega) = A_V \cos(\alpha + \phi) + s_1\,\omega + s_0 \ , \qquad\qquad (eq.\ 5\text{-}4)$$

where $A_V = \frac{I_0 R_0}{2} MR\, A_{MR}$ (cf. eq. 5-2) is the oscillation amplitude, $\phi$ is the harmonic phase lag, $\alpha$ is the residual phase, and $s_1\,\omega + s_0$ is a residual fitting function [13]. The amplitude and phase are given by:

$$A_V = \frac{D}{\omega\sqrt{4\,\zeta^2\,\omega_0^2 + \frac{(\omega_0^2 - \omega^2)^2}{\omega^2}}} \qquad \phi = \tan^{-1}\!\left(\frac{2\,\zeta\,\omega\,\omega_0}{\omega_0^2 - \omega^2}\right) \qquad (eq.\ 5\text{-}5)$$

where $\zeta$ is the harmonic damping ratio, and D is proportional to the driving force (and absorbs the $\frac{I_0 R_0}{2} MR$ factor), and $\tan^{-1}$ is the arctangent to the interval $(0,\pi)$. The fitting parameters are then [14]: $\omega_0$, $\zeta$, $\alpha$, D, $s_1$, and $s_0$.

## 5-2.2. Single domain FMR response

The measurement of a 110 nm wide track, with a T trap on the bottom, in the single domain state is shown in FIG. 5-16. The homodyne voltage ($V_{DC}$) is plotted as a function of the injected RF signal frequency (applied RF power was -2dBm), for two magnetisation orientations (±x). The model fitted well to the data, with a large peak observed at 7.5 GHz for both domain orientations. The peak width was also similar for both configurations (the FWHM [15] was 0.45 GHz for the leftward domain, and 0.48 GHz for the rightwards). A smaller peak at 3.7 GHz was also observed in the leftward domain spectrum, to which we will return further below.

---

[13] This function serves to eliminate other broad features in the spectrum (such as a second FMR peak).

[14] $\omega_0$ and $s_1$ will be presented in frequency units and not in angular frequency units ($2\pi\,f_0 = \omega_0$).

[15] FWHM = $2\,\zeta\,f_0$ .

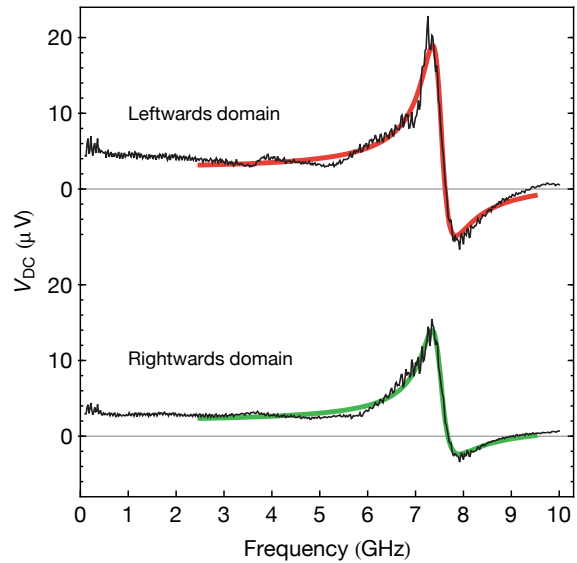

| fit parameter | leftwards | rightwards |
|---|---|---|
| $f_0$ (GHz) | 7.511±0.005 | 7.488±0.005 |
| $\zeta$ ($10^{-3}$) | 29.8 ±0.6 | 32.3 ±0.6 |
| $\alpha$ (rad) | -0.38 ±0.02 | -0.54 ±0.02 |

**FIG. 5-16 Single domain FMR.** Structure with a T trap on the bottom. **Top.** $V_{DC}$ vs. frequency (black lines) and fitted peak function (thick coloured lines) for both domain orientations. **Bottom.** Fitting parameters. The track was 110 nm wide, applied RF power was -2dBm, $R_0$ = 1876 Ω, no applied field.

**Resonance frequency versus applied field**

The same experiment was repeated under a horizontal externally applied (DC) field $H_x$, in a single domain state for both domain orientations, FIG. 5-17. The frequency of the main peak shifts linearly with the applied field, towards higher frequencies when the field and the magnetisation are parallel and towards lower frequencies when they are anti-parallel. The smaller peak also shifts with external field, and with negative fields it is no longer observable.

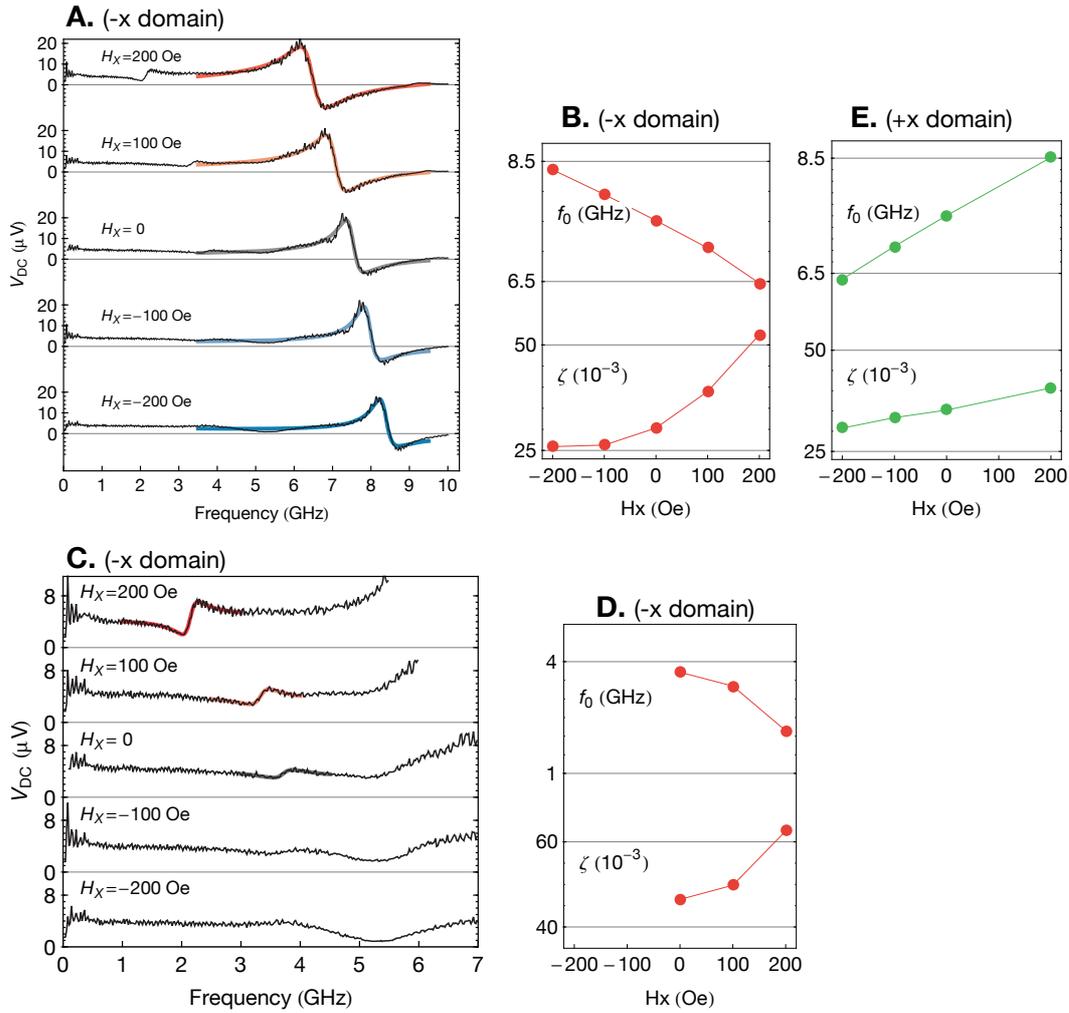

**FIG. 5-17 Single domain FMR vs. H_X.** Structure with a T trap on the bottom magnetised in −x (A–D) or +x (E). **A.** V_DC vs. frequency (black lines) and fitted peak function (coloured thick lines) for different external static fields. **B.** Main fitting parameters of the large peak. **C.** Replot showing the low-frequency peak, which was also fitted (coloured lines in C, and main parameters in **D**). **E.** The fitting parameters of the main peak with the magnetisation rightwards (+x). The track was 110 nm wide, applied RF power was -2dBm, R_0 = 1876 Ω.

Two other structures were also measured, of larger width (200 nm), one with the T trap on the bottom and the other on the top. In these other structures, a large peak was equally observed, and the fitted f₀ versus applied DC external field is shown in FIG. 5-18. No secondary peaks were observed.

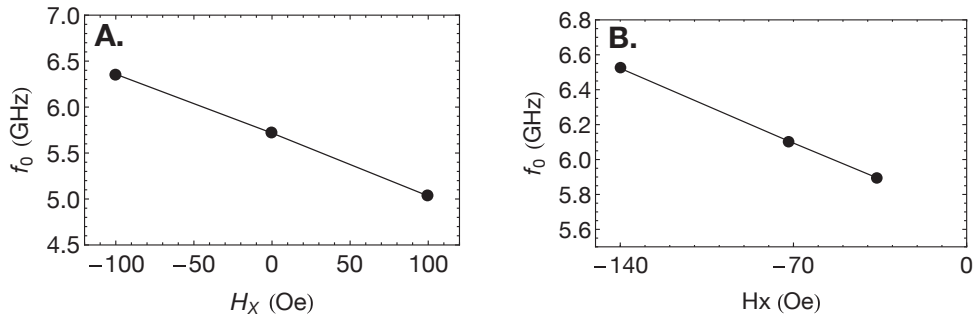

**Fig. 5-18 FMR vs applied field in 200 nm wide structures.** The fitted frequency of the resonance peak of two 200 nm structures magnetised in the −x direction. Structure **A** had a ⊤ trap on the top, and **B** on the bottom.

## Analysis: Kittel law

The linear shift of the large resonant peak suggests it corresponds to the uniform FMR mode. To verify if this hypothesis is correct, we will analyse our experimental results using Kittel's FMR theory and, later, we will compare our findings to micromagnetic simulations.

The uniform mode is a precessing mode where the magnetisation is spatially uniform (in contrast with other magnetic excitations, such as spin-waves, or DW resonance). The Kittel law [Kittel 1948] predicts that, in a strip, the FMR frequency of the uniform mode is given by:

$$\omega_0 = \gamma \, \mu_0 \, \sqrt{H_0 + \mathcal{N}_Y M_S} \; \sqrt{H_0 + (1 - \mathcal{N}_Y) M_S} \qquad (eq. 5\text{-}6)$$

where $H_0$ is the applied DC field, $N_Y$ is the demagnetisation factor along Y ($N_X = 0$, and $N_Z = 1 - N_Y$), $\gamma = g \, \mu_B \, / \, \hbar$ is the gyromagnetic ratio, and g is the spectroscopic splitting factor ($\approx 2.15$ for Py) [Kittel 1948; 2005].

The dependence of frequency shift with the external field is approximately linear, which is consistent with the observed results. Using $M_S = 800$ kA/m (typical value for bulk Py [Miyajima & Sato 1976]) and the demagnetisation factor of an infinite track [Aharoni 1998], the predicted resonant frequency at $H_0 = 0$ is 8.9 GHz (for the 110 nm wide track) and 7.2 GHz (200 nm wide), which are greater than the observed values, shown in Fig. 5-19. This may be due to a lower effective saturation magnetisation, as eq. 5-6 does predict $\omega_0 \propto M_S$.

Another way to analyse these results is to extract the effective $M_S$ and $N_Y$ and then compare the obtained values among different structures and to the expected values.

The product of the slope and intersect of the frequency vs. field line yields the value of the effective saturation magnetisation:

$$\omega_0 \frac{\partial \omega_0}{\partial H_0} \Big|_{H_0 = 0} = M_S \left( \gamma \mu_0 \right)^2$$

and the inverse of the slope yields the effective demagnetisation factor:

$$(\frac{\partial \omega_0}{\partial H_0})^{-1} \Big|_{H_0 = 0} = \frac{2}{\gamma \mu_0} \sqrt{N_Y} \sqrt{1 - N_Y}$$

These values thus obtained are only dependent on well-known constants ($\gamma$ and $\mu_0$), the DC external field calibration (correct within 5–10%), and, for the effective $M_S$, on the field zero calibration [16]. The obtained values can be seen in FIG. 5-19. The obtained effective $M_S$ ranged 604–692 kA/m, a value lower than expected for Py (~800 kA/m; e.g. [Miyajima & Sato 1976]). A lower effective $M_S$ indicates either an actual lower material $M_S$, or that not all the spins are precessing but that a portion is being pinned at the track border (caused by a non-uniform demagnetisation field).

The obtained demagnetisation factors agree well with the expected value for long tracks [Aharoni 1998], with an average difference of 12%. This difference is well within the expected error of the calculated $N_Y$. This factor was calculated from the track width, measured using SEM (typical error ±10 nm in 110 or 200 nm), and the free layer thickness, measured indirectly from the deposition time (typical error less than ±.5 nm, in 8 nm).

| Structure | Width (nm) | Domain | $f_0$ @H=0 (GHz) | $\partial f_0/\partial H_X$ (MHz/Oe) | Eff. $M_S$ (kA/m) | Eff. $N_Y$ |
|---|---|---|---|---|---|---|
| FIG. 5-17 | 110 | -x | 7.47 | -4.67 | 613 | 0.118 |
| | | +x | 7.48 | 5.26 | 692 | 0.090 |
| FIG. 5-18A | 200 | -x | 5.70 | -6.62 | 664 | 0.055 |
| FIG. 5-18B | 200 | -x | 5.67 | -6.05 | 604 | 0.066 |

**FIG. 5-19 Frequency vs. applied field: fit and Kittel law.** A linear fit was applied to the $f_0$ vs $H_0$ curves shown before. The effective $M_S$ and demagnetisation factor were calculated with the Kittel law for stripes (see text). The theoretical $N_Y$ are 0.095 (for 110 nm wide) and 0.060 (for 200 nm).

---

[16] The field zero calibration may be off by ~10 Oe due to inter-layer magnetostatic interactions. This has only a small effect on the obtained effective $M_S$ of ~1%.

In summary, our observations of a large resonant peak are in excellent quantitative agreement with Kittel law for the uniform resonant mode. The demagnetisation factor extracted from the measurements is very close to the theoretical value calculated from the geometrical dimensions of two different track sizes. The $M_S$ value, though, is significantly smaller than the expected value. To understand if this difference is caused by an actual difference in the material parameter, or instead by pinned magnetisation at the border of the resonating volume, we will perform micromagnetic simulations.

## Micromagnetic simulation

The response of a straight track, without a T-trap, to an oscillating Oersted field was simulated using the OOMMF code [Donahue & Porter 1999]. The Oersted field was scaled so its maximum amplitude was 1 Oe at the Py layer. The track dimensions were smaller than the experiment ($8 \times 50 \times 300$ nm$^3$) due to computation time limitations. The simulation was repeated with a static applied field in the x direction, and the effective $M_S$ and $N_Y$ were calculated as described before.

FIG. 5-20 shows the obtained results. FIG. 5-20A shows the amplitude of the change in normalised horizontal magnetisation ($m_X$) versus the frequency of the applied RF Oersted field, at $H_{DC} = 0$, revealing a clear resonant peak at 10.34 GHz. The same simulation was repeated for two other values of $H_{DC}$ ($\pm 100$ Oe), and the peaks were fitted to a Lorentz function (fitted $f_0$ and $\zeta$ are shown in FIG. 5-20B). The calculated effective $M_S$ and $N_Y$ are shown in FIG. 5-20D.

The resonance frequency of the simulated track is higher than experiment (10.34 GHz vs. 7.5 or 5.7 GHz), which can be attributed to the difference in track width. The calculated effective $M_S$ is lower than the material $M_S$ (690.1 vs. 800 kA/m), which is in excellent agreement with the experimentally obtained value. A snapshot of the oscillation speed ($\partial m / \partial t$) at resonance, FIG. 5-20C, also showed that the spins near the borders did not oscillate as much as those in the middle of the volume. The effective $N_Y$ is higher than the theoretical value for an infinite track (0.195 vs. 0.170), possibly due to the finite length of the simulated track.

This result supports the hypothesis that the **reduced effective $M_S$ is due to the pinning effect of the track borders** and not to an actually reduced material $M_S$.

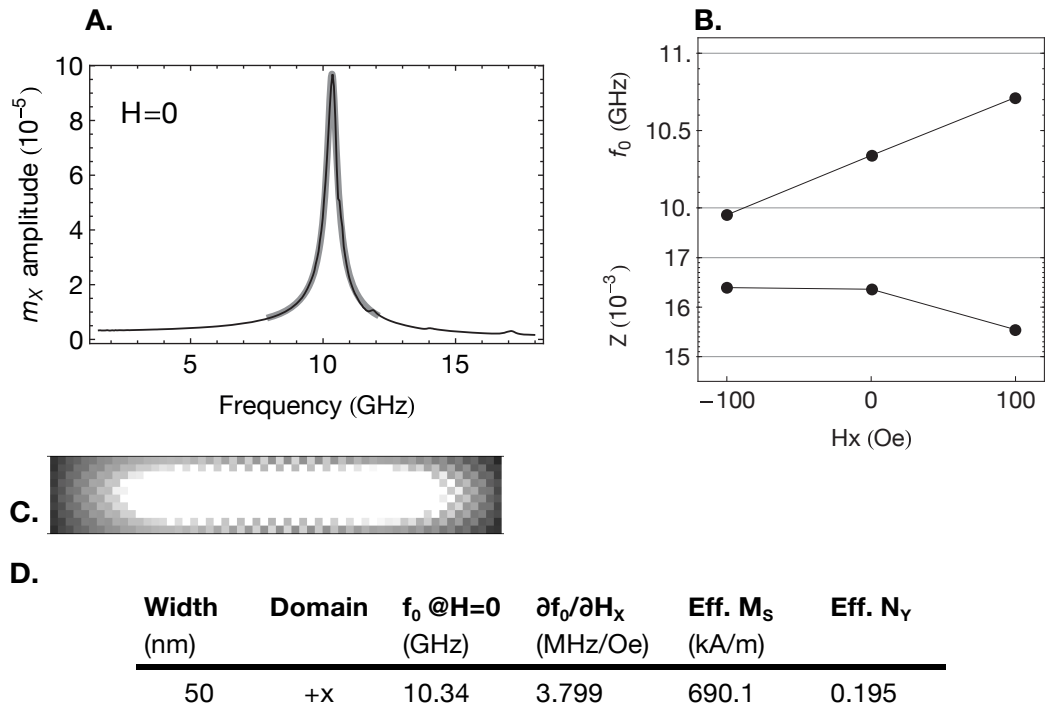

**A.**

**B.**

**C.**

**D.**

| Width (nm) | Domain | $f_0$ @H=0 (GHz) | $\partial f_0/\partial H_X$ (MHz/Oe) | Eff. $M_S$ (kA/m) | Eff. $N_Y$ |
|---|---|---|---|---|---|
| 50 | +x | 10.34 | 3.799 | 690.1 | 0.195 |

**FIG. 5-20 Track FMR** (micromagnetic simulation). **A.** Amplitude of oscillation of the normalised $m_x$ vs. applied Oersted field frequency (black line) and a fit to a harmonic resonance peak (red). **B.** Variation of $f_0$ and $\zeta$ with static applied field $H_X$. **C.** Snapshot of $\partial m/\partial t$ at 10 GHz in the simulation volume, white corresponding to $10^9$ rad/s, black to 0. **D.** Calculated effective $M_S$ and $N_Y$. The theoretical $N_Y$ for a track of infinite length is 0.170. Simulated volume = $8\times50\times300$ nm³, cell size = $5\times5\times8$ nm³, $M_S$ = 800 kA/m, α = 0.01. Simulated using the OOMMF code [Donahue & Porter 1999].

## Simulation of a T trap

The same simulation was performed on a track with a T-trap, magnetised in a single domain state. The mX change amplitude response is shown as a function of Oersted field frequency, for two values of $H_{DC}$, in FIG. 5-21. Comparing the spectrum obtained with the previous straight track, several new peaks are visible. This is due to the existence of regions with different demagnetisation factors, which resonate at different frequencies, especially the region of the junction with the T-trap. This could be the source of the smaller peak observed in some of the previous measurements. Not all this peaks may be visible in an actual experiment. Different peaks correspond to different regions of the structure, which have different resting angles and different current densities, and so can produce very different homodyne voltage peaks.

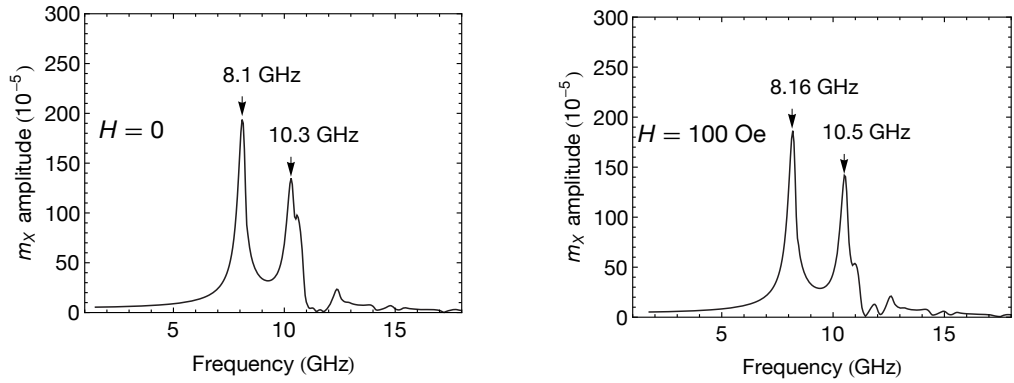

**FIG. 5-21 Track with T-trap FMR** (micromagnetic simulation). The plots represent the $m_x$ oscillation amplitude vs. Oersted field frequency, at two different values of applied field. Same simulation parameters as FIG. 5-20. The track was 300 nm long with a 150 nm long stub at the midpoint.

## 5-2.3. Tracks with two domains

The same track as in FIG. 5-18A was magnetised in a two domain state, with a DW pinned at the T-trap in the *P-wide* configuration. The resonance spectrum was obtained as before and fitted to a double peak function (i.e. the sum of two peak functions defined in eqs. 5–4 and 5–5), results shown in FIG. 5-22. From this peak, it is seen that the signal is composed of the uniform mode response of the two domains. As in the single domain mode, the frequency shifts linearly with applied static field. As there are two domains, one right and the other leftwards, the two peaks shift in opposite directions. The linear fit slopes are similar to those found before (here: -6.13 and +6.51 MHz/Oe, versus -6.6 MHz/Oe in the single domain state). The 0-intersect frequencies of the two peaks are approximately equal (5.71 and 5.72 GHz). The amplitudes are very different, with the fitted amplitude factor (*D* in eq. 5–5) differing by a factor of 5 times. These results are in close agreement with the single domain observations. We can then conclude that we observe here the uniform mode of the track in the two domains, before and after the T-trap.

It would be expectable that the amplitudes of the two peaks to be proportional to the lengths of the respective track segments, i.e. the same oscillation in a longer segment should produce a larger change in resistance. However, the signal amplitudes of the two peaks differ by far more than the relative resistances of the two segments.

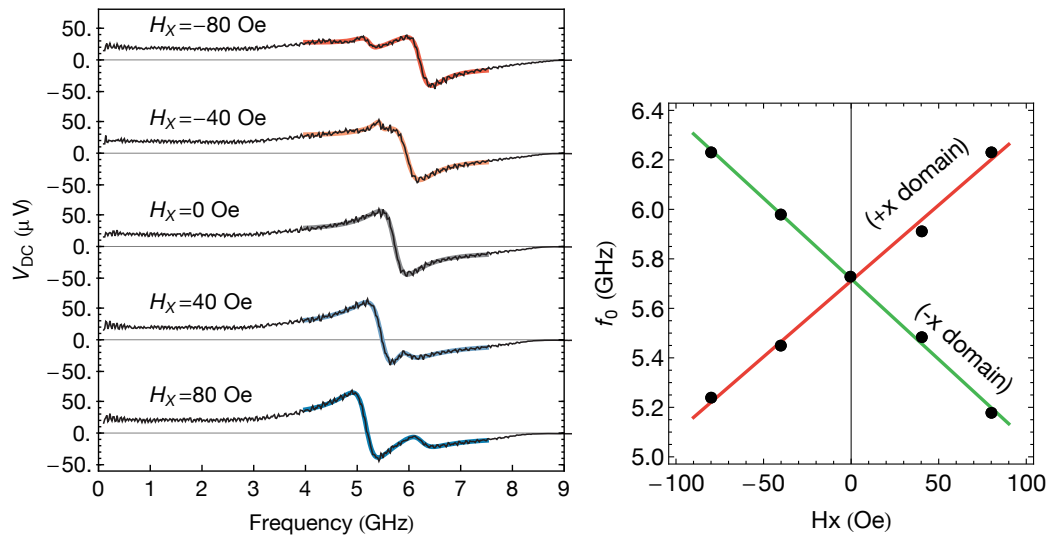

**Fig. 5-22 Double domain fmr vs. H_X.** Structure with a т trap on the bottom, 200 nm wide, magnetised in the *P-wide* configuration. **Left.** $V_{DC}$ vs. frequency (black lines) and fitted double peak function (coloured thick lines) for different external static fields. **Right.** $f_0$ parameters of the fitted function vs. applied static field (black circles) and linear fit (coloured lines).

As explained before, in an ideal track, these resonant modes should produce no signal. We proposed before that they indeed do as a consequence of a deviation from the symmetry of the ideal case, probably due to a small error in the angle of the reference layer. The observed variation between peak amplitudes of different segments suggests that this deviation from perfect symmetry may not be uniform in the whole structure.

## 5-2.4. Summary

We have shown the **self-excited, self-homodyne measurement of fmr of sv tracks** under moderate currents. The observed mode was in excellent agreement with Kittel's theory for uniform mode resonance. Using a simple model for the self-homodyne peak, basic parameters were extracted from the measurements, allowing the calculation of the effective $M_S$ and demagnetisation factor. We observed a reduced effective $M_S$, which we attributed to the pinning effect of the demagnetisation field at the track borders. We then performed micromagnetic simulations of fmr excitation by the induced Oersted field, which were in excellent agreement with the measured spectra and values, indicating that the Oersted field is the main excitation driver.

# 5-3. Conclusion

We studied the **effects of DC currents on the DW depinning from a T trap**. By measuring different DW polarities, different pinning configurations, and different depinning directions, we were able to quantify the effects on depinning of three phenomena: Joule heating, Oersted field, and spin-transfer torque.

The influence of the **spin-transfer torque** was determined to be in the range of a few Oe (0.4–4 Oe) at a current density in the Permalloy layer of $\approx 2.8 \times 10^{11}$ A·m$^{-2}$. The strength of this effect is similar to what was reported in monolayer tracks. We have not observed an enhanced effect in SV tracks, as was observed in some, but not all, reported studies.

Though the origin of the enhanced spin-transfer effect torque in SV tracks is still unknown, some authors (referenced before) have suggested that it is caused by perpendicular-to-plane spin currents, which may be further investigated by testing different SV compositions. Reducing the Oersted field may also be advantageous to making the effect of spin-transfer torque clearer. This could be achieved by adding current shunting layers to the SV stack, so that the current density profile is centred with the Py layer, with a tolerable reduction in MR signal.

The present study shows that the T trap structure presents a good model system to investigate current induced DW depinning and propagation effects, due to its well-characterised pinning mechanisms, known DW structure, the ability to separate different current effects with different pinning configurations, and the ability to create very weakly pinned DWs (H$_{PULL}$ in the *AP-wide* case).

We then studied the **effect of RF currents**, and demonstrated the **self-excited, self-homodyne measurement of the FMR of a single nanoscaled track**. The observed resonance peaks were in excellent agreement with the uniform mode predicted by Kittel's theory, and with micromagnetic simulations.

The method shown here could be adapted in the future to study other FMR effects, such as resonance of pinned DWs or of multiple coupled DWs [Bedau et al. 2007; O'Brien 2010]. Different modes couple differently with the excitation Oersted field and, in order to excite these different modes, the direction and amplitude of the Oersted field could be tailored by changing the position and number of the electrical contacts. The overall magnitude of the Oersted field can also be adjusted by changing the SV composition.

# 5-4. References

# [6] Conclusion

We have studied the propagation and pinning of DWs in SV nanotracks, in straight tracks and in track circuits with artificial traps, using electrical measurements to determine the DW position.

We started by investigating how to obtain a SV nanotrack that behaved as a good DW conduit: a track wherein DWs propagated with minimal pinning, driven by external fields much smaller than the nucleation field. We found that the composition of the reference layer was a crucial parameter, as well as the patterning technique. By using a SAF reference layer, patterned by an ion milling etch process, we were able to demonstrate the conduit behaviour on tracks down to 33 nm width, the narrowest DW conduits ever measured to this date. The propagation field, a measure of the significance of natural pinning defects, was lower than what has been reported in SV tracks (e.g. [Grollier et al. 2003; Himeno et al. 2003; Lacour et al. 2004; Uhlir et al. 2010]), though higher than what has been achieved in monolayer Py tracks. By using L-shaped tracks, we controlled the internal structure of the injected DW, as subsequently confirmed by studies on T gates. This control greatly simplifies the analysis of DW pinning, but is often unaddressed in studies of SV tracks (e.g. [Briones et al. 2008]). Moreover, the MR signal we measured enabled the determination of the DW position with great resolution (tens of nm) even in the narrowest tracks. To the best of our knowledge, these are the narrowest DW conduits, either in SV or Py, ever measured to date, and the SV tracks with best DW conduit properties.

Using these SV DW conduits, we could then demonstrate and study four families of DW digital devices: the spiral track, the T gate, the NOT gate (including two designs for high density gates), and the DW interaction parallel tracks. We compared their behaviour with earlier studies on Py structures, and found that they worked analogously to their Py counterparts, though often with a smaller operating margin (the range of applied fields producing the desired behaviour), which we attributed to a higher density of natural pinning defects. Using these structures, we demonstrated several digital functions on SV tracks: digital counting of field rotations, a controllable DW valve with a T gate, the NOT gate, a 7-gate shift register, and DW depinning induced by another DW. All but the first [Mattheis et al. 2006] had only been demonstrated in monolayer systems [Allwood et al. 2004; Petit et al. 2008], and the last had never been observed.

The single-shot resistance measurements also enabled us to characterise these logic structures, measuring some phenomena which are either new or only predicted in simulation to date. These included, among others, the magnetostatic interaction of the DW with the T gate, the generation, pinning, and separation of 360° DWs in a T gate, the multiplicity of effects induced by the interaction of two DWs in adjacent tracks, and the progress of a DW inside a NOT gate. These systems have clearly showed the usefulness of studying DW logic devices in SV nanotracks, as well as having demonstrated that complex DW logic circuits can be implemented and electrically measured in SV nanotracks, an important step towards the electronic integration of future DW logic devices.

Finally, we studied the effect of injected electrical currents on the track magnetisation and on DW depinning. Using the symmetry of DW depinning from a T gate, we determined the magnitude of different interaction mechanisms of a DC current on DW depinning (Joule heating, Oersted field, and spin-transfer torque). Of these, the spin-transfer torque incites a special technological interest, as it can be applied to electrically propagate DWs. We found that the influence of the spin-transfer torque was equivalent to a few Oe (0.4–4 Oe) at a current density of $\approx 3 \times 10^{11}$ A·m$^{-2}$, similar to what has been reported in studies of monolayer Py tracks (e.g. [Vernier et al. 2004]) and in some studies of SV track [Jiang et al. 2011; Mihai et al. 2011]. We have not observed an enhanced effect in SV tracks observed in some other SV studies [Grollier et al. 2004; Ravelosona et al. 2007; Pizzini et al. 2009]. This discrepancy, of great technological importance, remains to be determined. We suggest that the T gate used here, by virtue of its multiple and well characterised pinning configurations and its ability to create very weakly pinned DWs, is a useful model system to study current induced depinning.

We also studied the effect of RF currents, and demonstrated the self-excited, self-homodyne on-chip measurement of the FMR of a single nanoscaled track. Our measurements were in excellent agreement to Kittel's theory, and with micromagnetic simulations.

We believe that the SV nanotracks will be of great use in future experiments. On the particular subjects of this thesis, we think that it would be useful to demonstrate electronic data input, storage, and output in a shift register, by joining the shift register demonstrated here with the injection of DWs with current lines demonstrated in [Himeno et al. 2005]. It would also be interesting to study the effect of the pinning potential to the 'drive-by' depinning mechanism (§4-4.2), and in particular to demonstrate the depinning of a series of DWs by 'drive-by' depinning (also proposed in §4-4.2). The FMR measurement of current induced DW resonance (demonstrated in [Bedau et al. 2007]) would also be of use in the study of the pinning potentials brought about by the artificial traps magnetostatic interactions studied in this thesis. In addition, it would also enable the observation of the oscillation mechanisms of multiple coupled DWs in adjacent tracks [O'Brien 2010], with potential technological applications to RF oscillators. On the subject of current induced DW depinning and propagation, as stated before, we believe that more experimental observations on SV tracks of different compositions will be useful to understand the origin of the enhanced STT effect in SVs reported in literature.

Overall, our results have shown that DW logic structures, as well as fundamental studies of DW propagation and pinning, can be successfully implemented with SV tracks, with great advantages for their measurement and characterisation. The single-shot measurements possible in SV tracks are of great use when studying stochastic phenomena, as is the case of DW depinning and even of DW propagation at a smaller scale. The precise determination of DW position is a precious tool in the study of DW interactions with artificial structures or other DWs, as done in this thesis, and also for the study of DW dynamics (e.g. [Glathe et al. 2008]). It also provides a way of integrating DW logic devices with electronics, of great technological interest.

We would like to point out that many of these results, regarding the implementation of DW conduits and logic circuits in SV tracks, are of interest also in the wider class of magnetic multilayer tracks. Indeed, many of the problems addressed here, such as magnetostatic inter-layer interactions, are common to all multi-layer systems. These include proposed stacked multi-conduit devices [Cowburn et al. 2010], and DW conduits implemented in magnetic tunnel junctions. This latter class has the advantage over the SV track of higher signal, and thus finer measurements and easier electronic integration, and of high efficiency spin-transfer torque DW propagation, which has been recently demonstrated [Chanthbouala et al. 2011].

[6] Conclusion

# [A] Sample details

We list here the fabrication details of the samples mentioned in this thesis. For all presented SV stacks, thickness in nms, Py = $Ni_{19}Fe_{81}$, CoFe = $Co_{80}Fe_{20}$, MnIr=$Mn_{76}Ir_{24}$.

## A-1. Sample HM01

**SV Stack**

Ion beam assisted sputtering (at INESC-MN, Lisbon).

(Si/SiO$_2$)// Ta 2/Py 8/Cu 2/CoFe 2.2/Ru 0.8/CoFe 2.2/MnIr 6/Ta 5

### Patterning of SV layer

1. **PMMA coating:** ➢Pre-ash (O$_2$ plasma etch) 10 W for 30″. ➢PMMA 950k 4% in anisole, spin-coated at 1.3 kRPM for 2′ (~300 nm). ➢Bake in hot plate at 120 °C for 2′.

2. **EBL:** ➢20 kV, 20 μm aperture, I = 97pA. ➢Tracks composed by 2, 6, or 10 parallel lines with dose 700 pC/cm. ➢Developed in MIBK:IPA 1′, washed in IPA. ➢Post-ash (O$_2$ plasma etch) 10 W for 2′.

3. **Hard mask deposition:** ➢Thermal evaporation of Ti 11.7 nm. ➢Lift-off in acetone (1 hr) + washing.

4. **Ion beam milling:** ➢beam parameters 600 V / 28 A. For 4′ (in 30″ steps with 30″ pauses). ➢Used calibration samples of SV on glass and Ti on glass.

### Patterning of contact layer

5. **PMMA coating:** ➢Pre-ash (O$_2$ plasma etch) 10 W for 1′. ➢PMMA 950k 4% in anisole, spin-coated at 1.5 kRPM for 2′ (~280 nm). ➢Bake in hot plate at 110 °C for 1′.

6. **EBL:** ➢20 kV, 60 μm. I = 930 pA. ➢Area dose 200 μC/cm$^2$. ➢Developed in MIBK:IPA 4′ (in two steps), washed in IPA. ➢Post-ash (O$_2$ plasma etch) 10 W for 1′.

7. **Contact deposition:** ➢Pre-etch: ion beam milling (600 V / 28 mA) for 29″ (~ 3 nm). ➢In-chamber magnetron sputtering of Ta 3 nm/Cu 80 nm/Ta 5 nm. ➢Lift-off in acetone + washing.

---

# A-2. Sample HM03

## SV Stack

Ion beam assisted sputtering (at INESC-MN, Lisbon).

(Si/SiO₂)// Ta 2/Py 8/Cu 2/CoFe 2.2/MnIr 6/Ta 5

## Patterning of SV layer

1. **PMMA coating:** ➢Pre-ash (O₂ plasma etch) 10 W for 1′. ➢PMMA 950k 4% in anisole, spin-coated at 1.5 kRPM for 2′00″ (~300 nm). ➢Bake in hot plate at 120 °C for 1′.

2. **EBL:** ➢20 kV, 20 μm aperture, I = 91 pA ➢Tracks composed by 2, 6, or 10 parallel lines with dose 800 pC/cm ➢Developed in MIBK:IPA 1′30″, washed in IPA. ➢Post-ash (O₂ plasma etch) 10 W for 1′.

3. **Hard mask deposition:** ➢Evaporation of Ti 22.0 nm. ➢Lift-off in acetone (1 hr) + washing.

4. **Ion beam milling:** ➢beam parameters 600 V / 28 A, for 8′ (in 30″ steps with 30″ pauses). ➢Used calibration samples of SV on glass and Ti on glass.

## Patterning of contact layer

5. **PMMA coating:** ➢Pre-ash (O₂ plasma etch) 10 W for 1′. ➢PMMA 950k 4% in anisole, spin-coated at 2 kRPM for 2′ (~280 nm). ➢Bake in hot plate at 140 °C for 1′.

6. **EBL:** ➢20 kV, 60 μm, I = 927 pA. ➢Area dose 210 μC/cm². ➢Developed in MIBK:IPA 1′30″, washed in IPA. ➢Post-ash (O₂ plasma etch) 10 W for 1′.

7. **Contact deposition:** ➢Pre-etch: ion beam milling (600 V / 28 mA) for 20″ (~ 2 nm). ➢In-chamber magnetron sputtering of Ta 3 nm/Cu 80 nm/Ta 6 nm. ➢Lift-off in acetone (1 hr) + washing.

## A-3. Sample HM14

### SV Stack

Ion beam assisted sputtering (at INESC-MN, Lisbon).

(Si/SiO$_2$)// Ta 3.5/ Py 8/ CoFe 2/ Cu 2.4/ CoFe 2/ Ru 0.8/ CoFe 2/ MnIr 7/ Ta 2

MR = 5.3%      H$_F$ = 10.0 Oe      H$_C$ = 1.0 Oe

### Patterning of SV layer

1. **PMMA coating:** ➤Pre-ash (O$_2$ plasma etch) 10 W for 1′. ➤PMMA 950k 4% in anisole, spin-coated at 3.3 kRPM for 1′30″ (~270 nm). ➤Bake in hot plate at 145 °C for 1′.

2. **EBL:** ➤20 kV, 20 μm aperture, I = 145 pA ➤Tracks composed by 3, 6, or 12 parallel lines with dose 1000 pC/cm. ➤Developed in MIBK:IPA 1′, washed in IPA. ➤Post-ash (O$_2$ plasma etch) 10 W for 1′.

3. **Hard mask deposition:** ➤Thermal evaporation of Ti 18 nm. ➤Lift-off in acetone (1 hr) + washing.

4. **Ion beam milling:** ➤beam parameters 600 V / 28 A. For 4′ (in 30″ steps with 30″ pauses). ➤Used calibration sample of SV on glass.

### Patterning of contact layer

5. **PMMA coating:** ➤Pre-ash (O$_2$ plasma etch) 10 W for 1′. ➤PMMA 950k 4% in anisole, spin-coated at 3 kRPM for 2′ (~280 nm). ➤Bake in hot plate at 130 °C for 1′.

6. **EBL:** ➤20 kV, 60 μm, I = 1350 pA ➤Area dose 280 μC/cm$^2$. ➤Developed in MIBK:IPA 1′, washed in IPA. ➤Post-ash (O$_2$ plasma etch) 10 W for 1′.

7. **Pre-milling:** ➤Ion beam milling (600 V / 28 mA) for 10″ (~ 1 nm). ➤In-chamber magnetron sputtering of Ta 15 nm.

8. **Contact deposition:** ➤Thermal evaporation of Ti 5 nm/Au 200 nm (imprecise calibration). ➤Lift-off in acetone (1 hr) + washing.

## A-4. Sample HM18

### SV Stack

Ion beam assisted sputtering (at INESC-MN, Lisbon).

$(Si/SiO_2)//$ Ta 3.5/ Py 8/ CoFe 2/ Cu 2.4/ CoFe 2/ Ru 0.8/ CoFe 2/ MnIr 7/ Ta 2

MR = 5.3%          $H_F$ = 10.0 Oe      $H_C$ = 1.0 Oe

## Patterning of sv layer

1. **PMMA coating:** ➢Pre-ash ($O_2$ plasma etch) 10 W for 1'. ➢PMMA 950k 4% in anisole, spin-coated at 5 kRPM for 1'30" (~250 nm). ➢Bake in hot plate at 130 °C for 1'.

2. **EBL:** ➢20 kV, 30 μm aperture, I = 345 pA. ➢Tracks composed by 3, 6, or 12 parallel lines with dose 800 pC/cm. ➢Developed in MIBK:IPA 1'15", washed in IPA. ➢Post-ash ($O_2$ plasma etch) 10 W for 1'.

3. **Hard mask deposition:** ➢Thermal evaporation of Ti 18 nm. ➢Lift-off in acetone (1 hr) + washing.

4. **Ion beam milling:** ➢beam parameters 600 V / 28 A. For 4' (in 30" steps with 30" pauses). ➢Used calibration sample of SV on glass.

## Patterning of contact layer

5. **PMMA coating:** ➢Pre-ash ($O_2$ plasma etch) 10 W for 1'. ➢PMMA 950k 4% in anisole, spin-coated at 3 kRPM for 2' (~280 nm). ➢Bake in hot plate at 130 °C for 1'.

6. **EBL:** ➢20 kV, Multiple apertures (30 and 120 μm), I = 350 pA (for 30 μm aperture). ➢Area dose 280 μC/cm². ➢Developed in MIBK:IPA 1'30", washed in IPA. ➢Post-ash ($O_2$ plasma etch) 10 W for 1'.

7. **Pre-milling:** ➢Ion beam milling (600 V / 28 mA) for 20" (~ 2 nm). ➢In-chamber magnetron sputtering of Ta 30 nm.

8. **Contact deposition:** ➢Thermal evaporation of Ti 5 nm/Au 200 nm (imprecise calibration). ➢Lift-off in acetone (1 hr) + washing.

# [B] Additional data (Chapter 5)

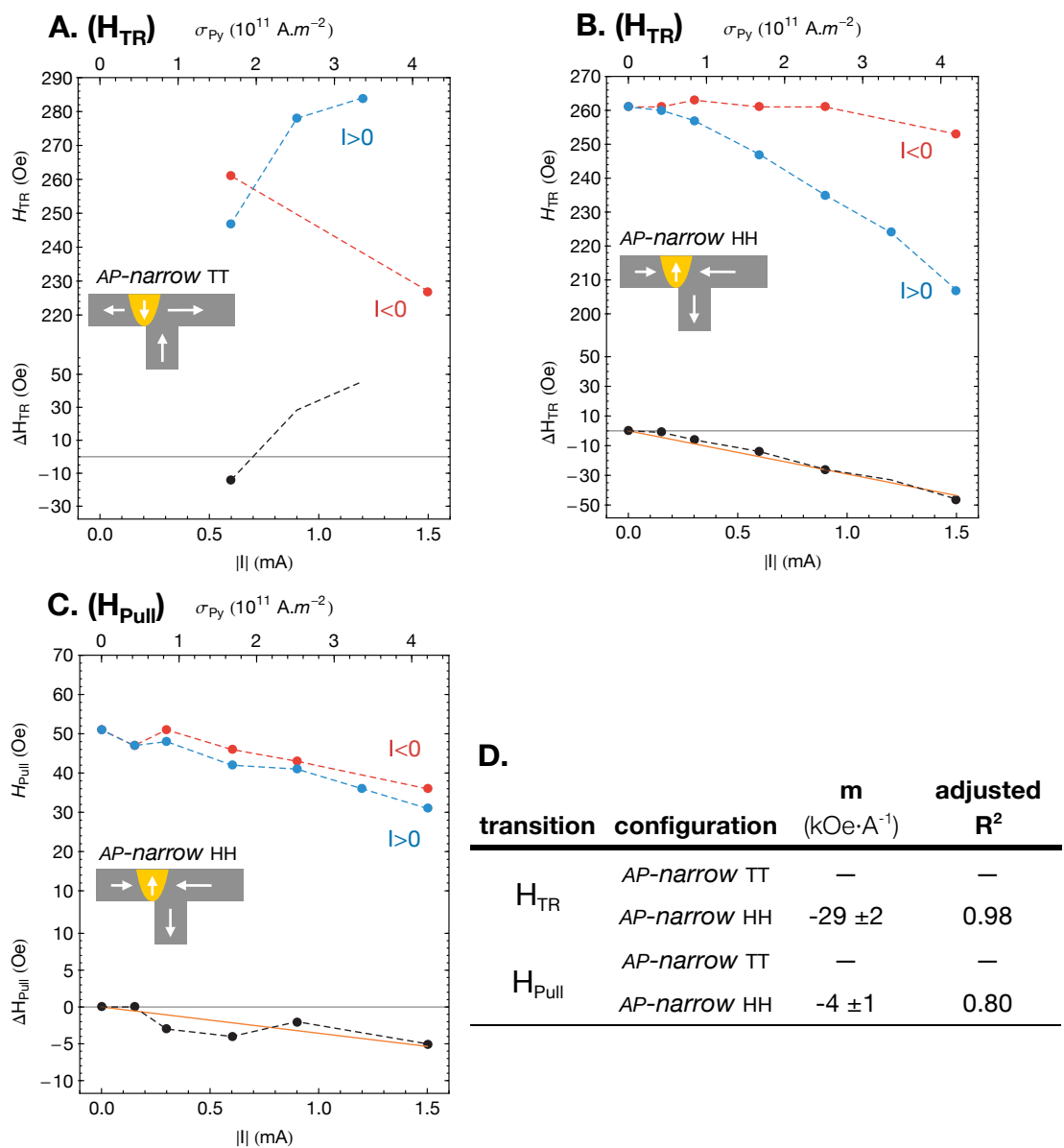

**FIG. B-1 Variation with current of H$_{TR}$ and H$_{Pull}$ (*AP-narrow* case).** The top plots show H$_{TR}$ (**A**, **B**) or H$_{Pull}$ I vs. current amplitude for negative and positive currents (red and blue points) and for TT and HH DWS. The bottom plots show ΔH$_{TR}$ or ΔH$_{Pull}$ (black points). The top axis is the estimated current density in the Py layer. The solid orange lines and the table (**D.**) are a linear fit with model $\Delta H\_ = m \cdot |I|$ .

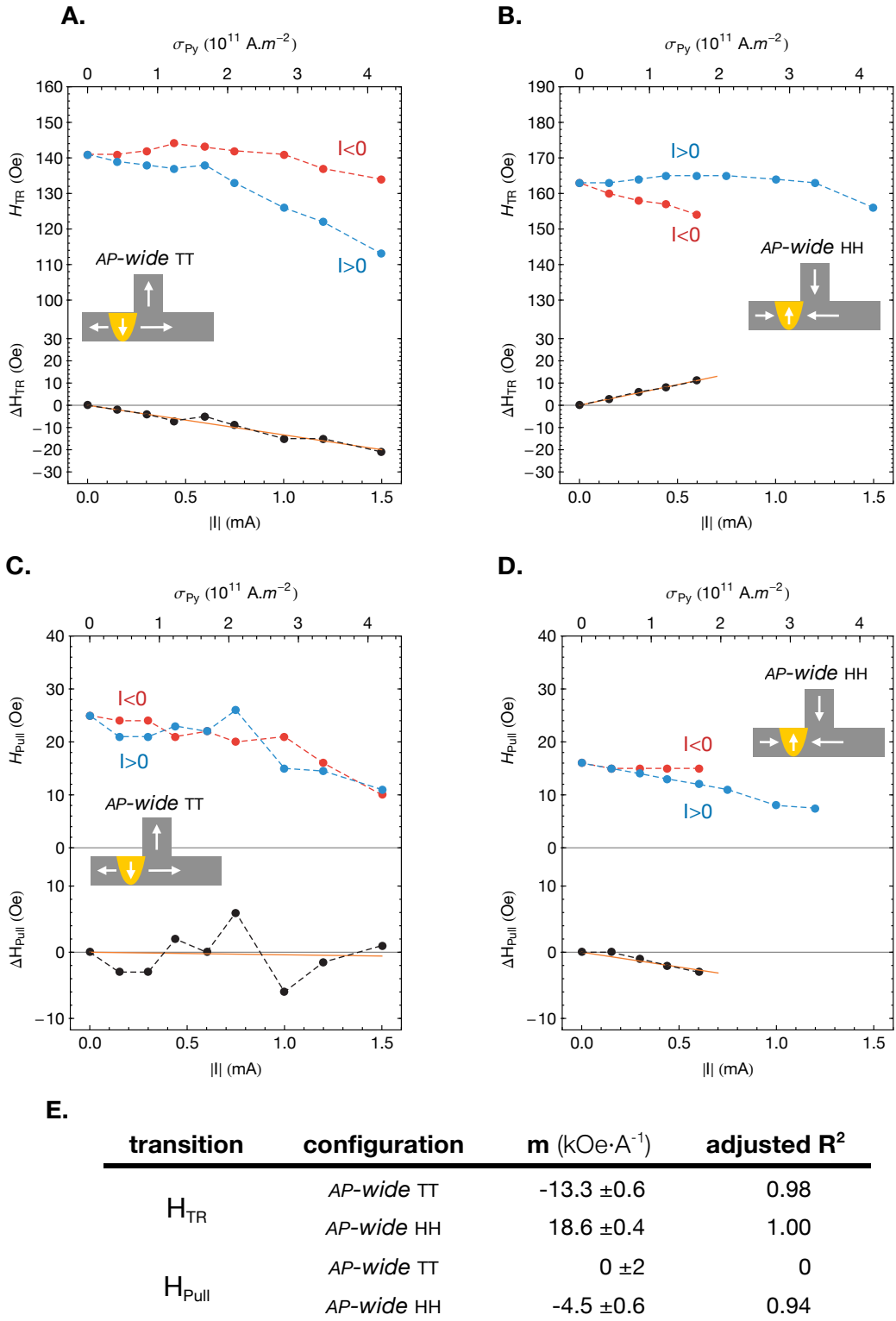

**Fɪɢ. B-2 Variation with current of $H_{TR}$ and $H_{Pull}$ (*AP-wide* case).** The top plots show $H_{TR}$ (**A.**, **B.**) or $H_{Pull}$ (**C.**, **D.**) vs. current amplitude for negative and positive currents (red and blue points) and for tt and hh ᴅᴡs. The bottom plots show $\Delta H_{TR}$ or $\Delta H_{Pull}$ (black points). The top axis is the estimated current density in the Py layer. The solid orange lines and the table (**E.**) are a linear fit with model $\Delta H\_\_ = m \cdot |I|$ .